\documentclass[onecolumn]{aastex631}

\usepackage{svg}
\usepackage{tikz}
\usepackage{amsmath}
\usepackage{xspace}
\usepackage{longtable}
\usepackage{outlines}
\usepackage{calligra}
\usepackage[T1]{fontenc}
\usepackage{tabularx}

\newcommand{\projname}{DESI Strong Lens Foundry\xspace}
\newcommand{\progname}{Spectroscopic Confirmation of DESI Lens Candidates with VLT/MUSE program\xspace}

\newcommand{\ang}{\AA\xspace}

\newcommand{\HST}{\emph{Hubble Space Telescope}\xspace}

\newcommand{\lamlam}{\ensuremath{\lambda\lambda}\xspace}

\extrafloats{110}
\begin{document}

\title{DESI Strong Lens Foundry~IV:\\
Spectroscopic Confirmation of DESI Lens Candidates with VLT/MUSE}

\shortauthors{Lin et al.}
\shorttitle{Spectroscopic Confirmation of DESI Lens Candidates}

\author[0009-0006-5989-4899]{Emerald~Lin}
\affiliation{Department of Physics, University of California, Berkeley, Berkeley, CA 94702, USA}

\author[0009-0006-3410-5531]{Ivonne Toro Bertolla}
\affiliation{Gemini Observatory / NSF's NOIRLab, Casilla 603, La Serena, Chile}
\author[0000-0001-7101-9831]{Aleksandar Cikota}
\affiliation{Gemini Observatory / NSF's NOIRLab, Casilla 603, La Serena, Chile}

\author[0000-0001-8156-0330]{Xiaosheng Huang}
\affiliation{Department of Physics \& Astronomy, University of San Francisco, 2130 Fulton Street, San Francisco, CA 94117-1080, USA}
\affiliation{Physics Division, Lawrence Berkeley National Laboratory, 1 Cyclotron Road, Berkeley, CA 94720, USA}

\author[[0000-0002-0385-0014]{Christopher Storfer}
\affiliation{Institute for Astronomy, University of Hawaii, Honolulu, HI 96822-1897}

\author[0009-0008-0518-8045]{Marcos Tamargo-Arizmendi}
\affiliation{Department of Physics and Astronomy and PITT PACC, University of Pittsburgh, Pittsburgh, PA 15260, USA}
\affiliation{Physics Division, Lawrence Berkeley National Laboratory, 1 Cyclotron Road, Berkeley, CA 94720, USA}

\author[0000-0002-5042-5088]{David J. Schlegel}
\affiliation{Physics Division, Lawrence Berkeley National Laboratory, 1 Cyclotron Road, Berkeley, CA 94720, USA}

\author[0000-0003-1889-0227]{William Sheu}
\affiliation{Department of Physics \& Astronomy, University of California, Los Angeles, 430 Portola Plaza, Los Angeles, CA 90095, USA}

\author[0000-0001-7266-930X]{Nao Suzuki}
\affiliation{Physics Division, Lawrence Berkeley National Laboratory, 1 Cyclotron Road, Berkeley, CA 94720, USA}
\affiliation{Kavli Institute for the Physics and Mathematics of the Universe, University of Tokyo, Kashiwa 277-8583, Japan}







\begin{abstract}
We present integral field spectroscopic observations of 76 strong gravitational lens candidates identified with a residual neural network in the DESI Legacy Imaging Surveys, obtained with the Multi Unit Spectroscopic Explorer (MUSE) on the ESO’s Very Large Telescope. These observations are part of an ongoing effort to build a large, spectroscopically confirmed sample of strong lensing systems for studies on dark matter, galaxy structure, and cosmology. 
Our MUSE program targets both lens and source redshifts, with particular emphasis on southern hemisphere systems. MUSE's wide spectral coverage and integral field capability allow for efficient identification of multiple sources, lens environments, and weak spectral features. 
Redshifts for lenses and sources were obtained via manual identification of spectral features in extracted 1D spectra. Our dataset includes systems with complex configurations, such as multiple source planes and group or cluster-scale environments. We extracted and analyzed 223 spectra, successfully determining both the lens and the source redshifts for 55 gravitational lens systems. For an additional 15 targets, we measured the redshifts of the lenses but were unable to determine the redshifts of the background sources. Six targets were confirmed to not be gravitational lenses.
The results presented here complement space-based imaging from our HST SNAPshot program and spectroscopic follow-up with DESI and Keck, and have lasting legacy value for identifying interesting high-redshift sources and complex lensing configurations.
\end{abstract}



\section{Introduction}
Strong gravitational lensing systems are a powerful tool for astrophysics and cosmology. 
They provide an effective way to measure dark matter (DM) mass distribution in the central regions of galaxies and galaxy clusters \citep[e.g.,][]{1991ApJ...373..354K, treu2002probing, bolton2006constraint, koopmans2006sloan, bolton2008sloan, bradavc2008dark, huang2009hubble, jullo2010cosmological, grillo2015clash, tessore2016lensed, shu2017sloan}
and are the only way to detect DM substructure at cosmological distances or low-mass halos \citep[e.g.,][]{vegetti2010detection, hezaveh2016detection, sengul2022substructure}. 
{Moreover, time-delay $H_{0}$ measurements from multiply imaged supernovae \citep[e.g.,][]{pierel2019turning, kelly2023constraints, Suyu_2024, 2025ApJ...979...13P, 2025arXiv250912319S, 2026ApJ...998..219P} and lensed quasars \citep[e.g.,][]{Wong_2019, 2025A&A...704A..63T}, 
combined with measurements from distance ladders \citep[e.g.,][]{freedman2020astrophysical, Riess_2022, 2025ApJ...985..203F, 2025ApJ...985..182L},
provide consistency checks and competitive constraints on $H_{0}$.

We found $\sim 3500$ new strong gravitational lensing candidates in the DESI Legacy Imaging Surveys.\footnote{The entire catalog of these candidates can be found on our project website \url{https://sites.google.com/usfca.edu/neuralens/}.} 
These consist of $\sim 3000$ new lenses candidates identified using residual neural networks \citep[ResNet;][hereafter H20, H21, and S24, respectively]{huang2020finding, huang2021discovering, storfer2024new}
and 436 new lensed quasar candidates, identified using an autocorrelation algorithm \citep{dawes2023finding}.
In the same dataset, we also found lensed supernova and lensed quasar candidates in targeted lensed transient searches using differencing image techniques \citep[]{Sheu_2023, sheu2024targeted}.

We follow up the gravitational lens system candidates using the Multi Unit Spectroscopic Explorer (MUSE, \citealt{bacon2010muse} installed the ESO’s Very Large Telescope (VLT).
MUSE is well suited for obtaining lens redshifts for targets in the southern hemisphere, where alternative spectroscopic facilities are limited.
The VLT and MUSE provide a significant advantage due to its sensitivity to weak absorption and emission features. 
This enables redshift determination beyond $z\sim$ 1.5, even in challenging cases \citep{cikota2023desi}. 
MUSE is particularly powerful in complex gravitational lens systems, such as those with multiple background sources \citep{sheu2024carousel}, LOS interlopers \citep{cikota2023desi}, or group cluster lenses, where its wide field of view and spatially resolved spectroscopy allow simultaneous characterization of multiple components. 
Our MUSE programs (PI: Cikota) have demonstrated the efficiency and scientific value of this approach.

For high resolution imaging, our \HST SNAP program (Go-15867, PI: Huang) observed 51 of our most promising candidates and confirmed all of them.
We report the results of this program in Paper~I of this series \citep[]{huang2025desistronglensfoundry}.  
Spectroscopic observations are also being carried out. 
The Dark Energy Spectroscopic Instrument  \citep[DESI;][]{aghamousa2016desia, aghamousa2016desib}
is uniquely suited for spectroscopic observations of such a large number of lensing systems. The first results will be reported in Paper~II in this series (Huang et al., in prep).
For a subset of the candidate systems in this program, the source redshifts are too high and the typical emission features (e.g., the [\ion{O}{2}] doublet at \lamlam 3726, 3729~\ang) are beyond the optical range of DESI.
We target these with an ongoing near-IR  spectroscopic program on the Keck~2 Telescope. 
The results for the first set of systems will be reported in Paper~III in this series (Agarwal et al., submitted).
In this paper --- Paper~IV in the \projname series --- we present the DESI \progname. 

Note that in our discovery paper series \citep{huang2020finding, huang2021discovering, storfer2024new}, the lens candidates are named following this convention: ``DESI-'' followed with RA and Dec, both in digital format with four decimal places. In 2024, DESI established an official naming convention that is very similar to that for our candidates, with a small difference --- the inclusion of ``J''. Thus, in our strong lens discovery paper series (including a fourth one, J.~Inchausti et al. in prep) and the DESI Strong Lens Foundry series (including this paper), all candidates and confirmed non-lenses will continue to be named without the ``J'' whereas fully confirmed systems will include the ``J''.

We describe the MUSE observations in Sect.~\ref{sect:MUSEspectra}, and the data reduction and methods in Sect.~\ref{sect:Reduction}. We present the results in Sect.~\ref{sect:results} and summarize and conclude in Sect.~\ref{sect:summary}.


\section{MUSE observations}
\label{sect:MUSEspectra}

The sample was observed between 2022 and 2024 as part of four ESO filler programs aimed at characterizing candidate galaxy-galaxy gravitational lensing systems using MUSE, installed on UT4 of ESO’s VLT at Cerro Paranal, Chile.

MUSE is an integral field spectrograph offering a $60''\times 60''$ field of view in Wide Field Mode, with a spatial sampling of 0.2'' per pixel. It covers a spectral range from 4750 to 9350 \AA, with a spectral resolution varying from R = 2000 to 4000 across the bandpass.

Each observation consisted of four 700 second exposures, and the data were reduced using standard procedures with version 2.2 of the MUSE pipeline \citep{2020A&A...641A..28W}, implemented via the ESO Recipe Execution Tool (ESOREX). Sky emission features were further mitigated using the Zurich Atmosphere Purge (ZAP) sky subtraction tool \citep{2016MNRAS.458.3210S}.

The program IDs are: 109.238W.004 (15h, PI: Bian), 111.24UJ.008 (10h, PI: Bian), 111.24P8.001 (50h, PI: Cikota), 112.2614.001 (50h, PI: Cikota), and 113.267Q.001 (50h, PI: Cikota). Because of the nature of the program, not all observing blocks were executed, and not all observations were successful; some had low SNR due to poor sky transparency and/or poor seeing. Out of the 175 observing blocks (OBs), 100 OBs were executed of 76 different targets (obseravtions of some targets were repeated in different semesters), and 92 observations  contained useful signal, while 8 were not useful due to poor weather.

The full observation log of successful observations is provided in Table~\ref{tab:obs_log}. Several targets were observed at two epochs in different semesters due to unsatisfactory data quality in earlier observations.

\begin{center}
\tiny
\begin{longtable*}{llllllll}
\caption{\label{tab:obs_log}Observations log.}\\
\hline
Name                       & UT Time              &     R.A.      &  Dec        &    Exposure  &   Airmass &  Seeing   & Prog. ID    \\
 & & hh:mm:ss & dd:mm:ss & s & & '' & \\
\hline
\endfirsthead
\multicolumn{8}{c}%
{\tablename\ \thetable\ -- \textit{Continued from previous page}} \\
\hline
Name                       & Start UT Time              &  Center R.A.      &  Center Dec        &    Exposure    &   Airmass &  Seeing   & Prog. ID    \\
 & & hh:mm:ss & dd:mm:ss & s & & '' & \\
\hline
\endhead
\hline \multicolumn{8}{r}{\textit{Continued on next page}} \\
\endfoot
\endlastfoot
DESI~J003.6745-13.5042     & 2022-06-18 08:35:47  &     00:14:42  &  -13:30:16  &   4 $\times$700 &        1.27  &       2.21 & 109.238W.004 \\
DESI-023.6758+04.5638     & 2024-01-20 01:43:29  &     01:34:42  &  +04:33:50  &   4 $\times$700 &        1.83  &       1.18 & 112.2614.001 \\
DESI-024.1634+00.1386     & 2023-12-27 01:25:41  &     01:36:39  &  +00:08:19  &   4 $\times$700 &        1.19  &       0.86 & 112.2614.001 \\
DESI-031.7781-27.4454     & 2023-07-23 06:56:28  &     02:07:07  &  -27:26:44  &   4 $\times$700 &         1.6  &       0.53 & 111.24UJ.008 \\
DESI~J033.8101-29.1565     & 2023-12-27 04:37:12  &     02:15:14  &  -29:09:23  &   4 $\times$700 &        1.67  &       1.13 & 112.2614.001 \\
DESI-043.5347-28.7960     & 2023-06-21 09:00:59  &     02:54:08  &  -28:47:46  &   4 $\times$700 &        2.09  &        0.9 & 111.24P8.001 \\
DESI-043.5347-28.7960     & 2024-08-15 07:06:38  &     02:54:08  &  -28:47:46  &   4 $\times$700 &        1.28  &        1.01 & 113.267Q.001 \\
DESI-043.5347-28.7960     & 2023-10-03 03:42:22  &     02:54:08  &  -28:47:46  &   4 $\times$700 &        1.34  &       1.47 & 112.2614.001 \\
DESI~J043.6663-04.3068     & 2023-07-19 08:03:00  &     02:54:40  &  -04:18:25  &   4 $\times$700 &        1.89  &       0.42 & 111.24P8.001 \\
DESI-044.9808+01.1384     & 2023-07-09 09:08:10  &     02:59:55  &  +01:08:18  &   4 $\times$700 &         1.8  &       1.72 & 111.24P8.001 \\
DESI~J053.6251-13.1869     & 2023-07-19 08:48:43  &     03:34:30  &  -13:11:13  &   4 $\times$700 &        1.67  &       0.45 & 111.24P8.001 \\
DESI~J055.0894-25.5581     & 2023-08-04 09:24:11  &     03:40:21  &  -25:33:29  &   4 $\times$700 &        1.14  &       0.71 & 111.24P8.001 \\
DESI-057.2074-10.2962     & 2023-09-27 05:58:17  &     03:48:50  &  -10:17:45  &   4 $\times$700 &        1.21  &       1.17 & 111.24UJ.008 \\
DESI-057.2074-10.2962     & 2023-12-26 02:16:26  &     03:48:50  &  -10:17:44  &   4 $\times$700 &        1.03  &        0.8 & 112.2614.001 \\
DESI-058.7923-18.5265     & 2023-08-08 07:20:52  &     03:55:10  &  -18:31:35  &   4 $\times$700 &        1.89  &       0.91 & 111.24P8.001 \\
DESI~J060.5238-22.0990     & 2023-08-18 06:51:39  &     04:02:06  &  -22:05:56  &   4 $\times$700 &         1.8  &       0.95 & 111.24P8.001 \\
DESI-064.8539-25.2245     & 2023-08-11 07:42:46  &     04:19:25  &  -25:13:28  &   4 $\times$700 &        1.71  &       0.73 & 111.24P8.001 \\
DESI~J065.6453-28.0646     & 2023-08-11 08:56:55  &     04:22:35  &  -28:03:53  &   4 $\times$700 &        1.27  &       0.75 & 111.24P8.001 \\
DESI-070.4130-09.7774     & 2023-08-18 07:54:31  &     04:41:39  &  -09:46:39  &   4 $\times$700 &        1.75  &       0.94 & 111.24P8.001 \\
DESI-070.4130-09.7774     & 2024-01-17 01:35:52  &     04:41:39  &  -09:46:40  &   4 $\times$700 &        1.03  &       1.03 & 112.2614.001 \\
DESI~J073.5286-10.2227     & 2023-08-18 08:55:58  &     04:54:07  &  -10:13:22  &   4 $\times$700 &        1.39  &       1.05 & 111.24P8.001 \\
DESI~J073.9027-25.5132     & 2023-08-11 09:28:07  &     04:55:37  &  -25:30:47  &   4 $\times$700 &        1.28  &       0.78 & 111.24P8.001 \\
DESI~J074.9646-30.7233     & 2023-08-19 09:38:38  &     04:59:52  &  -30:43:24  &   4 $\times$700 &        1.16  &       0.69 & 111.24P8.001 \\
DESI~J075.2793-24.4176     & 2023-08-20 09:19:09  &     05:01:07  &  -24:25:03  &   4 $\times$700 &        1.21  &       1.04 & 111.24P8.001 \\
DESI-078.3561-30.8433     & 2023-09-02 06:49:18  &     05:13:25  &  -30:50:36  &   4 $\times$700 &        1.86  &       1.29 & 111.24P8.001 \\
DESI-081.7544-18.9677     & 2023-09-12 08:43:28  &     05:27:01  &  -18:58:04  &   4 $\times$700 &        1.14  &       1.41 & 111.24P8.001 \\
DESI~J086.3072-26.5878     & 2024-01-21 02:02:55  &     05:45:14  &  -26:35:16  &   4 $\times$700 &        1.01  &       0.71 & 112.2614.001 \\
DESI~J087.1525-36.2427     & 2023-09-02 07:49:25  &     05:48:37  &  -36:14:34  &   4 $\times$700 &        1.61  &       1.73 & 111.24P8.001 \\
DESI~J090.9854-35.9683     & 2022-09-09 07:43:14  &     06:03:57  &  -35:58:06  &   4 $\times$700 &        1.56  &       1.17 & 109.238W.004 \\
DESI~J090.9854-35.9683     & 2023-09-30 07:45:10  &     06:03:57  &  -35:58:06  &   4 $\times$700 &         1.2  &       0.83 & 111.24UJ.008 \\
DESI-109.9018+27.9032     & 2023-12-25 04:23:24  &     07:19:36  &  +27:54:11  &   4 $\times$700 &         1.8  &       1.18 & 112.2614.001 \\
DESI~J122.0852+10.5284     & 2024-01-18 02:00:54  &     08:08:20  &  +10:31:42  &   4 $\times$700 &        1.82  &       0.59 & 112.2614.001 \\
DESI~J154.6975-01.3588     & 2024-01-20 03:20:28  &     10:18:47  &  -01:21:32  &   4 $\times$700 &        1.94  &       0.97 & 112.2614.001 \\
DESI~J157.4222+20.4043     & 2024-01-30 05:12:25  &     10:29:41  &  +20:24:16  &   4 $\times$700 &        1.53  &        0.9 & 112.2614.001 \\
DESI~J157.6135-06.6858     & 2023-04-25 03:38:14  &     10:30:27  &  -06:41:08  &   4 $\times$700 &        1.33  &       2.37 & 111.24P8.001 \\
DESI~J160.1719+18.8480     & 2023-05-17 01:13:37  &     10:40:41  &  +18:50:53  &   4 $\times$700 &         1.5  &       1.25 & 111.24P8.001 \\
DESI~J160.1719+18.8480     & 2024-01-23 04:30:03  &     10:40:41  &  +18:50:53  &   4 $\times$700 &        1.95  &       0.52 & 112.2614.001 \\
DESI-161.1162+31.2340     & 2024-01-23 05:39:40  &     10:44:28  &  +31:14:02  &   4 $\times$700 &        2.03  &       0.38 & 112.2614.001 \\
DESI~J161.4114-08.8358     & 2024-01-20 04:19:18  &     10:45:39  &  -08:50:08  &   4 $\times$700 &        1.51  &       0.67 & 112.2614.001 \\
DESI~J161.4114-08.8358     & 2023-04-22 05:42:53  &     10:45:39  &  -08:50:07  &   4 $\times$700 &        2.15  &       1.04 & 111.24P8.001 \\
DESI~J166.9974+04.1560     & 2024-03-01 05:44:49  &     11:08:00  &  +04:09:22  &   4 $\times$700 &        1.15  &       0.77 & 112.2614.001 \\
DESI~J166.9974+04.1560     & 2023-05-17 02:26:50  &     11:08:00  &  +04:09:24  &   4 $\times$700 &        1.37  &       0.85 & 111.24P8.001 \\
DESI~J168.7680+16.7604     & 2024-03-20 04:38:06  &     11:15:04  &  +16:45:38  &   4 $\times$700 &        1.34  &       0.71 & 112.2614.001 \\
DESI~J168.7680+16.7604     & 2023-05-17 03:25:44  &     11:15:05  &  +16:45:41  &   4 $\times$700 &        2.07  &        1.1 & 111.24P8.001 \\
DESI~J174.5484+14.7866     & 2023-06-30 00:31:39  &     11:38:12  &  +14:47:12  &   4 $\times$700 &        1.77  &       1.14 & 111.24P8.001 \\
DESI-178.0772+08.8167     & 2024-01-20 05:04:33  &     11:52:19  &  +08:49:00  &   4 $\times$700 &        2.12  &       0.81 & 112.2614.001 \\
DESI-180.2707-02.3681     & 2023-04-25 04:37:06  &     12:01:05  &  -02:22:04  &   4 $\times$700 &        1.26  &       1.91 & 111.24P8.001 \\
DESI~J186.4036-07.4200     & 2024-01-27 04:27:33  &     12:25:37  &  -07:25:12  &   4 $\times$700 &        2.24  &       0.85 & 112.2614.001 \\
DESI-189.9885+12.6693     & 2024-01-28 05:13:25  &     12:39:57  &  +12:40:10  &   4 $\times$700 &        2.41  &       0.63 & 112.2614.001 \\
DESI~J190.7935+21.3334     & 2023-07-15 01:03:58  &     12:43:10  &  +21:20:00  &   4 $\times$700 &        2.33  &       0.85 & 111.24P8.001 \\
DESI~J196.4575+22.9256     & 2023-05-01 01:41:24  &     13:05:50  &  +22:55:31  &   4 $\times$700 &        1.64  &       0.33 & 111.24P8.001 \\
DESI~J196.4575+22.9256     & 2024-01-28 07:09:53  &     13:05:50  &  +22:55:32  &   4 $\times$700 &        1.83  &       0.86 & 112.2614.001 \\
DESI~J197.5704+14.7474     & 2024-03-23 02:24:10  &     13:10:17  &  +14:44:51  &   4 $\times$700 &        2.23  &       1.08 & 112.2614.001 \\
DESI~J197.5704+14.7474     & 2023-04-30 02:23:30  &     13:10:17  &  +14:44:53  &   4 $\times$700 &        1.34  &       0.66 & 111.24P8.001 \\
DESI~J200.7678+03.7216     & 2023-04-30 00:57:11  &     13:23:04  &  +03:43:17  &   4 $\times$700 &        1.48  &       0.64 & 111.24P8.001 \\
DESI~J202.6690+04.6708     & 2024-01-28 06:11:41  &     13:30:41  &  +04:40:15  &   4 $\times$700 &        1.97  &       0.77 & 112.2614.001 \\
DESI~J218.2479-07.2268     & 2022-04-29 02:26:46  &     14:33:00  &  -07:13:36  &   4 $\times$700 &        1.27  &       0.63 & 109.238W.004 \\
DESI~J220.4549+14.6891     & 2023-04-25 07:19:25  &     14:41:49  &  +14:41:20  &   4 $\times$700 &        1.55  &       2.33 & 111.24P8.001 \\
DESI~J220.4549+14.6891     & 2024-01-29 08:02:22  &     14:41:49  &  +14:41:21  &   4 $\times$700 &        1.83  &       0.63 & 112.2614.001 \\
DESI~J234.4780+14.7229     & 2024-03-01 06:59:28  &     15:37:55  &  +14:43:22  &   4 $\times$700 &        1.78  &       0.67 & 112.2614.001 \\
DESI-234.8707+16.8379     & 2024-03-02 08:02:13  &     15:39:29  &  +16:50:16  &   4 $\times$700 &        1.49  &       1.31 & 112.2614.001 \\
DESI~J238.5690+04.7276     & 2024-03-20 05:23:02  &     15:54:17  &  +04:43:39  &   4 $\times$700 &        1.83  &       0.74 & 112.2614.001 \\
DESI~J245.7514+21.6226     & 2023-05-15 05:52:47  &     16:23:00  &  +21:37:21  &   4 $\times$700 &        1.45  &       0.74 & 111.24P8.001 \\
DESI~J246.0068+01.4842     & 2023-05-01 02:55:00  &     16:24:02  &  +01:29:02  &   4 $\times$700 &        1.91  &       0.38 & 111.24P8.001 \\
DESI~J246.0068+01.4842     & 2024-03-20 06:21:18  &     16:24:01  &  +01:29:01  &   4 $\times$700 &        1.53  &       0.92 & 112.2614.001 \\
DESI-252.2720+02.3993     & 2023-05-15 05:07:25  &     16:49:05  &  +02:23:57  &   4 $\times$700 &        1.15  &       0.96 & 111.24P8.001 \\
DESI~J253.2534+26.8843     & 2023-05-22 04:00:54  &     16:53:01  &  +26:53:04  &   4 $\times$700 &        1.81  &       0.53 & 111.24P8.001 \\
DESI~J260.8405+23.8442     & 2023-05-27 07:02:27  &     17:23:22  &  +23:50:40  &   4 $\times$700 &        1.61  &       0.63 & 111.24P8.001 \\
DESI~J304.0068-49.9067     & 2022-04-04 09:06:49  &     20:16:02  &  -49:54:25  &   4 $\times$700 &        1.37  &        0.8 & 109.238W.004 \\
DESI-306.4726-51.2868     & 2022-06-13 08:42:09  &     20:25:53  &  -51:17:12  &   4 $\times$700 &        1.14  &       2.72 & 109.238W.004 \\
DESI-311.4249-10.6762     & 2022-05-26 05:32:30  &     20:45:42  &  -10:40:32  &   4 $\times$700 &         1.7  &       0.51 & 109.238W.004 \\
DESI~J318.0376-01.7568     & 2022-05-26 06:35:00  &     21:12:09  &  -01:45:23  &   4 $\times$700 &        1.56  &        0.4 & 109.238W.004 \\
DESI-324.2094-62.6820     & 2022-05-30 04:19:13  &     21:36:50  &  -62:40:55  &   4 $\times$700 &        2.35  &       0.57 & 109.238W.004 \\
DESI-324.5073-60.1290     & 2022-05-21 07:39:53  &     21:38:01  &  -60:07:55  &   4 $\times$700 &        1.42  &       0.82 & 109.238W.004 \\
DESI~J326.0105-43.3965     & 2022-05-30 05:32:39  &     21:44:02  &  -43:23:48  &   4 $\times$700 &        1.79  &       0.49 & 109.238W.004 \\
DESI~J329.6820+02.9584     & 2023-07-18 05:28:19  &     21:58:44  &  +02:57:30  &   4 $\times$700 &        1.23  &       0.66 & 111.24P8.001 \\
DESI~J329.6820+02.9584     & 2023-10-01 02:15:21  &     21:58:44  &  +02:57:31  &   4 $\times$700 &        1.13  &       0.39 & 112.2614.001 \\
DESI~J331.8083-52.0487     & 2022-05-30 07:02:19  &     22:07:14  &  -52:02:55  &   4 $\times$700 &        1.44  &       0.49 & 109.238W.004 \\
DESI-333.3655-13.2491     & 2023-05-17 06:57:03  &     22:13:28  &  -13:14:57  &   4 $\times$700 &        2.13  &       1.41 & 111.24P8.001 \\
DESI-333.3655-13.2491     & 2023-10-04 02:42:19  &     22:13:28  &  -13:14:57  &   4 $\times$700 &        1.03  &       0.75 & 112.2614.001 \\
DESI~J335.5354+27.7596     & 2023-07-08 07:37:37  &     22:22:08  &  +27:45:35  &   4 $\times$700 &        1.65  &       0.77 & 111.24P8.001 \\
DESI-336.1611-01.8757     & 2022-07-05 04:28:29  &     22:24:39  &  -01:52:34  &   4 $\times$700 &        1.95  &       0.46 & 109.238W.004 \\
DESI~J339.8883-04.4880     & 2023-10-04 03:54:28  &     22:39:33  &  -04:29:19  &   4 $\times$700 &        1.14  &       0.79 & 112.2614.001 \\
DESI~J339.8883-04.4880     & 2023-06-19 05:56:51  &     22:39:33  &  -04:29:17  &   4 $\times$700 &        1.78  &       1.17 & 111.24P8.001 \\
DESI-340.2310-00.0123     & 2023-07-12 09:09:12  &     22:40:55  &  -00:00:44  &   4 $\times$700 &        1.15  &       1.38 & 111.24P8.001 \\
DESI~J341.0212+27.9883     & 2023-07-09 08:35:37  &     22:44:05  &  +27:59:18  &   4 $\times$700 &        1.65  &       0.82 & 111.24P8.001 \\
DESI-341.8012-02.0939     & 2023-07-16 04:40:44  &     22:47:12  &  -02:05:38  &   4 $\times$700 &        1.62  &       0.37 & 111.24P8.001 \\
DESI~J342.9290-03.4136     & 2022-07-27 03:30:37  &     22:51:43  &  -03:24:49  &   4 $\times$700 &        1.89  &       0.66 & 109.238W.004 \\
DESI~J343.0402-04.2187     & 2022-07-27 04:31:00  &     22:52:10  &  -04:13:07  &   4 $\times$700 &         1.4  &       0.76 & 109.238W.004 \\
DESI~J344.6262-58.6910     & 2022-05-26 06:53:49  &     22:58:30  &  -58:41:28  &   4 $\times$700 &        1.87  &       0.34 & 109.238W.004 \\
DESI-345.0725+22.2254     & 2023-07-16 07:18:52  &     23:00:17  &  +22:13:32  &   4 $\times$700 &        1.51  &        0.4 & 111.24P8.001 \\
DESI~J345.8606+23.4757     & 2023-07-16 07:50:28  &     23:03:27  &  +23:28:33  &   4 $\times$700 &        1.51  &        0.4 & 111.24P8.001 \\
 
\hline

\end{longtable*}

\end{center}

\section{Data Reduction}
\label{sect:Reduction}

For each gravitational lens system in our MUSE sample, we generated synthetic images in the Johnson–Cousins $V$ (yellow), $R$ (magenta), and $I$ (cyan) bands by convolving the MUSE data cubes with their respective passbands. These images provide a broad-band view of the lens and its multiple lensed source images. Representative color-composite images were created to visually identify and label  the individual components of the lensing configuration.

We obtained the positions and magnitudes of both the lensing galaxies and the lensed source images by cross-matching their locations in the MUSE field with sources listed in the DESI Legacy Imaging Surveys. The astrometric and photometric information was taken directly from the DESI Legacy Survey catalogs, providing DESI $g$, $r$, $i$, and $z$-band magnitudes and accurate celestial coordinates. Instead of performing independent photometric calibration on the MUSE data, these catalog values were used to construct the photometric catalogs for our sample of gravitational lens systems and are given in Table~\ref{tab:sourcespositions}.

To analyze the spectral properties, we extracted integrated 1D spectra of the lens and the multiple images of the background source directly from the MUSE data cubes. 
Spectra were extracted using circular apertures, or lists of manually selected spaxels for non circular targets (e.g., lensed arcs), and were averaged within each aperture. For manual selection of the spaxels we used a self-written python tool, which displays an flux-average 2D image of the MUSE 3D data cube and prints tuples of the (x,y) coordinates of the selected spaxels. Figure~\ref{fig:countourexample} shows an example of the circular apertures, and manually selected spaxels for a lensed arc.}

\begin{figure}[!ht]
\centering
\includegraphics[width=0.4\textwidth]{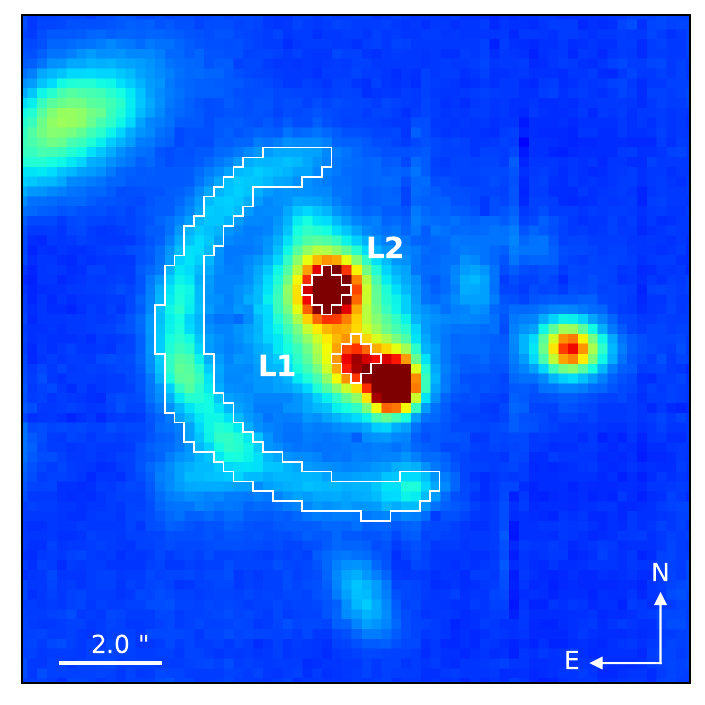}
\label{fig:countourexample}
\caption{Circular apertures and a contour denote the regions from which spectra were extracted for DESI~J245.7514+21.6226. Note that the object southwest of L1 is a star.}
\end{figure}

The averaged spectra were visually inspected, and initial redshifts for both the lens and the source components were determined through manual identification of prominent emission and absorption features in the extracted spectra. 

Thereafter, redshifts were determined by fitting Gaussian models to prominent spectral features in each observed spectrum. The spectra were lightly smoothed using a Savitzky--Golay filter and we then performed least-squares fitting using \texttt{curve\_fit} \citep{2020SciPy-NMeth}, adopting an initial redshift guess from our visual determination and modeling key emission or absorption features with Gaussian profiles tied to their rest-frame wavelengths. 

For most sources we focused on the Ca H\&K absorption lines or the [O\,II] $\lambda\lambda3726,3729$ doublet, fitting it with a two-component Gaussian model with a shared redshift parameter. For some spectra, alternative features (e.g., C\,IV, C\,III], Mg\,II, or Ly$\alpha$) were used in an analogous manner, depending on wavelength coverage and redshift. 
The best-fit redshift and its statistical uncertainty were obtained from the covariance matrix of the fit, and the resulting redshifts were visually inspected by overplotting the fitted model on the data.


Moreover, we adopt a quality flag, $Q_z$, for the determined redshifts. The quality flag is determined from the visual inspection of the spectra. For spectra with clearly recognizable and unambiguous spectral features we assign `1' = Robust --- this is the case for the vast majority of the spectra. A few spectra have a quality flag of `2' = Probable (in that case there are spectral features recognizable, but the signal-to-noise ratio may be low and the features may not be unambiguous) or `3' = Possible (in that case the signal-to-noise ratio is usually too low to fit the features with gaussian models, but there are some weak features to estimate a rough redshift).

The results are given in Table~\ref{tab:sourcespositions}.

\begin{center}
\tiny
\setlength{\tabcolsep}{4pt}
\begin{longtable}{llllllccccp{1cm}}

\caption{\label{tab:sourcespositions}: Confirmed lens systems}\\
\hline
Name & Object & R.A. & Dec & Redshift & $Q_z$ & DESI $g$ & DESI $r$ & DESI $i$ & DESI $z$ & Notes \\
& & deg & deg & & & mag & mag & mag & mag  \\
\hline
\endfirsthead
\multicolumn{9}{c}%
{\tablename\ \thetable\ -- \textit{Continued from previous page}} \\
\hline
Name & Object & R.A. & Dec & Redshift & $Q_z$ & DESI $g$ & DESI $r$ & DESI $i$ & DESI $z$ & Notes \\
& & deg & deg & & & mag & mag & mag & mag  \\
\hline
\endhead
\hline \multicolumn{9}{r}{\textit{Continued on next page}} \\
\endfoot
\hline 
\multicolumn{9}{c}{\textbf{Notes.} Redshift quality ($Q_z$); 1 = Robust, 2 = Probable, 3 = Possible.} \\
\multicolumn{9}{c}{$^{\rm a}$ Based on the fitted \ion{Ca}{2} H and K lines.} \\
\multicolumn{9}{c}{$^{\rm b}$ Based on the fitted $\lambda$3727\,\AA\ + $\lambda$3729\,\AA\ [\ion{O}{2}] doublet.} \\
\multicolumn{9}{c}{$^{\rm c}$ Based on the fitted $\lambda$1526.7\,\AA\ \ion{Si}{2} and the $\lambda$1548.2\,\AA\ + $\lambda$1550.8\,\AA\ [\ion{C}{4}] doublet. } \\
\multicolumn{9}{c}{$^{\rm d}$ Based on the fitted $\lambda$2795.5\,\AA\ + $\lambda$2802.7\,\AA\ \ion{Mg}{2} doublet.} \\    
\multicolumn{9}{c}{$^{\rm e}$ Based on the fitted $\lambda$1215.67\,\AA\ Ly$\alpha$ line.} \\    
\multicolumn{9}{c}{$^{\rm f}$ Based on the fitted $\lambda$1548.2\,\AA\ + $\lambda$1550.8\,\AA\ [\ion{C}{4}] doublet.} \\
\multicolumn{9}{c}{$^{\rm g}$ Based on the fitted $\lambda$6562.81\,\AA\ H$\alpha$ line.} \\
\multicolumn{9}{c}{$^{\rm h}$ Based on the fitted $\lambda$1906.68 \,\AA\ + $\lambda$1908.73\,\AA\ \ion{C}{3}] doublet.} \\
\multicolumn{9}{c}{$^{\rm i}$ Based on the fitted $\lambda$2586.65\,\AA\ \ion{Fe}{2} and $\lambda$2600.17\,\AA\ \ion{Fe}{2} lines.} \\
\multicolumn{9}{c}{$^{\rm j}$ Based on the fitted $\lambda$1526.7\,\AA\ \ion{Si}{2} line. } \\

\multicolumn{9}{c}{$^{\rm *}$ Based on visual line matching. Low SNR spectrum.} \\  
\multicolumn{9}{c}{$^{\rm \dagger}$ Based on a combined spectrum of multiple sources.} \\

\endlastfoot

DESI~J003.6745-13.5042 & Lens 1 & 3.6745 & -13.5042 & 0.43119 $\pm$ 0.00013$^{\rm a}$ & 1 & 20.20 & 18.37 & 17.76 & 17.41 \\

& Lens 2  & 3.6757 & -13.5034 & 0.42944 $\pm$ 0.00044$^{\rm a}$ & 1 & 24.14 & 22.32 & 22.02 & 24.72 \\

& Source  & 3.6744  & -13.5029 & 0.90999 $\pm$ 0.00023$^{\rm a}$ & 1 & 21.97 & 20.23 & 19.46 & 19.00 \\

DESI~J031.7781-27.4454 & Lens & 31.7778 & -27.4456 & 0.35451 $\pm$ 0.00016$^{\rm a}$ & 1 & 18.66 & 17.06 & - & 16.17 \\

 & Source  &  31.7787 & -27.4464 & 1.67711 $\pm$ 0.00005$^{\rm i}$ & 1 & 20.93 & 20.90 & - & 21.06\\

DESI~J033.8095-29.1570 & Lens & 33.8096 & -29.1574 & 0.9$^{\rm *}$ & 3 & - & - & - & - \\

 & Source  &   33.8100 & -29.1572 &  1.85757 $\pm$ 0.00011$^{\rm i}$ & 1 & 23.86 & 23.55 & 23.23 & 23.49\\

DESI~J043.6663-04.3068 & Lens 1  & 43.6660 & -4.3072 & 0.34477 $\pm$ 0.00016$^{\rm a}$ & 1 & 19.42 & 17.77 & 17.19 & 16.88 \\

 & Lens 2  & 43.6663 & -4.3068 & 0.34356 $\pm$ 0.00014$^{\rm a}$ & 1 & 23.87 & 21.74 & 21.12 & 20.75 \\

 & Source  & 43.6661 & -4.3060 & 2.4493 $\pm$ 0.0002$^{\rm c}$ & 1 & 21.55 & 21.12 & 21.02 & 20.99 \\

DESI~J053.6251-13.1869 & Lens  & 53.6251 & -13.1869 & 0.3866 $\pm$ 0.00032$^{\rm a}$ & 1 & 19.76 & 17.98 & 17.38 & 17.05 \\

 & Source A  &  53.6274 & -13.1852 & 2.30117 $\pm$ 0.00019$^{\rm c}$ & 1 & 21.51 & 21.07 & 20.93 & 20.78 \\
 
 & Source B & 53.6240 & -13.1868 & 2.30137 $\pm$ 0.00045$^{\rm c}$  & 1 & 22.15 & 22.06 & 22.15 & 22.61\\

DESI~J055.0894-25.5581 & Lens  & 55.0894 & -25.5581 & 0.65708 $\pm$ 0.00018$^{\rm a}$ & 1 & 21.06 & 19.34 & 18.30 & 17.89 \\

 & Source  &  55.0892 & -25.5573 & 2.6802 $\pm$ 0.0005$^{\rm d}$ & 2 & 23.68 & 23.70 & 24.24 & 23.88 \\

DESI~J057.2077-10.2959 & Lens  & 57.2074 & -10.2962 & 0.74493 $\pm$ 0.00013$^{\rm a}$ & 1 & 21.95 & 20.17 & - & 18.45 \\

 & Source  &  57.2069 & -10.2954 & 1.60376 $\pm$ 0.00014$^{\rm i}$ & 1 & 21.08 & 20.87 & - & 20.47 \\

DESI~J060.5238-22.0990 & Lens 1 & 60.5238 & -22.0990 & 0.4669 $\pm$ 0.0005$^{\rm a}$ & 1 & 21.16 & 19.34 & 18.64 & 18.28 \\

 & Lens 2  &   60.5233 & -22.0995 &  0.46506 $\pm$ 0.00013$^{\rm a}$ & 1 & 21.35 & 19.70 & 19.04 & 18.70 \\

 & Source  &  60.5234 & -22.1004 &  0.81998 $\pm$ 0.00028$^{\rm b}$  & 1 & 22.95 & 22.86 & 23.04 & 22.64 \\

DESI~J065.6453-28.0646 & Lens & 65.6453 & -28.0646 & 0.62107 $\pm$ 0.00018$^{\rm a}$ & 1 & 21.45 & 19.67 & 18.65 & 18.17 \\

 & Source  &  65.6441 & -28.0649 &  1.17482 $\pm$ 0.00018$^{\rm a}$ & 1 & 25.02 & 23.70 & 22.51 & 21.31 \\

DESI~J070.4130-09.7774 & Lens 1 & 70.4148 & -9.7743 & 0.45467 $\pm$ 0.00015$^{\rm a}$ & 1 & 20.21 & 18.31 & - & 17.22 \\
 & Lens 2  & 70.4130 & -9.7751 &  0.45707 $\pm$ 0.00014$^{\rm a}$ & 1 & 20.78 & 18.87 & - & 17.80 \\
 
 & Lens 3  & 70.4130 & -9.7772 &  0.35942 $\pm$ 0.00013$^{\rm a}$ & 1 & 20.80 & 18.96 & - & 17.98 \\

 & Source  & 70.4114 & -9.7837 &  1.80133 $\pm$ 0.00012$^{\rm i}$  & 1 & 21.62 & 21.32 & - & 21.03 \\

DESI~J073.5286-10.2227 & Lens & 73.5286 & -10.2227 & 0.24820 $\pm$ 0.00016$^{\rm a}$ & 1 & 7.80 & 16.29 & 15.79 & 15.47 \\

 & Source  &  73.5294 & -10.2213 &  1.04384 $\pm$ 0.00009$^{\rm b}$ & 1 & 21.59 & 20.98 & 20.36 & 19.87 \\

DESI~J073.9027-25.5132 & Lens & 73.9027 & -25.5132 & 0.3795 $\pm$ 0.00017$^{\rm a}$ & 1 & 19.62 & 17.99 & 17.41 & 17.03 \\

 & Source A &  73.9075 & -25.5109 & 2.8230 $\pm$ 0.0003$^{\rm c}$ & 1 & 22.36 & 21.61 & 21.48 & 21.37 \\

 & Source B &  73.9019 & -25.5074 &  0.70821 $\pm$ 0.00017$^{\rm a}$ & 1 & 23.42 & 22.33 & 21.55 & 21.07 \\

DESI~J074.9646-30.7233 & Lens & 74.9646 & -30.7233 & 0.44137 $\pm$ 0.00016$^{\rm a}$ & 1 & 20.32 & 18.48 & 17.84 & 17.48 \\

 & Source  & 74.9641 & -30.7228 & 1.44884 $\pm$ 0.00002$^{\rm b}$ & 1 & 21.90 & 21.55 & 21.30 & 20.88 \\

DESI~J075.2793-24.4176 & Lens & 75.2793 & -24.4176 & 0.32059 $\pm$ 0.00013$^{\rm a}$ & 1 & 20.86 & 19.23 & 18.68 & 18.36 \\

 & Source  & 75.2756 & -24.4177 &  2.8290 $\pm$ 0.0004$^{\rm c}$ & 1 & 22.63 & 22.21 & 22.27 & 22.34 \\

DESI~J086.3072-26.5877 & Lens 1 & 86.3066 & -26.5884 & 0.27534 $\pm$ 0.00018$^{\rm a}$ & 1 & 19.09 & 17.48 & 16.97 & 16.66 \\

 & Lens 2 & 86.3062 & -26.5892 & 0.27426 $\pm$ 0.00021$^{\rm a}$ & 1 & 18.22 & 16.67 & 16.15 & 15.82 \\
 
 & Source A & 86.3092 & -26.5854 &  2.17441 $\pm$ 0.00037$^{\rm c \dagger}$ & 2 & 21.83 & 21.57 & 21.41 & 21.25 \\

 & Source B & 86.3065 & -26.5848 & 2.17441 $\pm$ 0.00037$^{\rm c \dagger}$  & 1 & 20.13 & 19.81 & 19.63 & 19.54\\

 & Source C & 86.3054 & -26.5847 & 2.17441 $\pm$ 0.00037$^{\rm c \dagger}$  & 1 & 22.22 & 22.07 & 22.00 & 22.11 \\

DESI~J087.1525-36.2427 & Lens & 87.1525 & -36.2427 & 0.3011 $\pm$ 0.00018$^{\rm a}$ & 1 & 18.31 & 16.75 & 16.24 & 15.91 \\

 & Source A & 87.1510 & -36.2435 &  0.84611 $\pm$ 0.00005$^{\rm b}$ & 1 & - & - & - & - \\

 & Source B & 87.1547 & -36.2424 & 0.84661 $\pm$ 0.00006$^{\rm b}$ & 1 & 22.13 & 21.87 & 21.32 & 21.24 \\

DESI~J090.9854-35.9683  & Lens 1  & 90.9854 & -35.9682 & 0.4883 $\pm$ 0.00019$^{\rm a}$ & 1 & 20.26 & 18.43 & 17.68 & 17.29 \\

& Lens 2  & 90.9845 & -35.9677 &  0.48835 $\pm$ 0.00016$^{\rm a}$ & 1 & 22.96 & 20.92 & 20.18 & 19.77 \\

& Source A & 90.9807 & -35.9701 & 1.43211 $\pm$ 0.00012$^{\rm b}$ & 1 & 21.15 & 20.83 & 20.58 & 20.21 \\

& Source B & 90.9832 & -35.9658 & 1.43192 $\pm$ 0.00017$^{\rm b}$ & 1 & 21.33 & 20.89 & 20.65 & 20.19 \\

& Source C & 90.9911 & -35.9668 & 1.43209 $\pm$ 0.00008$^{\rm b}$ & 1 & 22.01 & 21.74 & 21.51 & 21.06 \\

& Source D & 90.9855 & -35.9714 & 1.43225 $\pm$ 0.00012$^{\rm b}$ & 1 & 21.64 & 21.35 & 21.09 & 20.66 \\

DESI~J122.0852+10.5284 & Lens & 122.0852 & 10.5284 &  0.47529 $\pm$ 0.00007$^{\rm a}$ & 1 & 19.96 & 18.16 & 17.47 & 17.11 \\

 & Source A & 122.0835 & 10.5283 & 1.23669 $\pm$ 0.00004$^{\rm b}$  & 1 & 23.96 & 23.24 & 22.47 & 21.77 \\

 & Source B & 122.0845 & 10.5272 & 1.45292 $\pm$ 0.0001 $^{\rm b}$ & 1 & - & - & - & \\

DESI~J154.6975-01.3588 & Lens & 154.6975 & -1.3588 & 0.38827 $\pm$ 0.00007$^{\rm a}$ & 1 & 20.11 & 18.32 & 17.73 & 17.38 \\

 & Source A & 154.6978 & -1.3587 & 1.431 $\pm$ 0.00012 $^{\rm b}$ & 1 & 23.49 & 23.18 & 22.93 & 22.32 \\

 & Source B & 154.6963 & -1.3590 & 1.43056 $\pm$ 0.00007$^{\rm b}$  & 1 & 22.13 & 21.67 & 21.19 & 20.74 \\

DESI~J157.4222+20.4043 & Lens 1 & 157.4224 & 20.4047 & 0.39203 $\pm$ 0.00008$^{\rm a}$ & 1 & 24.13 & 21.73 & - & 20.99  \\ 

 & Lens 2 & 157.4222 & 20.4043 & 0.39098 $\pm$ 0.00009$^{\rm a}$  & 1 & 19.86 & 18.01 & - & 17.03 \\
 
 & Source & 157.4235 & 20.4050 & 1.30729 $\pm$ 0.00015$^{\rm b}$  & 1 & 23.10 & 21.54 &    -   & 19.29 \\

DESI~J157.6135-06.6858 & Lens 1 & 157.6135 & -6.6858 & 0.46717 $\pm$ 0.00013$^{\rm a}$ & 1 & 20.44 & 18.64 & 17.93 & 17.54  \\ 

 & Source & 157.6120 & -6.6848 & 1.5806 $\pm$ 0.0001$^{\rm i}$ & 1 &   21.15    &    20.95   &    -   &20.52& \\

DESI~J160.1719+18.8480 & Lens & 160.1719 & 18.8480 & 0.31331 $\pm$ 0.00013$^{\rm a}$ & 1 & 18.50 & 16.91 & 16.37 & 16.03 \\

 & Source A & 160.1731 & 18.8493 & 0.88096 $\pm$ 0.00019$^{\rm a}$ & 1 & 23.12 & 21.64 & 20.48 & 19.88 \\

 & Source B1 & 160.1725 & 18.8481 & 0.87779 $\pm$ 0.00026$^{\rm a}$ & 1 & 24.22 & 22.27 & 20.91 & 20.08 \\
 
 & Source B2 & 160.1705 & 18.8464 & 0.8778 $\pm$ 0.00024$^{\rm a}$  & 2 & 24.25 & 23.48 & 22.39 & 22.22 \\

 & Source C & 160.1706 & 18.8456 & 0.87566 $\pm$ 0.00013$^{\rm a}$ & 2 & 23.70 & 22.84 & 21.90 & 21.38 \\

DESI~J161.1162+31.2340 & Lens 1 &  161.1159 & 31.2337  & 0.69819 $\pm$ 0.00027$^{\rm a}$  & 1 & 22.90 & 21.03 & - & 19.45 \\ 
& Lens 2 &  161.1146 &  31.2343 & 0.70127 $\pm$ 0.0008$^{\rm b}$ & 1 & - &  & - & - \\ 
& Member &  161.1149 & 31.2338 &  0.69957 $\pm$ 0.0001$^{\rm b}$  & 1 & 20.18 & 19.91 & - & 19.35\\
& Source & 161.1137 & 31.2344 & 1.51637 $\pm$ 0.00014$^{\rm i}$ & 1 & 20.18 & 19.91 & - & 19.35\\



DESI~J161.4114-08.8358 & Lens & 161.4114 &  -8.8358  &  0.82754 $\pm$ 0.00018$^{\rm a}$ & 1 & 22.57 & 21.25 & - & 19.48 \\  
 & Source  & 161.4114 & -08.8355 & 2.079 $\pm$ 0.002$^{\rm d}$ & 3 &  22.00     &   21.90    &    -   &   22.08    & Noisy \\

DESI~J166.9974+04.1560 & Lens 1 & 166.9975 & 4.1565 & 0.35153 $\pm$ 0.00017$^{\rm a}$ & 1 & 22.74 & 20.93 & 20.32 & 19.94 \\

 & Lens 2 & 166.9972 & 4.1559 & 0.35271 $\pm$ 0.00016$^{\rm a}$ & 1 & 19.31 & 17.55 & 16.94 & 16.61 \\

 & Source & 166.9965 & 4.1543 & 1.28023 $\pm$ 0.00013$^{\rm b}$ & 1 & 23.16 & 23.45 & 23.63 & 21.74 \\

DESI~J168.7680+16.7604 & Lens & 168.7680 & 16.7604 & 0.53714 $\pm$ 0.0002$^{\rm a}$ & 1 & 20.99 & 19.08 & 18.26 & 17.86 \\

 & Foreground & 168.7676 & 16.7589 & 0.53503 $\pm$ 0.00016$^{\rm a}$ & 1 & 21.67 & 21.58 & 21.49 & 21.46 \\

 & Source A & 168.7687 & 16.7590 & 1.71573 $\pm$ 0.00012$^{\rm i}$ & 1 & 22.49 & 21.27 & 20.48 & 20.12\\

 & Source B & 168.7686 & 16.7576 & 3.46264 $\pm$ 0.00004$^{\rm e}$ & 1 & 22.13 & 22.31 & 22.67 & 22.62 \\

DESI~J174.5481+14.7863 & Lens & 174.5481 & 14.7863 &  0.56486 $\pm$ 0.00019$^{\rm a}$ & 1 & 21.49 & 19.79 & 18.94 & 18.55 \\

 & Source  & 174.5473 & 14.7866 & 2.41864 $\pm$ 0.00065$^{\rm c}$  & 1 & 21.52 & 20.73 & 20.50 & 20.13 \\

DESI~J186.4036-07.4200 & Lens & 186.4036 & -07.420 & 0.40484 $\pm$ 0.00037$^{\rm a}$ & 1 & 20.85 & 19.21 & 18.63 & 18.31 \\

 & Source A & 168.7687 & 16.7590 & 1.08376 $\pm$ 0.00004$^{\rm b}$ & 1 & 22.71 & 22.41 & 22.07 & 21.73 \\

 & Source B & 168.7686 & 16.7576 & 1.1001 $\pm$ 0.00004$^{\rm b}$ & 1 & 23.74 & 23.83 & 23.35 & 22.83 \\

DESI~J190.7935+21.3334 & Lens & 190.7935 & 21.3334 & 0.34832 $\pm$ 0.00006$^{\rm a}$  & 1 & 19.80 & 18.07 & 17.50 & 17.18 \\

 & Source & 190.7924 & 21.3318 & 1.48152 $\pm$ 0.00005 $^{\rm b}$ & 1 & 21.58 & 21.49 & 21.50 & 21.08 \\

DESI~J196.4575+22.9256 & Lens & 196.4575 & 22.9256 & 0.39426 $\pm$ 0.00017$^{\rm a}$ & 1 & 19.91 & 18.32 & 17.75 &17.42  \\ 

& Source & 196.4580 & 22.9251 & 0.713$^{\rm *}$  & 3 &   21.81    &   21.96    &    -   &   23.11   & Noisy \\

DESI~J197.5704+14.7474 & Lens & 197.5704 & 14.7474 & 0.26144 $\pm$ 0.00007$^{\rm a}$ & 1 & 18.71 & 17.17 & 16.68 & 16.37 \\

 & Source & 197.5721 & 14.7463 & 0.85834 $\pm$ 0.00015$^{\rm a}$ & 1 & 21.39 & 20.50 & 19.72 & 19.34 \\

DESI~J200.7678+03.7216 & Lens 1 & 200.7689 & 3.7218 & 0.35402 $\pm$ 0.00015$^{\rm a}$ & 1 & 19.43 & 17.65 & 17.08 & 16.75 \\

 & Lens 2 & 200.7672 & 3.7221 & 0.35262 $\pm$ 0.00018$^{\rm a}$ & 1 & 19.30 & 17.50 & 16.92 & 16.58 \\

 & Source A & 200.7682 & 3.7236 &1.01586 $\pm$ 0.00017$^{\rm a}$ & 1 & 23.13 & 21.17 & 20.03 & 19.15 \\

& Source B & 200.7664 & 3.7227 & 1.015 $\pm$ 0.00022$^{\rm a}$ & 1 & - & 23.59 & 21.65 & 20.59 \\

& Source C & 200.7666 & 3.7207 & 1.01537 $\pm$ 0.00017$^{\rm a}$ & 1 & 22.87 & 21.13 & 20.09 & 19.25 \\

DESI~J202.6690+04.6708 & Lens & 202.6690 & 4.6708 & 0.3364 $\pm$ 0.0001$^{\rm a}$ & 1 & 18.59 & 16.94 & 16.38 & 16.062 \\

 & Foreground & 202.6692 & 4.6696 & 0.03650 $\pm$ 0.00002$^{\rm g}$ & 1 & 21.03 & 20.84 & 20.66 & 20.40 \\

 & Source & 202.6701 & 4.6698 & 1.16828 $\pm$ 0.00001$^{\rm b}$  & 1 & 22.31 & 22.16 & 21.83 & 21.35 \\

DESI~J218.2479-07.2268 & Lens 1 & 218.2479 & -7.2268  &0.33851 $\pm$ 0.00013$^{\rm a}$  & 1 & 18.47 & 16.81 & - & 15.89& \\ 

& Lens 2 & 218.2474 & -7.2262 &  0.243$^{\rm *}$   & 2 & 24.21 & 23.48 &   -   & 23.45 &\\

&Source A  &  218.2466 & -7.2265 & 1.104$^{\rm *}$  & 2 &   -    &   -    &    -   &     -  & Noisy \\

&Source B & 218.2480 & -7.2279 & 1.104$^{\rm *}$  & 2 &   -    &    -   &    -   &  -     & Noisy \\

&Source C & 218.2485 & -7.2275 & 1.104$^{\rm *}$  & 2 &   20.47    &   20.01,    &  -     &    19.54   & Noisy \\
 
DESI~J220.4549+14.6891 & Lens & 220.4549 & 14.6891 & 0.7421 $\pm$ 0.0002$^{\rm a}$ & 1 & 22.14 & 20.22 & - & 18.43 \\

 & Source A & 220.4557 & 14.6893 & 1.43353 $\pm$ 0.00005$^{\rm b}$  & 1 & 21.47 & 21.57 & - & 21.54 \\

 & Source B & 220.4552 & 14.6897 & 1.43344 $\pm$ 0.00005$^{\rm b}$ & 1 & 21.80 & 21.80 & - & 21.81 \\

 & Source C & 220.4539 & 14.6895 & 1.43357 $\pm$ 0.00006$^{\rm b}$ & 1 & 22.13 & 22.33 & - & 22.46 \\

 & Source D & 220.4549 & 14.6885 & 1.43338 $\pm$ 0.00014$^{\rm b}$  & 1 & 22.00 & 21.98 & - & 21.91 \\

DESI~J234.4780+14.7229 & Lens & 234.4780 & 14.7229 & 0.73066 $\pm$ 0.00016$^{\rm a}$   & 1 & 22.39 & 21.16 & - & 19.52 \\

 & Source A & 234.4785 & 14.7235 & 2.47831 $\pm$ 0.00123$^{\rm h}$ & 3 & - & - & - & - \\

 & Source B &  234.4783 & 14.7226 & 2.4734 $\pm$ 0.00063$^{\rm h}$ & 1 & 21.92 & 21.65 & - & 21.92 \\

DESI~J238.5690+04.7276 & Lens & 238.5690 & 4.7276 & 0.77591 $\pm$ 0.00014$^{\rm a}$ & 1 & 21.38 & 20.62 & 19.75 & 19.20 \\

 & Source A & 238.5695 & 4.7275 & 1.717$^{\rm *}$ & 1 & 24.26 & 25.17 & 19.75 & 19.20 \\

 & Source B & 238.5684 & 4.7274 & 1.717$^{\rm *}$ & 1 & 23.62 & 23.73 & 25.86 & - \\

DESI~J245.7514+21.6226 & Lens & 245.7512 & 21.6223  & 0.75824 $\pm$ 0.00015$^{\rm a}$ & 1 & 21.66  & 20.11 & 18.31 &  - &    \\ 

&Source & 245.7523 & 21.6224  &  1.7254 $\pm$ 0.0001$^{\rm i}$ & 1 &  21.07 & 20.67 &-  & 20.01  &  \\

DESI~J246.0068+01.4842 & Lens & 246.0068 & 1.4842 & 1.09308 $\pm$ 0.00008$^{\rm a}$ & 1 & 24.22 & 22.25 & 21.12 & 20.18 \\

 & Source &  246.0059 & 1.4829 & 2.36583 $\pm$ 0.00018$^{\rm i}$ & 1 & 24.26 & 25.17 & 19.75 & 19.20 \\
 
 & Galaxy 1 & 246.0070 & 1.4836 & 1.06936 $\pm$ 0.00017$^{\rm b}$ & 3 & 22.05 & 21.54 & 21.38 & 21.07 \\
 
 & Galaxy 2 & 246.0058 & 1.4841 & 1.4413 $\pm$ 0.00017$^{\rm b}$ & 1 & 22.92 & 22.56 & 22.42 & 21.91 \\
 
 & Galaxy 3 & 246.0054 & 1.4835 & 2.3664 $\pm$ 0.00036$^{\rm j}$ & 3 & 23.05 & 22.37 & 22.10 & 21.63 \\

 & Galaxy 4 & 246.0063 & 1.4822 & 1.0813 $\pm$ 0.0002$^{\rm a}$ & 1 & 29.04 & 24.64 & 22.97 & 22.11 \\

DESI~J253.2534+26.8843 & Lens & 253.2534 & 26.8843 & 0.63632 $\pm$ 0.00013$^{\rm a}$ & 1 & 23.64 & 21.86 & - & 20.47 \\

 & Source A & 253.2528 & 26.8846 & 2.5964 $\pm$ 0.0006$^{\rm c}$  & 1 & 22.63 & 22.18 & - & 21.85 \\

 & Source B & 253.2542 & 26.8840 & 2.5957 $\pm$ 0.0004$^{\rm c}$  & 1 & 22.53 & 21.69 & - & 21.69 \\

 & Source C & 253.2531 & 26.8838 & 2.5946 $\pm$ 0.0005$^{\rm c}$  & 1 & 22.15 & 21.56 & - & 21.24 \\

 & Source D & 253.2538 & 26.8847 & 2.5945 $\pm$ 0.0008$^{\rm c}$ & 1 & 23.49 & 22.96 & - & 22.66 \\

DESI~J260.8405+23.8442 & Lens & 260.8405 & 23.84423 & 0.2296 $\pm$ 0.00015$^{\rm a}$  & 1 &  17.93 & 16.57 & - & 15.74 \\

 & Source A & 260.8441 & 23.8448 & 0.97503 $\pm$ 0.00005$^{\rm b}$ & 1 &  20.53 & 20.16 & - & 19.25 \\
 
 & Source B & 260.8426 & 23.8472 & 0.97505 $\pm$ 0.00004$^{\rm b}$ & 1 &  20.74 & 20.46 & - & 19.73  \\

DESI~J304.0068-49.9067 & Lens 1 & 304.0067 & -49.9066 & 0.27811 $\pm$ 0.00018$^{\rm a}$ & 1  & 18.45 & 16.88 & 16.36 & 16.04 \\

 & Lens 2 & 304.0079 & -49.9065 & 0.26604 $\pm$ 0.00016$^{\rm a}$ & 1  & 21.77 & 20.41 & 19.94 & 19.63 \\

 & Lens 3 & 304.0060 & -49.9074 & 0.27939 $\pm$ 0.00031$^{\rm a}$ & 1  & - & - & - & - & \\

 & Source A & 304.0081 & -49.9049 & 1.23186 $\pm$ 0.00004$^{\rm b}$  & 1 & 21.90 & 21.69 & 21.42 & 21.31    \\
 
 & Source B & 304.0016 & -49.9066 & 1.2319 $\pm$ 0.00005$^{\rm b}$  & 1 & 22.42 & 22.61 & 22.31 & 22.31  \\

 & Source C & 304.0131 & -49.9070 & 1.2320 $\pm$ 0.00003$^{\rm b}$ & 1 & 21.70 & 21.67 & 21.39 & 21.25 \\

DESI~J318.0376-01.7568 & Lens  & 318.0376 & -1.7567 & 0.22406 $\pm$ 0.00013$^{\rm a}$ & 1 & 18.23 & 16.82 & 16.34 & 16.02  \\

 & Source & 318.0367 & -1.7561 &  1.10902 $\pm$ 0.00002$^{\rm b}$ & 1 & 20.20 & 19.83 & 19.44 & 19.12 &  \\

DESI~J326.0105-43.3965 & Lens 1 & 326.0105 & -43.3965 &  0.27394 $\pm$ 0.00016$^{\rm a}$ & 1 & 18.52 & 16.97 & 16.46 & 16.15    \\

& Lens 2 & 326.0109 & -43.3957 & 0.27081 $\pm$ 0.00011$^{\rm a}$ & 1 & 21.49 & 20.02 & 19.54 & 19.27\\

& Lens 3 & 326.0100 & -43.3972 & 0.27225 $\pm$ 0.00025$^{\rm a}$ & 1 & 22.77 & 21.45 & 20.96 & 20.71 \\

&Source   & 326.0129 & -43.3975 & 1.08206 $\pm$ 0.00038$^{\rm b}$ & 2 &   22.96    &  22.54     &    -   &   21.18      & Noisy \\

DESI~J329.6820+02.9584 & Lens & 329.6820 & 2.9584 & 0.28677 $\pm$ 0.00006$^{\rm a}$ & 1 & 18.83 & 17.28 & - & 16.42 \\

 & Source A & 329.6824 & 2.9595 & 2.07591 $\pm$ 0.00151$^{\rm f}$ & 1 & 21.70 & 21.45 & - & 21.38 \\

 & Source B & 329.6830 & 2.9582 & 2.07975 $\pm$ 0.00124$^{\rm f}$ & 1 & 21.77 & 21.76 & - & 22.18 \\

 & Source C & 329.6825 & 2.9574 & 2.07971 $\pm$ 0.00015$^{\rm f}$ & 1 & 22.05 & 21.91 & - & 22.10 \\

 & Source D & 329.6813 & 2.9585 & 2.0781 $\pm$ 0.00353 $^{\rm f}$ & 2 & - & - & - & - \\

DESI~J331.8083-42.0487 & Lens 1 & 331.8083 & -52.0487 &0.35494 $\pm$ 0.00017$^{\rm a}$ & 1 & 19.08 & 17.39 & 16.82 & 16.52   \\

&Lens 2 & 331.8079 & -52.0490 &  0.35204 $\pm$ 0.00018$^{\rm a}$ & 1 & 22.23 & 20.56 & 19.98 & 19.69  \\

&Source A & 331.8052 & -52.0503 & 1.07$^{\rm *\dagger}$  & 2 &   22.77    &   22.84    & 22.86 & 22.76 & Noisy  \\

&Source B & 331.8049 & -52.0485 & 1.07$^{\rm *\dagger}$   & 2 &  23.57     &   22.78    &  -  & 22.49& Noisy  \\

&Source C & 331.8095 & -52.0491 & 1.07$^{\rm *\dagger}$   & 2 &   22.77    &  22.84    &    22.86   &   22.76      & Noisy  \\

DESI~J335.5354+27.7596 & Lens 1 & 335.5357 & 27.7598 & 0.49016 $\pm$ 0.00012$^{\rm a}$ & 1 & 20.43 & 18.57 & - & 17.40 \\

 & Lens 2 & 335.5364 & 27.7592 & 0.49278 $\pm$ 0.00016$^{\rm a}$ & 1 & 21.89 & 20.06 & - & 18.90 \\

 & Lens 3 & 335.5359 & 27.7583 & 0.4910 $\pm$ 0.0001$^{\rm a}$  & 1 & 22.01 & 20.30 & - & 19.14 \\

 & Source & 335.5358 & 27.7569 & 2.29167 $\pm$ 0.00029$^{\rm c}$ & 1 & 21.37 & 21.10 & - & 20.85 \\

 & Quasar A & 335.5378 & 27.7606 & 2.79791 $\pm$ 0.0004$^{\rm f}$ & 1 & 20.65 & 20.42 & - & 20.61 \\

 & Quasar B & 335.5366 & 27.7611 & 2.80515 $\pm$ 0.00041$^{\rm f}$ & 1 & 20.97 & 20.58 & - & 20.48 \\

 & Quasar C &  335.5330 & 27.7605 & 2.79884 $\pm$ 0.00041$^{\rm f}$ & 1 & 21.36 & 20.95 & - & 20.94 \\


DESI~J339.8883-04.4880 & Lens & 339.8883 & -4.4880 & 0.55752 $\pm$ 0.00011$^{\rm a}$  & 1 & 21.39 & 19.53 & 18.65 & 18.24 \\

 & Source & 339.8873 & -4.4890 & 1.11577 $\pm$ 0.00001$^{\rm b}$ & 1 & 21.55 & 21.32 & 20.99 & 20.80 \\

DESI~J341.0212+27.9883 & Lens & 341.0212 & 27.9883 & 0.34281 $\pm$ 0.00007$^{\rm a}$ & 1 & 19.35 & 17.55 & - & 16.60 \\

 & Source & 341.0212 & 27.9860 & 0.96045 $\pm$ 0.00006$^{\rm a}$  & 1 & 20.82 & 19.50 & - & 17.77 \\

DESI~J341.8012-2.0939 & Lens & 341.8012 & -2.0939 & 0.33414 $\pm$ 0.00015$^{\rm a}$ & 1 & 19.88 & 18.22 & - & 17.34\\

 & Source &  341.7995 & -2.0922 & 1.76782 $\pm$ 0.0001$^{\rm i}$ & 1 & 21.20 & 20.99 & - & 20.83 \\

DESI~J342.9290-03.4136 & Lens 1 & 342.9290 & -03.4136 & 0.24592 $\pm$ 0.00018$^{\rm a}$ & 1 & 18.06 & 16.55 & 16.04 & 15.71\\

 & Lens 2 & 342.9283 & -3.4123 &  0.25016 $\pm$ 0.00016$^{\rm a}$ & 1 & 20.13 & 18.69 & 18.19 & 17.87 \\

 & Lens 3 & 342.9262 & -3.4117 &0.25136 $\pm$ 0.00010$^{\rm a}$ & 1 & 19.98 & 18.51 & 18.02 & 17.72 \\

 & Source & 342.9287 & -3.41 & 1.05956 $\pm$ 0.00008$^{\rm b}$  & 1 & 21.48 & 20.88 & 20.28 & 19.89 &  \\

DESI~J343.0402-04.2187 & Lens 1 & 343.0402 & -04.2187 & 0.43519 $\pm$ 0.00014$^{\rm a}$ & 1 & 19.78 & 17.94 & 17.30 & 16.94\\

 & Lens 2 & 343.0410 & -4.2179 & 0.47$^{\rm *}$ & 3 & 24.44 & 22.80 & 22.19 & 21.83 \\

 & Lens 3 & 343.0395 & -4.2175 & 0.43182 $\pm$ 0.0002$^{\rm a}$ & 1 & 23.88 & 22.04 & 21.41 & 21.07 \\

 & Source A & 343.0395 & -4.2208 & 0.58$^{\rm *}$ & 3 & 22.99 & 22.93 & 22.81 & 22.76  \\

 & Source B & 343.0414 & -4.2204 & 0.58$^{\rm *}$ & 3 & 23.52 & 23.27 & 23.03 & 22.60  \\

 & Source C & 343.0421 & -4.2198 & 0.58$^{\rm *}$ & 3 & 23.34 & 23.40 & 23.72 & 23.34  \\

 & Source D & 343.0433 & -4.2183 & 0.58$^{\rm *}$ & 3 & 24.26 & 23.96 & 24.42 & 24.22  \\

DESI~J344.6262-58.6910 & Lens 1 & 344.6261 & -58.6909 & 0.29435 $\pm$ 0.00017$^{\rm a}$ & 1 & 21.98 & 20.15 & 19.58 & 19.30 \\

 & Lens 2 & 344.6267  & -58.6913 & 0.37$^{\rm *}$ & 2 & 18.32 & 16.84 & 16.30 & 15.97 \\

 & Lens 3 & 344.6282  & -58.6924 &  0.29597 $\pm$ 0.00016$^{\rm a}$ & 1 & 20.84 & 19.24 & 18.73 & 18.43  \\

 & Source A & 344.6278 & -58.6898 &  0.88023 $\pm$ 0.00005$^{\rm b}$ & 1 & 19.96 & 19.06 & 18.45 & 18.20 & \\

 & Source B & 344.6313 & -58.6918 &0.88064 $\pm$ 0.00005$^{\rm b}$  & 1 & 23.63 & 22.53 & 22.13 & 21.91  \\

DESI~J345.0725+22.2254 & Lens 1 &  345.0749 & 22.2239 & 0.43738 $\pm$ 0.00012$^{\rm a}$ & 1 & 21.55 & 19.52 & - & 18.41 \\

 & Source &  345.0772 & 22.2240 & 1.9273 $\pm$ 0.0001$^{\rm i}$ & 1 & 22.65 & 22.27 & - & 21.84 \\

 & Foreground &  345.0769 & 22.2226 & 0.4$^{\rm *}$ & 3 & 22.03 & 21.19 & - & 20.35 \\

DESI~J345.8606+23.4757 & Lens & 345.8606 & 23.4757 & 0.27653 $\pm$ 0.00007$^{\rm a}$ & 1 & 18.61 & 16.96 & - & 15.99 \\

 & Object X &  345.8617 & 23.4750 & 0.28914 $\pm$ 0.00032$^{\rm a}$ & 3 & - & - & - & - \\

 & Source A1 & 345.8600 & 23.4765 & 1.29593 $\pm$ 0.00003$^{\rm b}$  & 1 & 23.53 & 24.10 & - & 23.28 \\

 & Source A2 &  345.8606 & 23.4742 & 1.29552$\pm$ 0.00004$^{\rm b}$ & 1 & - & - & - & - \\

 & Source B1 & 345.8582 & 23.4775 & 3.01007 $\pm$ 0.00024$^{\rm c}$ & 1 & 23.59 & 23.27 & - & 23.73 \\

 & Source B2 & 345.8584 & 23.4765 & 3.01024 $\pm$ 0.00025$^{\rm c}$ & 1 & 22.33 & 21.80 & - & 21.97 \\

 & Source B3 & 345.8586 & 23.4750 & 3.01134 $\pm$ 0.00048$^{\rm c}$ & 2 & 23.84 & 23.06 & - & 22.95 \\

 & Source B4 & 345.8591 & 23.4740 & 3.01081 $\pm$ 0.00039$^{\rm c}$ & 1 & 23.09 & 22.50 & - & 22.62 \\

\hline
\end{longtable}

\end{center}


\section{Results}
\label{sect:results}


\subsection{Confirmed Lenses}
\label{sect:confirmed}

In this section we will briefly describe all of the gravitational lens systems with fully confirmed redshifts, listed in Table~\ref{tab:sourcespositions}. Unless otherwise stated, the quality flag is $Q_z=1$ (``Robust'').
For each system, we present an accompanying figure showing the MUSE color image alongside the extracted spectra of the lenses and sources.

The spectra for the lens galaxies in our survey typically contain the following spectral features: the 4000\,\AA\ break seen in the spectral energy distribution and the Fraunhofer absorption lines. The most commonly seen and prominent lines are \ion{Ca}{2} H and K at $\lambda$3968\,\AA\ and $\lambda$3934\,\AA\ respectively, the G band at $\lambda$4308\,\AA, H$\beta$ at $\lambda$4862\,\AA, Mg B at $\lambda$5175\,\AA, and \ion{Na}{1} D at $\lambda$5892\,\AA. These features are typical of passive galaxies. If not otherwise stated, the redshift for the lensing galaxies were determined by identifying at least three of these characteristics in the spectra.

\begin{enumerate}

     \item \textit{DESI~J003.6745-13.5042} is presented in Figure \ref{fig:MUSEspectra16_big}. Both lens galaxies, labeled L1 and L2, have similar redshifts of $z=0.43119 \pm 0.00013$ and $z=0.42944 \pm 0.00044$ respectively. The lensed source (red arc) is slightly affected by an object in the middle of the arc, but it did not significantly influence our extraction of the source image near it. We identified several absorption lines, namely the H and K at $\lambda$3969\,\AA, and $\lambda$3934\,\AA, as well as G band at $\lambda$4308\,\AA. Furthermore, we detected the $\lambda$3727\,\AA\ + $\lambda$3729\,\AA\ [\ion{O}{2}] and $\lambda$4364\,\AA\ [\ion{O}{3}] emission lines. These spectral features allow us to confidently determine the source redshift of $z=0.90999 \pm 0.00023$. \label{Ref:lens16_1}

     \item \textit{DESI~J031.7781-27.4454} is a stunning system, with the source lensed into a bright blue arc which almost completes a full ring. It is displayed in Fig.~\ref{fig:MUSEspectra2_031}. The yellow lens galaxy in the center has redshift $z=0.35451 \pm 0.00016$. The source arc has a bright knot on the west-most end, so we extracted the spectra for the knot and arc separately. Both spectra place the source at redshift $z=1.67711 \pm 0.00005$, but interestingly, the intensity of the spectral features they display differ. Most notably, the knot spectra has more prominent $\lambda$1808\,\AA\ \ion{Si}{2}; $\lambda$1855\,\AA\ and $\lambda$1863\,\AA, \ion{Al}{3}; $\lambda$2344\,\AA, $\lambda$2374\,\AA, $\lambda$2382\,\AA, $\lambda$2587\,\AA, and $\lambda$2600\,\AA\ \ion{Fe}{2}; $\lambda$2382\,\AA\ and $\lambda$2587\,\AA\ \ion{Mg}{2} absorption lines. Meanwhile, the arc spectra have stronger emission lines ($\lambda$2324\,\AA\ + $\lambda$2325\,\AA\ [\ion{C}{2}] and $\lambda$1909\,\AA\ [\ion{C}{3}]).  
     \label{Ref:lens2_031}

     \item \textit{DESI~J033.8095-29.1570} is displayed in \ref{fig:MUSEspectra112}. Though the image data is noisy, we identified a red lens galaxy in the center with a multiply imaged blue source. This was done by referencing data from our \HST SNAP program (GO-15867), which can be seen in Figure 1 of \cite{huang2025desistronglensfoundry}. In our RGB image, the east and west sides of the source images appear to be different colors even though they belong to the same source. This can be explained by the bad quality of our data, as a colored line runs through the east side of the source, shifting the color. The lens redshift assigned to the lens of $z=0.9$ was determined purely through getting the DESI photo-z redshifts of surrounding red galaxies, as the spectrum was too noisy to identify any features. As for the source, a redshift of $z=1.85757 \pm 0.00011$ was determined through the $\lambda$1855\,\AA, and $\lambda$1863\,\AA\ \ion{Al}{3}; $\lambda$2344\,\AA, $\lambda$2374\,\AA, $\lambda$2382\,\AA, $\lambda$2587\,\AA, and $\lambda$2600\,\AA\ \ion{Fe}{2}; $\lambda$2382\,\AA\ and $\lambda$2587\,\AA\ \ion{Mg}{2} absorption lines. This makes it one of the sources with the highest redshifts recorded in this paper.
     \label{Ref:lens112}

    \item \textit{DESI~J043.6663-04.3068} is presented in Figure \ref{fig:MUSEspectra16}. Both lenses, labeled L1 and L2, have spectra which place their redshifts at $z=0.34477 \pm 0.00016$ and $z=0.34356 \pm 0.00014$, respectively. The lensed source exhibits a "tadpole" morphology, with a distinct knot and a tail-like arc extending to the west. The extracted spectra from both sections have a high signal-to-noise ratio and prominent spectral features (Figure \ref{fig:MUSEspectra16}). We identified the $\lambda$1392\,\AA\ \ion{Si}{4}, $\lambda$1402\,\AA\ \ion{Si}{4} + \ion{O}{4}, $\lambda$1526\,\AA\ \ion{Si}{2}, $\lambda$1549\,\AA\ \ion{C}{4}, $\lambda$1608\,\AA\ \ion{Fe}{2}, $\lambda$1671\,\AA\ \ion{Al}{1} absorption lines and the $\lambda$1909\,\AA\ [\ion{C}{3}] emission line in the spectra. These features allow us to confidently place the source redshift at $z=2.4493 \pm 0.0002$. 
    \label{Ref:lens16_2}

    \item \textit{DESI~J053.6251-13.1869} is presented in Figure \ref{fig:MUSEspectra19}. We determined the redshift of the lens galaxy to be $z=0.3866 \pm 0.00032$. The source has two images, with arc A in the north-east and arc B in the south-west. Both spectra have consistent features, with the absorption lines \ion{Si}{2} at $\lambda$1526\,\AA, \ion{C}{4} at $\lambda$1549\,\AA, \ion{Fe}{2} at $\lambda$1608\,\AA, and \ion{Al}{2} at $\lambda$1671\,\AA. We also see the $\lambda$1909\,\AA\ [\ion{C}{3}] emission line. Based on these features, we determine the source redshifts to be $z=2.30117 \pm 0.00019$ (Source A) and $z=2.30137 \pm 0.00045$ (Source B). Though the \ion{Fe}{2} and [\ion{C}{3}] lines are less prominent in source B, we are confident in our redshift determination based on the other spectral features. Additionally, the MUSE color image shows that both arcs are the same color, further supporting that these are two images of the same source.
    \label{Ref:lens19}

    \item \textit{DESI~J055.0894-25.5581} The spectrum for the lens galaxy is displayed in Figure \ref{fig:MUSEspectra20} at a redshift of $z=0.65708 \pm 0.00018$. The lensed arc of the source is quite faint in the MUSE data cube, so we extracted the topmost section of the arc. The extracted spectrum exhibits absorption lines of [\ion{O}{2}] + \ion{Si}{2} at $\lambda$1303,\AA\ and \ion{C}{2} at $\lambda$1334,\AA. Together with the [\ion{C}{3}] emission at $\lambda$1909\,\AA, these features yield a source redshift of $z = 2.6802 \pm 0.0005$. However, because all emission lines are fairly weak, we assigned a source redshift quality flag of $Q_z=2$ (``Probable'').
    \label{Ref:lens20}

    \item \textit{DESI~J057.2077-10.2959} is a single arc system displayed in Figure \ref{fig:MUSEspectra4_057} with a lens galaxy redshift of $z=0.74493 \pm 0.00013$ and a source redshift of $z=1.60376 \pm 0.00014$. The source spectrum displays many low ionization absorption lines, such as \ion{Fe}{2} at $\lambda$2344\,\AA, $\lambda$2374\,\AA, $\lambda$2382\,\AA, $\lambda$2587\,\AA, and $\lambda$2600\,\AA\; as well as \ion{Mg}{2} at $\lambda$2796\,\AA\ and $\lambda$2803\,\AA.
    \label{Ref:lens4_057}

    \item \textit{DESI~J060.5238-22.0990} This system has two lens galaxies, labeled L1 and L2. Their spectra are displayed in Figure \ref{fig:MUSEspectra22} at a redshift of $z=0.4669 \pm 0.0005$ and $z=0.46506 \pm 0.00013$, respectively. Both galaxy spectra have similar spectral features. The lensed arc spectra of the source has a very prominent [\ion{O}{2}] emission at $\lambda$3727\,\AA\ and $\lambda$3729\,\AA\ with a doublet shown in the cut-out, as well as weaker emission lines H$\gamma$ at $\lambda$4340 and [\ion{O}{3}] at $\lambda$4958\,\AA\ and $\lambda$5006\,\AA. This places the source galaxy redshift at $z=0.81998 \pm 0.00028$. A detailed inspection shows a slight velocity dispersion offset in the latter three emission lines, as the expected rest wavelength falls slightly red-ward of the observed H$\gamma$ line, while the expected rest wavelength of the [\ion{O}{3}] falls slightly blue-ward. The error spectra, plotted at the bottom, demonstrate that this offset is not due to skylines.
    \label{Ref:lens22}
   
    \item \textit{DESI~J065.6453-28.0646} is presented in Figure \ref{fig:MUSEspectra24}, showing a red lens galaxy, indicating relatively high redshift, and a red arc. We confidently determined the redshift of the lens galaxy to be $z=0.62107 \pm 0.00018$. Like the lens galaxy, the source displays \ion{Ca}{2} H and K absorption lines at $\lambda$3968 and $\lambda$3934\,\AA. In addition, the source spectrum exhibits a weak [\ion{O}{2}] emission line doublet (see the zoomed in panel). Based on these features, we determine the source galaxy to be at a redshift of $z=1.17482 \pm 0.00018$.
    \label{Ref:lens24}

     \item \textit{DESI~J070.4130-09.7774} consists of a large blue arc lensed around a cluster of orange lens galaxies. We extracted three of these, labeled L1 ($z=0.45467 \pm 0.00015$), L2 ($z=0.45707 \pm 0.00014$), and L3 ($z=0.35942 \pm 0.00013$) in \ref{fig:MUSEspectra25}. The source arc exhibits the same common low ionization absorption lines seen previously, which place it at a redshift of $z=1.80133 \pm 0.00012$. Namel, we see the $\lambda$2344\,\AA, $\lambda$2374\,\AA, $\lambda$2382\,\AA, $\lambda$2587\,\AA, and $\lambda$2600\,\AA\ \ion{Fe}{2} absorption lines; $\lambda$2382\,\AA\ and $\lambda$2587\,\AA\ \ion{Mg}{2} absorptions, and a $\lambda$1909\,\AA\ [\ion{C}{3}] emission line. 
     \label{Ref:lens25}

    \item \textit{DESI~J073.5286-10.2227} This gravitational lens system also features a single red lensed arc from the source galaxy, with the lens galaxy at a redshift of $z=0.24820 \pm 0.00016$. As seen in the MUSE color image (Figure \ref{fig:MUSEspectra26}), there are blue and yellow interloper galaxies in front of the arc. To extract the spectra of the source, we selected the red pixels around the interlopers, specifically in north-east and south-west regions of the arc. In the source spectra, we detected a prominent [\ion{O}{2}] emission with a double peak shown in the zoom-in, as well as Ca H and K absorption lines. These place the source galaxy at a redshift of $z=1.04384 \pm 0.00009$.\label{Ref:lens26}

    \item \textit{DESI~J073.9027-25.5132} We see a yellow foreground galaxy lensing a blue source galaxy into a very extended arc (A) and a red galaxy which appears as a smaller arc (B) in Figure \ref{fig:MUSEspectra27}. The spectrum of the lens galaxy displays some of the common absorption lines which were previously mentioned. More prominent, however, are the strong emission lines: [\ion{O}{2}], H$\beta$, [\ion{O}{3}], H$\alpha$, and the \ion{S}{2} doublet. Based on these, we confidently place the lens galaxy at a redshift of $z=0.3795 \pm 0.00017$. As for Source A, we see that the arc is split up into separate sections. To extract the spectra, we combined the pixels from all sections. We found a redshift of $z=2.8230 \pm 0.0003$ based on the strong metal absorption lines: \ion{Si}{4} at $\lambda$1392\,\AA, \ion{Si}{4} and \ion{O}{4} at $\lambda$1402\,\AA, \ion{Si}{2} at $\lambda$1526\,\AA, \ion{C}{4} at $\lambda$1549\,\AA, \ion{Fe}{2} at $\lambda$1608 \,\AA, and \ion{Al}{2} at $\lambda$1671\,\AA. The strong $\lambda$1909\,\AA\ [\ion{C}{3}] emission line confirms our redshift determination. For Source B, we found a redshift of $z=0.70821 \pm 0.00017$, based on the [\ion{O}{2}] emission and very strong \ion{Ca}{2} H and K absorptions at $\lambda$3968\,\AA\ and $\lambda$3934\,\AA. For source B, we didn't find any counter-image. Therefore, Source B may also be just an normal (not-lensed) galaxy. \label{Ref:lens27}

    \item \textit{DESI~J074.9646-30.7233} The system features a small blue arc around a yellow lens galaxy, presented in Figure \ref{fig:MUSEspectra28}. We determined the redshifts of the lens and source to be $z=0.44137 \pm 0.00016$, and $z=1.44884 \pm 0.00002$, respectively. The source redshift determination is solely based on the very strong [\ion{O}{2}] emission line at $\lambda$3727\,\AA. Upon zooming in, we see a beautiful double peak, which confirms the feature as [\ion{O}{2}].
    \label{Ref:lens28}

    \item \textit{DESI~J075.2793-24.4176} The gravitational lens system  features a background galaxy which has been lensed around a large galaxy cluster and forms a long blue arc (Figure \ref{fig:MUSEspectra29}). We found a lens galaxy redshift of $z=0.32059 \pm 0.00013$. The background source spectra exhibits the metal absorption lines we have previously seen: $\lambda$1303\,\AA\ [\ion{O}{2}] + \ion{Si}{2}, $\lambda$1334\,\AA\ \ion{C}{2}, $\lambda$1392\,\AA\ \ion{Si}{4}, $\lambda$1402\,\AA\ \ion{Si}{4} + \ion{O}{4}, $\lambda$1526\,\AA\ \ion{Si}{2}, $\lambda$1549\,\AA\ \ion{C}{4}, $\lambda$1608 \ion{Fe}{2}, and $\lambda$1671\,\AA\ \ion{Al}{2} absorption lines. We can also see a weak $\lambda$1909\,\AA\ [\ion{C}{3}] emission line. These lines allow us to assign the source spectra a redshift of $z=2.8290 \pm 0.0004$.
    \label{Ref:lens29}

    \item \textit{DESI~J086.3072-26.5877} This gravitational lens consists of two lens galaxies and one blue source galaxy (Figure \ref{fig:MUSEspectra111}). The two lens galaxies, L1 and L2, have redshifts of $z=0.27534 \pm 0.00018$ and $z=0.27426 \pm 0.00021$, respectively. We broke up the background source into three separate arcs, labeled A, B, and C. Arcs B and C exhibit at least three out of the following four spectral features: $\lambda$1526\,\AA\ \ion{Si}{2}, $\lambda$1549\,\AA\ \ion{C}{4}, $\lambda$1608\,\AA\ \ion{Fe}{2}, $\lambda$1671\,\AA\ \ion{Al}{2}, and $\lambda$1909\,\AA\ [\ion{C}{3}]. The spectra for arc A has a very low signal to noise ratio, so there are no prominent spectral features. However, the shape of the flux continuum and the color of the arcs match, leading us to believe arc A belongs to the same galaxy at a redshift of $z=2.17441 \pm 0.00037$. Because of this, we set the quality flag of A to be $Q_z=2$. The co-added spectra of all three arcs, displayed at the bottom, has a much higher SNR, showing all the spectral features mentioned above. 
    \label{Ref:lens111}

    \item \textit{DESI~J087.1525-36.2427} The doubly imaged source galaxy is displayed in Figure \ref{fig:MUSEspectra32}. The two orange arcs, labeled A and B, can be seen on either side of the lens galaxy cluster. The lens galaxy spectrum has a very high signal to noise ratio, and we confidently assign it a redshift of $z=0.3011 \pm 0.00018$. As for the background galaxy, both arcs exhibit a strong [\ion{O}{2}] emission line with a clear double peak, as seen in the zoom in. The spectra for source arc B features strong H$\beta$ and [\ion{O}{3}] emission lines, as well as weaker \ion{Ne}{3} and H$\gamma$ emission lines, which are seldom seen among the other gravitational lens systems covered in this paper. The source arc A spectrum has a lower SNR, but we can see the H$\beta$ and [\ion{O}{3}] emission lines. Additionally, velocity dispersion in the H$\beta$ and [\ion{O}{3}] lines can be seen in both arc spectra. Based on these features, we assign redshifts of $z=0.84611 \pm 0.00005$ (Source A) and $z=0.84661 \pm 0.00006$ (Source B).
    \label{Ref:lens32}

    \item \textit{DESI~J090.9854-35.9683} is an extraordinary gravitational lens system with a total of seven observed lensed sources centered around a cluster of lens galaxies. Here, we highlight just one of the sources and two of the lens galaxies. Further analysis of all seven sources and their cosmological constraints can be found in \cite{sheu2024carousel} and \cite{2026arXiv260216077U}, respectively. The four blue images of the source are arranged in an arc-like morphology, labeled A, B, C and D (Figure \ref{fig:MUSEspectra16repl_all}). Despite being faint, we clearly detect the $\lambda$3727\,\AA\ [\ion{O}{2}] doublet in the four images. The lens galaxies, labeled L1 and L2, have respective redshifts of $z=0.4883 \pm 0.00019$ and $z=0.48835 \pm 0.00016$. These spectral features allow us to confidently determine the source redshift components ranging closely around $z=1.432$ (e.g., Source A at $z=1.43211 \pm 0.00012$).  \label{ref:lens16repl}

    \item \textit{DESI~J122.0852+10.5284} has two different background source galaxies which have both been singly lensed about the same foreground galaxy. The foreground galaxy has a redshift of $z=0.47529 \pm 0.00007$. Source A has a redshift of $z=1.23669 \pm 0.00004$, identified with a strong [\ion{O}{2}] emission with a textbook double peak, as seen in the zoom in. Source A is clearly visible in Fig.~\ref{fig:MUSEspectra113} as a long orange arc. Source B is hard to see in the MUSE color image, but  displays a prominent [\ion{O}{2}] emission line. It is also well visible in observations from our \HST SNAP program (GO-15867). It has a higher redshift of $z=1.45292 \pm 0.0001$, based on an [\ion{O}{2}] emission line. 
    \label{Ref:lens113}

    \item \textit{DESI~J154.6975-01.3588} displays an arc-counterarc morphology. The foreground lens has a redshift of $z=0.38827 \pm 0.00007$. The counterarc and arc are labeled with A and B respectively, and they have consistent redshifts of $z=1.431 \pm 0.00012$ and $z=1.43056 \pm 0.00007$. Despite the faint appearance of the arcs in the color image (Fig.~\ref{fig:MUSEspectra107}), both images display a strong [\ion{O}{2}] emission line. The double peaks can be seen in the zoom in. 
    \label{Ref:lens107}

    \item \textit{DESI~J157.4222+20.4043} consists of a red background galaxy that has been lensed into a distinctive "tadpole" shape (Figure~\ref{fig:MUSEspectra115}). Its spectrum reveals a weak [\ion{O}{2}] emission peak alongside strong Ca H and K absorption lines, confirming its redshift at $z = 1.30729 \pm 0.00015$. The lens consists of two galaxies, labeled L1 and L2, with similar redshifts of $z = 0.39203 \pm 0.00008$ and $z=0.39098 \pm 0.00009$.
    \label{Ref:lens115}

    \item \textit{DESI~J157.6135-06.6858} is presented in Figure \ref{fig:MUSEspectra34}. The lens galaxy has a redshift of $z=0.46717 \pm 0.00013$.  The image of the lensed source appears as an arc west of the the lens galaxy, with a distinct point in the south part of the arc (Figure~\ref{fig:MUSEspectra34}). We identified absorption lines, specifically \ion{Fe}{2} $\lambda$2344\,\AA, $\lambda$2374\,\AA, $\lambda$2382\,\AA, $\lambda$2586\,\AA, $\lambda$2600\,\AA\ and \ion{Mg}{2} $\lambda$2796\,\AA. These spectral features allow us to determine the source redshift to be $z=1.5806 \pm 0.0001$. \label{Ref:lens34}

    \item \textit{DESI~J160.1719+18.8480} features a red background galaxy, which has been lensed around a foreground galaxy at $z=0.31331 \pm 0.00013$. There are four images labeled A, B1, B2, and C from north to south (Figure~\ref{fig:MUSEspectra35}).
    Based on the redshifts, the four images come from a trio of physically close galaxies: B1 and B2 correspond to the same source at redshifts of $z=0.87779 \pm 0.00026$ and $z=0.8778 \pm 0.00024$, while A is at a slightly higher redshift of $z=0.88096 \pm 0.00019$. C corresponds to a third source at a lower redshift of $z=0.87566 \pm 0.00013$. In the spectra, we can see that they all have an {\ion{O}{2}} peak, as well as Ca~H and K absorption lines (clearer for the brighter images A and B1). Because B2 and C have lower SNR, we assigned quality flags of $Q_z=2$ for those sources. 
    \label{Ref:lens35}

    \item \textit{DESI~J161.1162+31.2340} has multiple lens galaxies, with one at redshift $z=0.70$, labeled L1 in Figure \ref{fig:MUSEspectra36}. To the east, we see another at the same redshift, labeled L2. Its spectra differs from L1, as it exhibits a very strong \ion{O}{2} and \ion{O}{3} emission, indicating that it may be an AGN. There is another group member to the east, labeled L3. The lensed source arc has three prominent knots; we extracted all three separately, as well as the regions in between them (labeled as arc in the plot). Each spectrum prominently features $\lambda$2344\,\AA, $\lambda$2374\,\AA, $\lambda$2382\,\AA, $\lambda$2587\,\AA, and $\lambda$2600\,\AA\ \ion{Fe}{2} absorption lines and $\lambda$2382\,\AA\ and $\lambda$2587\,\AA\ \ion{Mg}{2} absorptions. This sets the source redshift at $z=1.516$. Additionally, we see a foreground object to the southeast with a redshift of $z=0.7$, determining that it is another group member, separate from the source. \label{ref:lens36}

    \item \textit{DESI~J161.4114-08.8358} is presented in Figure \ref{fig:MUSEspectra37}. The lens galaxy is characterized by a redshift of $z=0.82754 \pm 0.00018$. The lensed source is visible as an arc in the upper region. We identified several absorption lines (Figure \ref{fig:MUSEspectra37}), including \ion{Fe}{2} $\lambda$2344\,\AA, $\lambda$2374\,\AA, $\lambda$2382\,\AA, $\lambda$2586\,\AA, $\lambda$2600\,\AA\ and \ion{Mg}{2} $\lambda$2796\AA. Despite the presence of noise (which led us to assign a quality flag of $Q_z=3$), the \ion{Fe}{2} absorption lines are distinctly observed, enabling the determination of the source redshift as $z=2.079 \pm 0.002$. \label{ref:lens37}

    \item \textit{DESI~J166.9974+04.1560} We see two lens galaxies (labeled L1 and L2) in Figure \ref{fig:MUSEspectra38}, which are at a redshift of $z=0.35153 \pm 0.00017$ and $z=0.35271 \pm 0.00016$. As for the background source, we see that it is singly lensed with a spectra exhibiting a very strong [\ion{O}{2}] doublet. We can also see slight dips in the flux continuum corresponding to Ca H and K absorptions. Therefore, we confidently assign a redshift of $z=1.28023 \pm 0.00013$. 
    \label{Ref:lens38}

    \item \textit{DESI~J168.7680+16.7604} In this system, we observe one lens galaxy (L1) with two different imaged background galaxies A and B (Figure \ref{fig:MUSEspectra39}). Galaxy L1 has redshift $z=0.53714 \pm 0.0002$. Between the two arcs we see an "interloper" foreground galaxy (L2), which has a slightly lower redshift of $z=0.53503 \pm 0.00016$. The spectra for source A has unexpectedly low SNR, considering how bright it appears in the MUSE color image. We see a few weak spectral features, specifically the  $\lambda$1402\,\AA\ \ion{Si}{4}, $\lambda$1808\,\AA\ \ion{Si}{2}, $\lambda$1549\,\AA\ \ion{C}{4}, and $\lambda$1862\,\AA\ \ion{Al}{2} absorption lines, as well as the $\lambda$1882\,\AA\ [\ion{Si}{3}] and $\lambda$1909\,\AA\ [\ion{C}{3}] emission lines. Additionally, the slightly downward sloping flux continuum matches the spectra of other galaxies with similar spectral features. With this, we assign a tentative redshift of $z=1.71573 \pm 0.00012$. Though source B appears fainter, the textbook Ly-$\alpha$ emission, characterized by the steep rise on the blue end and the gentler drop on the red end, in its spectra allows us to confidently calculate a redshift of $z=3.46264 \pm 0.00004$. Based on the spectral reduction, as well as the slight color difference between arcs A and B, we can confirm that this system has two different lensed sources. 
    \label{Ref:lens39}

    \item \textit{DESI~J174.5481+14.7863} is a gravitational lens system shown in Figure \ref{fig:MUSEspectra41} featuring a main lens galaxy labeled L, with a redshift of $z=0.56486 \pm 0.00019$. The lensed source is observed to the west of the lens galaxy, displaying an arc-like morphology. Although the spectrum has a low SNR, we identified absorption lines such as \ion{Si}{2} $\lambda$1526\,\AA, \ion{C}{4} $\lambda$1549\,\AA, \ion{Al}{2} $\lambda$1671\,\AA, \ion{O}{2} $\lambda$3727\,\AA\ and \ion{Fe}{2} $\lambda$2344\,\AA, $\lambda$2374\,\AA, $\lambda$2382\,\AA. These spectral features \citep{2017A&A...605A.118F} allow us to confidently determine the source redshift to be $z=2.41864 \pm 0.00065$. \label{ref:lens41}

    \item \textit{DESI~J186.4036-07.4200} has two different singly imaged background source galaxies, labeled A and B and shown in Figure \ref{fig:MUSEspectra106}. We found a lens redshift of $z=0.40484 \pm 0.00037$. The two arcs both show prominent [\ion{O}{2}] emission lines with clearly visible double peaks. We found slightly different redshifts for each arc, with $z=1.08376 \pm 0.00004$ for A and $z=1.1001 \pm 0.00004$ for B. Because of this, we conclude that arcs A and B belong to different galaxies. 
    \label{Ref:lens106}  

    \item \textit{DESI~J190.7935+21.3334} This system features a very large and bright lens galaxy at redshift $z=0.34832 \pm 0.00006$ (Figure \ref{fig:MUSEspectra47}). The lensed arc is also very bright, so the SNR of the source spectra is extremely high. We see a very bright [\ion{O}{2}] emission with a slight double peak, placing the background galaxy at a redshift of $z=1.48152 \pm 0.00005$. 
    \label{Ref:lens47}

    \item \textit{DESI~J196.4575+22.9256} has a lens galaxy at a redshift of $z=0.39426 \pm 0.00017$. The lensed source appears as a small arc south east relative to the lens (Figure \ref{fig:MUSEspectra48}). We identify the following absorption lines: \ion{Ca}{2} K and H $\lambda$3934\, \AA\  $\lambda$3969\,\AA, G band at $\lambda$4308\,\AA, Mg b at $\lambda$5176\,\AA, along with emission lines like  $H{\gamma}$ at $\lambda$4341\,\AA, \ion{O}{3} at $\lambda$4960\,\AA, and $H{\beta}$ at $\lambda$4862\,\AA.\ The source spectra is noisy (note the quality flag $Q_z=3$ -- ``Possible''), but these lead us to assign a source redshift of $z=0.713$.
    \label{ref:lens48}

    \item \textit{DESI~J197.5704+14.7474} is shown in Figure \ref{fig:MUSEspectra49}. This system consists of a red background galaxy lensed by a large yellow galaxy at a redshift of $z=0.26144 \pm 0.00007$. In the arc spectrum, we identified a weak [\ion{O}{2}] doublet, as well as the Ca H and K absorption lines, allowing us to determine a redshift of $z=0.85834 \pm 0.00015$.
    \label{Ref:lens49}

    \item \textit{DESI~J200.7678+03.7216} has two lens galaxies, labeled L1 and L2, each at a redshift of $z=0.35402 \pm 0.00015$ and $z=0.35262 \pm 0.00018$, respectively. The blue object in the center is a star. The images of the lensed source are arranged in three knots north west of the galaxies, labeled as A, B, and C (Figure \ref{fig:MUSEspectra50}). The images  A and B are connected by a faint arc. We identified absorption lines in all three sources, specifically the \ion{Ca}{2} H and K doublet $\lambda$3969\,\AA\ and $\lambda$3934\,\AA\ and G Band at $\lambda$4308\,\AA.\ Based on these spectral features, the source redshifts were confidently identified to be around $z=1.015$ (Source A at $z=1.01586 \pm 0.00017$, Source B at $z=1.015 \pm 0.00022$, and Source C at $z=1.01537 \pm 0.00017$).  \label{ref:lens50}

    \item \textit{DESI~J202.6690+04.6708} has a singly lensed background galaxy around a bright foreground galaxy at redshift $z=0.3364 \pm 0.0001$. There is also a blue foreground galaxy which overlaps with the lensed arc (Figure \ref{fig:MUSEspectra119}). We extracted its spectra and found a very low redshift of $z=0.03650 \pm 0.00002$, based on the following prominent emission lines: H$\beta$ at $\lambda$4862\,\AA, [\ion{O}{3}] at $\lambda$4958\,\AA\ and $\lambda$5006\,\AA, H$\alpha$ at $\lambda$6564\,\AA, and the \ion{S}{2} doublet at $\lambda$6718\,\AA\ and $\lambda$6732\,\AA. As for the lensed source, we extracted the spectra of the arc from areas around the foreground galaxy and determined a redshift of $z=1.16828 \pm 0.00001$, based solely on the very prominent [\ion{O}{2}] doublet (see the zoomed-in panel). 
    \label{Ref:lens119}

    \item \textit{DESI~J218.2479-07.2268} is shown in Figure \ref{fig:MUSEspectra1}. Two galaxies with different redshifts are observed: Lens 1 at $z=0.33851 \pm 0.00013$ and Lens 2 at $z=0.243$. We observed three lensed images of the source. Image A is located in the north-west with an arc-like shape, while images B and C appear as spots in the south-east. All three exhibit the \ion{Ca}{2} H and K absorption lines at $\lambda$3969\,\AA\ and $\lambda$3934\,\AA, though image C has the lowest intensity. Based on these absorption lines, we determined a redshift of $z=1.104$ for the source components. Because of the low signal-to-noise ratio, we assigned a quality flag of $Q_z=2$ for all three spectra. \label{ref:lens1}

    \item \textit{DESI~J220.4549+14.6891} This system features a textbook Einstein cross formation, with the quadruply lensed source labeled A, B, C, and D (Figure \ref{fig:MUSEspectra52}). It was first observed in 2009 in the CAmbridge Sloan Survey Of Wide ARcs in the skY (CASSOWARY) and named CSWA 20 \citep{pettini2010cassowary}. They secured the redshifts of two out of the four images. We secured redshifts for all four images. This system was additionally observed by DESI DR1 \citep[see][]{huang2025desistronglensfoundry}. The lens galaxy was found to have a redshift of $z = 0.7421 \pm 0.0002$, and the lensed source components have respective redshifts around $z = 1.433$ (Source A is $1.43353 \pm 0.00005$, Source B is $1.43344 \pm 0.00005$, Source C is $1.43357 \pm 0.00006$, Source D is $1.43338 \pm 0.00014$). In the MUSE data we see that all four images display [\ion{O}{2}] emission lines, with all double peaks aligning perfectly.
    \label{Ref:lens52p112}
 
    \item \textit{DESI~J234.4780+14.7229} The gravitational lens has one source with an arc-counterarc morphology, shown in Figure~\ref{fig:MUSEspectra120}. The arc is labeled B and the counterarc is labeled A. Arc A is very small and not easily noticed at first glance, but we observe a small blue spot above the red lens galaxy. The lens galaxy was found to have a redshift of $z=0.73066 \pm 0.00016$. The source spectra are mostly featureless, with the most notable feature being a prominent emission line in spectrum B. At first glance, it looks like another [\ion{O}{2}] emission, but it does not have a double peak upon zooming in. With support from Keck data (Storfer, priv. communication), we determined that it is actually a [\ion{C}{3}] emission. Looking at spectrum A, we do not see any notable lines (note the quality flag of $Q_z=3$), except one small peak at the same wavelength of the [\ion{C}{3}] emission line in spectrum B. This is supported by the weak emission lines \ion{He}{2} at $\lambda$1640\,\AA\ and \ion{O}{3} at $\lambda$1660\,\AA\ and $\lambda$1666\,\AA. Based on these, we determine a redshift for the source components of $z=2.47831 \pm 0.00123$ (Source A) and $z=2.4734 \pm 0.00063$ (Source B).
    \label{Ref:lens120}

    \item \textit{DESI~J238.5690+04.7276} The system features another galaxy lensed into an arc–counter-arc configuration, labeled A and B in Figure~\ref{fig:MUSEspectra122}. The foreground lens galaxy has a redshift of $z = 0.77591 \pm 0.00014$. As in the previous system, the spectra of both arcs are largely featureless, except for a notable peak in spectrum B. Based on Keck observations, we identify this feature as the [\ion{C}{3}] emission line at $\lambda$1909\,\AA. Spectrum A also shows a possible peak at the same wavelength, though it is significantly less pronounced than in the DESI~J234.4780+14.7229 system. Additionally, we identified other weak absorption lines, specifically the \ion{Al}{3} at $\lambda$1855\,\AA\ and $\lambda$1853\,\AA\ and the \ion{Fe}{2} at $\lambda$2344\,\AA, $\lambda$2375\,\AA, and $\lambda$2383\,\AA. These detections allow us to assign a source redshift of $z = 1.717$. Although our MUSE observations are noisy, because the lines are cross-identified in our Keck observations, the redshift is robust and we set the quality flag to $Q_z=1$. 
    \label{Ref:lens122}

    \item \textit{DESI~J245.7514+21.6226} The system is presented in Figure \ref{fig:MUSEspectra54}. Two galaxies, L1 and L2, are observed very close to each other with a redshift of $z=0.75824 \pm 0.00015$. The image of the lensed source forms a blue arc that almost completely surrounds them in a circle. The source is at a redshift of $z=1.7254 \pm 0.0001$ with a quality flag of $Q_z=1$, determined using absorption lines of  \ion{C}{3} at $\lambda$2382\,\AA, \ion{Fe}{3} at $\lambda$2344\,\AA, $\lambda$2374\,\AA, $\lambda$2382\,\AA, $\lambda$2586\,\AA, $\lambda$2600\,\AA\ and \ion{Mg}{3} at $\lambda$2796\,\AA, $\lambda$2803\,\AA, using the spectral features presented in \citep{2017A&A...605A.118F}. 
    \label{ref:lens54}

    \item \textit{DESI~J246.0068+01.4842}
    This system has a very high lens galaxy redshift of $z=1.09308 \pm 0.00008$. 
    South of the lens is another galaxy within the same group with a redshift of $z=1.0813 \pm 0.0002$, labelled G4.
    We observed this system twice and co-added the data to get higher quality spectra. The co-added RGB image is shown in Figure~\ref{fig:MUSEspectra123}. There are two blue source arcs (identified via Hubble images), one on top and one on bottom. There are also three galaxies (labeled G1, G2, and G3) which coincide with the source arc. To get the source redshift, we extracted three areas of the blue source arc: the top (avoiding the areas around G2 and G3), the bottom, and the left side of the bottom arc surrounding G1 but not including it. All three of these areas conform to the redshift of $z = 2.36583 \pm 0.00018$, with the most prominent absorption lines being \ion{Si}{2} at $\lambda$ 1526\,\AA\ and $\lambda$1808\,\AA; \ion{C}{4} at $\lambda$ 1549\,\AA; \ion{Fe}{2} at $\lambda$1608\,\AA, $\lambda$2344\,\AA, $\lambda$2375\,\AA, and $\lambda$2383\,\AA; \ion{Al}{2} at $\lambda$1671\,\AA; \ion{Al}{3} at $\lambda$1855\,\AA and $\lambda$1863\,\AA; and \ion{Mg}{1} at $\lambda$2027\,\AA. This redshift value is also supported by Keck data. Based on this, we determine that the two arcs are from the same source. In addition, we extracted spectra from each of the three galaxies. For G1, we found a redshift of $z = 1.06936 \pm 0.00017$ based on the \ion{O}{2} doublet. Though we are confident that G1 is not part of the source arc because of how much higher the flux is compared to the arc, we have lower confidence in this redshift value because the doublet we see in the zoom-in is not a perfect match with what we expect from an \ion{O}{2} emission. Because of this, we assign a quality flag of $Q_z = 3$. The redshift of G2 was also identified by an \ion{O}{2} emission to have redshift $z = 1.4413 \pm 0.00017$, and we have high confidence in this value based on the zoom-in. Finally, the extracted spectra of G3 exhibited many of the same features as that of the source arc, giving it a measured $z = 2.3664 \pm 0.00036$. However, we are certain that G3 is not part of the source because it is a different color, so we believe these features are due to contamination from the source, hence the quality flag $Q_z = 3$. 
    \label{ref:lens123}   

    \item \textit{DESI~J253.2534+26.8843}, shown in Figure \ref{fig:MUSEspectra57} is another textbook Einstein Cross, which was studied and published in \citet{2023ApJ...953L...5C}. At the center of the system, a galaxy with a redshift of $z=0.63632 \pm 0.00013$ is identified, encircled by four distinct blue images of the source (A, B, C, and D). Spectroscopic analysis reveals absorption lines attributed to \ion{C}{2} at $\lambda$1335\,\AA, \ion{Si}{4} at $\lambda$1397\,\AA, \ion{S}{2} + \ion{O}{4} at $\lambda$1399\,\AA,  \ion{C}{4} at $\lambda$1549\,\AA, and \ion{Fe}{2} at $\lambda$1908\,\AA, enabling the determination of a redshift of $z=2.5964 \pm 0.0006$ for Source A, with B, C, and D yielding matching components. 
    \label{ref:lens57}

    \item \textit{DESI~J260.8405+23.8442} is shown in Figure \ref{fig:MUSEspectra58}. The redshift of the predominant galaxy, labeled L, was determined to be $z=0.2296 \pm 0.00015$. The sources A and B, which appear to be elongated, are visible in the east. They show prominent $\lambda$3727\,\AA\ \ion{O}{2} emission lines, where the two peaks can be clearly distinguished, along with $H\gamma$ at $\lambda$4341\,\AA.  These features facilitate the determination of the redshift of the source at $z=0.97503 \pm 0.00005$ (Source A) and $z=0.97505 \pm 0.00004$ (Source B). 
    \label{ref:lens58}

    \item \textit{DESI~J304.0068-49.9067} A system with multiple galaxies is presented in Figure~\ref{fig:MUSEspectra304}, where L1, L2, and L3 are at respective redshifts of $z=0.27811 \pm 0.00018$, $z=0.26604 \pm 0.00016$, and $z=0.27939 \pm 0.00031$. Surrounding them are the blue images of the source (labeled A, B, and C). Their spectra exhibit a single, well-defined \ion{O}{2} at $\lambda$3727\,\AA\ emission line, allowing for a redshift determination of the background galaxy of $z=1.23186 \pm 0.00004$ for Source A, with components B and C sharing identical values. We extracted spectra for the blue arc south from the lens, and we find that it is at a much higher redshift than the source. It displays no strong spectral features and we could not determine the exact redshift.\label{ref:lens304}

    \item \textit{DESI~J318.0376-01.7568} is shown in Figure \ref{fig:MUSEspectra4}. The bright yellow lens galaxy has a redshift of $z=0.22406 \pm 0.00013$. The arc lies to the west and is also very bright, exhibiting a strong $\lambda$3727\,\AA\ \ion{O}{2} emission line and \ion{Fe}{2} and \ion{Mg}{2} absorption lines, confirming the redshift of the background object at $z=1.10902 \pm 0.00002$.
    \label{ref:lens4}

    \item \textit{DESI~J326.0105-43.3965} is presented in Figure \ref{fig:MUSEspectra10}. The redshifts of L1 and L2 were calculated as $z=0.27394 \pm 0.00016$ and $z=0.27081 \pm 0.00011$, and L3 as $z=0.27225 \pm 0.00025$. The lensed background galaxy appears as a faint arc south-east from the lens, slightly contaminated by nearby objects. Despite the low signal-to-noise ratio, an $\lambda$3727\,\AA\ \ion{O}{2} emission line was identified, allowing us to determine a redshift of $z=1.08206 \pm 0.00038$, with a quality flag of $Q_z=2$.
    \label{ref:lens10}

    \item \textit{DESI~J329.6820+02.9584} Here, the background galaxy is lensed into another arc-counterarc morphology, with the arc split into 3 sections (labeled A, B, and C) and counterarc D directly across towards the west. For the lens galaxy, we found a redshift of $z=0.28677 \pm 0.00006$. We display all four source spectra in the bottom panel of \ref{fig:MUSEspectra59}. All four display absorption lines $\lambda$1549\,\AA\, \ion{C}{4}; $\lambda$1608\,\AA\, \ion{Fe}{2}; $\lambda$1671\,\AA\, \ion{Al}{2}; $\lambda$1855\,\AA\ and $\lambda$1863\,\AA\, \ion{Al}{3}; $\lambda$2344\,\AA,$\lambda$2375\,\AA\  and $\lambda$2383\,\AA\, \ion{Fe}{2}; and $\lambda$2796\,\AA\ and $\lambda$2804\,\AA\, \ion{Mg}{2}. In addition, arc C exhibits a weak [\ion{C}{3}] emission line. Arc D is less bright and smaller, so the spectra have lower SNR and we assigned a quality flag of $Q_z=2$. However, it still has enough features to identify that arc D is at the same redshift. Based on this, we can confidently place the background galaxy at a redshift ranging around $z=2.07591 \pm 0.00151$ (Source A). 
    \label{Ref:lens59}

    \item \textit{DESI~J331.8083-52.0487} The system in Figure \ref{fig:MUSEspectra12} is composed of two galaxies, L1 and L2, with respective redshifts of $z=0.35494 \pm 0.00017$ and $z=0.35204 \pm 0.00018$. The sources A, B, and C are faintly observed, with source A showing a higher flux than B and C. Despite the low signal (hence the quality flag $Q_z=2$), several absorption lines of \ion{Fe}{2} at $\lambda$2344\,\AA, \ion{Ca}{2} $K$ and $H$ at $\lambda$3934\,\AA\ and at $\lambda$3969\,\AA, respectively, as well as emission lines such as \ion{Mg}{2} at $\lambda$2796\,\AA, \ion{O}{2} at $\lambda$3727\,\AA, and $H\gamma$ at $\lambda$4341\,\AA\ were recognized, allowing for the determination of a redshift for the source components of $z=1.07$. 
    \label{ref:lens12}

    \item \textit{DESI~J335.5354+27.7596} This system has been observed multiple times before \citep[e.g.][]{dahle2013sdss, acebron2022new}. There are three main lens galaxies, labeled L1, L2, and L3. They are displayed in the top panel of Figure~\ref{fig:MUSEspectra64} with respective redshifts of $z=0.49016 \pm 0.00012$, $z=0.49278 \pm 0.00016$, and $z=0.4910 \pm 0.0001$. There is also a blue triply lensed quasar, with the images marked as A, B, and C. The spectra are shown in the bottom panel of \ref{fig:MUSEspectra64}. The perfectly matched \ion{C}{4} and [\ion{C}{3}] in all 3 spectra show that the quasar components undoubtedly have a redshift ranging around $z=2.79791 \pm 0.0004$ (Quasar A). Finally, we see a lensed arc towards the south, labeled in \ref{fig:MUSEspectra64}, which was not spectroscopically confirmed by the other papers mentioned above. The main spectral features are the \ion{Si}{2}, \ion{C}{4}, and \ion{Fe}{2} absorptions as well as [\ion{O}{3}] and [\ion{C}{3}] emission lines, placing the source galaxy redshift at $z=2.29167 \pm 0.00029$. 
    \label{Ref:lens64}

    \item \textit{DESI~J339.8883-4.4880} In Figure~\ref{fig:MUSEspectra61}, we see the yellow foreground lens and a blue arc in the south. The lens galaxy has a redshift $z=0.55752 \pm 0.00011$. We confidently determine that the background galaxy has redshift $z=1.11577 \pm 0.00001$ based purely on the strong [\ion{O}{2}] emission. It has a high flux and an obvious double peak in the zoom in. 
    \label{Ref:lens61}

    \item \textit{DESI~J341.0212+27.9883} The system features another red background galaxy with a "tadpole" morphology, shown in Figure~\ref{fig:MUSEspectra65}. We found a lens galaxy redshift of $z=0.34281 \pm 0.00007$. As in the case of previous red lensed sources, the redshift is identified based on the [\ion{O}{2}] emission and \ion{Ca}{2} H and K absorption lines. These features are especially strong, and the SNR is sufficiently high for even the $\lambda$4101\,\AA\ H$\delta$ absorption line to be visible. Though the [\ion{O}{2}] zoom in doesn't show an obvious double peak, the other spectral features are sufficient to confidently assign a source redshift of $z=0.96045 \pm 0.00006$.
    \label{Ref:lens65}

    \item \textit{DESI~J341.8012-2.0939} has a singly lensed source around a lens galaxy with redshift $z=0.33414 \pm 0.00015$ (Figure \ref{fig:MUSEspectra63}). We see in the source spectrum the absorption lines \ion{Fe}{2} at $\lambda$2344\,\AA, $\lambda$2374\,\AA, $\lambda$2382\,\AA, $\lambda$2587\,\AA, and $\lambda$2600\,\AA; and \ion{Mg}{2} at $\lambda$2796\,\AA\ and $\lambda$2803\,\AA, which unquestionably place the source redshift at $z=1.76782 \pm 0.0001$.
    \label{Ref:lens63}

    \item \textit{DESI~J342.9290-03.4136} is presented in Figure \ref{fig:MUSEspectra7}. There are 3 galaxies at similar redshifts of $z=0.24592 \pm 0.00018$ (L1), $z=0.25016 \pm 0.00016$ (L2), and $z=0.25136 \pm 0.00010$ (L3). The background galaxy A, faintly visible in the north, is red and shows a prominent $\lambda$3727\,\AA\ \ion{O}{2} emission line. Upon zooming in on the spectrum, we can confirm the two peaks, providing a redshift of $z=1.05956 \pm 0.00008$ for the source.
    \label{ref:lens7}

    \item \textit{DESI~J343.0402-04.2187} In Figure \ref{fig:MUSEspectra8}, the lens galaxies are at different redshifts, identified as L1 at $z=0.43519 \pm 0.00014$, L2 at $z=0.47$, and L3 at $z=0.43182 \pm 0.0002$. The spectrum shown corresponds only to L1. In the south, an arc-shaped structure is observed, composed of blue knots labeled A, B, C, and D. While other objects with similar morphology are visible, they are not associated with the source. A faint emission line is detected in the four components of the arc, exhibiting a double-peaked structure consistent with [\ion{O}{2}] emission. This suggests a redshift of $z = 0.58$ for the background galaxy components, with a quality flag of $Q_z=3$ due to the low SNR.
    \label{ref:lens8}

    \item \textit{DESI~J344.6262-58.6910} is displayed in Figure \ref{fig:MUSEspectra9}. Redshifts were calculated for the three lens galaxies: L1 at $z=0.29435 \pm 0.00017$, L2 at $z=0.37$, and L3 at $z=0.29597 \pm 0.00016$, all with clearly identified spectral lines. Among these, only the spectrum of lens L1 is shown in figure \ref{fig:MUSEspectra9}. Two sources were identified: one in the north (A), with an arc shape and a reddish hue, and another in the east (B), with a point-like appearance. The redshift of this background galaxy was determined to be $z=0.88023 \pm 0.00005$ (Source A) and $z=0.88064 \pm 0.00005$ (Source B), based on the clear detection of strong emission lines from  \ion{O}{2} at $\lambda$3727\,\AA, $H\gamma$ at $\lambda$4341\,\AA, and $H\beta$ at $\lambda$4341\,\AA.  
    \label{ref:lens9}

    \item \textit{DESI~J345.0725+22.2254}, seen in Figure \ref{fig:MUSEspectra66}, consists of a blue source arc lensed around a cluster of lens galaxies ($z=0.43738 \pm 0.00012$). We also see yellow foreground galaxies which perfectly coincide with the blue source arc. Though we were not able to find a secure redshift, the shape of the continuum alone is enough to determine that they aren't at the same redshift at the source. We determined an approximate redshift of $z=0.4$ for them based on neighboring galaxies and what seems like a 4000\AA\ break. The source galaxy has a secure redshift of $z=1.9273 \pm 0.0001$, determined by the absorption lines \ion{Fe}{2} at $\lambda$2344\,\AA, $\lambda$2374\,\AA, $\lambda$2382, $\lambda$2587\,\AA, and $\lambda$2600\,\AA; and \ion{Mg}{2} at $\lambda$2796\,\AA\ and $\lambda$2803\,\AA.
    \label{Ref:lens66}

    \item \textit{DESI~J345.8606+23.4757} This gravitational lens system features two background sources, one doubly lensed (A) and one quadruply lensed (B), shown in Figure \ref{fig:MUSEspectra67}. The images of both sources are observed west of the lens galaxy, which has a redshift of $z=0.27653 \pm 0.00007$. There is another arc southeast of the lens, which we have named "Object X". Though the spectra has no immediately identifiable features, its shape seems to mirror that of other lens galaxy, with a clear rise in the flux continuum. Therefore, we have placed it at a measured redshift of $z=0.28914 \pm 0.00032$, with quality flag $Q_z=3$. This is similar enough to that of the lens galaxy to determine that the arc is not due to gravitational lensing. As for our lensed sources, source A has two arcs, labeled A1 and A2. Both spectra have very strong [\ion{O}{2}] doublets. They have slightly different redshifts, with A1 being higher, giving measured values of $z=1.29593 \pm 0.00003$ and $z=1.29552 \pm 0.00004$. The other background galaxy, with arcs B1, B2, B3, and B4, is much bluer in color than source A. The spectra display features which we have not seen in any other system discussed in this paper. Namely, we see a strong  Ly-$\alpha$ absorption line at $\lambda$1215\,\AA. In addition, the metal absorption lines are [\ion{O}{2}] + \ion{Si}{2}, \ion{C}{2}, \ion{Si}{4}, \ion{Si}{4} + \ion{O}{4}, \ion{Si}{2}, \ion{C}{4}, \ion{Fe}{2}, and \ion{Al}{1} incredibly strong. Though we often identify redshifts from these lines, seeing so many of them in one spectrum with such high SNR is very rare. Some arc spectra, specifically B1 and B3, also display a clear [\ion{C}{3}] emission line. These features suggest that source B could be a Lyman break galaxy \citep[see][]{shapley2003rest}, where the breadth of the Lyman absorption line is likely due to huge hydrogen column in front of radiation source. Alternatively, these features match the spectra of a gamma-ray burst \citep[]{Rau_2010} or a quasar \citep[]{bordoloi2022resolving}. All four B spectra have much higher measured redshifts centering around $z=3.010$ (e.g., Source B1 at $z=3.01007 \pm 0.00024$), though B3 has a quality flag of $Q_z=2$ due to its weaker features. The morphology of the arcs of both sources is fairly unexpected. That is, instead of bending around the lens galaxy as is common, they lie in almost a straight line stretching from north to south.
    \label{Ref:lens67}
    
\end{enumerate}


\subsection{Confirmed Non-Lenses}


In this section, we briefly describe all DESI targets that were initially identified as gravitational lens candidates, but which we have ruled out as lensing systems based on their spectroscopic properties. The systems are listed in Table~\ref{tab:non_lens_systems}.

\begin{table}[!ht]
\centering
\caption{\label{tab:non_lens_systems} Non-Lens Systems}
\begin{tabular}{llllllccccc} 
\hline
Name & Object & R.A. & Dec & Redshift & $Q_z$ & DESI $g$ & DESI $r$ & DESI $i$ & DESI $z$ & Notes \\
& & deg & deg & & & mag & mag & mag & mag & \\
\hline

DESI-180.2707-02.3681  & Galaxy 1  & 180.2707 & -2.3681 & 0.385 & 1 & 21.03 & 19.23 & 18.64 & 18.35 \\

 & Galaxy 2  & 180.2707 & -2.3672 & 0.39 & 1 & 21.09 & 20.44 & 20.20 & 19.70  \\ 

 & Galaxy 3  & 180.2716 & -2.3666 & 0.1 & 1 & 20.23 & 19.81 & 19.60 & 19.54  \\ 

  & Arc & 180.2703 & -2.3666 & 0.166  & 2  &   21.41    &   20.83    &   20.59    & 20.47 \\

DESI-189.9885+12.6693 & Galaxy 1  & 189.9900 & 12.6688 & 0.221 & 1 & 20.25 & 18.84 & 17.49 & 18.07 \\

 & Galaxy 2  & 189.9885 & 12.6693 & 0.221 & 1 & 19.23 & 17.91 & 17.49 & 17.16 \\

 & Arc  & 189.9896 & 12.6674 & 0.219 & 3 & 21.90 & 21.01 & 20.54 & 20.62 \\

 DESI-252.2720+02.3993 & Galaxy & 252.2720 & 02.3993  & 0.495 & 1 & 20.27 & 19.18 & 18.65 & 18.22  \\ 

 & Arc  & 252.2725 & 2.3995 & 0.495 & 1 & 22.20 & 22.29 & 23.20 & 23.33   \\

DESI-311.4249-10.6762 & Galaxy & 311.4249 & -10.6762 & 0.108 & 1 & 17.30 & 16.24 & 15.79 & 15.47   \\ 

 &Arc & 252.2725 & 2.3995 & 0.08 & 1 & 20.94 & 20.53 & 20.40 & 20.45   \\ 
 
DESI-333.3655-13.2491 & Galaxy & 333.3655 & -13.2491 & 0.111 & 1 & 17.57 & 16.54 & 16.07 & 15.77 \\

 & Arc  & 333.3661 & -13.2474 & 0.111 & 1 & 19.37 & 18.95 & 18.72 & 18.62 \\

DESI-340.2310-00.0123 & Galaxy & 340.2304 &	-0.0128 & 0.441 & 1 & 20.07 & 18.56 & 24.79 & 17.57 \\

 & Arc  & 340.2303 & -0.0134 & 0.442 & 1 & 22.56 & 22.27 & 23.02 & 23.64 \\
 
\hline
\end{tabular} \\
{Note --- Redshift quality ($Q_z$); 1 = Robust, 2 = Probable, 3 = Possible.}
\end{table}

\begin{enumerate}

    \item \textit{DESI~J180.2707-2.3681} In the system shown in Figure \ref{fig:MUSEspectra45}, three galaxies were observed. Two of them (L1 and L2) similar redshifts of $z=0.385$ and $z=0.39$, while L3 has a redshift of $z=0.1$. Only the spectrum of L1 is displayed here. Despite the noisy spectrum, it was possible to calculate the spectral lines of the suspected source. The emission lines $H_\gamma$, \ion{O}{2}, and $H_\gamma$ were clearly identified, as seen in spectrum \ref{fig:MUSEspectra45}. The redshift calculated from the arc is $z=0.166$, indicating that the galaxy is not in the background. \label{ref:lens45}
   
    \item \textit{DESI~J189.9885+12.6693} Here we see two galaxies, both at redshift $z=0.221$ with a very large arc between the two. Though there are no identifiable features in the arc spectra, the shape of the flux continuum seems to demonstrate the 4000\,\AA\ break, which would place the arc at a similar redshift to that of the galaxies. We have displayed the arc spectra in the bottom panel of Figure~\ref{fig:MUSEspectra118} at an approximate redshift of $z=0.219$. As the redshifts of all objects are similar to each other, this is not an example of gravitational lensing. Alternatively, it may be a huge tidal tail between the two merging galaxies L1 and L2.
    \label{ref:lens118}
    
    \item \textit{DESI~J252.2720+02.3993} In Figure \ref{fig:MUSEspectra56}, at first glance, the candidate source appears as a large blue arc, located to the east of the galaxy labeled L1, which has a redshift of $z=0.495$. A star overlaps with the arc. Due to its arc-shaped structure, it was initially suspected to be a background galaxy. However, its spectrum shows strong emission lines with high signal-to-noise ratio (Figure \ref{fig:MUSEspectra56}), clearly revealing \ion{O}{2} at $\lambda$3727\,\AA, $H_\gamma$ at $\lambda$4341\,\AA, $H_\beta$ at $\lambda$4862\,\AA, \ion{O}{3} at $\lambda$4960\,\AA,  and \ion{O}{3} at $\lambda$5008\,\AA. Based on this,  we calculate the same redshift $z=0.495$ as  L1, indicating that it is not a gravitationally lensed source. 
    \label{ref:lens56}
    
     \item \textit{DESI~J311.4249-10.6762} In Figure \ref{fig:MUSEspectra11}, galaxy L1 has a redshift of $z=0.108$, as shown in spectrum number \ref{fig:MUSEspectra11}. The supposed source, which appears as an arc south-west of L1, has a calculated redshift of $z=0.08$, with clear emission lines such as \ion{O}{3} and $H_\alpha$.This places it closer to us than L1, ruling it out as a candidate.
     \label{ref:lens11}
    
    \item \textit{DESI~J333.3655-13.2491} On first glance, this object seems to have a very prominent lensed arc (labeled in Figure~\ref{fig:MUSEspectra60}). Upon inspecting the spectrum, we see that the "lens" galaxy and "lensed arc" both show same emission lines: [\ion{O}{3}] at $\lambda$4958 and $\lambda$5006, H-$\alpha$ at $\lambda$6564\,\AA\ and the \ion{S}{2} doublet at $\lambda$6718\,\AA\ and $\lambda$6732\,\AA. They also share some Fraunhofer absorption lines (the G band and Na D). The very high SNR of these spectra allow us to confidently assign redshift $z=0.111$ for both the "lens" and the "source", showing us that this is not a gravitational lens system. Based on the bright cloud surrounding the objects, it seems that the arc is an arm in a large spiral galaxy with the "lens" in the center. However, it is unusual for a spiral galaxy to have a single significantly brighter arm such as this. 
    \label{ref:lens60}
    
    \item \textit{DESI~J340.2310-00.0123} This also looks like a gravitational lens system, but spectral data demonstrates that it is not. Both the "lens" and the "source" share the [\ion{O}{2}] and [\ion{O}{3}] emission lines (Figure \ref{fig:MUSEspectra62}). The "lens" also displays Fraunhofer absorption lines, like the lens galaxies previously mentioned in this paper. Conversely, the "source" spectra has more emission lines, specifically \ion{Ne}{3}, H$\gamma$, and H-$\beta$. Based on these spectral features, we assign the "lens" galaxy a redshift of $z=0.441$ and the arc a redshift of $z=0.442$, which are too similar to belong to a gravitationally lensed object. Like in the previous system, we theorize that the arc is the arm of a spiral galaxy with the "lens" at the center.
    \label{ref:lens62}

\end{enumerate}


\subsection{Systems which require further observation}

In this section, we present 15 systems for which we were able to determine only the redshifts of the lens galaxies (Table~\ref{tab:lens_only_systems}). The spectra of the background sources had insufficient signal-to-noise to allow the identification of spectral features from line catalogs, or the sources were not visible in the MUSE cubes owing to unfavorable weather conditions. As the lens galaxies are typically much brighter than the background sources, their redshifts could be measured in from the Ca H \& K absorption features. Given the poorer overall data quality for this set, we do not present the corresponding MUSE images for these systems. Instead we extracted the cut-outs for these systems from the DESI Legacy Survey, shown in Fig.~\ref{fig:lens_only_systems_mosaic}. Furthermore, the spectra of the systems are shown in Fig.~\ref{fig:lens_only_systems_spectra}.


\begin{table*}[!ht]
\centering
\caption{\label{tab:lens_only_systems} Lens Only Systems}
\begin{tabular}{llllllccccc} 
\hline
Name & Object & RA & Dec & Redshift & $Q_z$ & DESI $g$ & DESI $r$ & DESI $i$ & DESI $z$ & Notes \\
& & deg & deg & & & mag & mag & mag & mag  \\
\hline

DESI-023.6758+04.5638 & Lens & 23.6766 & 4.5639 & 0.55128 $\pm$ 0.00018 & 1 & 20.70 & 18.78 & 17.88 & 17.47 \\

DESI-024.1634+00.1386 & Lens & 24.1634 & 0.1386 & 0.34423 $\pm$ 0.0002 & 1 & 19.33 & 17.68 & 17.12 & 16.80 \\


DESI-043.5347-28.7960 & Lens & 43.5347 & -28.7960 & 0.62217 $\pm$ 0.00022 & 1 & 22.03 & 20.15 & 19.11 & 18.68 \\

DESI-044.9808+01.1384 & Lens & 44.9808 & 1.1384 & 0.88994 $\pm$ 0.00031 & 1 & 21.58 & 21.19 & 20.91 & 16.88 \\



DESI-058.7923-18.5265 & Lens & 58.7923 & -18.5265 & 0.29913 $\pm$ 0.00014 & 1 & 21.77 & 20.07 & 19.56 & 19.29 \\

DESI-064.8539-25.2245 & Lens & 64.8539 & -25.2245 & 0.371 $\pm$ 0.00014 & 1 & 19.23 & 17.44 & 16.86 & 16.55 \\


DESI-078.3561-30.8433 & Lens & 78.3561 & -30.8433 & 0.40699 $\pm$ 0.00015 & 1 & 19.79 & 17.98 & 17.38 & 17.04 \\

DESI-081.7544-18.9677 & Lens & 81.7544 & -18.9677 & 0.4513 $\pm$ 0.00026 & 1 & 20.35 & 18.47 & 17.80 & 17.44 \\

DESI-109.9018+27.9032 & Lens & 109.9018 & 27.9032 & 0.55391 $\pm$ 0.0002 & 1 & 21.02 & 19.18 & - & 17.95 \\



DESI-178.0772+08.8167 & Lens & 178.0772 & 8.8167 & 0.38851 $\pm$ 0.00015 & 1 & 20.48 & 18.89 & 18.33 & 18.02 \\


DESI-234.8707+16.8379 & Lens & 234.8707 & 16.8379 & 0.40948 $\pm$ 0.0002 & 1 & 19.77 & 17.86 & - & 16.89 \\

DESI-306.4726-51.2868 & Lens & 306.2500 & -51.2833 & 0.2249 $\pm$ 0.00015 & 1 & 18.74 & 17.34 & 16.86 & 16.54 \\

DESI-324.2094-62.6820 & Lens & 324.2094 & -62.6820 & 0.2307 $\pm$ 0.00017 & 1 & 17.78 & 16.37 & 15.90 & 15.60  \\ 

DESI-324.5073-60.1290 & Lens 1 & 324.5073 & -60.1290 & 0.32043 $\pm$ 0.00019 & 1 & 18.55 & 16.89 & 16.35 & 16.03 \\ 

&Lens 2 & 324.5035 & -60.1317 & 0.3152 $\pm$ 0.00058  & 1 & 18.74 & 17.11 & 16.57 & 16.26 \\

DESI-336.1611-01.8757 & Lens  & 336.1611 & -1.8757 & 0.20793 $\pm$ 0.00016 & 1 & 17.61 & 16.22& 15.72 & 15.37 \\ 



\hline
\end{tabular}

{Note --- Redshift quality ($Q_z$); 1 = Robust, 2 = Probable, 3 = Guess.}
\end{table*}

\section{Summary and conclusions}
\label{sect:summary}

We have presented integral field spectroscopic observations of 76 candidate strong gravitational lens systems discovered in the DESI Legacy Imaging Surveys, using MUSE on the ESO VLT. 
These observations were conducted over multiple ESO filler programs from 2022 to 2024. MUSE's unique combination of angular and spectral coverage makes it particularly well-suited for identifying lens and source redshifts, even in challenging configurations such as multiple source planes, and group- or cluster-scale lensing environments (see, for example, our work on the Einstein cross DESI~J253.2534+26.8843 by \citealt{2023ApJ...953L...5C}, and on the Carousel Lens, by \citealt{sheu2024carousel}). 

MUSE offers a unique combination of advantages: its large field of view, continuous spectral coverage, fine spatial sampling, and unparalleled multiplexing power. While other IFUs may be preferable for very high spectral resolution or restricted wavelength ranges, MUSE remains unmatched for general-purpose, deep-field, and wide-area integral-field spectroscopy.

We obtained 92 pointings of 76 different targets with MUSE in total, and extracted the spectra of 223 objects in these MUSE fields. 
Redshifts for both the lens and the source were successfully measured in 55 cases, allowing us to consider these systems fully confirmed. In these 55 targets, we extracted and analyzed 192 spectra, comprising of the lenses, sources, and significant foreground and background objects.

Six targets were determined not to be gravitational lens systems. Although they initially appeared to exhibit lens-like features, such as arcs or multiple sources, further analysis revealed that these were actually spiral arms or unrelated foreground/background objects.
For the remaining 15 targets, only the redshift of the lens was measured. In these cases, the data quality was insufficient to analyze the source, or the spectra lacked prominent features. These systems require additional observations for confirmation.

The distribution of measured lens and source redshifts are summarized in Figures~\ref{fig:lens-source-redshift-histogram} and \ref{fig:lens-source-redshift-scatter}. 

Looking ahead, the value of this spectroscopically confirmed lens sample will continue to grow. Combined with future high-resolution imaging (e.g., from \textit{JWST} or \textit{Euclid}) and lens modeling, these systems can be used to probe dark matter halo profiles, measure the stellar mass-to-light ratio, and constrain the subhalo mass function. In parallel, time-delay cosmography from multiply imaged supernovae and quasars will provide independent measurements of the Hubble constant $H_0$ and test for consistency with other cosmological probes.

As wide-field surveys such as LSST/Rubin, \textit{Euclid}, and \textit{Roman Space Telescope} come online, the demand for flexible, high-throughput follow-up strategies will only increase. Our work demonstrates the power of combining machine learning-based discovery with targeted follow-up using instruments like MUSE. 
The strategies developed here lay the groundwork for future strong gravitational lens discoveries and parameterizations, and deliver high-value samples for dark matter and cosmology.


\section*{Acknowledgments}
\small
The work of A.C. is supported by NOIRLab, which is managed by the Association of Universities for Research in Astronomy (AURA) under a cooperative agreement with the National Science Foundation. 
X.H. acknowledges the University of San Francisco Faculty Development Fund.
This work was supported in part by the Director, Office of Science, Office of High Energy Physics of the US Department of Energy under contract No. DE-AC025CH11231.
This research used resources of the National Energy Research Scientific Computing Center (NERSC), a U.S. Department of Energy Office of Science User Facility operated under the same contract as above and the Computational HEP program in The Department of Energy's Science Office of High Energy Physics provided resources through the ``Cosmology Data Repository'' project (grant No. KA2401022).
This work is based on observations collected at the European Organisation for Astronomical Research in the Southern Hemisphere under ESO programs 109.238W.004, 111.24UJ.008, 111.24P8.001, 112.2614.001, and 113.267Q.001. The execution in the service mode of these observations by the VLT operations staff is gratefully acknowledged.


{\sl Data Availability Statement:} 
All data supporting this work is publicly available. The raw MUSE observations can be accessed through the ESO Science Archive (https://archive.eso.org/) using the corresponding Program IDs given in the acknowledgments. We aim to upload the results to the Strong Gravitational Lens database (SLED, https://sled.amnh.org/), and the extracted spectra to CSD/VizieR.

\facilities{The observations were obtained with the Very Large Telescope at the European Southern Observatory's La Silla Paranal Observatory in Chile.}

%







\bibliography{impolbib}{} 

@ARTICLE{1991ApJ...373..354K,
       author = {{Kochanek}, Christopher S.},
        title = "{The Implications of Lenses for Galaxy Structure}",
      journal = {\apj},
         year = 1991,
        month = jun,
       volume = {373},
        pages = {354},
          doi = {10.1086/170057},
       adsurl = {https://ui.adsabs.harvard.edu/abs/1991ApJ...373..354K}
}

@ARTICLE{2023ApJ...953L...5C,
       author = {{Cikota}, Aleksandar and {Bertolla}, Ivonne Toro and {Huang}, Xiaosheng and {Baltasar}, Saul and {Ratier-Werbin}, Nicolas and {Sheu}, William and {Storfer}, Christopher and {Suzuki}, Nao and {Schlegel}, David J. and {Cartier}, Regis and {Torres}, Simon and {Cikota}, Stefan and {Jullo}, Eric},
        title = "{DESI-253.2534+26.8843: A New Einstein Cross Spectroscopically Confirmed with Very Large Telescope/MUSE and Modeled with GIGA-Lens}",
      journal = {\apjl},
         year = 2023,
        month = aug,
       volume = {953},
       number = {1},
          eid = {L5},
        pages = {L5},
          doi = {10.3847/2041-8213/ace9da},
archivePrefix = {arXiv},
       eprint = {2307.12470},
 primaryClass = {astro-ph.GA},
       adsurl = {https://ui.adsabs.harvard.edu/abs/2023ApJ...953L...5C}
}

@ARTICLE{2017A&A...605A.118F,
       author = {{Finley}, Hayley and {Bouch{\'e}}, Nicolas and {Contini}, Thierry and {Epinat}, Beno{\^\i}t and {Bacon}, Roland and {Brinchmann}, Jarle and {Cantalupo}, Sebastiano and {Erroz-Ferrer}, Santiago and {Marino}, Raffaella Anna and {Maseda}, Michael and {Richard}, Johan and {Schroetter}, Ilane and {Verhamme}, Anne and {Weilbacher}, Peter M. and {Wendt}, Martin and {Wisotzki}, Lutz},
        title = "{Galactic winds with MUSE: A direct detection of Fe II* emission from a z = 1.29 galaxy}",
      journal = {\aap},
         year = 2017,
        month = sep,
       volume = {605},
          eid = {A118},
        pages = {A118},
          doi = {10.1051/0004-6361/201730428},
archivePrefix = {arXiv},
       eprint = {1701.07843},
 primaryClass = {astro-ph.GA},
       adsurl = {https://ui.adsabs.harvard.edu/abs/2017A&A...605A.118F}
}

@ARTICLE{2016MNRAS.458.3210S,
       author = {{Soto}, Kurt T. and {Lilly}, Simon J. and {Bacon}, Roland and {Richard}, Johan and {Conseil}, Simon},
        title = "{ZAP - enhanced PCA sky subtraction for integral field spectroscopy}",
      journal = {\mnras},
         year = 2016,
        month = may,
       volume = {458},
       number = {3},
        pages = {3210-3220},
          doi = {10.1093/mnras/stw474},
archivePrefix = {arXiv},
       eprint = {1602.08037},
 primaryClass = {astro-ph.IM},
       adsurl = {https://ui.adsabs.harvard.edu/abs/2016MNRAS.458.3210S}
}

@ARTICLE{2020A&A...641A..28W,
author = {{Weilbacher}, Peter M. and {Palsa}, Ralf and {Streicher}, Ole and {Bacon}, Roland and {Urrutia}, Tanya and {Wisotzki}, Lutz and {Conseil}, Simon and {Husemann}, Bernd and {Jarno}, Aur{\'e}lien and {Kelz}, Andreas and {P{\'e}contal-Rousset}, Arlette and {Richard}, Johan and {Roth}, Martin M. and {Selman}, Fernando and {Vernet}, Jo{\"e}l},
title = "{The data processing pipeline for the MUSE instrument}",
journal = {\aap},
year = 2020,
month = sep,
volume = {641},
eid = {A28},
pages = {A28},
doi = {10.1051/0004-6361/202037855},
archivePrefix = {arXiv},
eprint = {2006.08638},
primaryClass = {astro-ph.IM},
adsurl = {https://ui.adsabs.harvard.edu/abs/2020A&A...641A..28W}
}

@inproceedings{bacon2010muse,
  title={The MUSE second-generation VLT instrument},
  author={Bacon, R and Accardo, M and Adjali, L and Anwand, H and Bauer, S and Biswas, I and Blaizot, J and Boudon, D and Brau-Nogue, S and Brinchmann, J and others},
  booktitle={Ground-based and airborne instrumentation for astronomy III},
  volume={7735},
  pages={131--139},
  year={2010},
  organization={SPIE}
}

@ARTICLE{treu2002probing,
  title={Probing dark matter distribution with gravitational lensing and stellar dynamics},
  author={Treu, Tommaso and Koopmans, L{\'e}on VE},
  journal={arXiv preprint astro-ph/0205335},
  year={2002}
}

@article{bolton2006constraint,
  title={Constraint on the post-Newtonian parameter $\gamma$ on galactic size scales},
  author={Bolton, Adam S and Rappaport, Saul and Burles, Scott},
  journal={Physical Review D—Particles, Fields, Gravitation, and Cosmology},
  volume={74},
  number={6},
  pages={061501},
  year={2006},
  publisher={APS}
}

@article{koopmans2006sloan,
  title={The Sloan Lens ACS Survey. III. The structure and formation of early-type galaxies and their evolution since z≈ 1},
  author={Koopmans, Leon VE and Treu, Tommaso and Bolton, Adam S and Burles, Scott and Moustakas, Leonidas A},
  journal={The Astrophysical Journal},
  volume={649},
  number={2},
  pages={599},
  year={2006},
  publisher={IOP Publishing}
}

@article{bolton2008sloan,
  title={The Sloan Lens ACS Survey. VII. Elliptical galaxy scaling laws from direct observational mass measurements},
  author={Bolton, Adam S and Treu, Tommaso and Koopmans, L{\'e}on VE and Gavazzi, Rapha{\"e}l and Moustakas, Leonidas A and Burles, Scott and Schlegel, David J and Wayth, Randall},
  journal={The Astrophysical Journal},
  volume={684},
  number={1},
  pages={248},
  year={2008},
  publisher={IOP Publishing}
}

@article{bradavc2008dark,
  title={Dark matter and baryons in the x-ray luminous merging galaxy cluster rx j1347. 5--1145},
  author={Brada{\v{c}}, Maru{\v{s}}a and Schrabback, Tim and Erben, Thomas and McCourt, Michael and Million, Evan and Mantz, Adam and Allen, Steve and Blandford, Roger and Halkola, Aleksi and Hildebrandt, Hendrik and others},
  journal={The Astrophysical Journal},
  volume={681},
  number={1},
  pages={187},
  year={2008},
  publisher={IOP Publishing}
}

@article{huang2009hubble,
  title={HUBBLE SPACE TELESCOPE DISCOVERY OF A z= 3.9 MULTIPLY IMAGED GALAXY BEHIND THE COMPLEX CLUSTER LENS WARPS J1415. 1+ 36 AT z= 1.026},
  author={Huang, Xiaosheng and Morokuma, T and Fakhouri, HK and Aldering, G and Amanullah, Rahman and Barbary, K and Brodwin, M and Connolly, NV and Dawson, KS and Doi, M and others},
  journal={The Astrophysical Journal},
  volume={707},
  number={1},
  pages={L12},
  year={2009},
  publisher={IOP Publishing}
}

@article{jullo2010cosmological,
  title={Cosmological constraints from strong gravitational lensing in clusters of galaxies},
  author={Jullo, Eric and Natarajan, Priyamvada and Kneib, Jean-Paul and D’Aloisio, Anson and Limousin, Marceau and Richard, Johan and Schimd, Carlo},
  journal={Science},
  volume={329},
  number={5994},
  pages={924--927},
  year={2010},
  publisher={American Association for the Advancement of Science}
}

@article{grillo2015clash,
  title={CLASH-VLT: INSIGHTS ON THE MASS SUBSTRUCTURES IN THE FRONTIER FIELDS CLUSTER MACS J0416. 1- 2403 THROUGH ACCURATE STRONG LENS MODELING},
  author={Grillo, C and Suyu, SH and Rosati, Piero and Mercurio, AMATA and Balestra, I and Munari, Emiliano and Nonino, Mario and Caminha, GB and Lombardi, M and De Lucia, GABRIELLA and others},
  journal={The Astrophysical Journal},
  volume={800},
  number={1},
  pages={38},
  year={2015},
  publisher={IOP Publishing}
}

@article{tessore2016lensed,
  title={LENSED: a code for the forward reconstruction of lenses and sources from strong lensing observations},
  author={Tessore, Nicolas and Bellagamba, Fabio and Metcalf, R Benton},
  journal={Monthly Notices of the Royal Astronomical Society},
  volume={463},
  number={3},
  pages={3115--3128},
  year={2016},
  publisher={The Royal Astronomical Society}
}

@article{shu2017sloan,
  title={The Sloan Lens ACS Survey. XIII. Discovery of 40 New Galaxy-scale Strong Lenses∗},
  author={Shu, Yiping and Brownstein, Joel R and Bolton, Adam S and Koopmans, L{\'e}on VE and Treu, Tommaso and Montero-Dorta, Antonio D and Auger, Matthew W and Czoske, Oliver and Gavazzi, Rapha{\"e}l and Marshall, Philip J and others},
  journal={The Astrophysical Journal},
  volume={851},
  number={1},
  pages={48},
  year={2017},
  publisher={IOP Publishing}
}

@article{vegetti2010detection,
  title={Detection of a dark substructure through gravitational imaging},
  author={Vegetti, S and Koopmans, LVE and Bolton, A and Treu, T and Gavazzi, R},
  journal={Monthly Notices of the Royal Astronomical Society},
  volume={408},
  number={4},
  pages={1969--1981},
  year={2010},
  publisher={Blackwell Publishing Ltd Oxford, UK}
}

@article{hezaveh2016detection,
  title={Detection of lensing substructure using ALMA observations of the dusty galaxy SDP. 81},
  author={Hezaveh, Yashar D and Dalal, Neal and Marrone, Daniel P and Mao, Yao-Yuan and Morningstar, Warren and Wen, Di and Blandford, Roger D and Carlstrom, John E and Fassnacht, Christopher D and Holder, Gilbert P and others},
  journal={The Astrophysical Journal},
  volume={823},
  number={1},
  pages={37},
  year={2016},
  publisher={IOP Publishing}
}

@article{sengul2022substructure,
  title={Substructure detection reanalysed: dark perturber shown to be a line-of-sight halo},
  author={Seng{\"u}l, Atin{\c{c}} {\c{C}}agan and Dvorkin, Cora and Ostdiek, Bryan and Tsang, Arthur},
  journal={Monthly Notices of the Royal Astronomical Society},
  volume={515},
  number={3},
  pages={4391--4401},
  year={2022},
  publisher={Oxford University Press}
}

@article{pierel2019turning,
  title={Turning gravitationally lensed supernovae into cosmological probes},
  author={Pierel, JDR and Rodney, S},
  journal={The Astrophysical Journal},
  volume={876},
  number={2},
  pages={107},
  year={2019},
  publisher={IOP Publishing}
}

@article{kelly2023constraints,
  title={Constraints on the Hubble constant from supernova Refsdal’s reappearance},
  author={Kelly, Patrick L and Rodney, Steven and Treu, Tommaso and Oguri, Masamune and Chen, Wenlei and Zitrin, Adi and Birrer, Simon and Bonvin, Vivien and Dessart, Luc and Diego, Jose M and others},
  journal={Science},
  volume={380},
  number={6649},
  pages={eabh1322},
  year={2023},
  publisher={American Association for the Advancement of Science}
}

@article{Suyu_2024,
   title={Strong Gravitational Lensing and Microlensing of Supernovae},
   volume={220},
   ISSN={1572-9672},
   url={http://dx.doi.org/10.1007/s11214-024-01044-7},
   DOI={10.1007/s11214-024-01044-7},
   number={1},
   journal={Space Science Reviews},
   publisher={Springer Science and Business Media LLC},
   author={Suyu, Sherry H. and Goobar, Ariel and Collett, Thomas and More, Anupreeta and Vernardos, Giorgos},
   year={2024},
   month=feb }

@article{Wong_2019,
   title={H0LiCOW – XIII. A 2.4 per cent measurement of H0 from lensed quasars: 5.3σ tension between early- and late-Universe probes},
   volume={498},
   ISSN={1365-2966},
   url={http://dx.doi.org/10.1093/mnras/stz3094},
   DOI={10.1093/mnras/stz3094},
   number={1},
   journal={Monthly Notices of the Royal Astronomical Society},
   publisher={Oxford University Press (OUP)},
   author={Wong, Kenneth C and Suyu, Sherry H and Chen, Geoff C-F and Rusu, Cristian E and Millon, Martin and Sluse, Dominique and Bonvin, Vivien and Fassnacht, Christopher D and Taubenberger, Stefan and Auger, Matthew W and Birrer, Simon and Chan, James H H and Courbin, Frederic and Hilbert, Stefan and Tihhonova, Olga and Treu, Tommaso and Agnello, Adriano and Ding, Xuheng and Jee, Inh and Komatsu, Eiichiro and Shajib, Anowar J and Sonnenfeld, Alessandro and Blandford, Roger D and Koopmans, Léon V E and Marshall, Philip J and Meylan, Georges},
   year={2019},
   month=sep, pages={1420–1439} }

@article{freedman2020astrophysical,
  title={Astrophysical distance scale. II. Application of the JAGB method: a nearby galaxy sample},
  author={Freedman, Wendy L and Madore, Barry F},
  journal={The Astrophysical Journal},
  volume={899},
  number={1},
  pages={67},
  year={2020},
  publisher={IOP Publishing}
}

@article{Riess_2022,
   title={A Comprehensive Measurement of the Local Value of the Hubble Constant with 1 km s−1 Mpc−1 Uncertainty from the Hubble Space Telescope and the SH0ES Team},
   volume={934},
   ISSN={2041-8213},
   url={http://dx.doi.org/10.3847/2041-8213/ac5c5b},
   DOI={10.3847/2041-8213/ac5c5b},
   number={1},
   journal={The Astrophysical Journal Letters},
   publisher={American Astronomical Society},
   author={Riess, Adam G. and Yuan, Wenlong and Macri, Lucas M. and Scolnic, Dan and Brout, Dillon and Casertano, Stefano and Jones, David O. and Murakami, Yukei and Anand, Gagandeep S. and Breuval, Louise and Brink, Thomas G. and Filippenko, Alexei V. and Hoffmann, Samantha and Jha, Saurabh W. and D’arcy Kenworthy, W. and Mackenty, John and Stahl, Benjamin E. and Zheng, WeiKang},
   year={2022},
   month=jul, pages={L7} }

@article{huang2020finding,
  title={Finding strong gravitational lenses in the DESI DECam legacy survey},
  author={Huang, Xiaosheng and Storfer, Christopher and Ravi, V and Pilon, A and Domingo, M and Schlegel, DJ and Bailey, S and Dey, A and Gupta, RR and Herrera, D and others},
  journal={The Astrophysical Journal},
  volume={894},
  number={1},
  pages={78},
  year={2020},
  publisher={IOP Publishing}
}

@article{huang2021discovering,
  title={Discovering new strong gravitational lenses in the DESI legacy imaging surveys},
  author={Huang, Xiaosheng and Storfer, Christopher and Gu, A and Ravi, V and Pilon, A and Sheu, W and Venguswamy, R and Banka, S and Dey, A and Landriau, M and others},
  journal={The Astrophysical Journal},
  volume={909},
  number={1},
  pages={27},
  year={2021},
  publisher={IOP Publishing}
}

@article{storfer2024new,
  title={New Strong Gravitational Lenses from the DESI Legacy Imaging Surveys Data Release 9},
  author={Storfer, C and Huang, X and Gu, A and Sheu, W and Banka, S and Dey, A and Reyes, J Inchausti and Jain, A and Kwon, KJ and Lang, D and others},
  journal={The Astrophysical Journal Supplement Series},
  volume={274},
  number={1},
  pages={16},
  year={2024},
  publisher={IOP Publishing}
}

@article{dawes2023finding,
  title={Finding Multiply Lensed and Binary Quasars in the DESI Legacy Imaging Surveys},
  author={Dawes, C and Storfer, C and Huang, X and Aldering, G and Cikota, Aleksandar and Dey, Arjun and Schlegel, DJ},
  journal={The Astrophysical Journal Supplement Series},
  volume={269},
  number={2},
  pages={61},
  year={2023},
  publisher={IOP Publishing}
}

@article{Sheu_2023,
   title={Retrospective Search for Strongly Lensed Supernovae in the DESI Legacy Imaging Surveys},
   volume={952},
   ISSN={1538-4357},
   url={http://dx.doi.org/10.3847/1538-4357/acd1e4},
   DOI={10.3847/1538-4357/acd1e4},
   number={1},
   journal={The Astrophysical Journal},
   publisher={American Astronomical Society},
   author={Sheu, William and Huang, Xiaosheng and Cikota, Aleksandar and Suzuki, Nao and Schlegel, David J. and Storfer, Christopher},
   year={2023},
   month=jul, pages={10} }

@article{sheu2024targeted,
  title={A Targeted Search for Variable Gravitationally Lensed Quasars},
  author={Sheu, William and Huang, Xiaosheng and Cikota, Aleksandar and Suzuki, Nao and Palmese, Antonella and Schlegel, David J and Storfer, Christopher},
  journal={The Astrophysical Journal},
  volume={973},
  number={1},
  pages={24},
  year={2024},
  publisher={IOP Publishing}
}

@article{aghamousa2016desia,
  title={The DESI experiment part I: science, targeting, and survey design},
  author={Aghamousa, Amir and Aguilar, Jessica and Ahlen, Steve and Alam, Shadab and Allen, Lori E and Prieto, Carlos Allende and Annis, James and Bailey, Stephen and Balland, Christophe and Ballester, Otger and others},
  journal={arXiv preprint arXiv:1611.00037},
  year={2016}
}

@article{aghamousa2016desib,
  title={The DESI experiment part II: instrument design},
  author={Aghamousa, Amir and Aguilar, Jessica and Ahlen, Steve and Alam, Shadab and Allen, Lori E and Prieto, Carlos Allende and Annis, James and Bailey, Stephen and Balland, Christophe and Ballester, Otger and others},
  journal={arXiv preprint arXiv:1611.00037},
  year={2016}
}

@article{Rau_2010,
   title={A VERY METAL-POOR DAMPED LYMAN-α SYSTEM REVEALED THROUGH THE MOST ENERGETIC GRB 090926A},
   volume={720},
   ISSN={1538-4357},
   url={http://dx.doi.org/10.1088/0004-637X/720/1/862},
   DOI={10.1088/0004-637x/720/1/862},
   number={1},
   journal={The Astrophysical Journal},
   publisher={American Astronomical Society},
   author={Rau, A. and Savaglio, S. and Krühler, T. and Afonso, P. and Greiner, J. and Klose, S. and Schady, P. and McBreen, S. and Filgas, R. and Olivares E., F. and Rossi, A. and Updike, A.},
   year={2010},
   month=aug, pages={862–871} }

@article{bordoloi2022resolving,
  title={Resolving the H i in damped Lyman $\alpha$ systems that power star formation},
  author={Bordoloi, Rongmon and O’Meara, John M and Sharon, Keren and Rigby, Jane R and Cooke, Jeff and Shaban, Ahmed and Matuszewski, Mateusz and Rizzi, Luca and Doppmann, Greg and Martin, D Christopher and others},
  journal={Nature},
  volume={606},
  number={7912},
  pages={59--63},
  year={2022},
  publisher={Nature Publishing Group UK London}
}

@article{shapley2003rest,
  title={Rest-frame ultraviolet spectra of z~ 3 Lyman break galaxies},
  author={Shapley, Alice E and Steidel, Charles C and Pettini, Max and Adelberger, Kurt L},
  journal={The Astrophysical Journal},
  volume={588},
  number={1},
  pages={65},
  year={2003},
  publisher={IOP Publishing}
}

@article{cikota2023desi,
  title={DESI-253.2534+ 26.8843: a new Einstein Cross spectroscopically confirmed with very large telescope/MUSE and modeled with GIGA-Lens},
  author={Cikota, Aleksandar and Bertolla, Ivonne Toro and Huang, Xiaosheng and Baltasar, Saul and Ratier-Werbin, Nicolas and Sheu, William and Storfer, Christopher and Suzuki, Nao and Schlegel, David J and Cartier, Regis and others},
  journal={The Astrophysical journal letters},
  volume={953},
  number={1},
  pages={L5},
  year={2023},
  publisher={IOP Publishing}
}

@ARTICLE{2025arXiv250912319S,
       author = {{Suyu}, S.~H. and {Acebron}, A. and {Grillo}, C. and {Bergamini}, P. and {Caminha}, G.~B. and {Cha}, S. and {Diego}, J.~M. and {Ertl}, S. and {Foo}, N. and {Frye}, B.~L. and {Fudamoto}, Y. and {Granata}, G. and {Halkola}, A. and {Jee}, M.~J. and {Kamieneski}, P.~S. and {Koekemoer}, A.~M. and {Meena}, A.~K. and {Newman}, A.~B. and {Nishida}, S. and {Oguri}, M. and {Rosati}, P. and {Schuldt}, S. and {Zitrin}, A. and {Ca{\~n}ameras}, R. and {Hayes}, E.~E. and {Larison}, C. and {Mamuzic}, E. and {Millon}, M. and {Pierel}, J.~D.~R. and {Tortorelli}, L. and {Wang}, H.},
        title = "{Cosmology with supernova Encore in the strong lensing cluster MACS J0138-2155: Lens model comparison and H0 measurement}",
      journal = {arXiv e-prints},
         year = 2025,
        month = sep,
          eid = {arXiv:2509.12319},
        pages = {arXiv:2509.12319},
          doi = {10.48550/arXiv.2509.12319},
archivePrefix = {arXiv},
       eprint = {2509.12319},
 primaryClass = {astro-ph.CO},
       adsurl = {https://ui.adsabs.harvard.edu/abs/2025arXiv250912319S}
}

@ARTICLE{2026ApJ...998..219P,
       author = {{Pierel}, J.~D.~R. and {Hayes}, E.~E. and {Millon}, M. and {Larison}, C. and {Mamuzic}, E. and {Acebron}, A. and {Agrawal}, A. and {Bergamini}, P. and {Cha}, S. and {Dhawan}, S. and {Diego}, J.~M. and {Frye}, B.~L. and {Gilman}, D. and {Granata}, G. and {Grillo}, C. and {Jee}, M.~J. and {Kamieneski}, P.~S. and {Koekemoer}, A.~M. and {Meena}, A.~K. and {Newman}, A.~B. and {Oguri}, M. and {Padilla-Gonzalez}, E. and {Poidevin}, F. and {Rosati}, P. and {Schuldt}, S. and {Strolger}, L.~G. and {Suyu}, S.~H. and {Thorp}, S. and {Zitrin}, A.},
        title = "{Cosmology with Supernova Encore in the Strong Lensing Cluster MACS J0138─2155: Time Delays and Hubble Constant Measurement}",
      journal = {\apj},
         year = 2026,
        month = feb,
       volume = {998},
       number = {2},
          eid = {219},
        pages = {219},
          doi = {10.3847/1538-4357/ae3159},
archivePrefix = {arXiv},
       eprint = {2509.12301},
 primaryClass = {astro-ph.CO},
       adsurl = {https://ui.adsabs.harvard.edu/abs/2026ApJ...998..219P}
}

@ARTICLE{2025ApJ...979...13P,
       author = {{Pascale}, Massimo and {Frye}, Brenda L. and {Pierel}, Justin D.~R. and {Chen}, Wenlei and {Kelly}, Patrick L. and {Cohen}, Seth H. and {Windhorst}, Rogier A. and {Riess}, Adam G. and {Kamieneski}, Patrick S. and {Diego}, Jos{\'e} M. and {Meena}, Ashish K. and {Cha}, Sangjun and {Oguri}, Masamune and {Zitrin}, Adi and {Jee}, M. James and {Foo}, Nicholas and {Leimbach}, Reagen and {Koekemoer}, Anton M. and {Conselice}, C.~J. and {Dai}, Liang and {Goobar}, Ariel and {Siebert}, Matthew R. and {Strolger}, Lou and {Willner}, S.~P.},
        title = "{SN H0pe: The First Measurement of H$_{0}$ from a Multiply Imaged Type Ia Supernova, Discovered by JWST}",
      journal = {\apj},
         year = 2025,
        month = jan,
       volume = {979},
       number = {1},
          eid = {13},
        pages = {13},
          doi = {10.3847/1538-4357/ad9928},
archivePrefix = {arXiv},
       eprint = {2403.18902},
 primaryClass = {astro-ph.CO},
       adsurl = {https://ui.adsabs.harvard.edu/abs/2025ApJ...979...13P}
}

@ARTICLE{2025ApJ...985..182L,
       author = {{Lee}, Abigail J. and {Freedman}, Wendy L. and {Madore}, Barry F. and {Jang}, In Sung and {Owens}, Kayla A. and {Hoyt}, Taylor J.},
        title = "{The Chicago{\textendash}Carnegie Hubble Program: The JWST J-region Asymptotic Giant Branch Extragalactic Distance Scale}",
      journal = {\apj},
         year = 2025,
        month = jun,
       volume = {985},
       number = {2},
          eid = {182},
        pages = {182},
          doi = {10.3847/1538-4357/adc8a1},
archivePrefix = {arXiv},
       eprint = {2408.03474},
 primaryClass = {astro-ph.GA},
       adsurl = {https://ui.adsabs.harvard.edu/abs/2025ApJ...985..182L}
}

@ARTICLE{2025A&A...704A..63T,
       author = {{Tdcosmo Collaboration} and {Birrer}, Simon and {Buckley-Geer}, Elizabeth J. and {Cappellari}, Michele and {Courbin}, Fr{\'e}d{\'e}ric and {Dux}, Fr{\'e}d{\'e}ric and {Fassnacht}, Christopher D. and {Frieman}, Joshua A. and {Galan}, Aymeric and {Gilman}, Daniel and {Huang}, Xiang-Yu and {Knabel}, Shawn and {Langeroodi}, Danial and {Lin}, Huan and {Millon}, Martin and {Morishita}, Takahiro and {Motta}, Veronica and {Mozumdar}, Pritom and {Paic}, Eric and {Shajib}, Anowar J. and {Sheu}, William and {Sluse}, Dominique and {Sonnenfeld}, Alessandro and {Spiniello}, Chiara and {Stiavelli}, Massimo and {Suyu}, Sherry H. and {Tan}, Chin Yi and {Treu}, Tommaso and {van de Vyvere}, Lyne and {Wang}, Han and {Wells}, Patrick and {Williams}, Devon M. and {Wong}, Kenneth C.},
        title = "{TDCOSMO 2025: Cosmological constraints from strong lensing time delays}",
      journal = {\aap},
         year = 2025,
        month = dec,
       volume = {704},
          eid = {A63},
        pages = {A63},
          doi = {10.1051/0004-6361/202555801},
archivePrefix = {arXiv},
       eprint = {2506.03023},
 primaryClass = {astro-ph.CO},
       adsurl = {https://ui.adsabs.harvard.edu/abs/2025A&A...704A..63T}
}

@ARTICLE{2026arXiv260216077U,
       author = {{Urcelay}, Felipe and {Huang}, Xiaosheng and {Sheu}, William and {O'Donnell}, Jackson H. and {Jeltema}, Tesla and {Williams}, Demetrius Y. and {Xu}, Sean and {Agarwal}, Shrihan and {Aldering}, Greg and {{\'A}lvarez-Garc{\'\i}a}, David and {Ambardekar}, Harsh and {Barone}, Tania M. and {Bian}, Fuyan and {Bolton}, Adam S. and {Cikota}, Aleksandar and {Farren}, Gerrit S. and {Glazebrook}, Karl and {Hoyt}, Taylor and {Jain}, Aniket and {Jones}, Tucker and {Kacprzak}, Glenn G. and {Lin}, Emerald and {Perlmutter}, Saul and {Rubin}, David and {Schlegel}, David J. and {Silver}, Ethan and {Storfer}, Christopher J. and {Suzuki}, Nao and {Truong}, Jannik and {{\'U}beda}, M{\'o}nica and {C}, Keerthi Vasan G.},
        title = "{The Carousel Lens II: Cosmological Constraints with GIGA-Lens}",
      journal = {arXiv e-prints},
         year = 2026,
        month = feb,
          eid = {arXiv:2602.16077},
        pages = {arXiv:2602.16077},
          doi = {10.48550/arXiv.2602.16077},
archivePrefix = {arXiv},
       eprint = {2602.16077},
 primaryClass = {astro-ph.CO},
       adsurl = {https://ui.adsabs.harvard.edu/abs/2026arXiv260216077U}
}

@ARTICLE{2025ApJ...985..203F,
       author = {{Freedman}, Wendy L. and {Madore}, Barry F. and {Hoyt}, Taylor J. and {Jang}, In Sung and {Lee}, Abigail J. and {Owens}, Kayla A.},
        title = "{Status Report on the Chicago-Carnegie Hubble Program (CCHP): Measurement of the Hubble Constant Using the Hubble and James Webb Space Telescopes}",
      journal = {\apj},
         year = 2025,
        month = jun,
       volume = {985},
       number = {2},
          eid = {203},
        pages = {203},
          doi = {10.3847/1538-4357/adce78},
archivePrefix = {arXiv},
       eprint = {2408.06153},
 primaryClass = {astro-ph.CO},
       adsurl = {https://ui.adsabs.harvard.edu/abs/2025ApJ...985..203F}
}

@article{sheu2024carousel,
  title={The Carousel Lens: A well-modeled strong lens with multiple sources spectroscopically confirmed by VLT/MUSE},
  author={Sheu, William and Cikota, Aleksandar and Huang, Xiaosheng and Glazebrook, Karl and Storfer, Christopher and Agarwal, Shrihan and Schlegel, David J and Suzuki, Nao and Barone, Tania M and Bian, Fuyan and others},
  journal={The Astrophysical Journal},
  volume={973},
  number={1},
  pages={3},
  year={2024},
  publisher={IOP Publishing}
}

@misc{huang2025desistronglensfoundry,
      title={DESI Strong Lens Foundry I: HST Observations and Modeling with GIGA-Lens}, 
      author={X. Huang and S. Baltasar and N. Ratier-Werbin and C. Storfer and W. Sheu and S. Agarwal and M. Tamargo-Arizmendi and D. J. Schlegel and J. Aguilar and S. Ahlen and G. Aldering and S. Banka and S. BenZvi and D. Bianchi and A. Bolton and D. Brooks and A. Cikota and T. Claybaugh and A. de la Macorra and A. Dey and P. Doel and J. Edelstein and A. Filipp and J. E. Forero-Romero and E. Gaztanaga and S. Gontcho A Gontcho and A. Gu and G. Gutierrez and K. Honscheid and E. Jullo and S. Juneau and R. Kehoe and D. Kirkby and T. Kisner and A. Kremin and K. J. Kwon and A. Lambert and M. Landriau and D. Lang and L. Le Guillou and J. Liu and A. Meisner and R. Miquel and J. Moustakas and A. D. Myers and S. Perlmutter and I. Perez-Rafols and F. Prada and G. Rossi and D. Rubin and E. Sanchez and M. Schubnell and Y. Shu and E. Silver and D. Sprayberry and N. Suzuki and G. Tarle and B. A. Weaver and H. Zou},
      year={2025},
      eprint={2502.03455},
      archivePrefix={arXiv},
      primaryClass={astro-ph.CO},
      url={https://arxiv.org/abs/2502.03455}, 
}

@article{dahle2013sdss,
  title={SDSS J2222+ 2745: A GRAVITATIONALLY LENSED SEXTUPLE QUASAR WITH A MAXIMUM IMAGE SEPARATION OF 15.″1 DISCOVERED IN THE SLOAN GIANT ARCS SURVEY},
  author={Dahle, H{\aa}kon and Gladders, MD and Sharon, K and Bayliss, MB and Wuyts, E and Abramson, LE and Koester, BP and Groeneboom, N and Brinckmann, TE and Kristensen, MT and others},
  journal={The Astrophysical Journal},
  volume={773},
  number={2},
  pages={146},
  year={2013},
  publisher={IOP Publishing}
}

@article{acebron2022new,
  title={New strong lensing modelling of SDSS J2222+ 2745 enhanced with VLT/MUSE spectroscopy},
  author={Acebron, Ana and Grillo, CLAUDIO and Bergamini, PIETRO and Caminha, GB and Tozzi, Paolo and Mercurio, Amata and Rosati, Piero and Brammer, G and Meneghetti, MASSIMO and Nonino, Mario and others},
  journal={Astronomy \& Astrophysics},
  volume={668},
  pages={A142},
  year={2022},
  publisher={EDP Sciences}
}

@article{pettini2010cassowary,
  title={CASSOWARY 20: a wide separation Einstein Cross identified with the X-shooter spectrograph},
  author={Pettini, Max and Christensen, Lise and D'Odorico, Sandro and Belokurov, Vasily and Evans, N Wyn and Hewett, Paul C and Koposov, Sergey and Mason, Elena and Vernet, Jo{\"e}l},
  journal={Monthly Notices of the Royal Astronomical Society},
  volume={402},
  number={4},
  pages={2335--2343},
  year={2010},
  publisher={Blackwell Publishing Ltd Oxford, UK}
}

@article{2020SciPy-NMeth,
  author  = {Virtanen, P. and others},
  title   = {{{SciPy} 1.0: Fundamental Algorithms for Scientific Computing in Python}},
  journal = {Nature Methods},
  year    = {2020},
  volume  = {17},
  pages   = {261--272},
  doi     = {10.1038/s41592-019-0686-2}
}
\bibliographystyle{aasjournal}



\begin{figure}[!ht]
\centering
\includegraphics[width=1.0\textwidth]{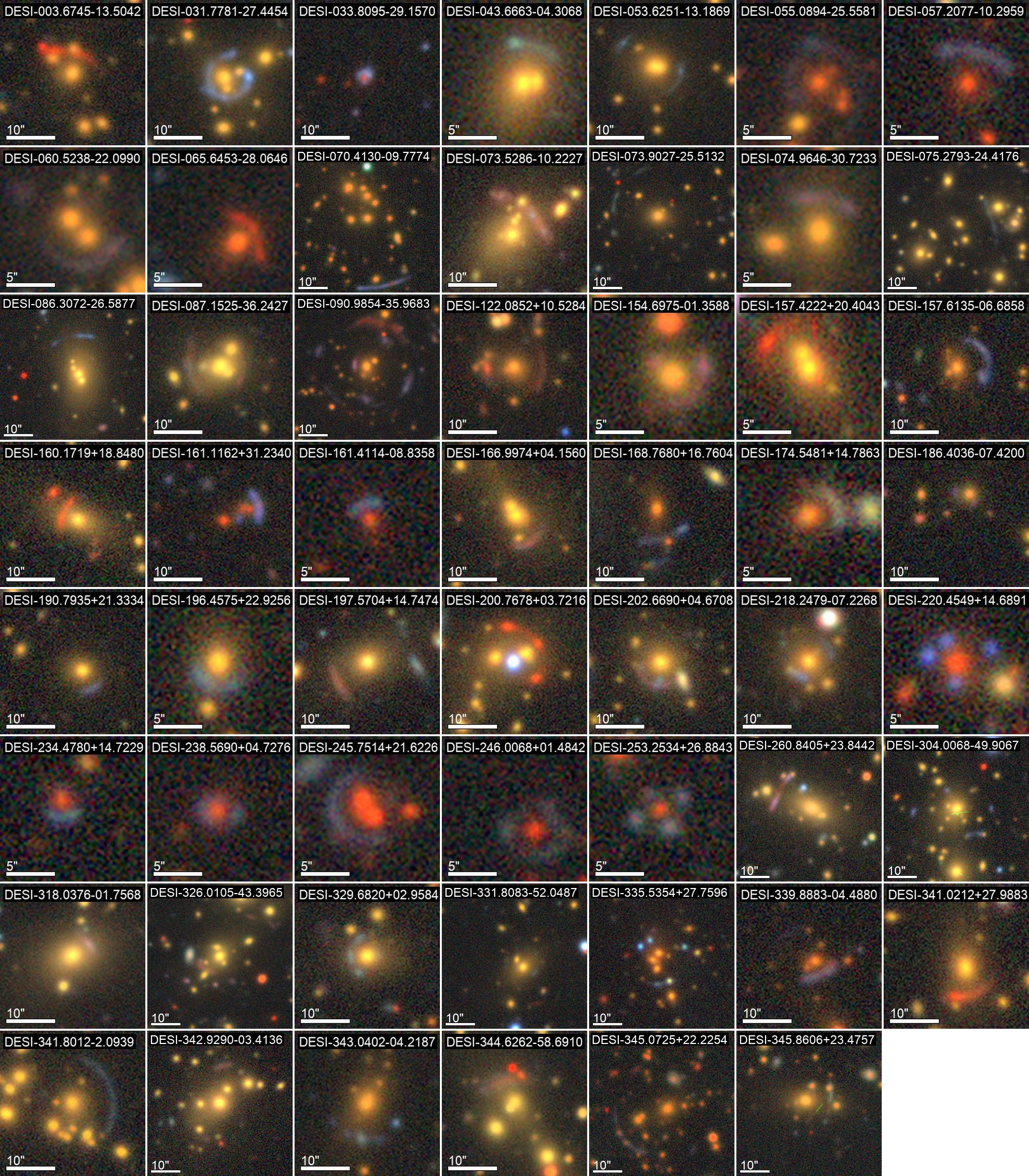}
\label{fig:mosaic_confirmed}
\caption{Mosaic of the 55 confirmed DESI gravitational lens candidates. The cutouts are taken from the DESI Legacy Imaging Surveys DR9. North is up and east is to the left. The scale bars in the lower-left corner of each cutout indicate the angular scale.}
\end{figure}

\begin{figure*}[!ht]
\centering

\begin{minipage}{1.0\textwidth}
\centering
\includegraphics[width=0.4\textwidth]{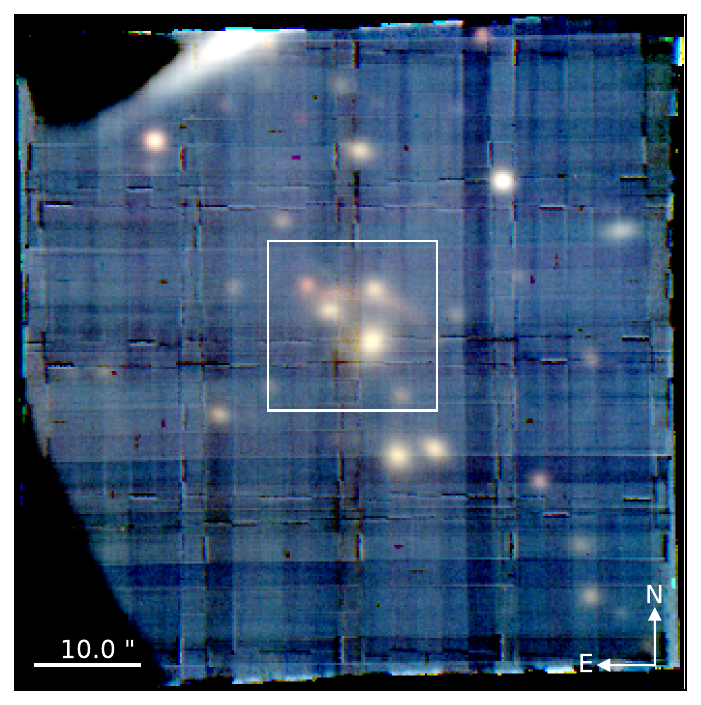}
\includegraphics[width=0.4\textwidth]{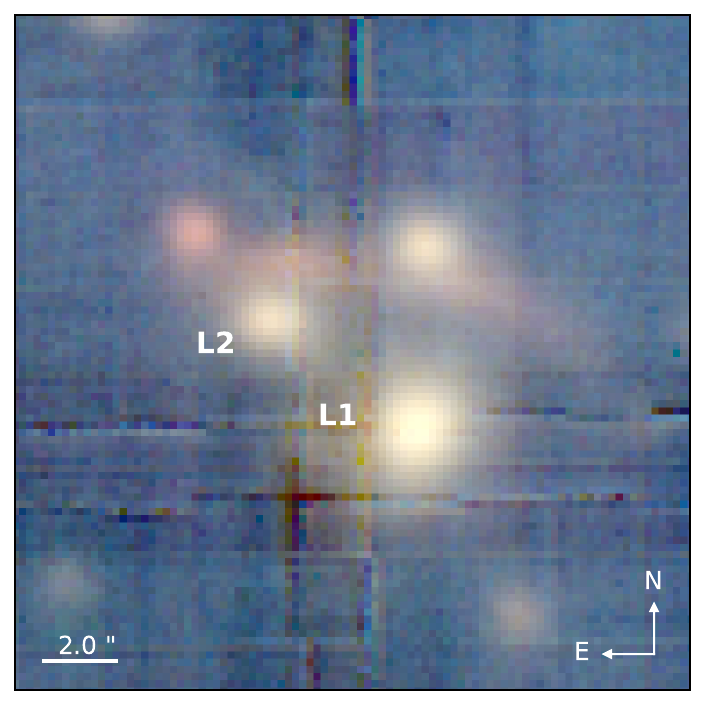}
\includegraphics[width=\textwidth]{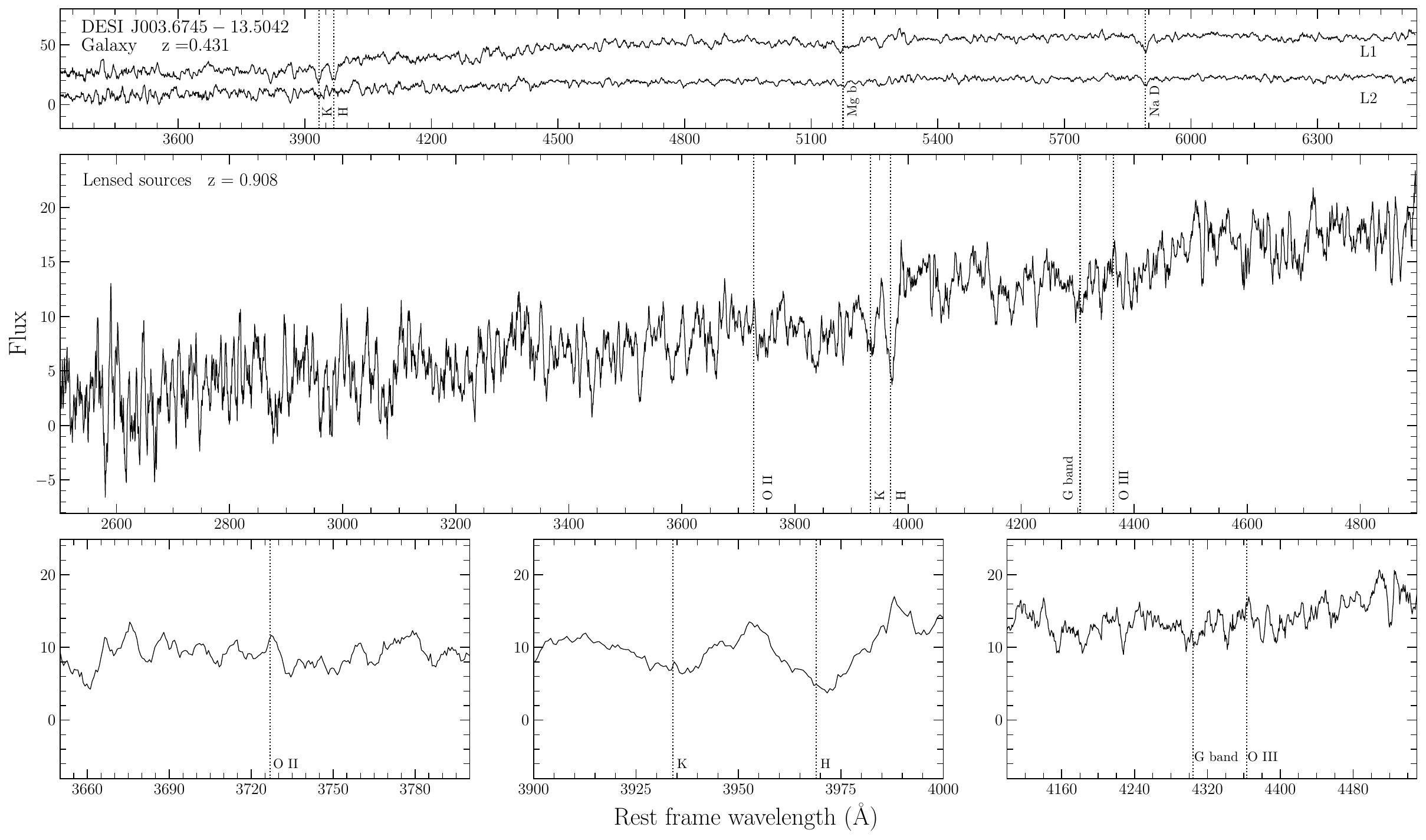}
\caption{\textit{Top:} RGB image of gravitational lens system DESI~J003.6745-13.5042 observed with MUSE. The image is a combination of the Johnson $V$ (yellow), Cousins $R$ (magenta), and Cousins $I$ (cyan) images generated from the MUSE data cube. The white square indicates the cutout shown in the right panel. \textit{Bottom:} MUSE spectra of DESI~J003.6745-13.5042 in rest frame wavelength, with the host galaxy above and the lensed source below. Redshifts are written at the top of each panel and in the lower panels, the emission and absorption lines are shown in more detail. For more information on the system, see Desc. \ref{Ref:lens16_1}.}
\label{fig:MUSElens16}

\label{fig:MUSEspectra16_big}
\end{minipage}

\end{figure*}

\begin{figure*}[!ht]
\centering
\begin{minipage}{1.0\textwidth}
\centering
\includegraphics[width=0.4\textwidth]{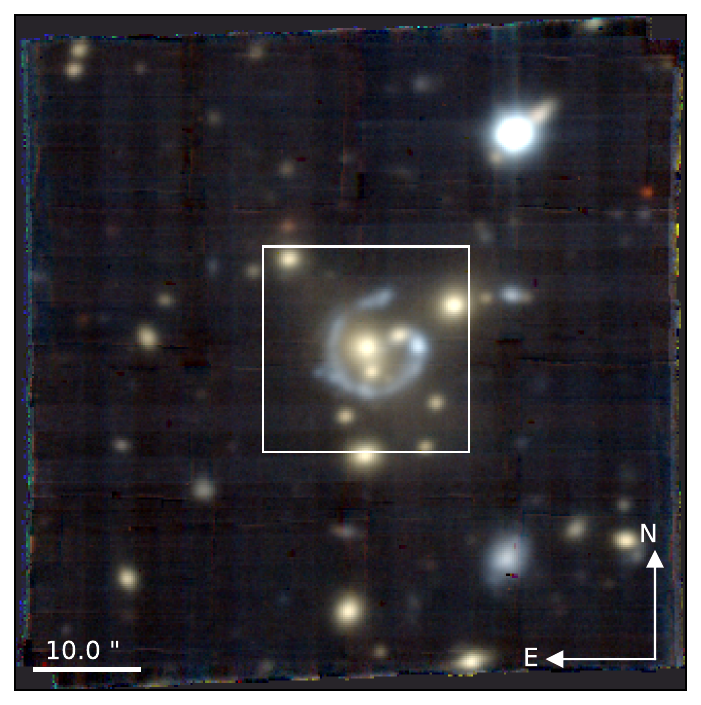}
\includegraphics[width=0.404\textwidth]{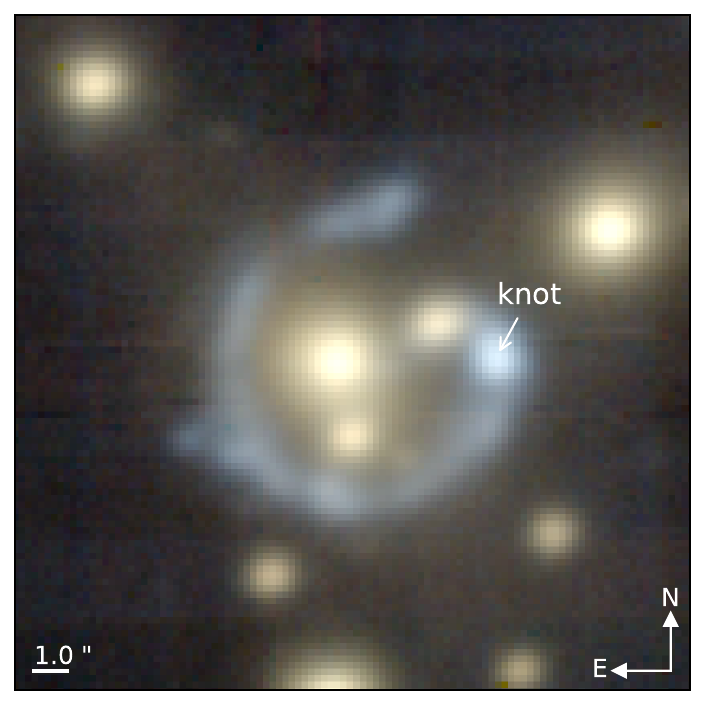}
\label{fig:MUSElens2_031img}
\includegraphics[width=\textwidth]{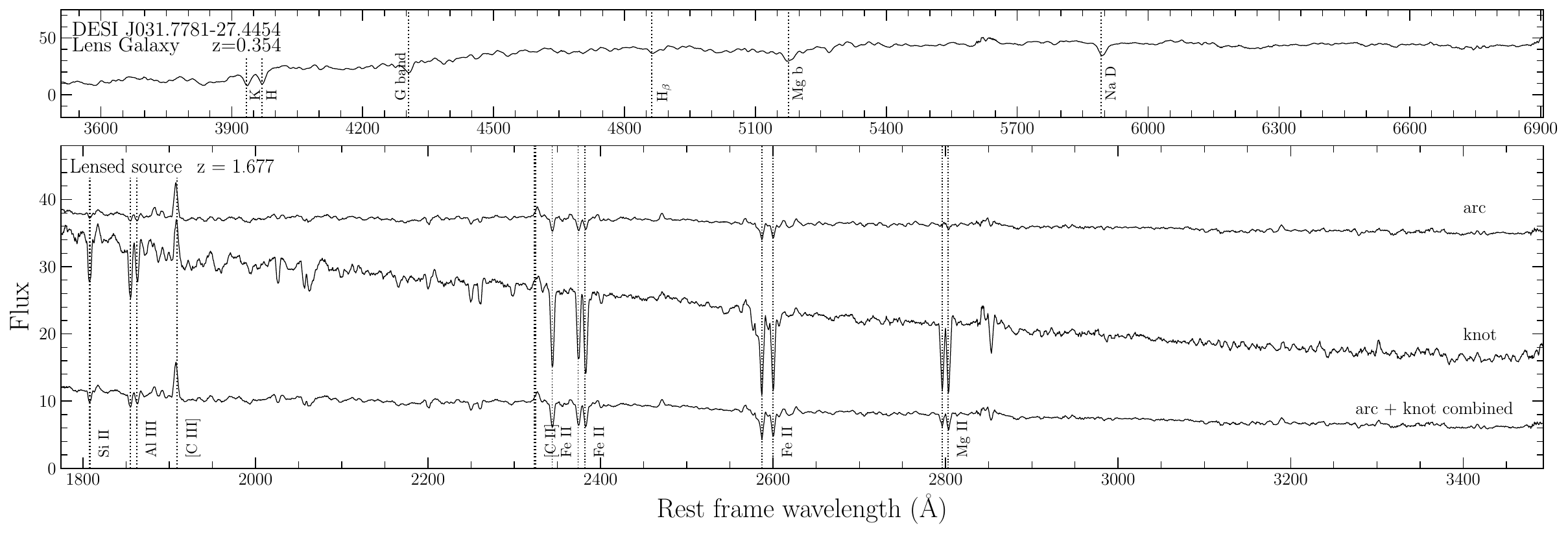}
\caption{\textit{Top:} RGB image of gravitational lens system DESI~J031.7781-27.4454 observed with MUSE. \textit{Bottom:}MUSE spectra of DESI~J031.7781-27.4454. For more information on the system, see Desc. \ref{Ref:lens2_031}.}
\label{fig:MUSEspectra2_031}
\end{minipage}
\end{figure*}



\begin{figure*}[!ht]
\centering
\begin{minipage}{1.0\textwidth}
\centering
\includegraphics[width=0.4\textwidth]{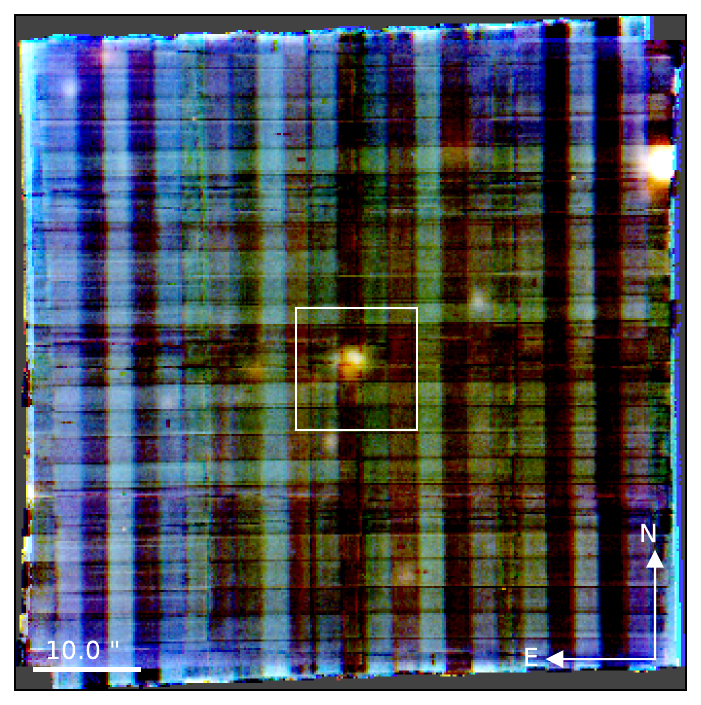}
\includegraphics[width=0.404\textwidth]{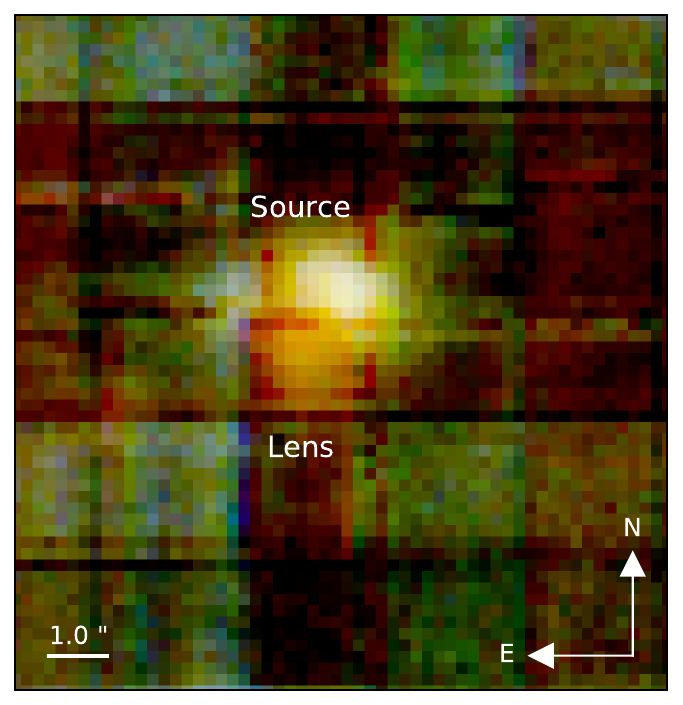}
\label{fig:MUSElens112img}
\includegraphics[width=\textwidth]{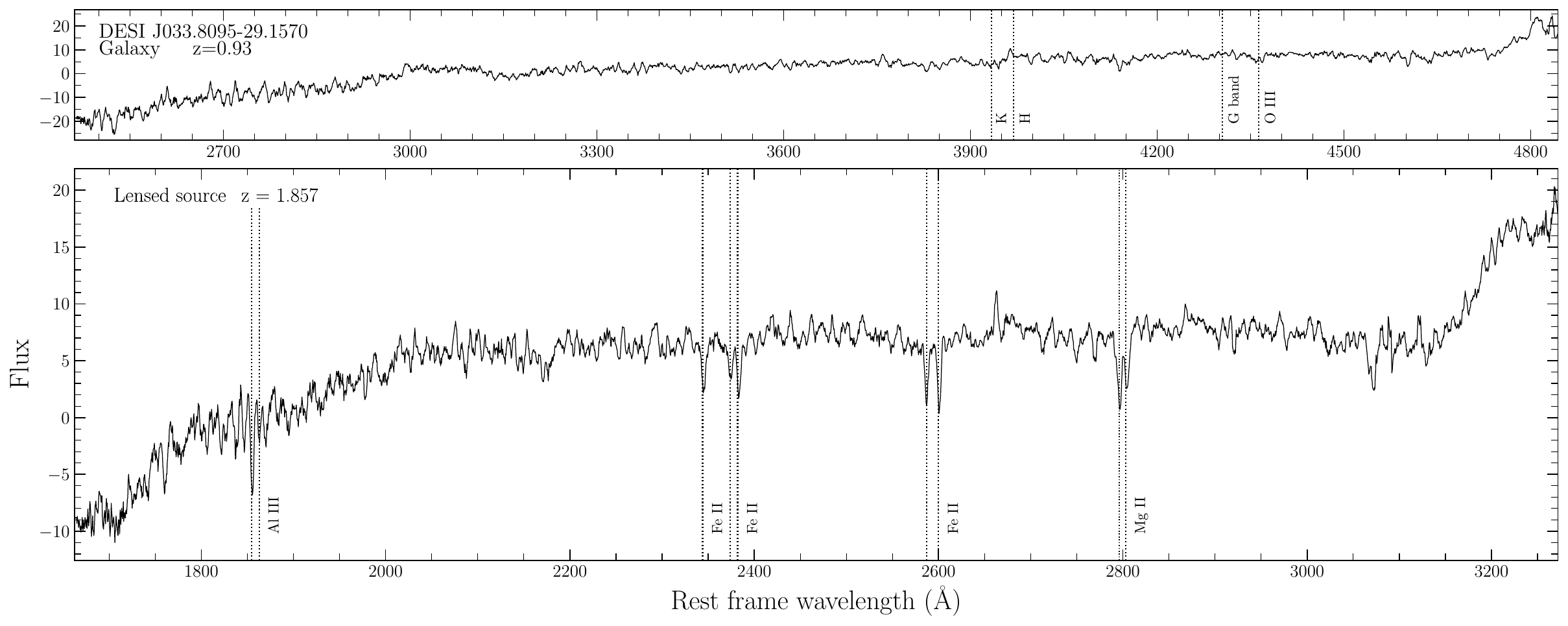}
\caption{\textit{Top:} RGB image of gravitational lens system DESI~J033.8095-29.1570 observed with MUSE. \textit{Bottom:}MUSE spectra of DESI~J033.8095-29.1570. Note that the lens quality flag is $Q_z=3$. For more information on the system, see Desc. \ref{Ref:lens112}.}
\label{fig:MUSEspectra112}
\end{minipage}
\end{figure*}

\begin{figure*}[!ht]
\centering
\begin{minipage}{1.0\textwidth}
\centering
\includegraphics[width=0.4\textwidth]{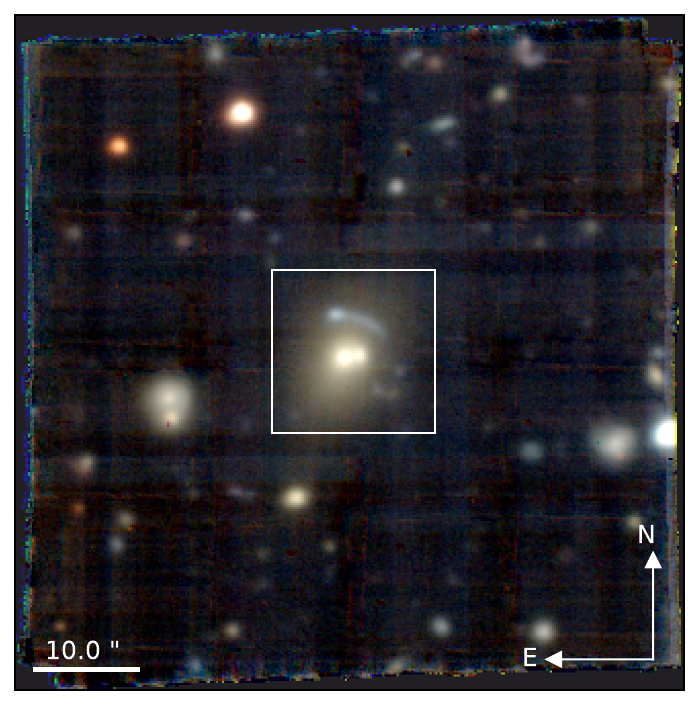}
\includegraphics[width=0.404\textwidth]{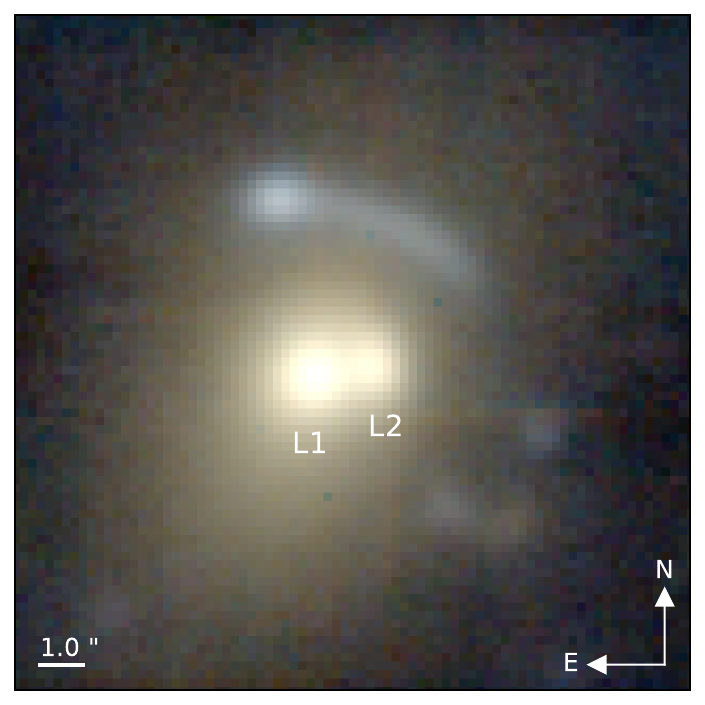}
\label{fig:MUSElens16img}
\includegraphics[width=\textwidth]{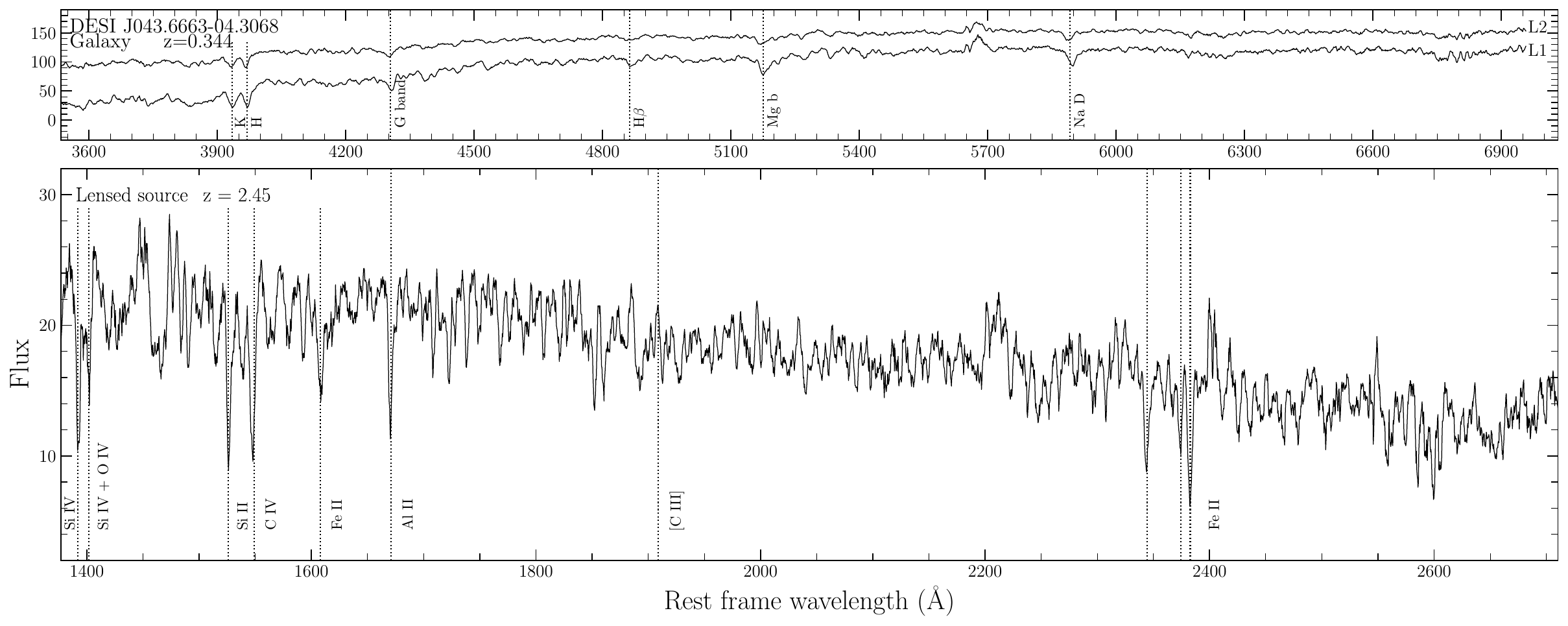}
\caption{\textit{Top:} RGB image of gravitational lens system DESI~J043.6663-04.3068 observed with MUSE. \textit{Bottom:}MUSE spectra of DESI~J043.6663-04.3068. For more information on the system, see Desc. \ref{Ref:lens16_2}.}
\label{fig:MUSEspectra16}
\end{minipage}
\end{figure*}

\begin{figure*}[!ht]
\centering
\begin{minipage}{1.0\textwidth}
\centering
\includegraphics[width=0.4\textwidth]{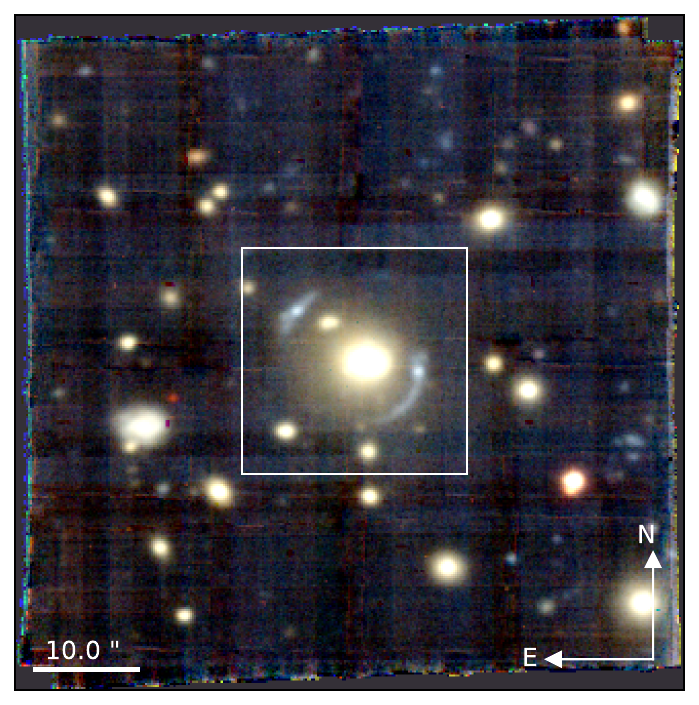}
\includegraphics[width=0.404\textwidth]{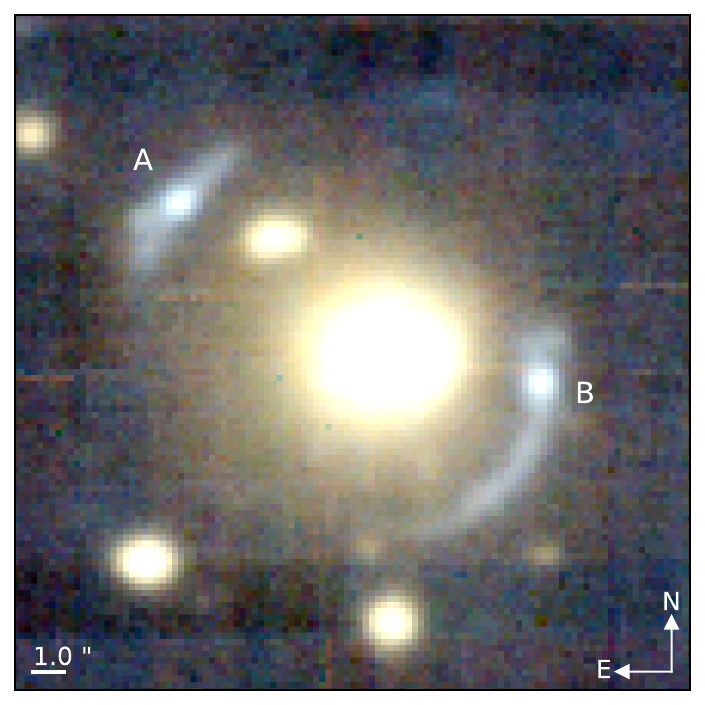}
\label{fig:MUSElens19img}
\includegraphics[width=\textwidth]{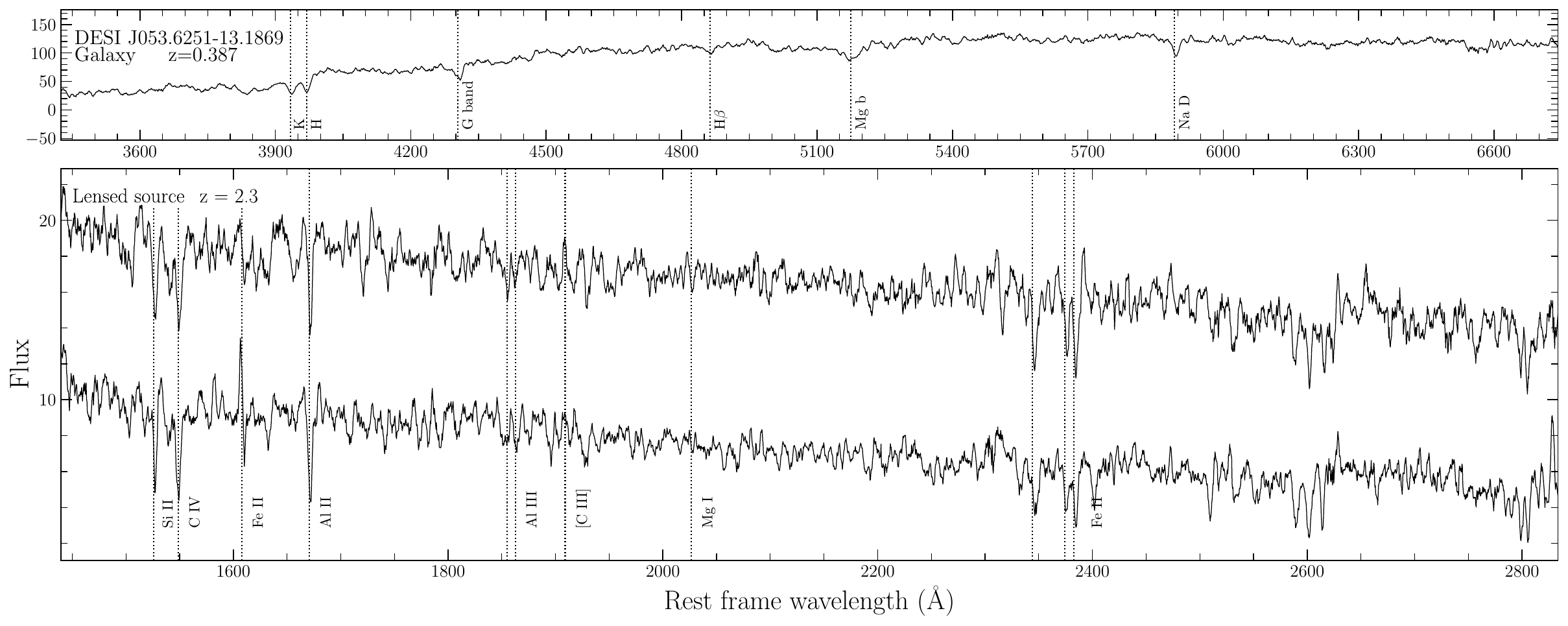}
\caption{\textit{Top:} RGB image of gravitational lens system DESI~J053.6251-13.1869 observed with MUSE. \textit{Bottom:} MUSE spectra of DESI~J053.6251-13.1869. For more information on the system, see Desc. \ref{Ref:lens19}.} 
\label{fig:MUSEspectra19}
\end{minipage}
\end{figure*}

\begin{figure*}[!ht]
\centering
\begin{minipage}{1.0\textwidth}
\centering
\includegraphics[width=0.4\textwidth]{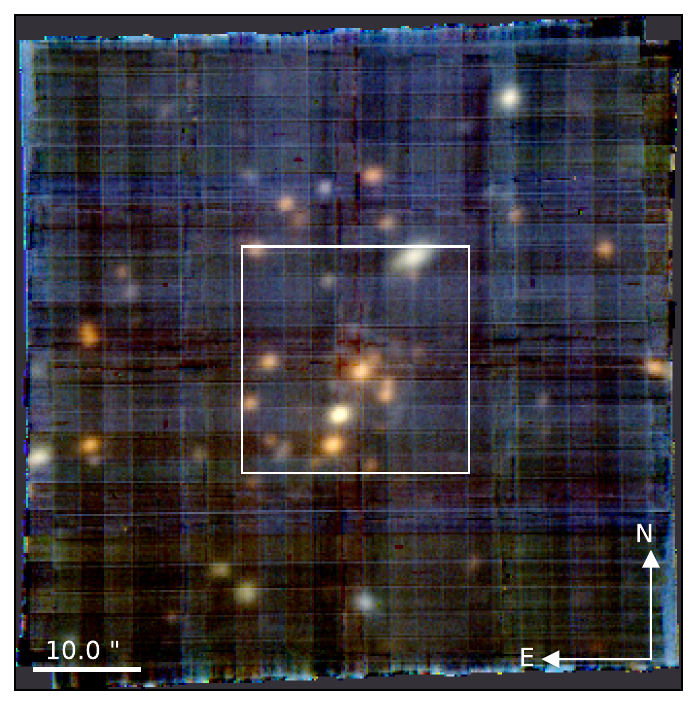}
\includegraphics[width=0.404\textwidth]{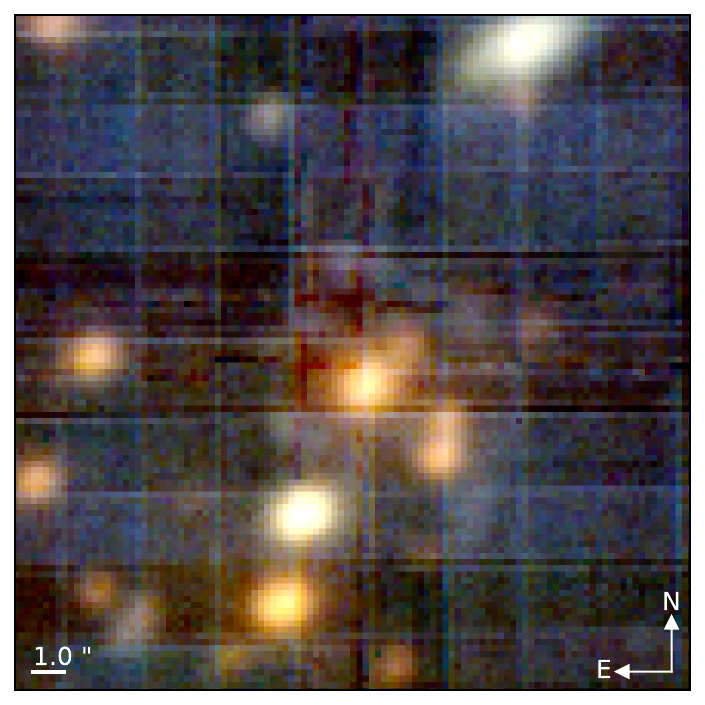}
\label{fig:MUSElens20img}
\includegraphics[width=\textwidth]{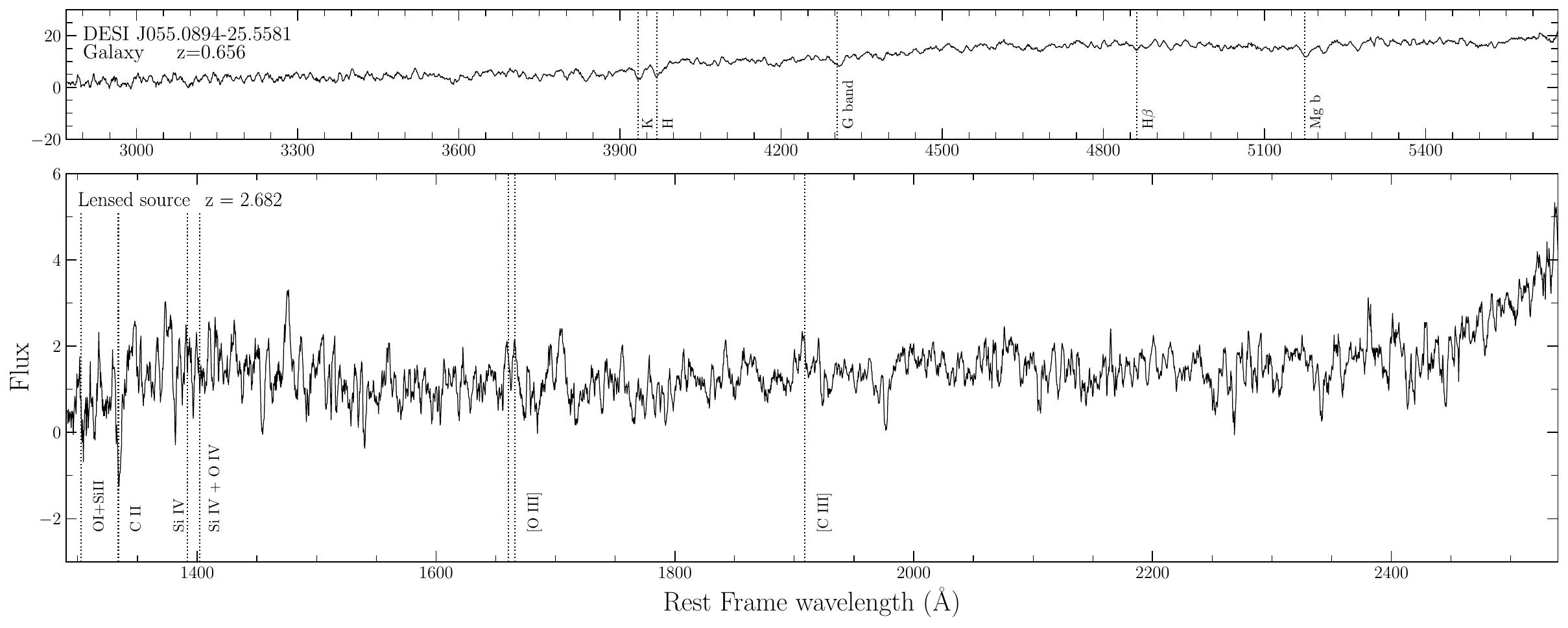}
\caption{\textit{Top:} RGB image of gravitational lens system DESI~J055.0894-25.5581 observed with MUSE. \textit{Bottom:} MUSE spectra of DESI~J055.0894-25.5581. Note that the source quality flag is $Q_z=2$. For more information on the system, see Desc. \ref{Ref:lens20}.}
\label{fig:MUSEspectra20}
\end{minipage}
\end{figure*}

\begin{figure*}[!ht]
\centering
\begin{minipage}{1.0\textwidth}
\centering
\includegraphics[width=0.4\textwidth]{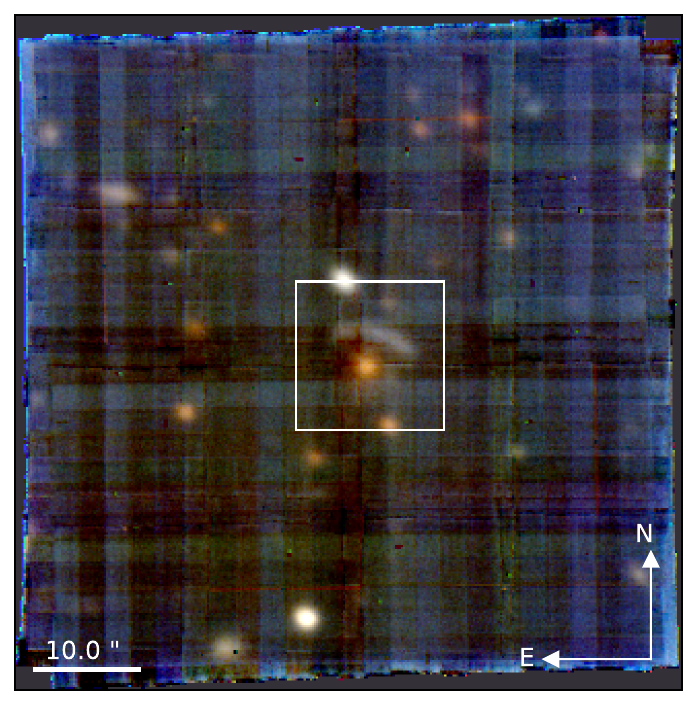}
\includegraphics[width=0.404\textwidth]{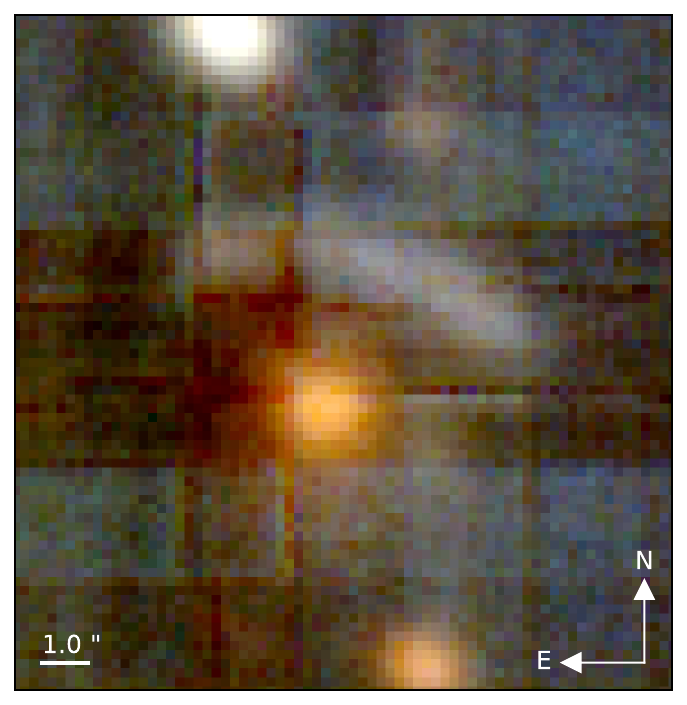}
\label{fig:MUSElens4_057img}
\includegraphics[width=\textwidth]{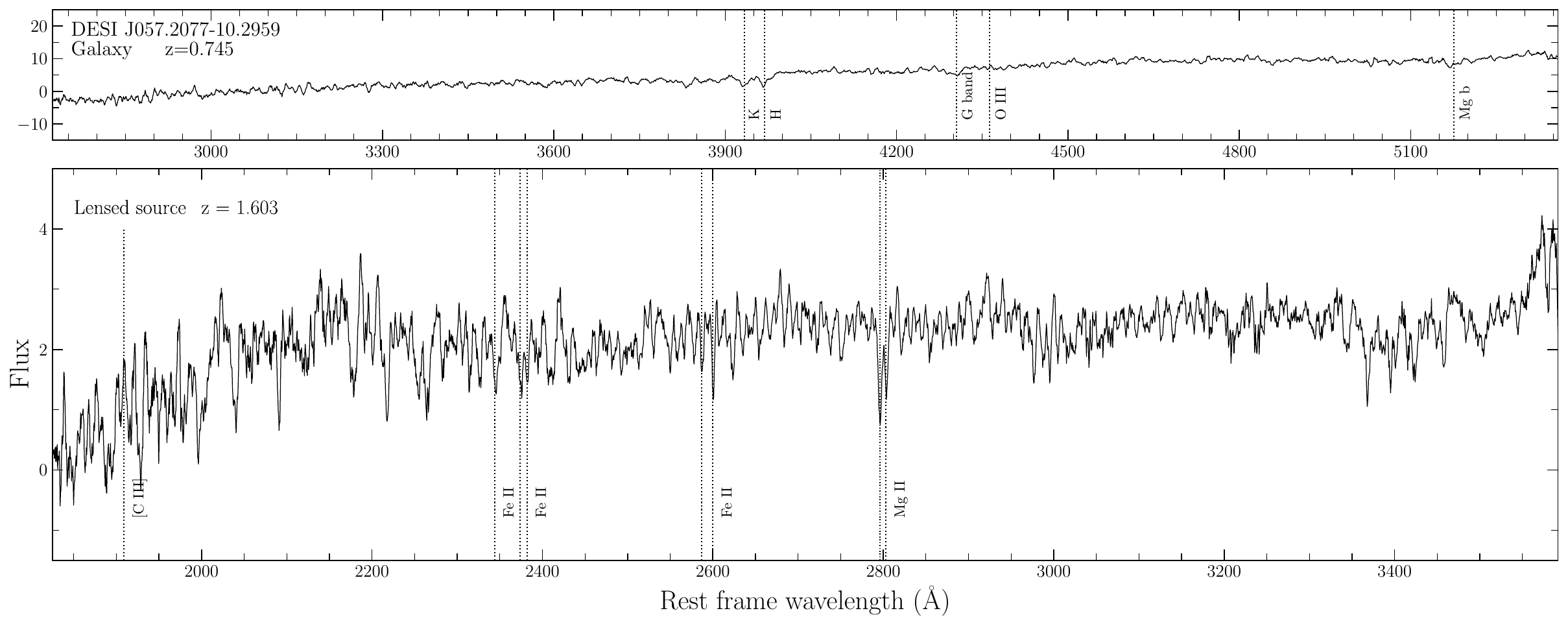}
\caption{\textit{Top:} RGB image of gravitational lens system DESI~J057.2077-10.2959 observed with MUSE. \textit{Bottom:} MUSE spectra of DESI~J057.2077-10.2959. Note that the source quality flag is $Q_z=2$. For more information on the system, see Desc. \ref{Ref:lens4_057}.}
\label{fig:MUSEspectra4_057}
\end{minipage}
\end{figure*}

\begin{figure*}[!ht]
\centering
\begin{minipage}{1.0\textwidth}
\centering
\includegraphics[width=0.4\textwidth]{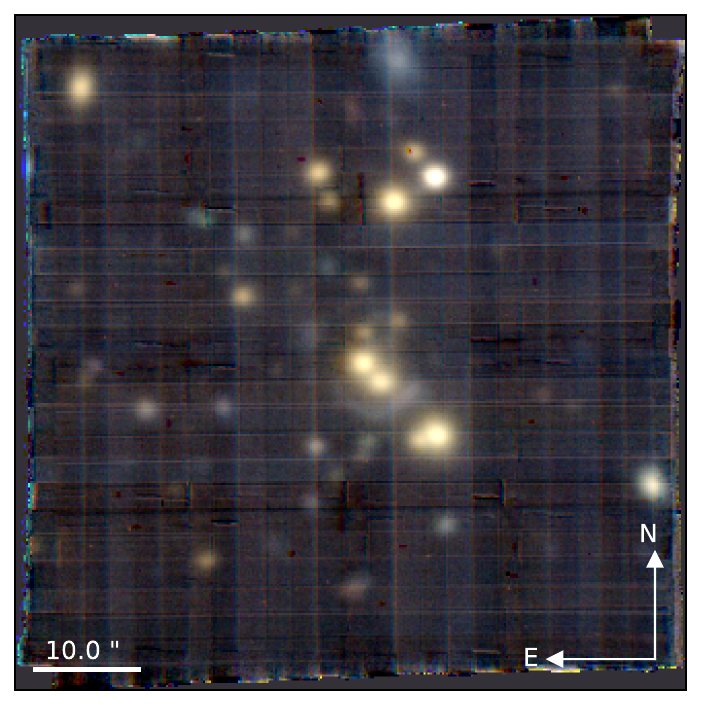}
\includegraphics[width=0.404\textwidth]{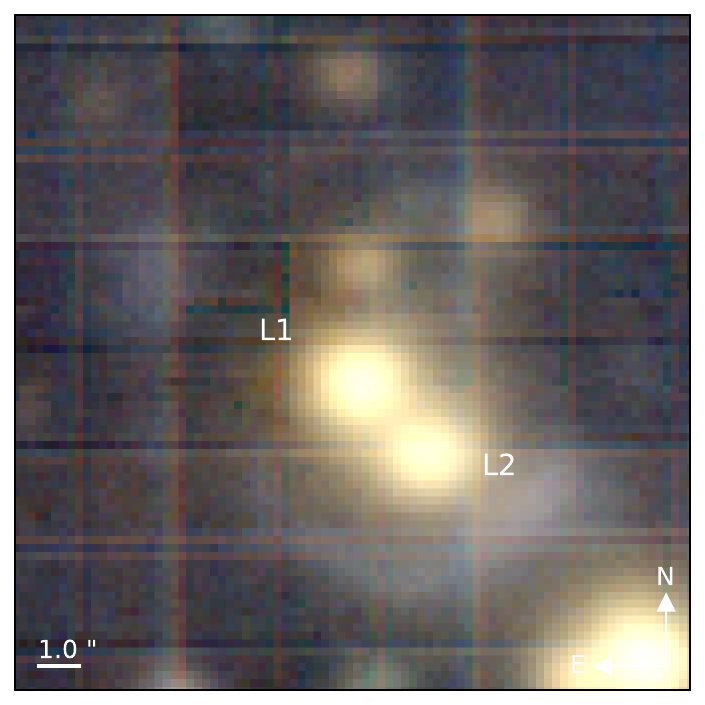}

\includegraphics[width=\textwidth]{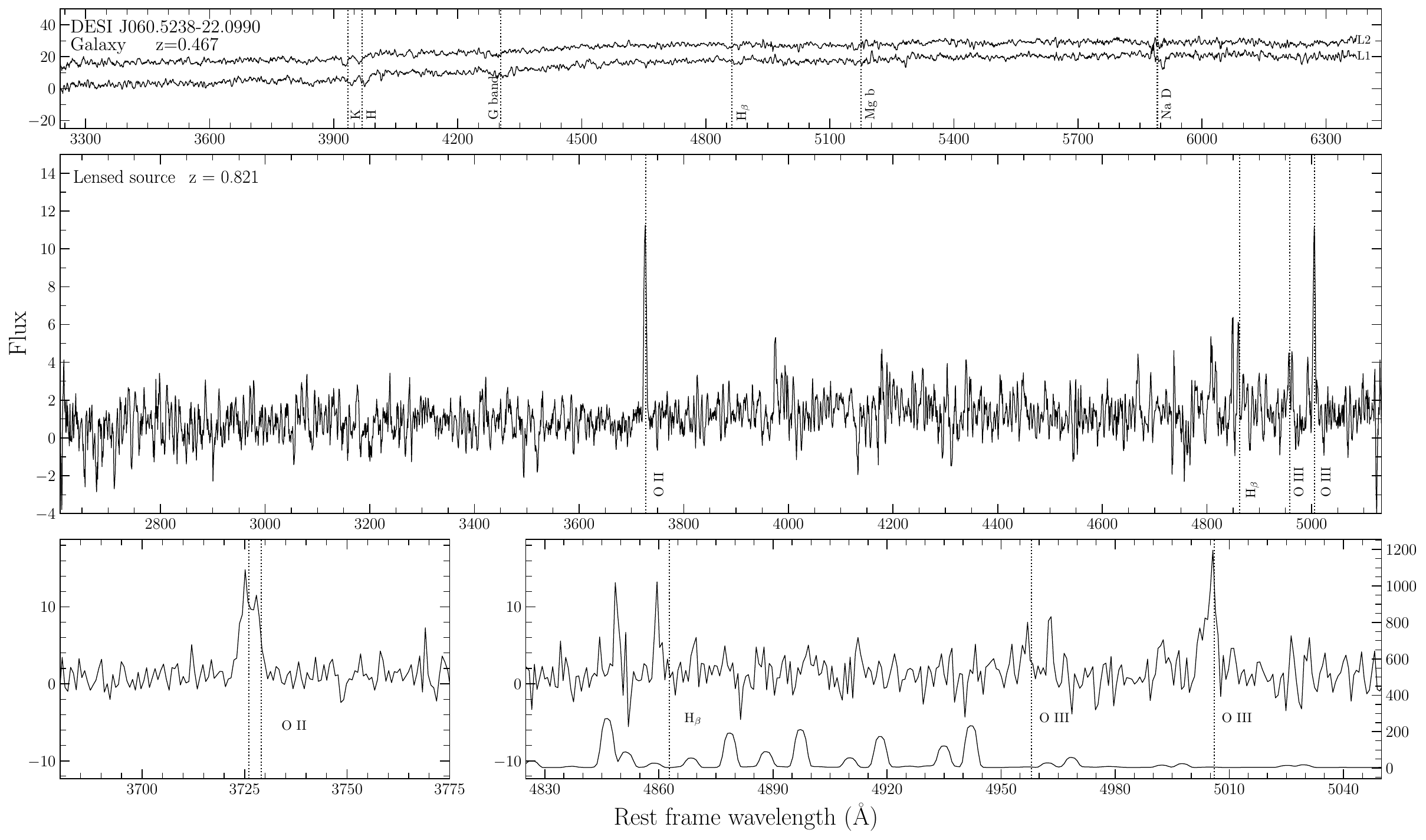}
\caption{\textit{Top:} RGB image of gravitational lens system DESI~J060.5238-22.0990 observed with MUSE. \textit{Bottom:} MUSE spectra of DESI~J060.5238-22.0990. For more information on the system, see Desc. \ref{Ref:lens22}. }
\label{fig:MUSEspectra22}
\end{minipage}

\end{figure*}

\begin{figure*}[!ht]
\centering
\begin{minipage}{1.0\textwidth}
\centering
\includegraphics[width=0.4\textwidth]{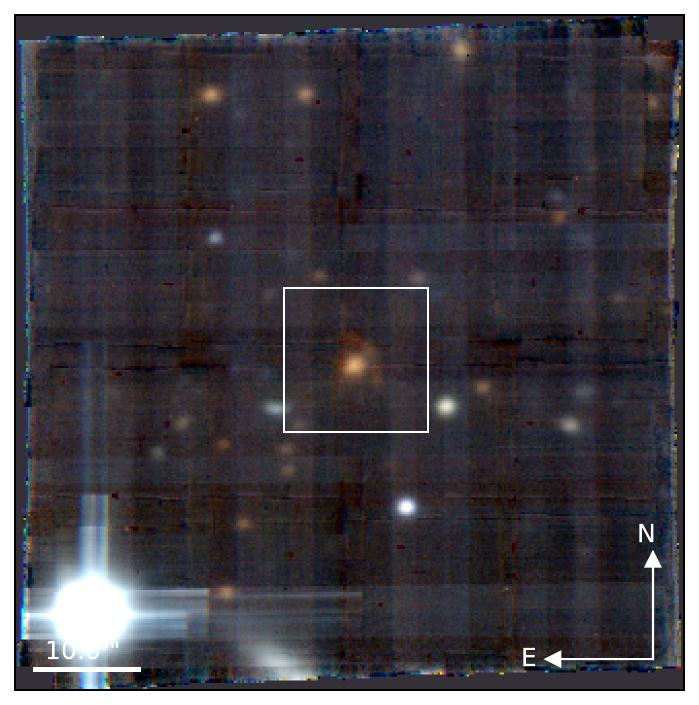}
\includegraphics[width=0.404\textwidth]{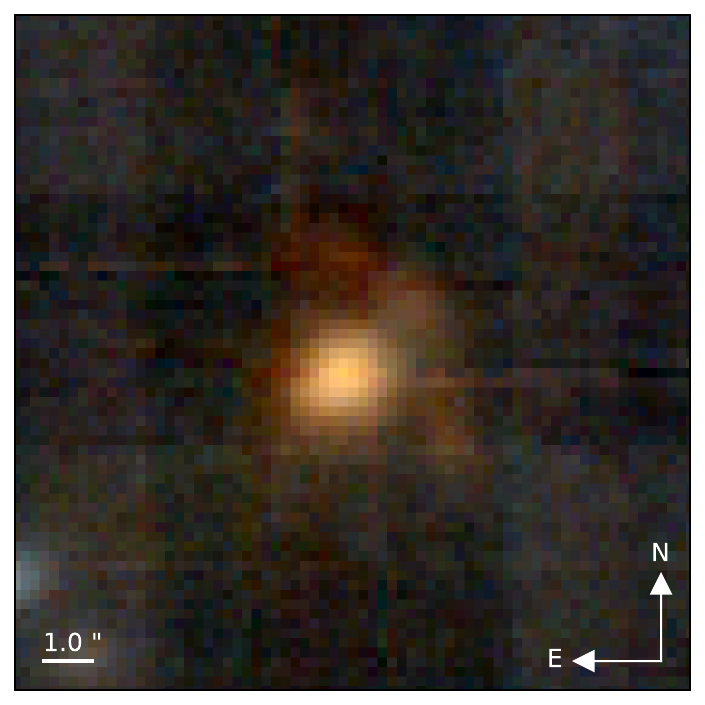}

\includegraphics[width=\textwidth]{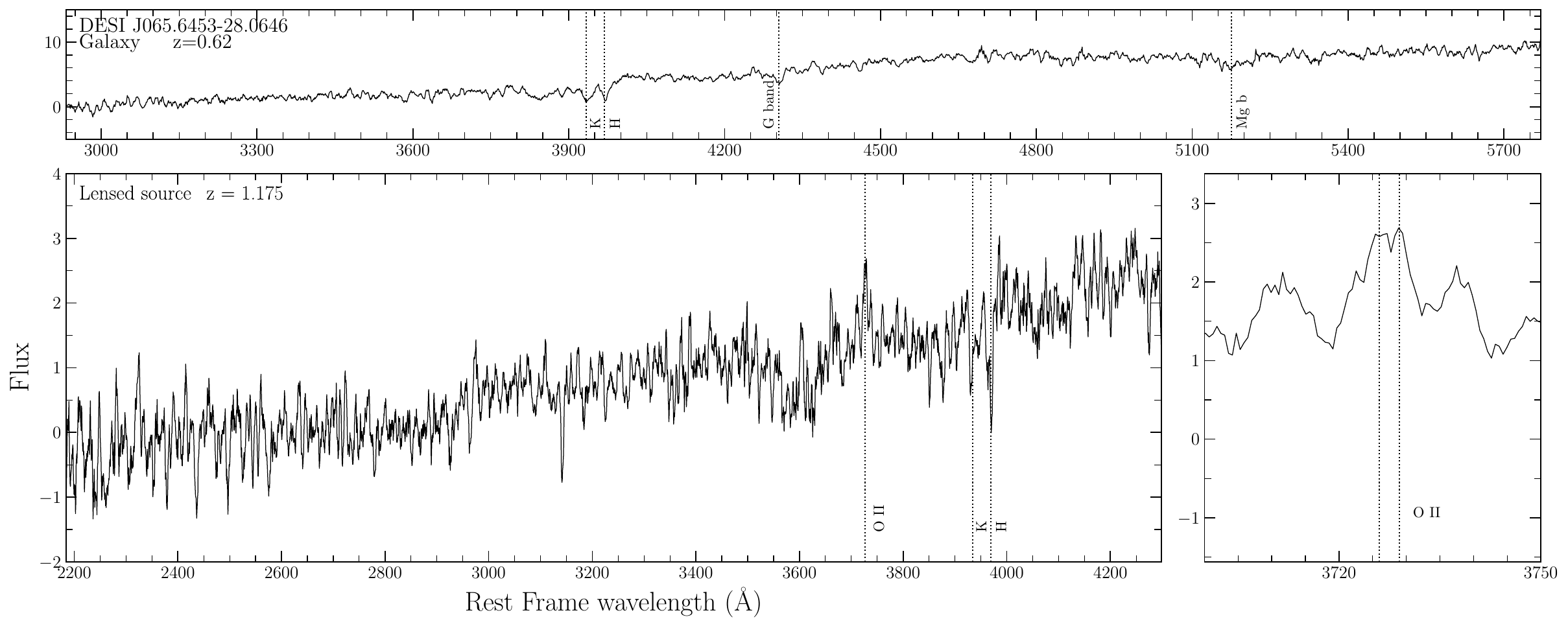}
\caption{\textit{Top:} RGB image of gravitational lens system DESI~J065.6453-28.0646 observed with MUSE. \textit{Bottom:} MUSE spectra of DESI~J065.6453-28.0646. For more information on the system, see Desc. \ref{Ref:lens24}. }
\label{fig:MUSEspectra24}
\end{minipage}
\end{figure*}

\begin{figure*}[!ht]
\centering
\begin{minipage}{1.0\textwidth}
\centering
\includegraphics[width=0.4\textwidth]{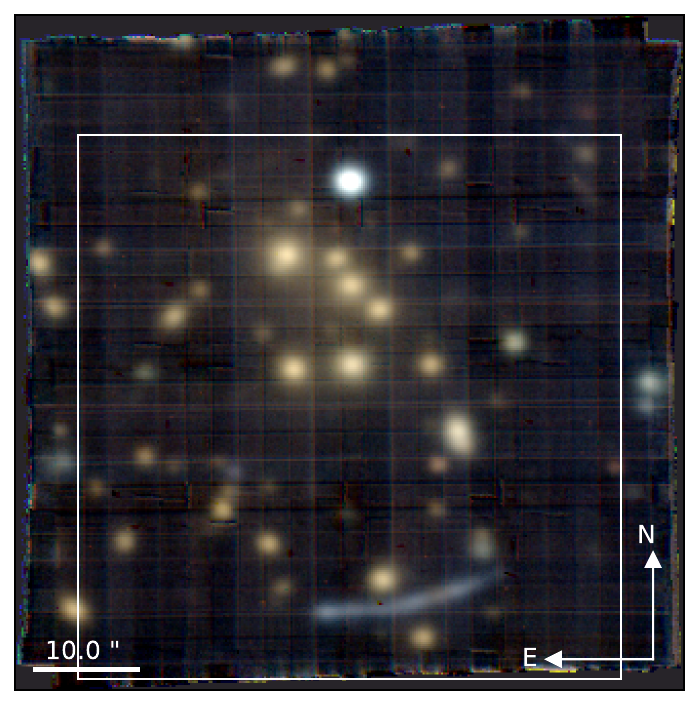}
\includegraphics[width=0.404\textwidth]{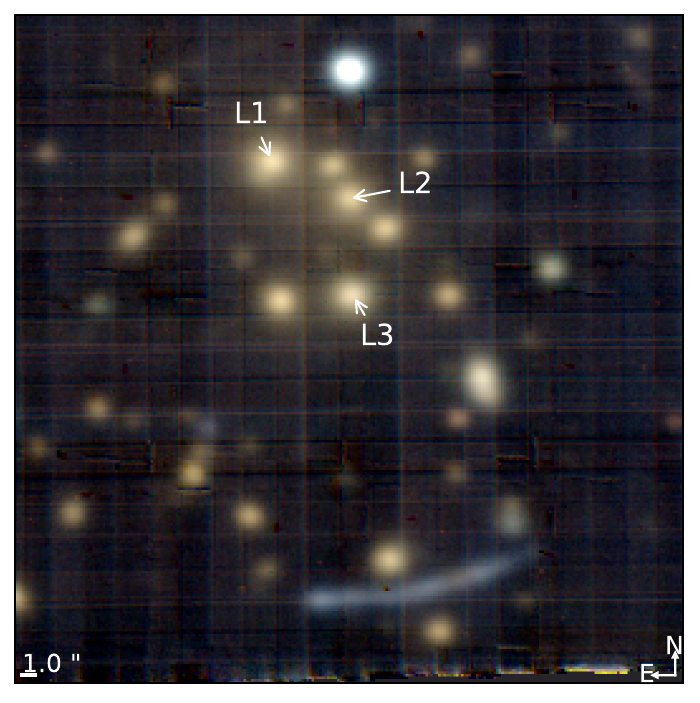}

\includegraphics[width=\textwidth]{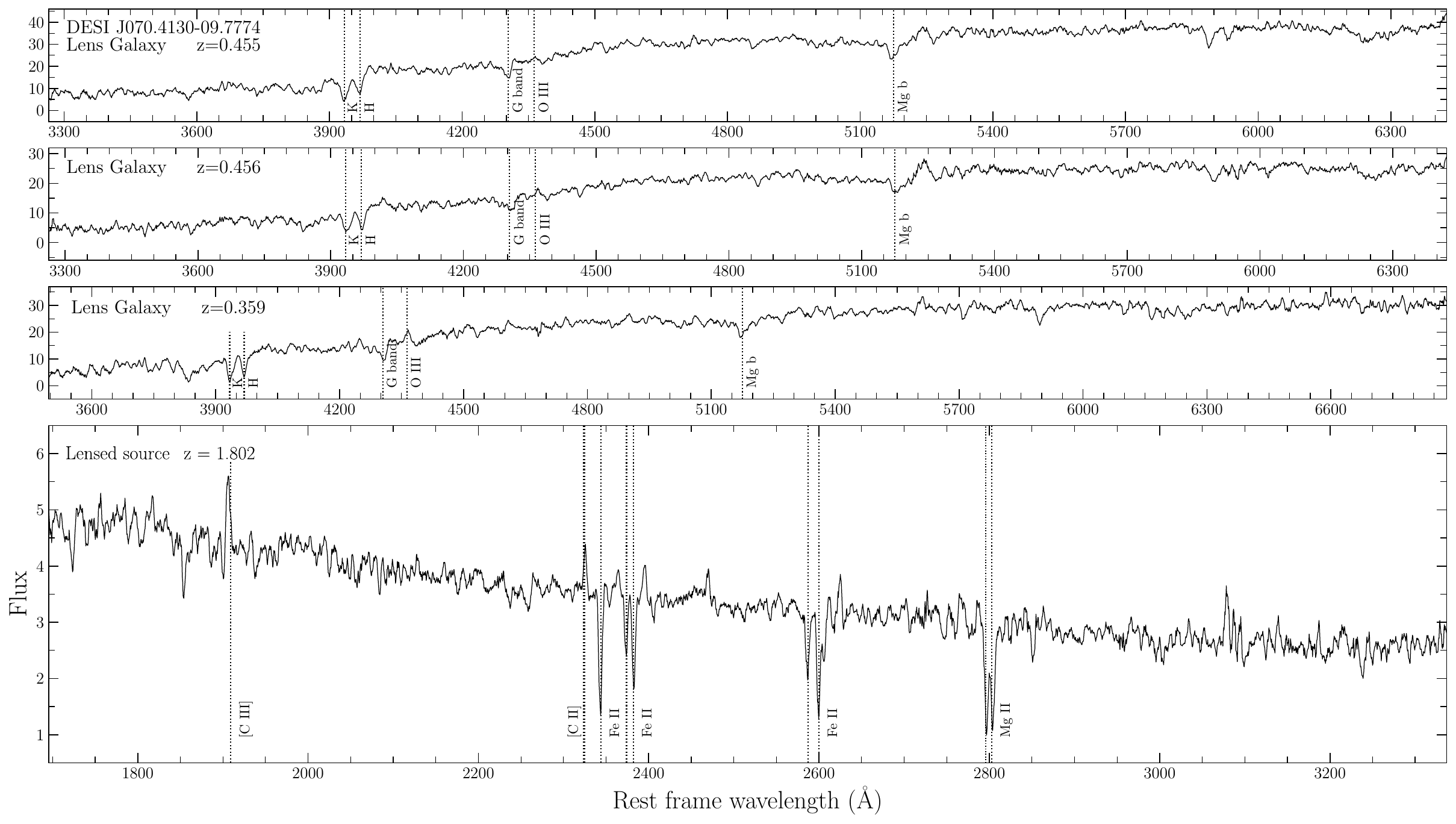}
\caption{\textit{Top:} RGB image of gravitational lens system DESI~J070.4130-09.7774 observed with MUSE. \textit{Bottom:} MUSE spectra of DESI~070.4130-09.7774. For more information on the system, see Desc. \ref{Ref:lens25}. }
\label{fig:MUSEspectra25}
\end{minipage}
\end{figure*}

\begin{figure*}[!ht]
\centering
\begin{minipage}{1.0\textwidth}
\centering
\includegraphics[width=0.4\textwidth]{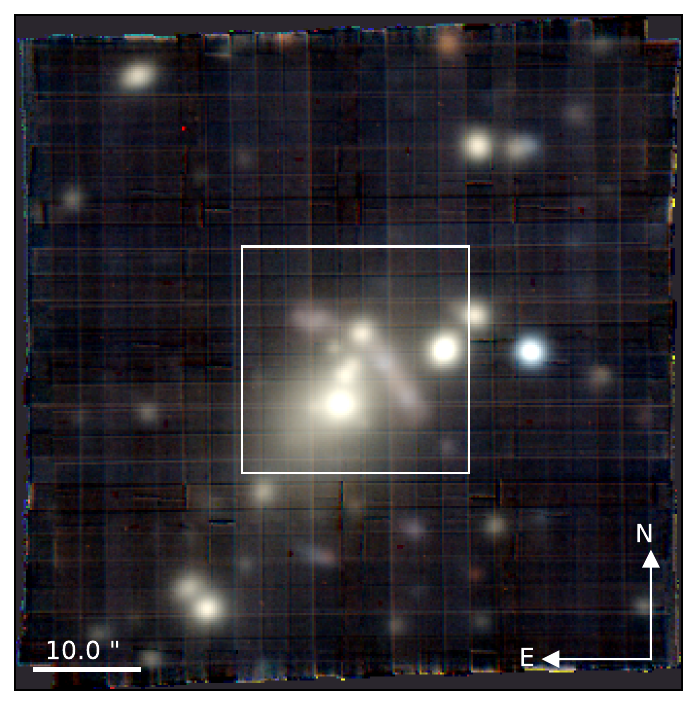}
\includegraphics[width=0.404\textwidth]{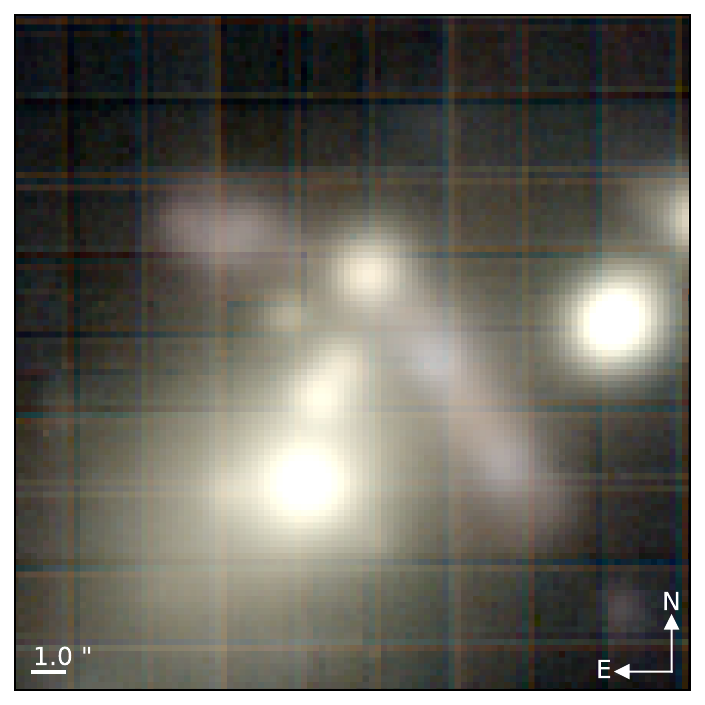}

\includegraphics[width=\textwidth]{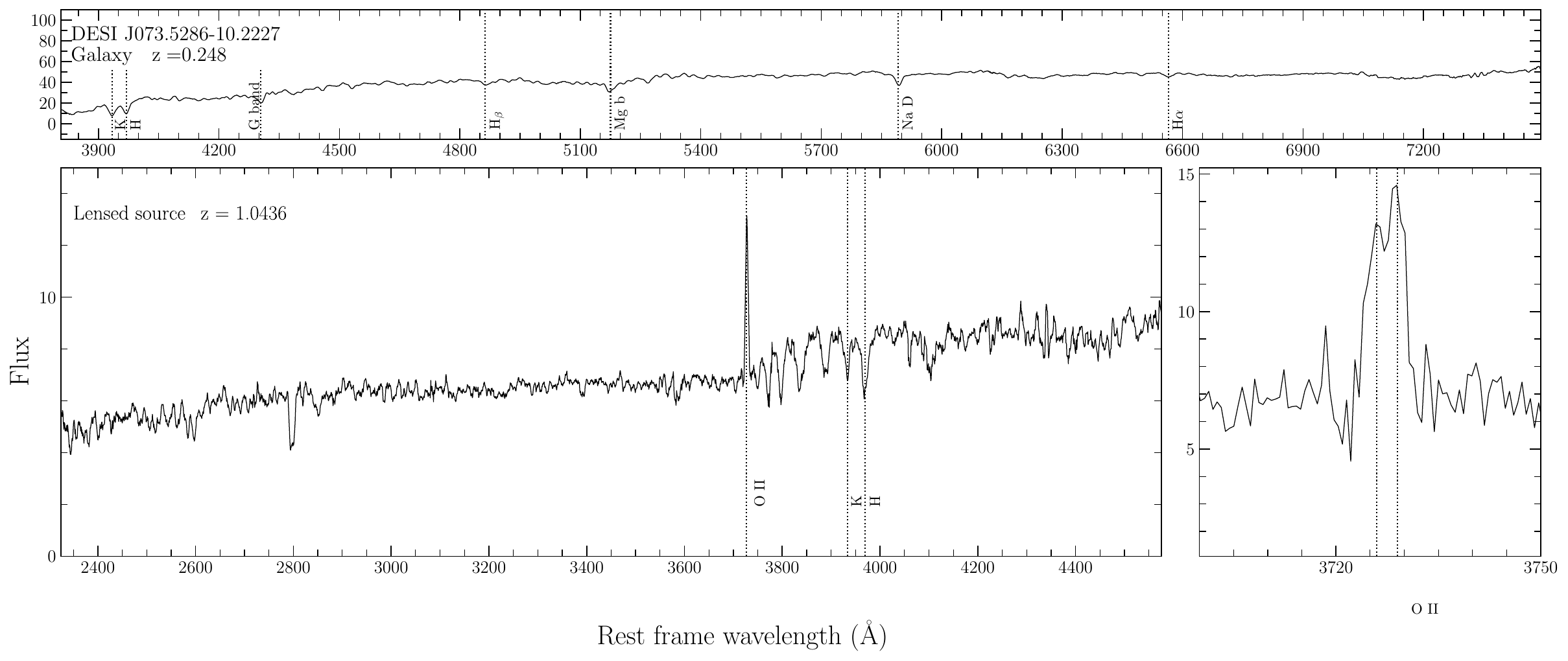}
\caption{\textit{Top:} RGB image of gravitational lens system DESI~J073.5286-10.2227 observed with MUSE. \textit{Bottom:} MUSE spectra of DESI~J073.5286-10.2227. For more information on the system, see Desc. \ref{Ref:lens26}.}
\label{fig:MUSEspectra26}
\end{minipage}
\end{figure*}

\begin{figure*}[!ht]
\centering
\begin{minipage}{1.0\textwidth}
\centering
\includegraphics[width=0.4\textwidth]{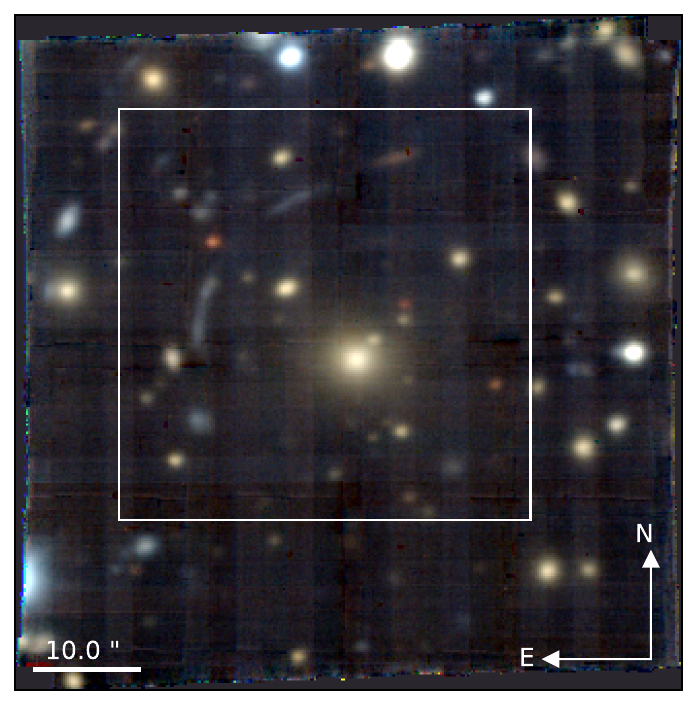}
\includegraphics[width=0.404\textwidth]{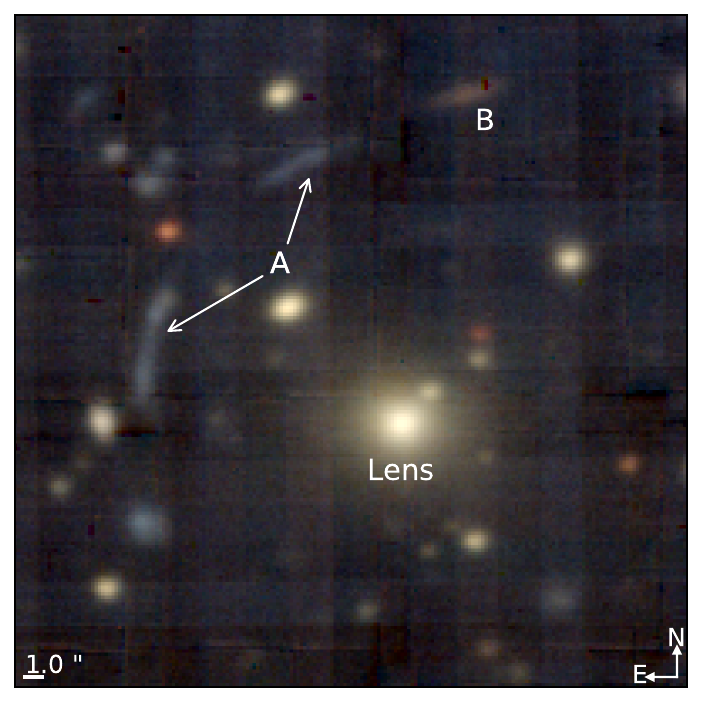}

\includegraphics[width=\textwidth]{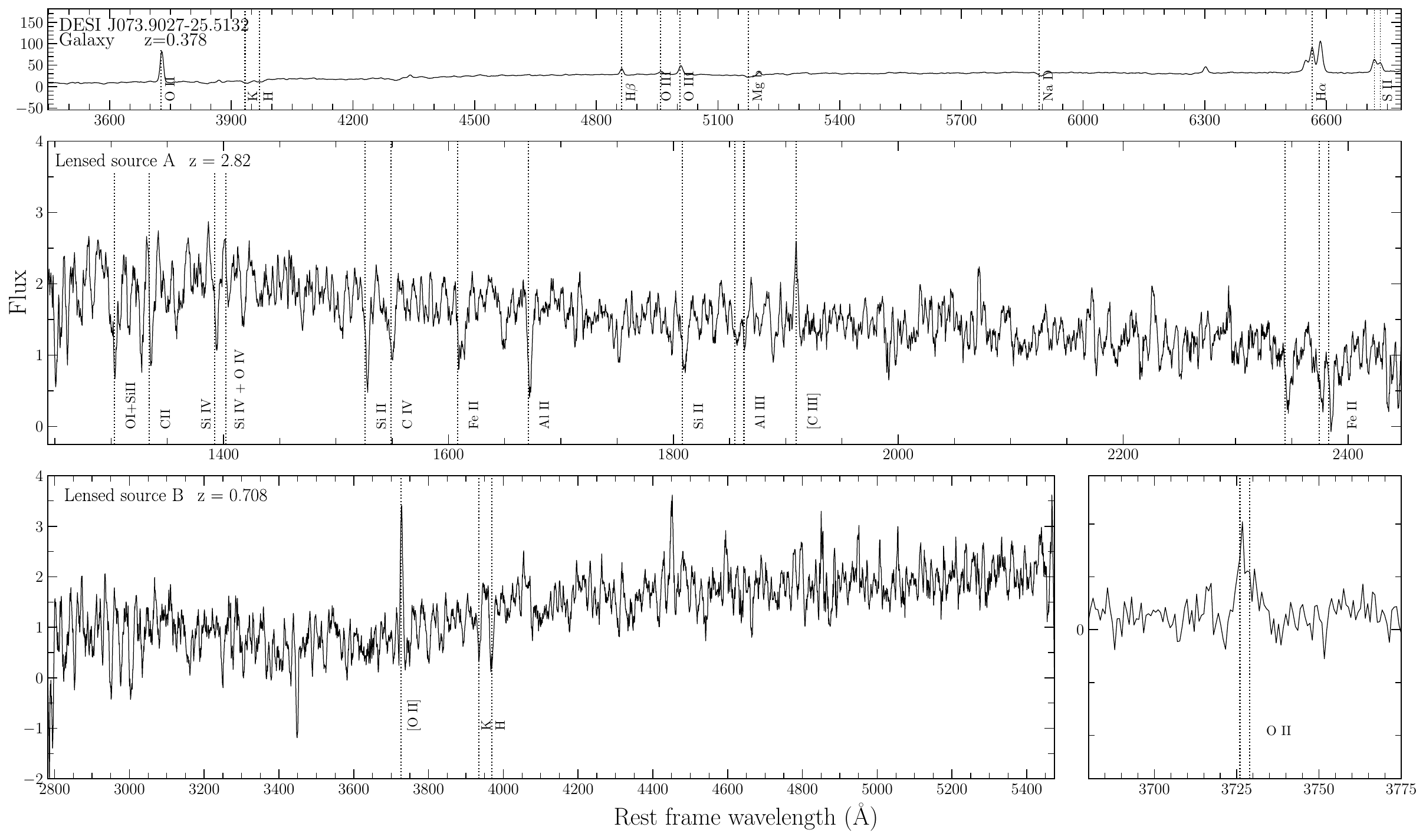}
\caption{\textit{Top:} RGB image of gravitational lens system DESI~J073.9027-25.5132 observed with MUSE. \textit{Bottom:} MUSE spectra of DESI~J073.9027-25.5132. For more information on the system, see Desc. \ref{Ref:lens27}.}
\label{fig:MUSEspectra27}
\end{minipage}
\end{figure*}

\begin{figure*}[!ht]
\centering
\begin{minipage}{1.0\textwidth}
\centering
\includegraphics[width=0.4\textwidth]{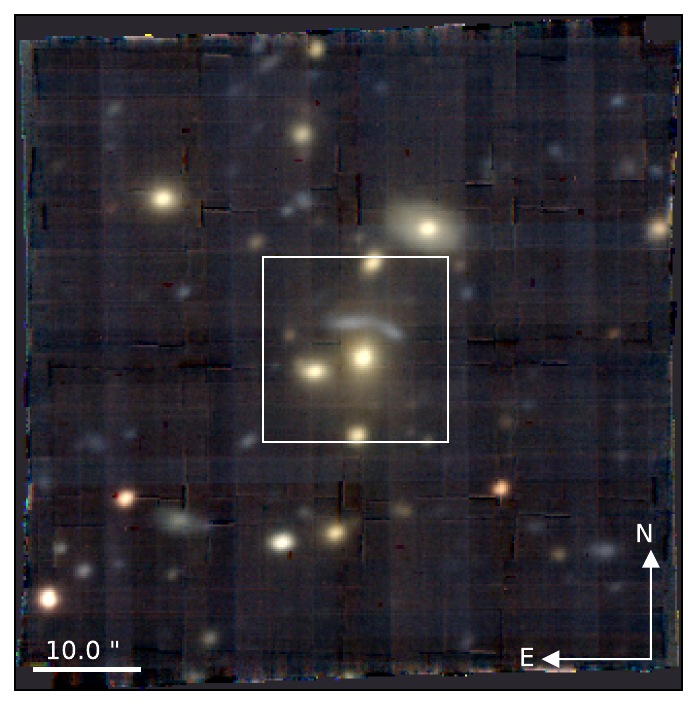}
\includegraphics[width=0.404\textwidth]{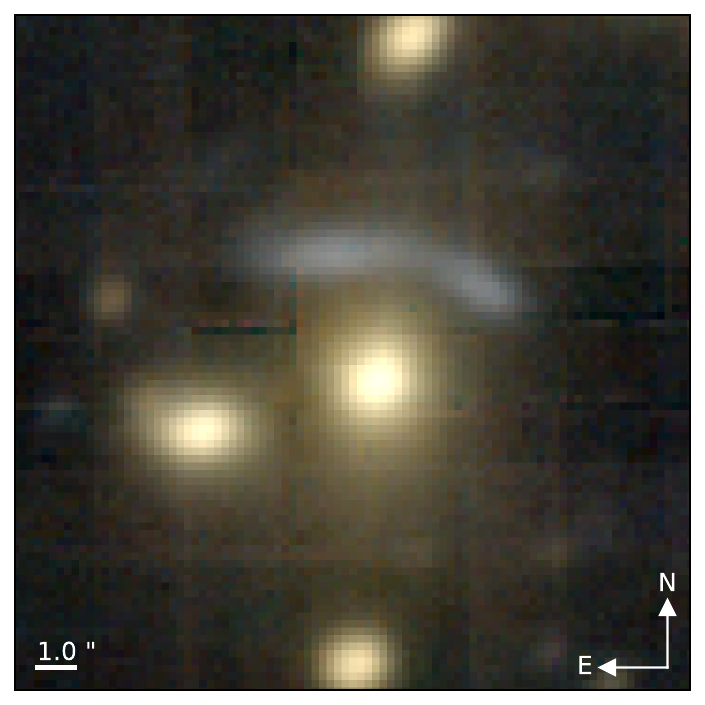}

\includegraphics[width=\textwidth]{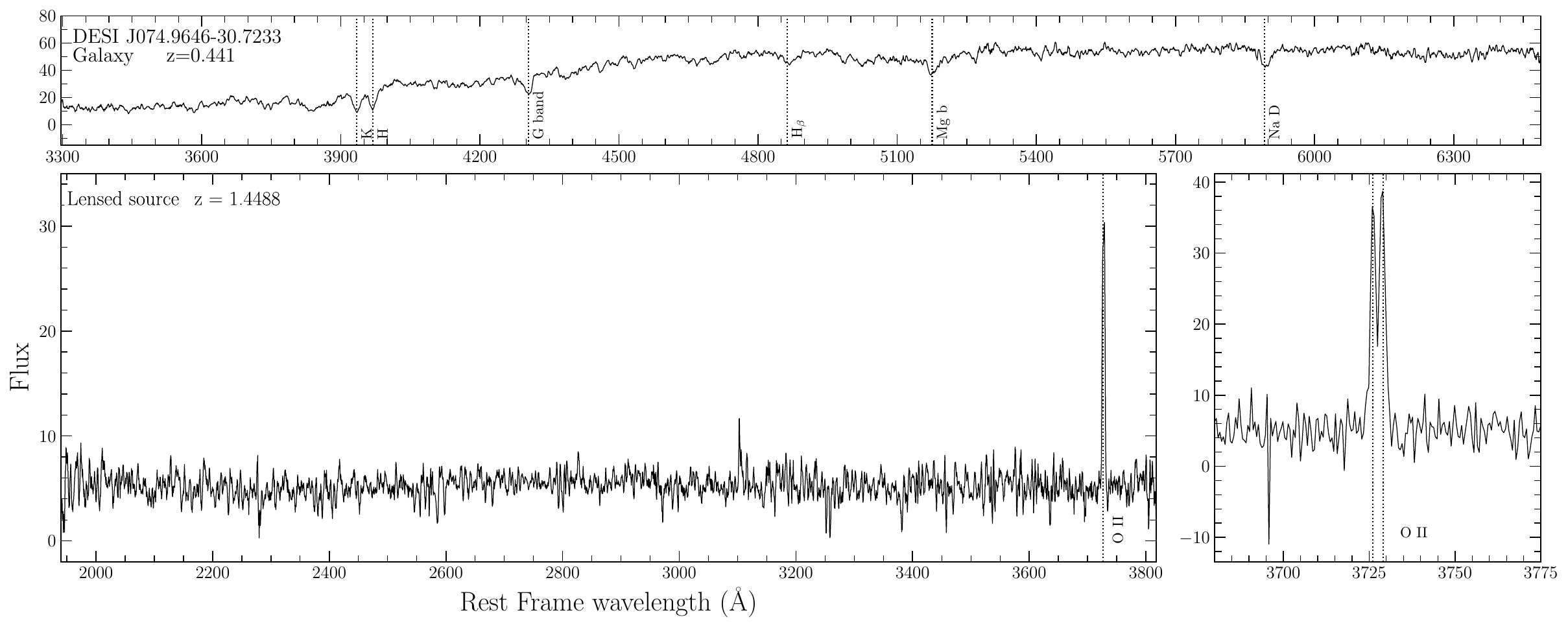}
\caption{\textit{Top:} RGB image of gravitational lens system DESI~J074.9646-30.7233 observed with MUSE. \textit{Bottom:} MUSE spectra of DESI~J074.9646-30.7233. For more information on the system, see Desc. \ref{Ref:lens28}. }
\label{fig:MUSEspectra28}
\end{minipage}
\end{figure*}

\begin{figure*}[!ht]
\centering

\begin{minipage}{1.0\textwidth}
\centering
\includegraphics[width=0.4\textwidth]{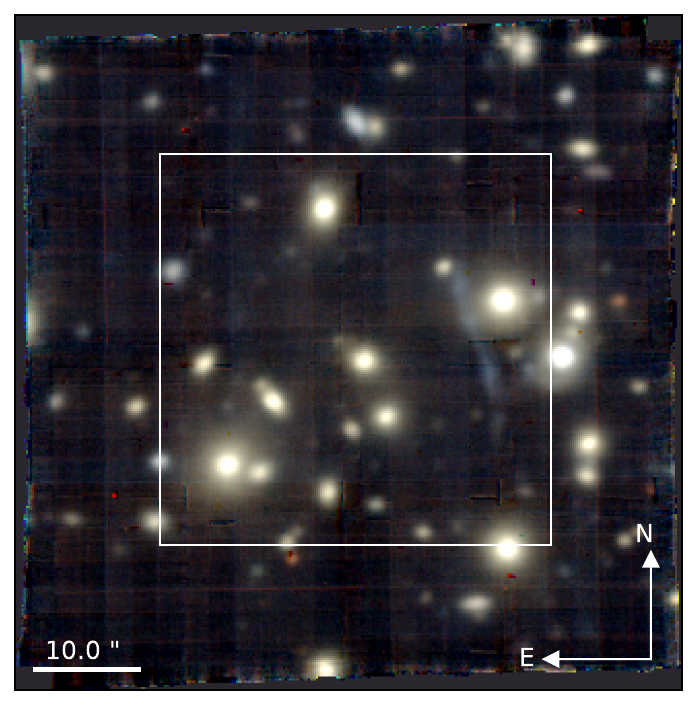}
\includegraphics[width=0.404\textwidth]{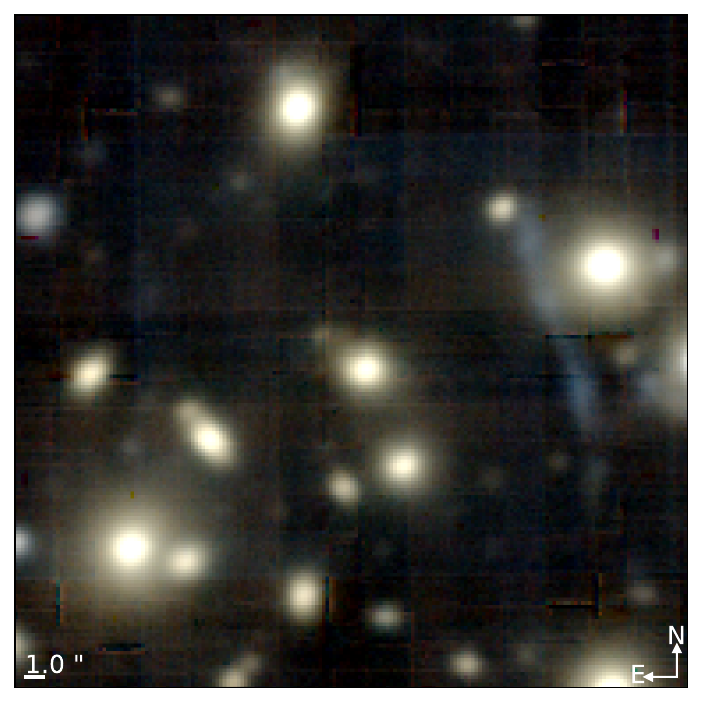}

\includegraphics[width=\textwidth]{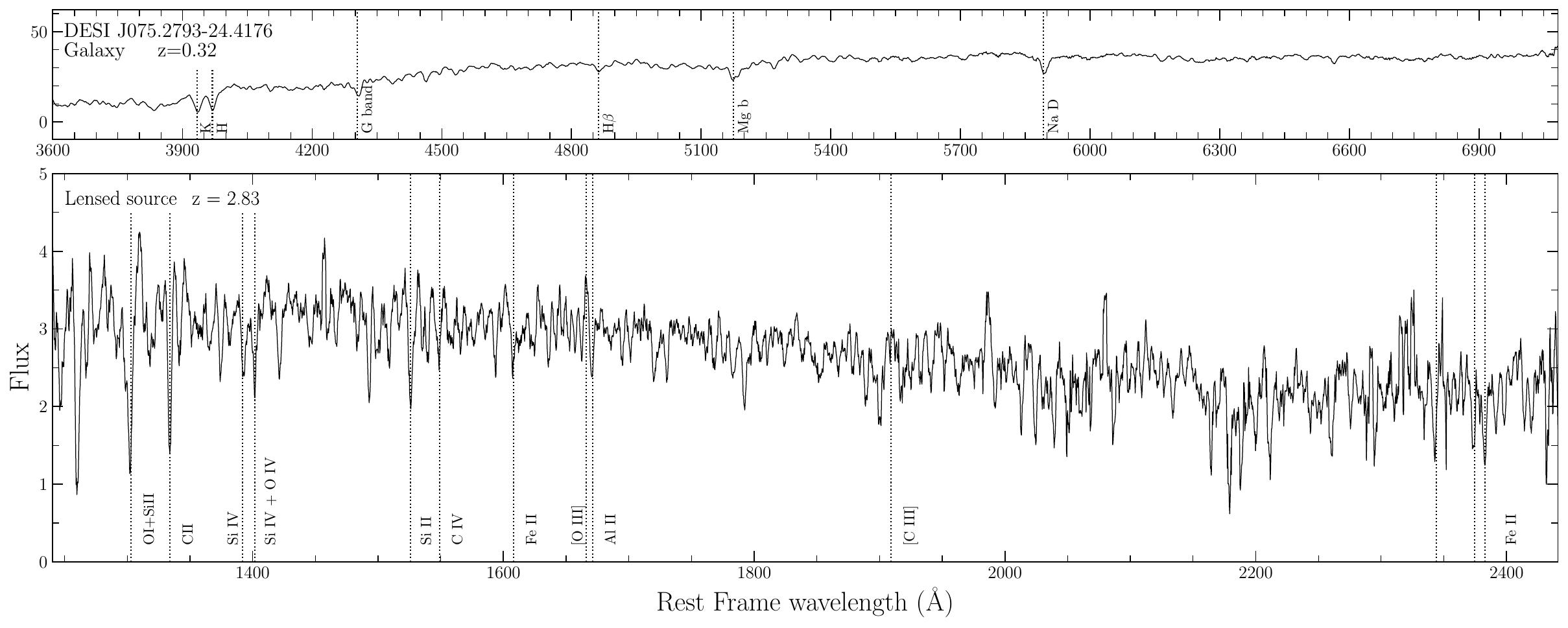}
\caption{\textit{Top:} RGB image of gravitational lens system DESI~J075.2793-24.4176 observed with MUSE. \textit{Bottom:} MUSE spectra of DESI~J075.2793-24.4176. For more information on the system, see Desc. \ref{Ref:lens29}.}
\label{fig:MUSEspectra29}
\end{minipage}
\end{figure*}

\begin{figure*}[!ht]
\centering
\begin{minipage}{1.0\textwidth}
\centering
\includegraphics[width=0.4\textwidth]{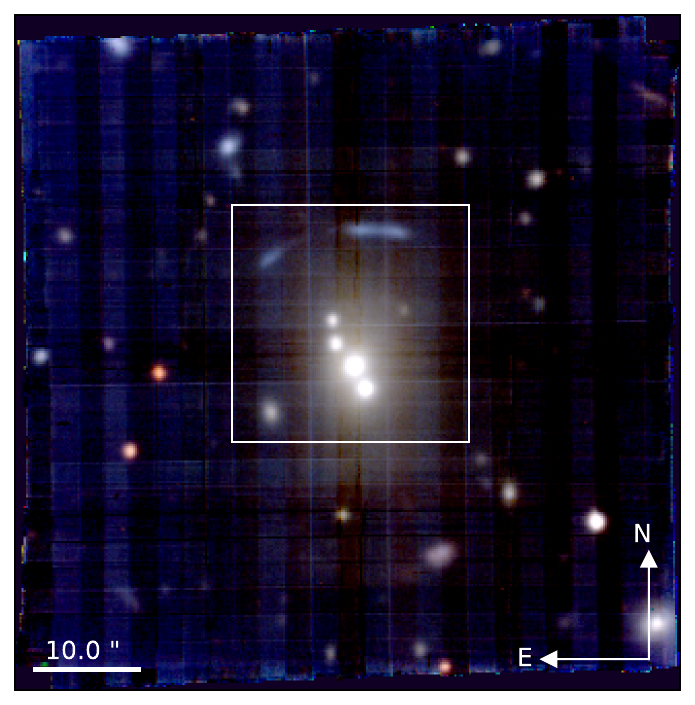}
\includegraphics[width=0.404\textwidth]{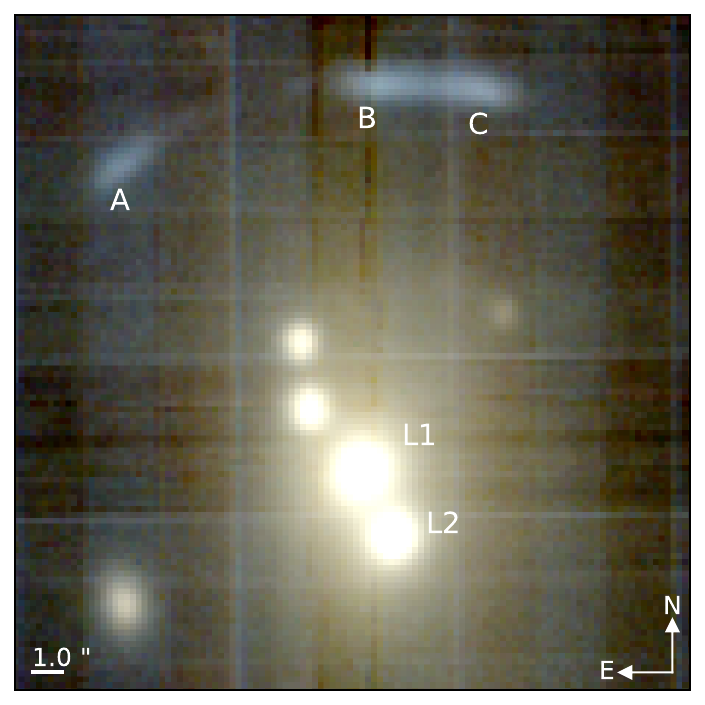}

\includegraphics[width=\textwidth]{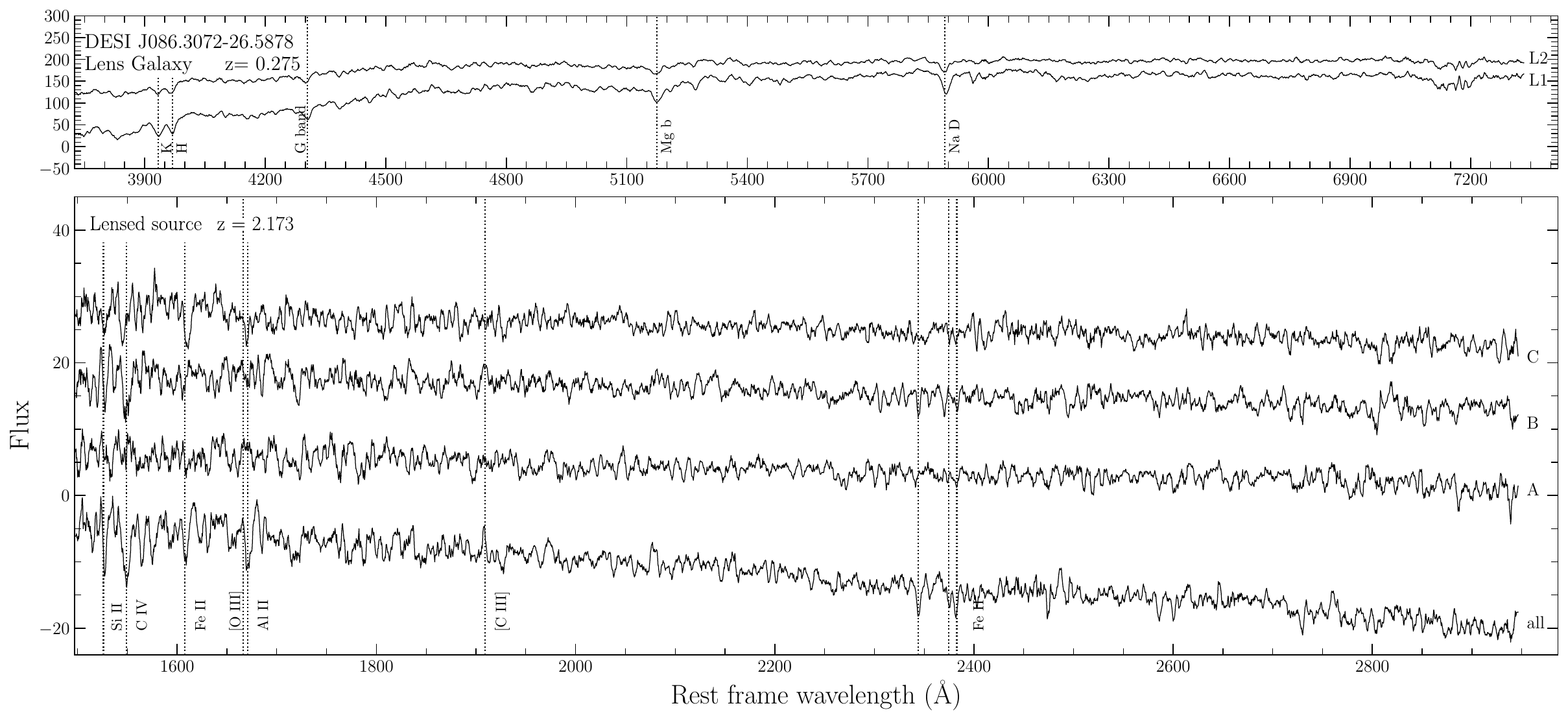}
\caption{\textit{Top:} RGB image of gravitational lens system DESI~J086.3072-26.5878 observed with MUSE. \textit{Bottom:} MUSE spectra of DESI~J086.3072-26.5878. Note that Source A's quality flag is $Q_z=2$. For more information on the system, see Desc. \ref{Ref:lens111}.}
\label{fig:MUSEspectra111}
\end{minipage}
\end{figure*}

\begin{figure*}[!ht]
\centering

\begin{minipage}{1.0\textwidth}
\centering
\includegraphics[width=0.4\textwidth]{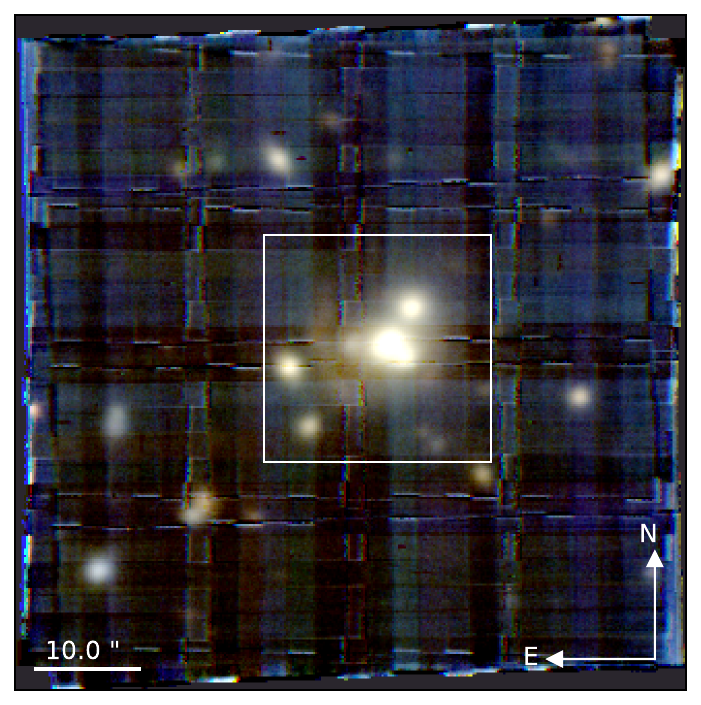}
\includegraphics[width=0.404\textwidth]{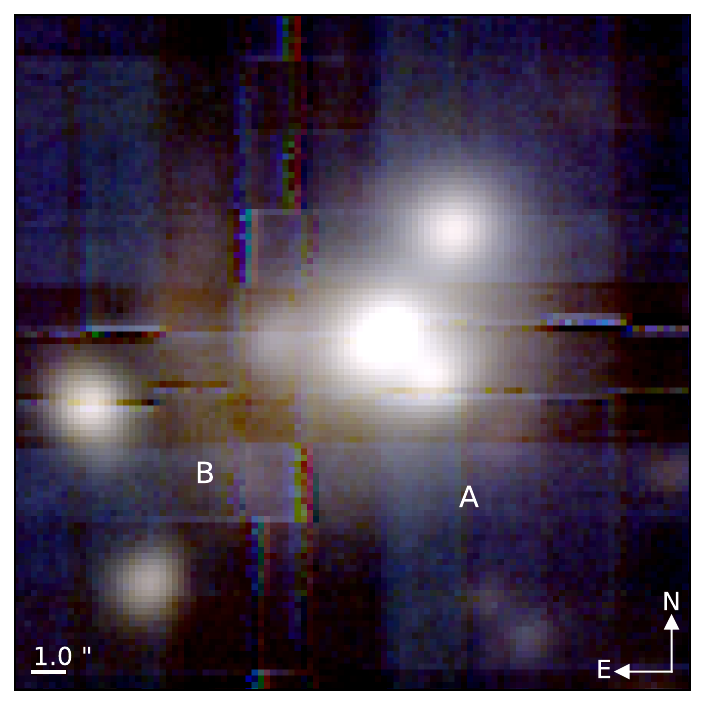}

\includegraphics[width=\textwidth]{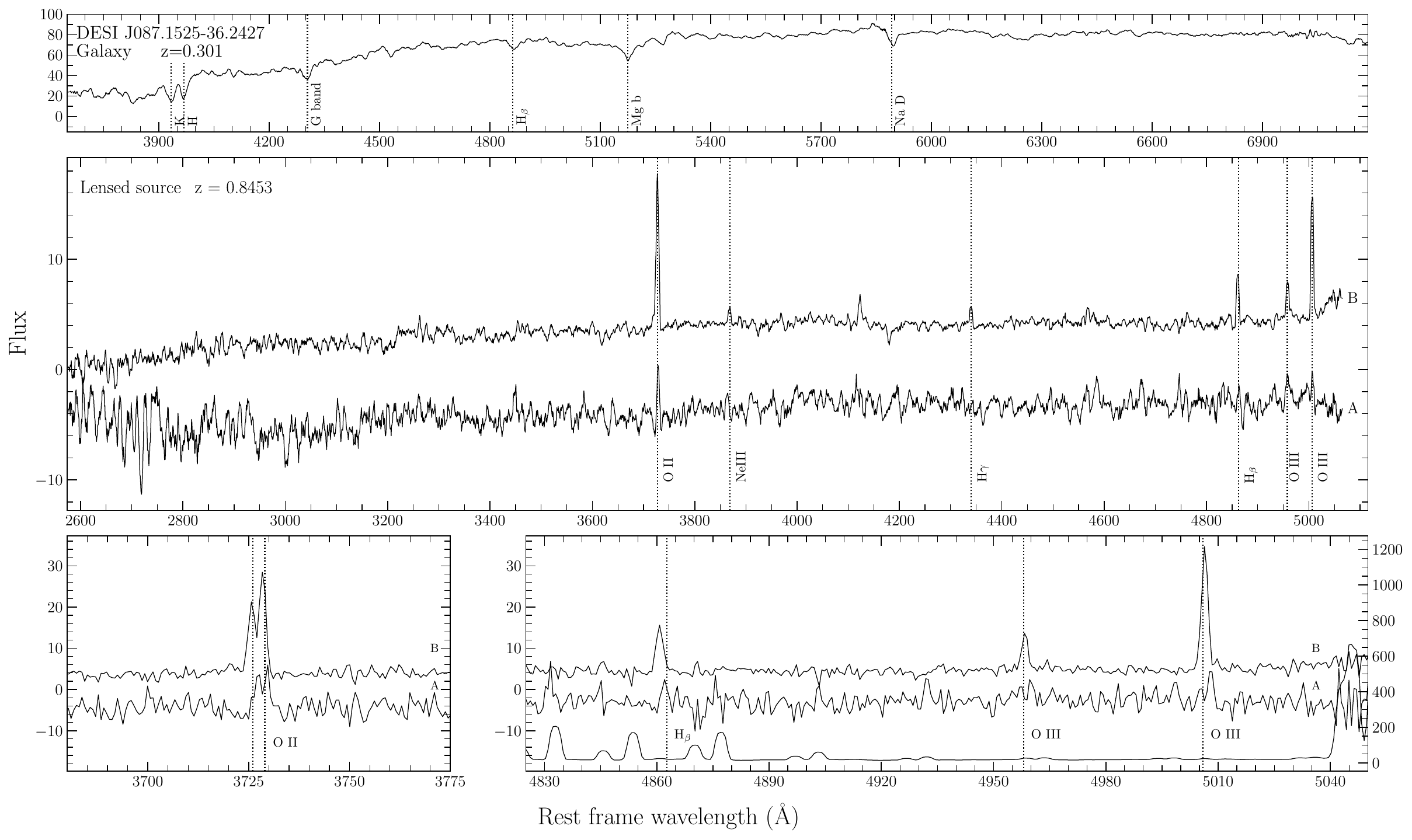}
\caption{\textit{Top:} RGB image of gravitational lens system DESI~J087.1525-36.2427 observed with MUSE. \textit{Bottom:} MUSE spectra of DESI~J087.1525-36.2427. For more information on the system, see Desc. \ref{Ref:lens32}. }
\label{fig:MUSEspectra32}
\end{minipage}
\end{figure*}

\begin{figure*}[!ht]
\centering
\begin{minipage}{1.0\textwidth}
\centering
\includegraphics[width=0.4\textwidth]{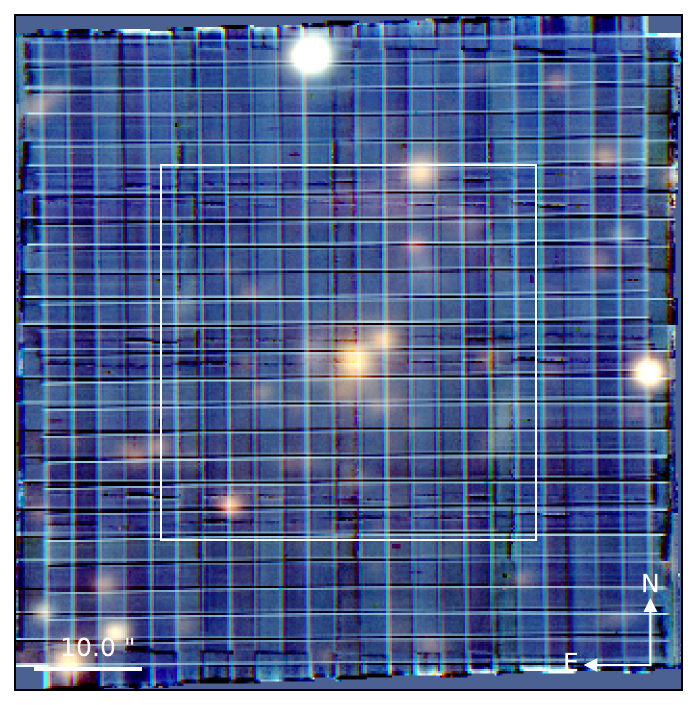}
\includegraphics[width=0.404\textwidth]{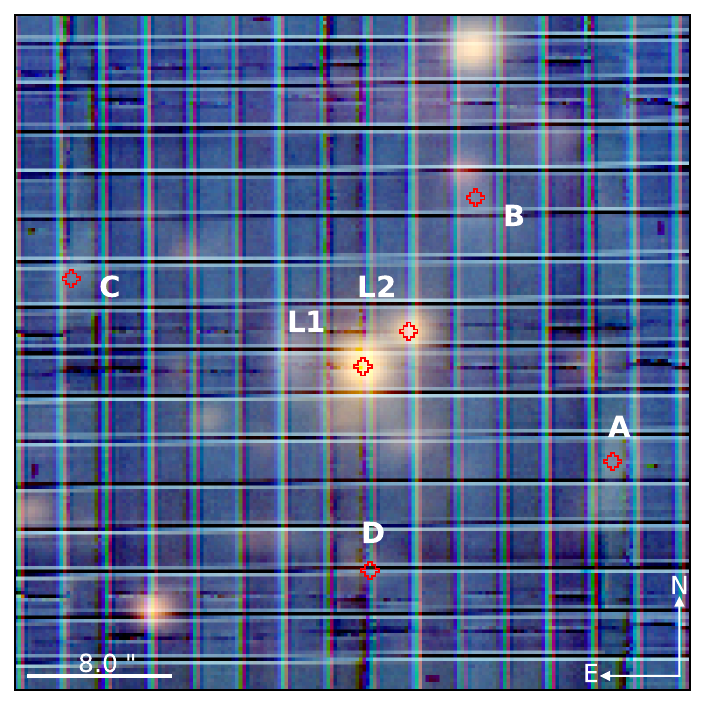}

\includegraphics[width=\textwidth]{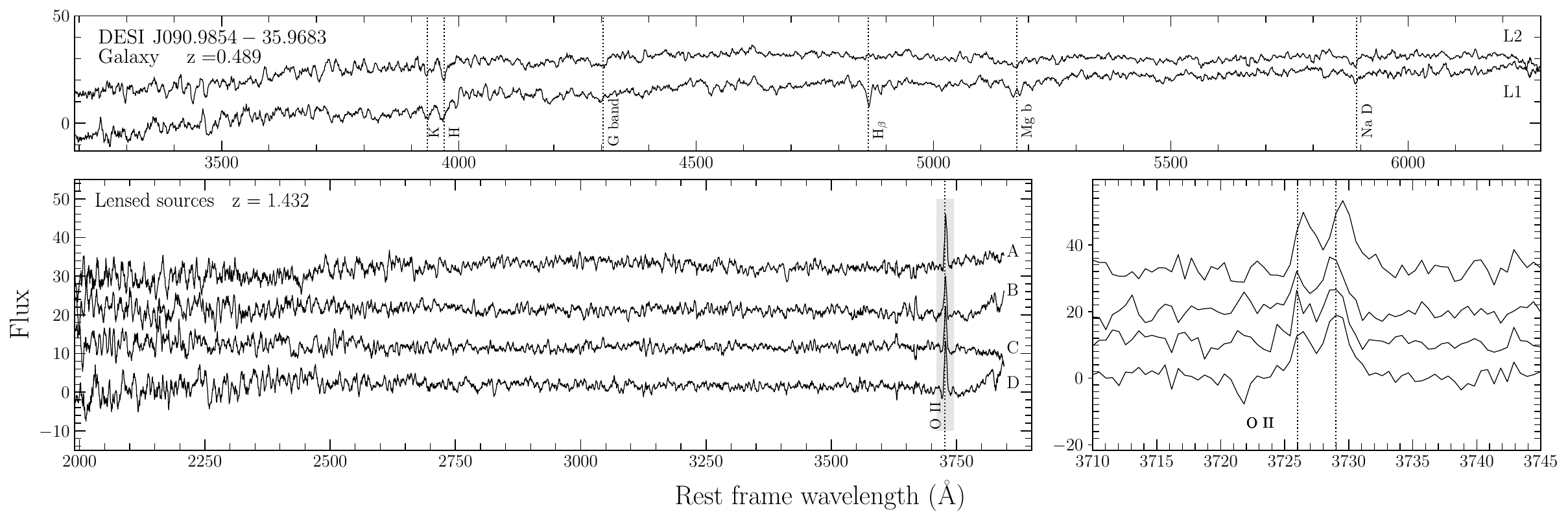}
\caption{\textit{Top:} RGB image of gravitational lens system DESI~J090.9854-35.9683 observed with MUSE. The red contours are extraction apertures of the spectra. \textit{Bottom:} MUSE spectra of DESI~J090.9854-35.9683. For more information on the system, see Desc. \ref{ref:lens16repl}. See \cite{sheu2024carousel} for a more detailed analysis of the Carousel Lens.}
\label{fig:MUSEspectra16repl_all}
\end{minipage}
\end{figure*}

\begin{figure*}[!ht]
\centering
\begin{minipage}{1.0\textwidth}
\centering
\includegraphics[width=0.4\textwidth]{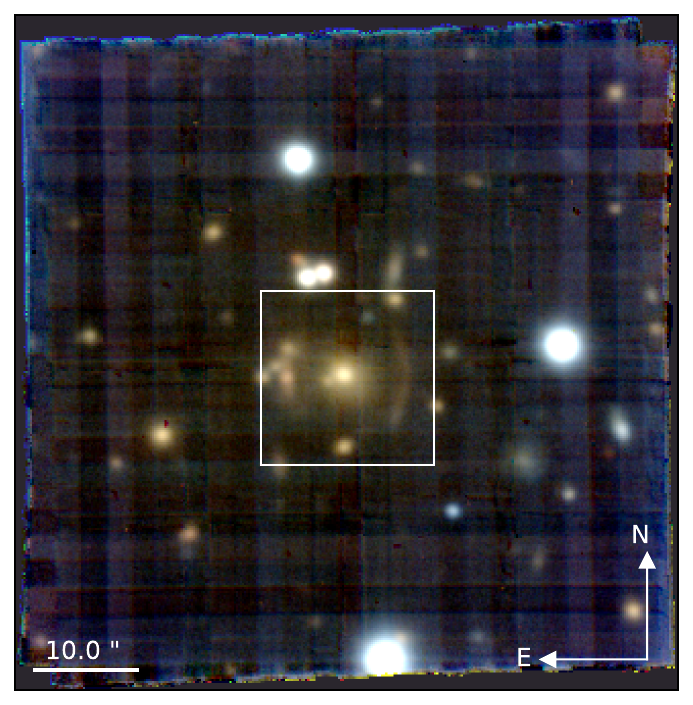}
\includegraphics[width=0.404\textwidth]{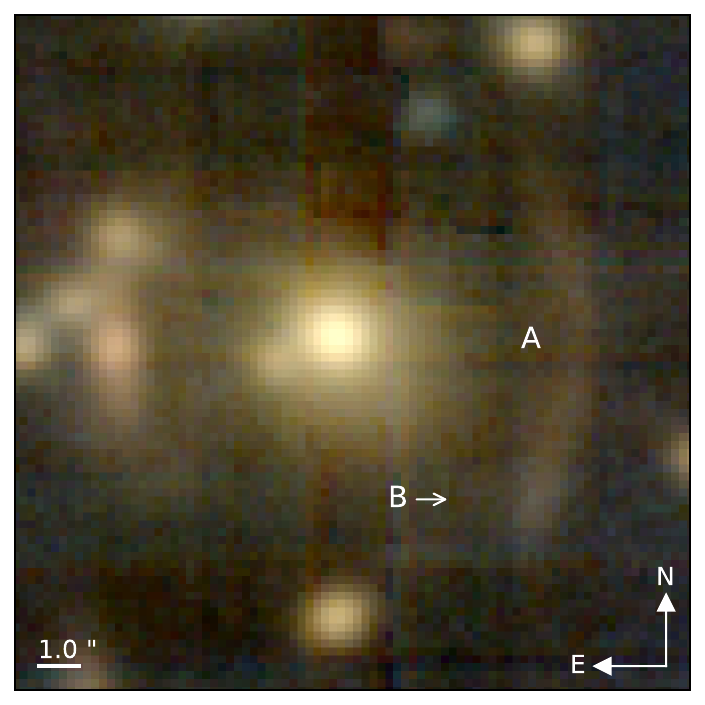}

\includegraphics[width=\textwidth]{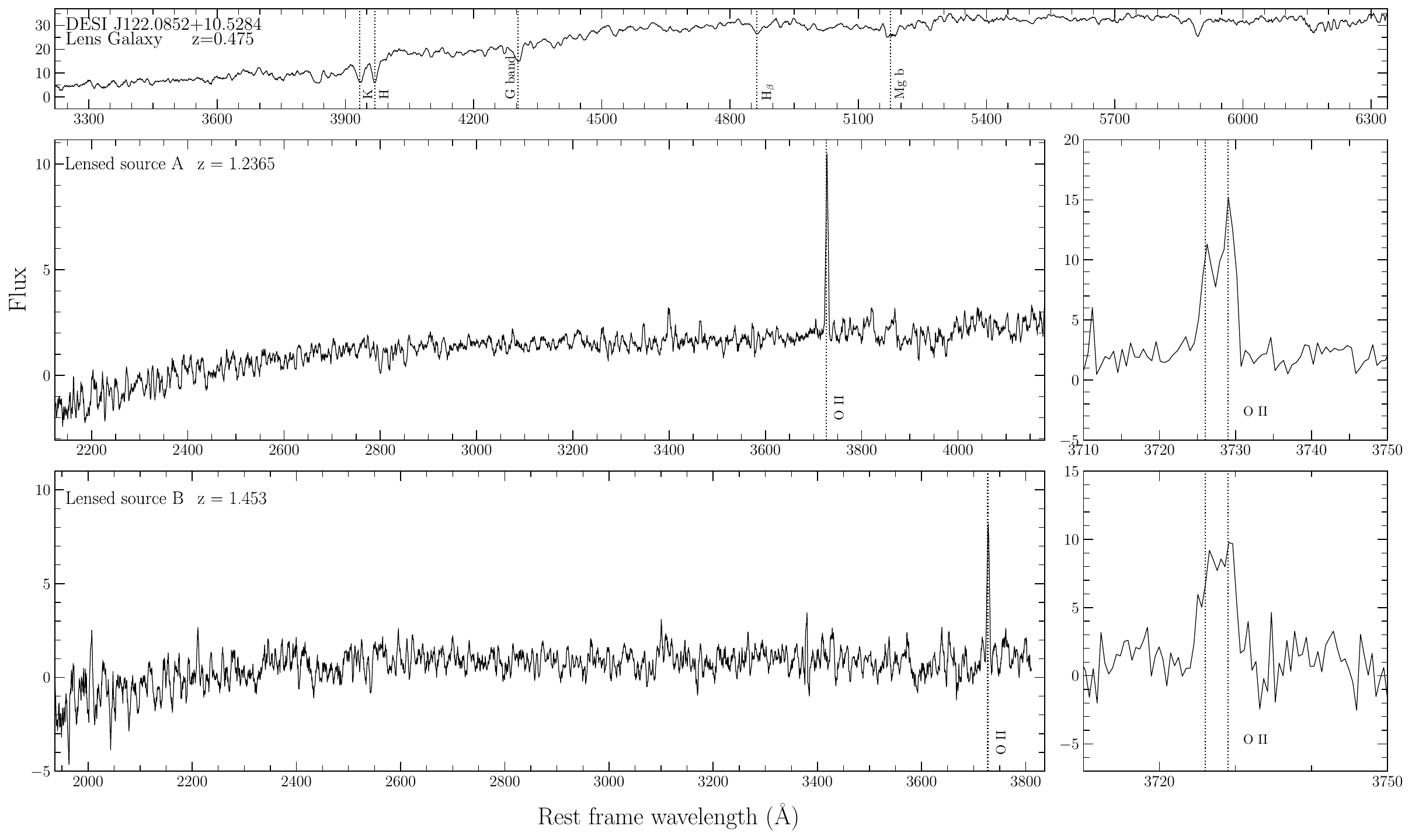}
\caption{\textit{Top:} RGB image of gravitational lens system DESI~J122.0852+10.5284 observed with MUSE. \textit{Bottom:} MUSE spectra of DESI~J122.0852+10.5284. For more information on the system, see Desc. \ref{Ref:lens113}.}
\label{fig:MUSEspectra113}
\end{minipage}
\end{figure*}

\begin{figure*}[!ht]
\centering
\begin{minipage}{1.0\textwidth}
\centering
\includegraphics[width=0.4\textwidth]{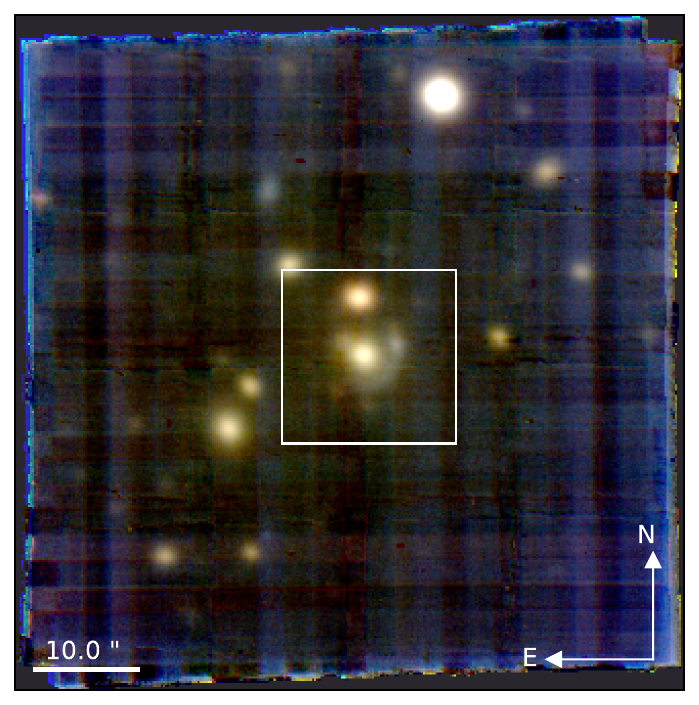}
\includegraphics[width=0.404\textwidth]{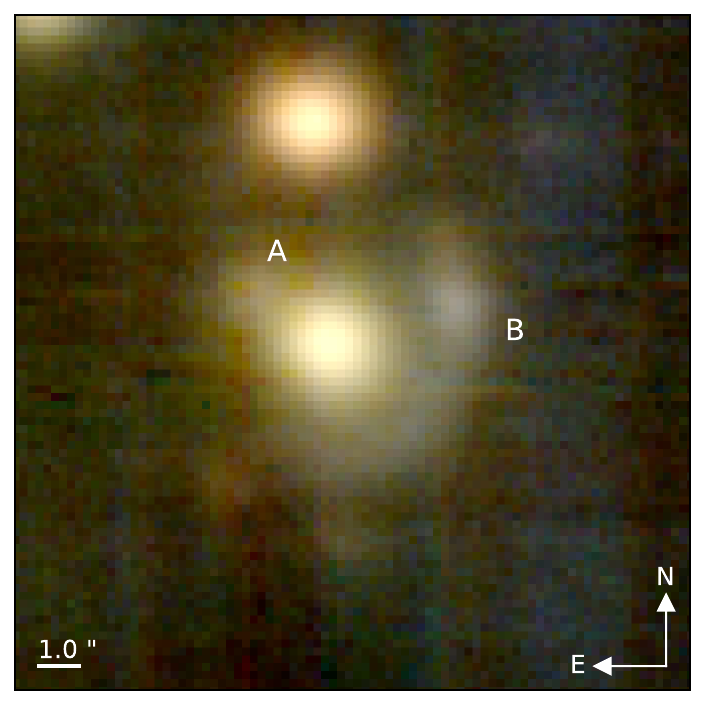}

\includegraphics[width=\textwidth]{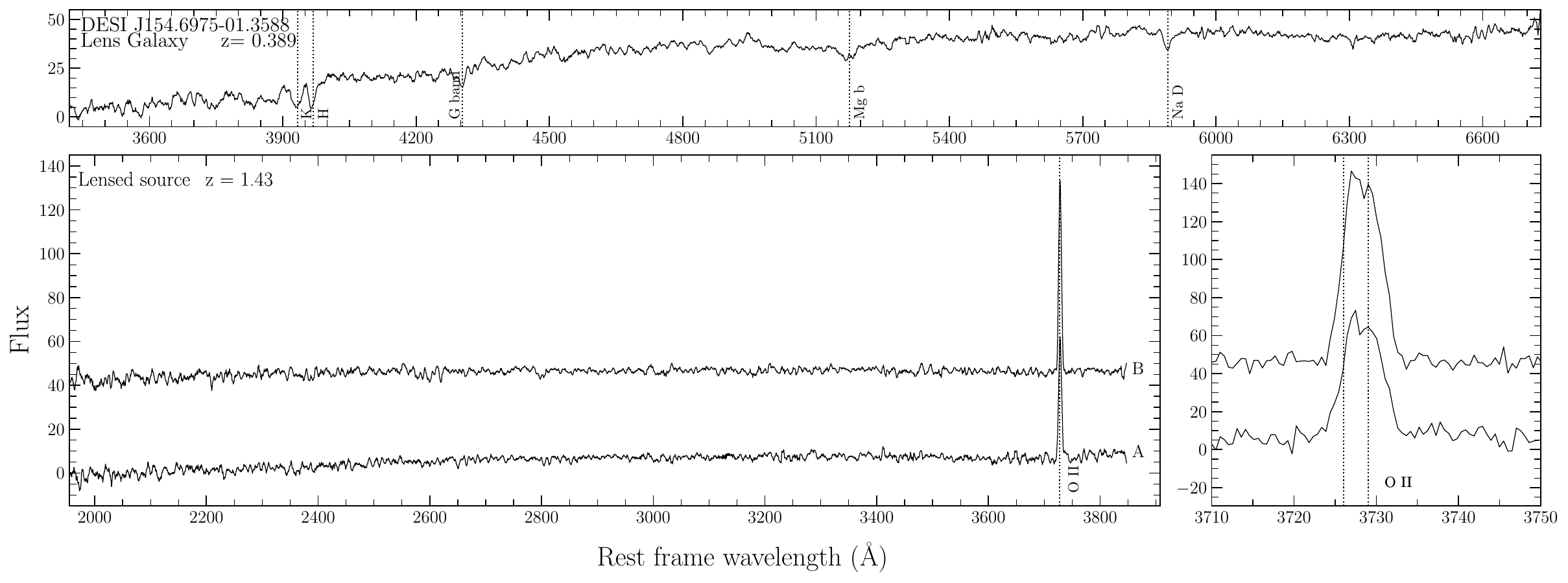}
\caption{\textit{Top:} RGB image of gravitational lens system DESI~J154.6975-01.3588 observed with MUSE. \textit{Bottom:} MUSE spectra of DESI~J154.6975-01.3588. For more information on the system, see Desc. \ref{Ref:lens107}. }
\label{fig:MUSEspectra107}
\end{minipage}
\end{figure*}

\begin{figure*}[!ht]
\centering
\begin{minipage}{1.0\textwidth}
\centering
\includegraphics[width=0.4\textwidth]{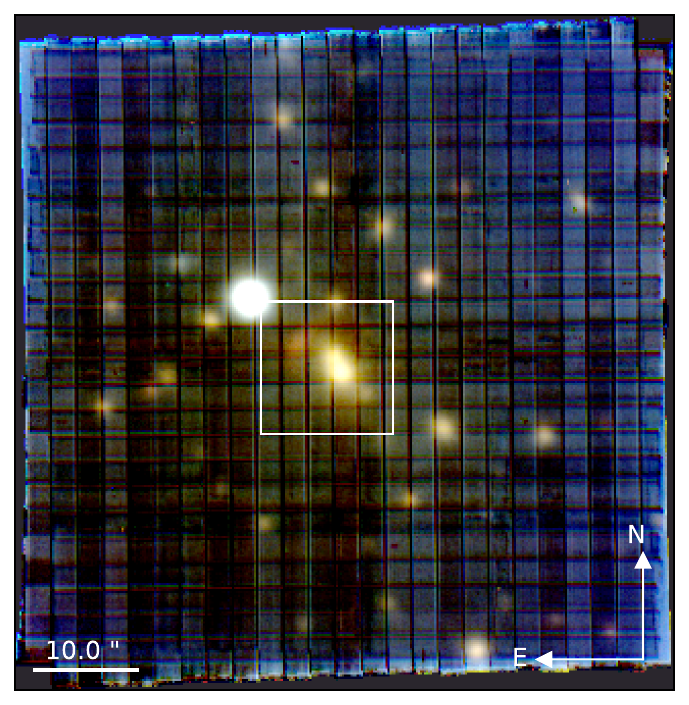}
\includegraphics[width=0.404\textwidth]{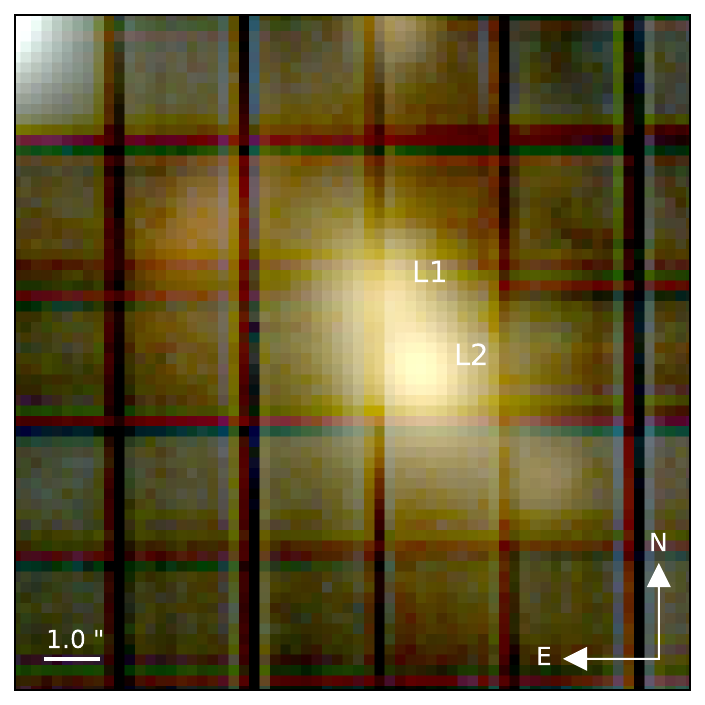}

\includegraphics[width=\textwidth]{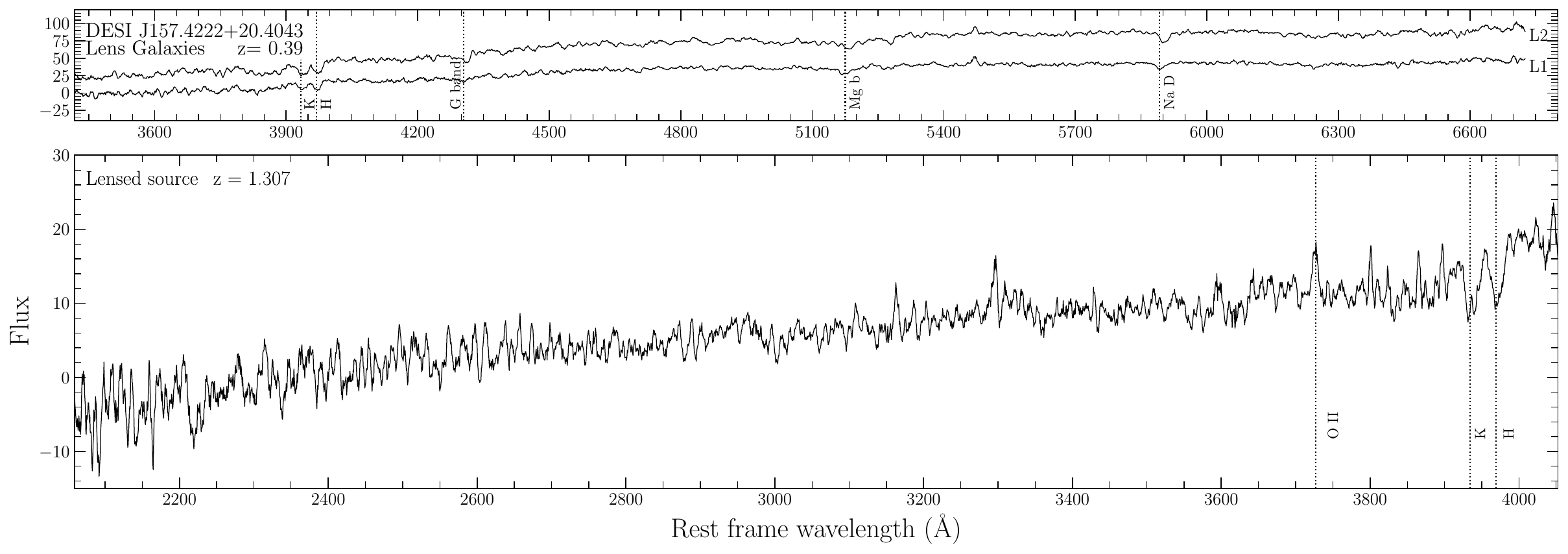}
\caption{\textit{Top:} RGB image of gravitational lens system DESI~J157.4222+20.4043 observed with MUSE. \textit{Bottom:} MUSE spectra of DESI~J157.4222+20.4043. For more information on the system, see Desc. \ref{Ref:lens115}. }
\label{fig:MUSEspectra115}
\end{minipage}
\end{figure*}

\begin{figure*}[!ht]
\centering
\begin{minipage}{1.0\textwidth}
\centering
\includegraphics[width=0.4\textwidth]{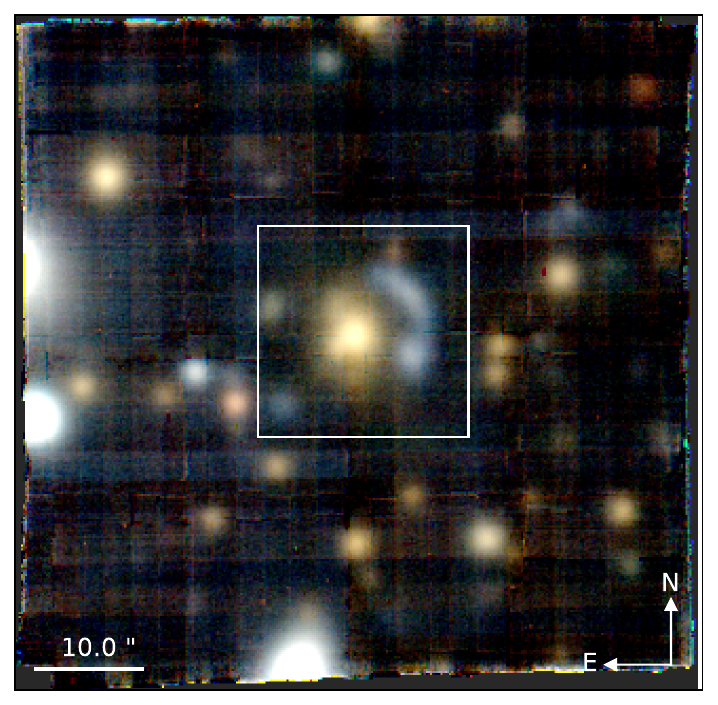}
\includegraphics[width=0.404\textwidth]{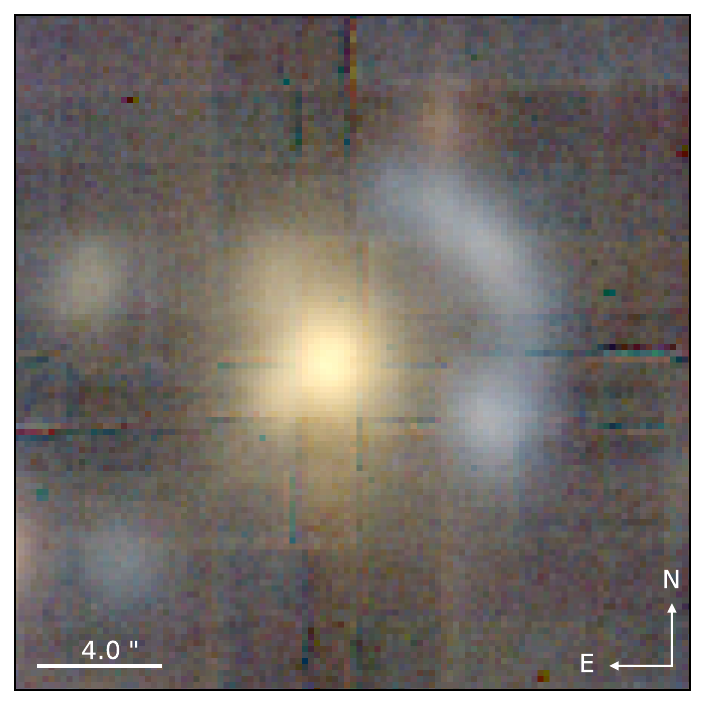}
\label{fig:MUSEspectra34img}
\includegraphics[width=\textwidth]{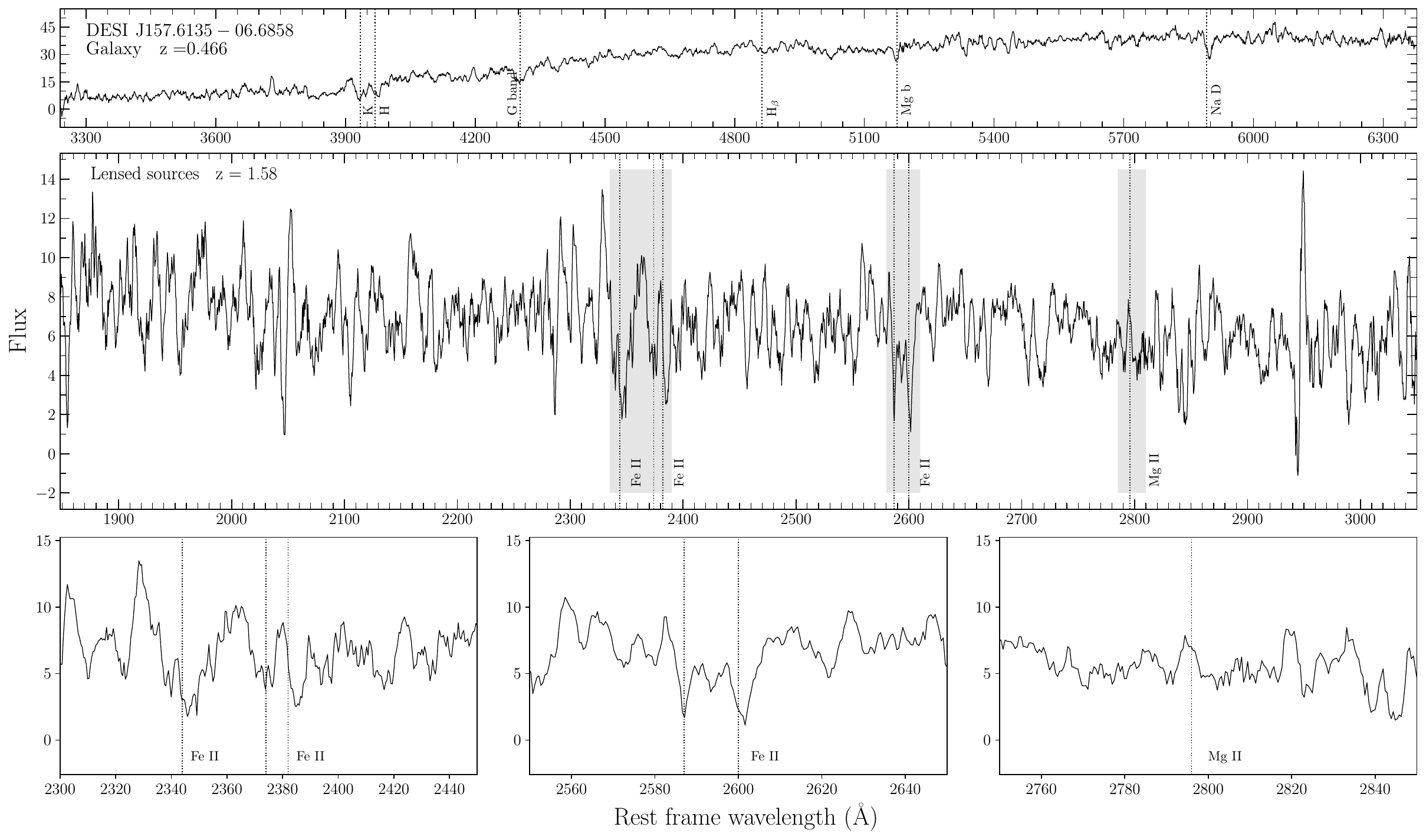}
\caption{\textit{Top:} RGB image of gravitational lens system DESI~J157.6135-06.6858 observed with MUSE. \textit{Bottom:} MUSE spectra of DESI~J157.6135-06.6858. For more information on the system, see Desc. \ref{Ref:lens34}. }
\label{fig:MUSEspectra34}
\end{minipage}
\end{figure*}

\begin{figure*}[!ht]
\centering
\begin{minipage}{1.0\textwidth}
\centering
\includegraphics[width=0.4\textwidth]{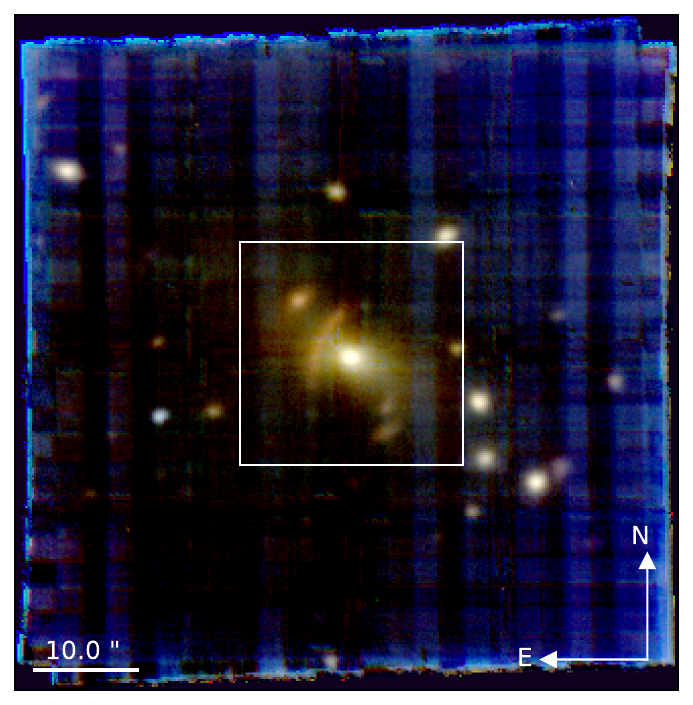}
\includegraphics[width=0.404\textwidth]{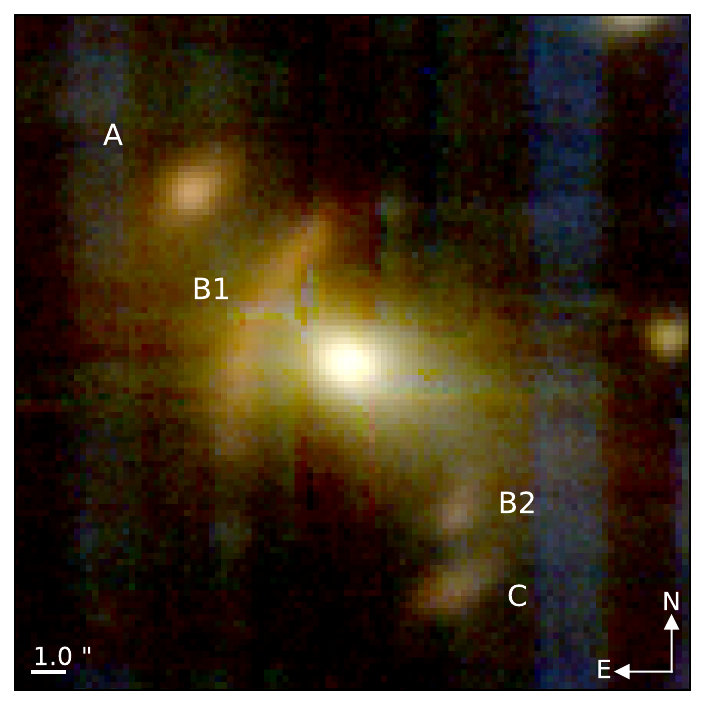}

\includegraphics[width=\textwidth]{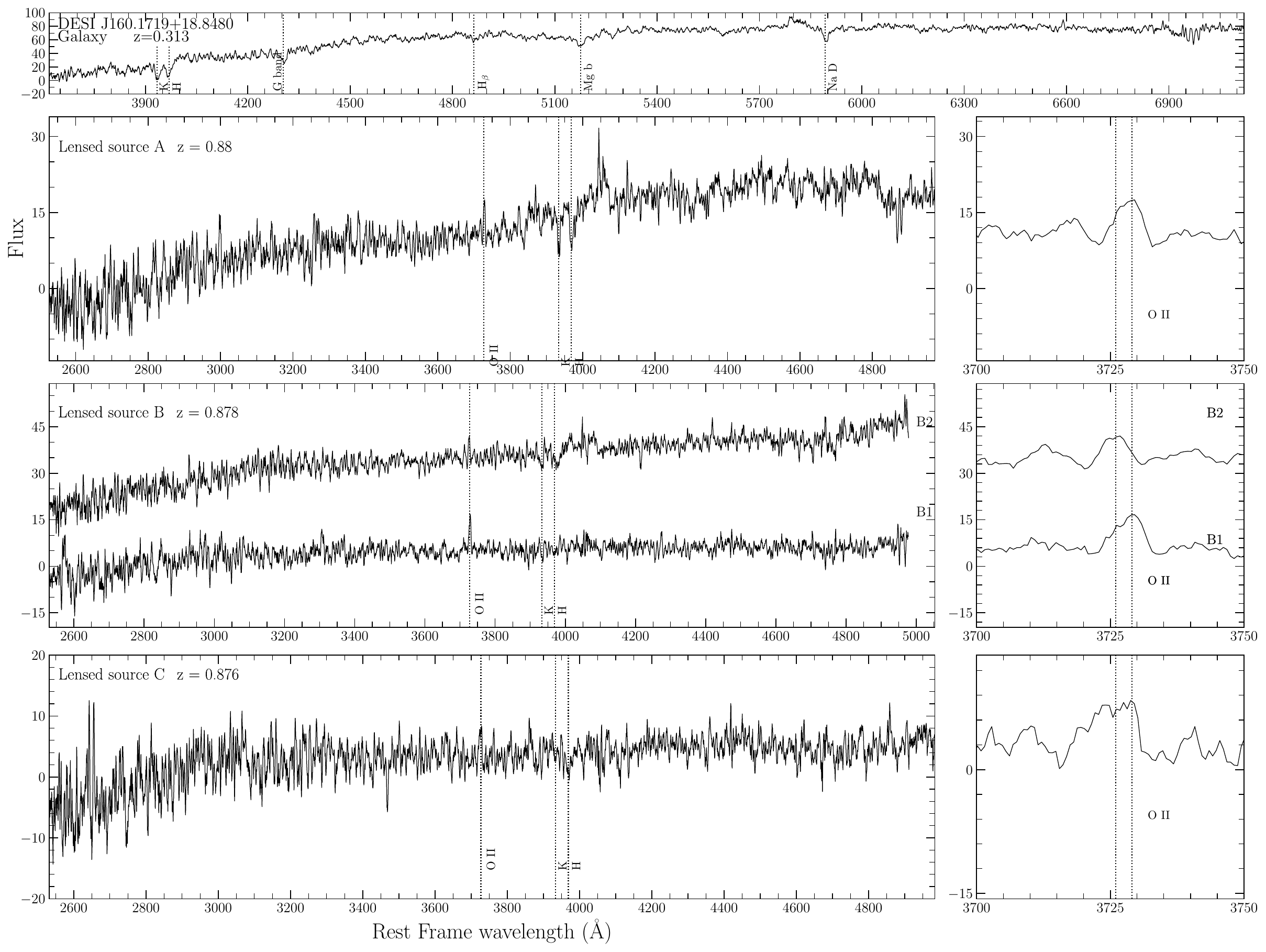}
\caption{\textit{Top:} RGB image of gravitational lens system DESI~J160.1719+18.8480 observed with MUSE. \textit{Bottom:} MUSE spectra of DESI~J160.1719+18.8480. For more information on the system, see Desc. \ref{Ref:lens35}.}
\label{fig:MUSEspectra35}
\end{minipage}
\end{figure*}

\begin{figure*}[!ht]
\centering
\begin{minipage}{1.0\textwidth}
\centering
\includegraphics[width=0.4\textwidth, trim=11mm 8mm 0mm 0mm, clip]{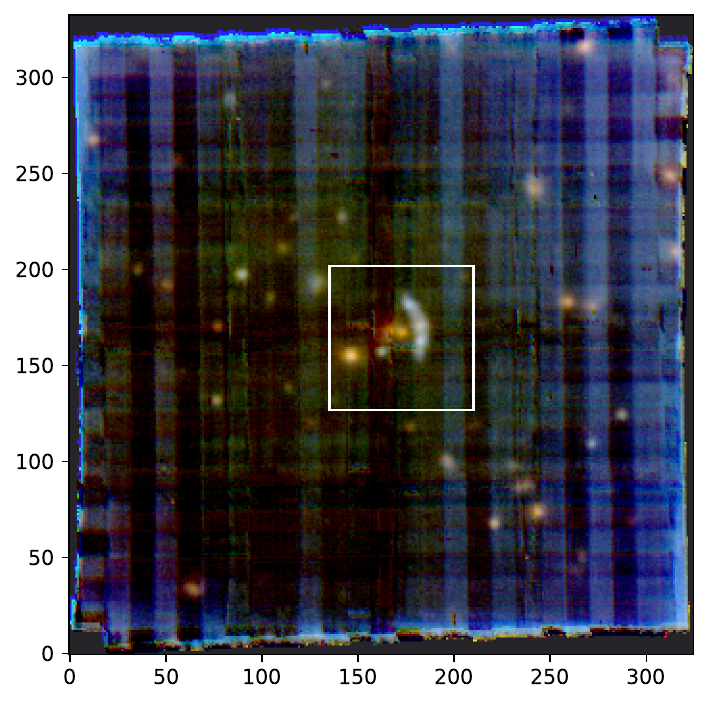}
\includegraphics[width=0.41\textwidth]{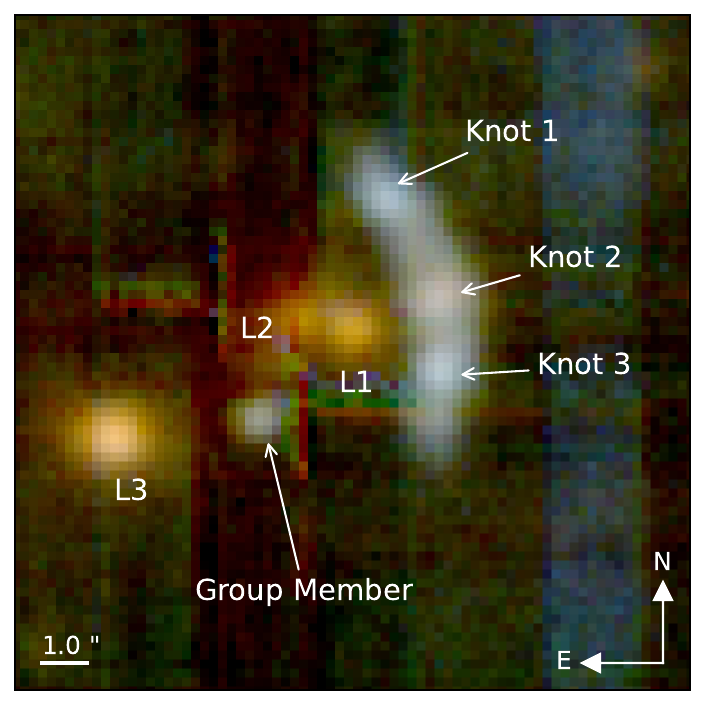}

\includegraphics[width=\textwidth]{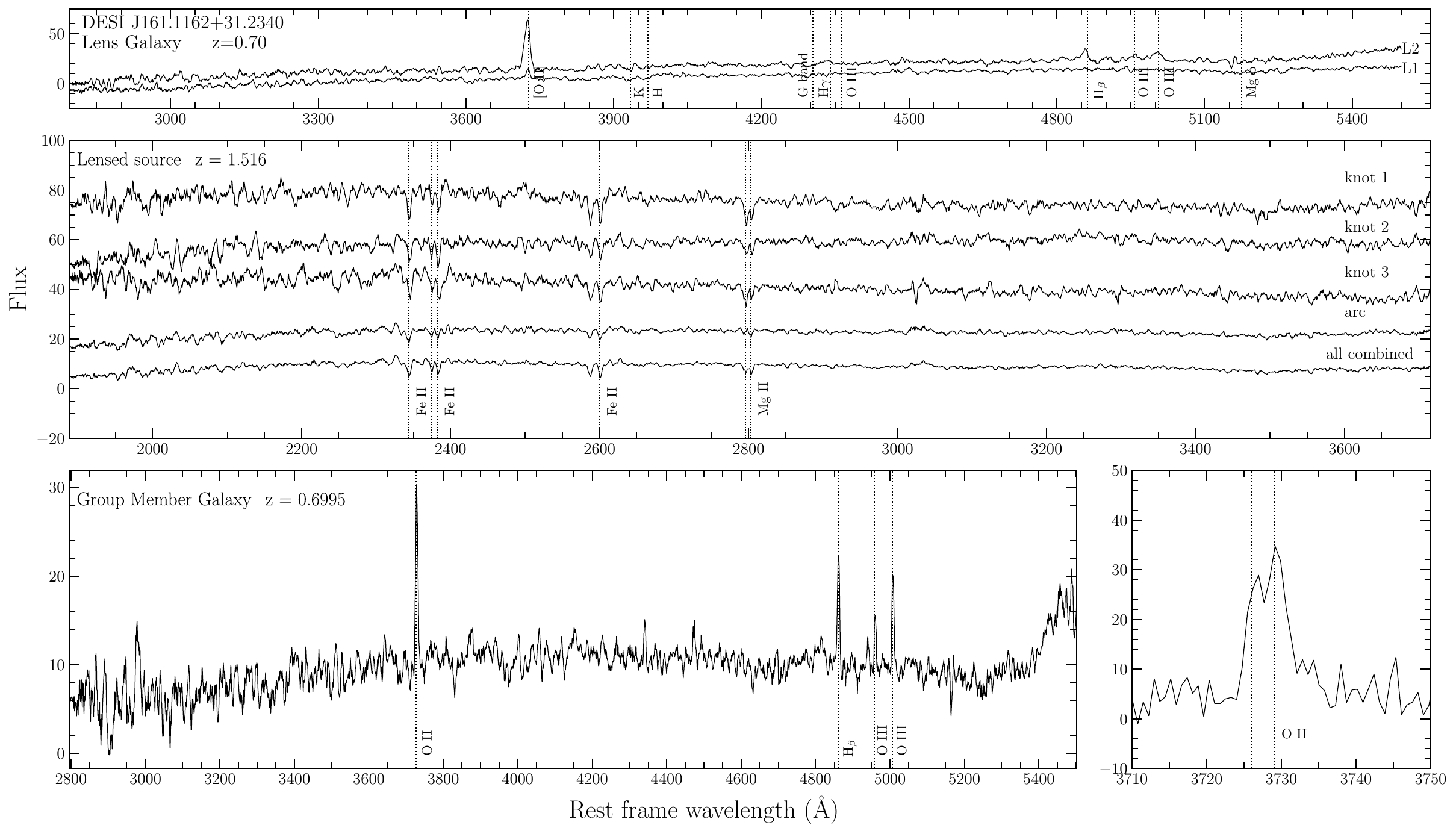}
\caption{\textit{Top:} RGB image of gravitational lens system DESI~J161.1162+31.2340 observed with MUSE. \textit{Bottom:} MUSE spectra of DESI~J161.1162+31.2340. Note that the source quality flag is $Q_z=3$. For more information on the system, see Desc. \ref{ref:lens36}.}
\label{fig:MUSEspectra36}
\end{minipage}
\end{figure*}

\begin{figure*}[!ht]
\centering
\begin{minipage}{1.0\textwidth}
\centering
\includegraphics[width=0.4\textwidth]{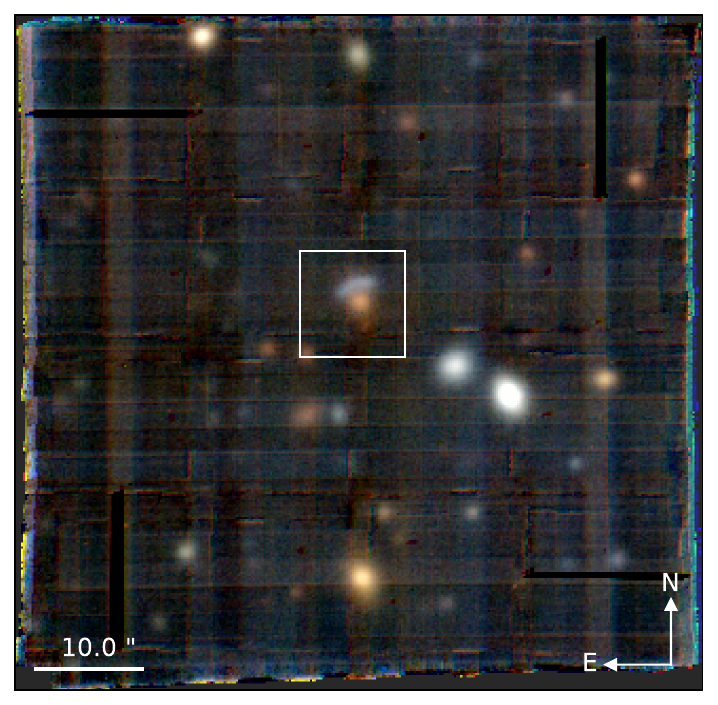}
\includegraphics[width=0.404\textwidth]{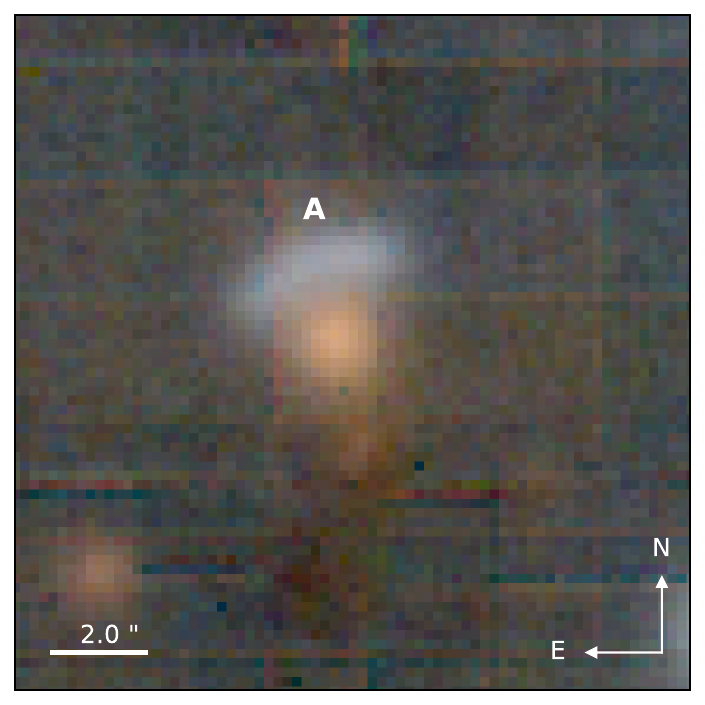}

\includegraphics[width=\textwidth]{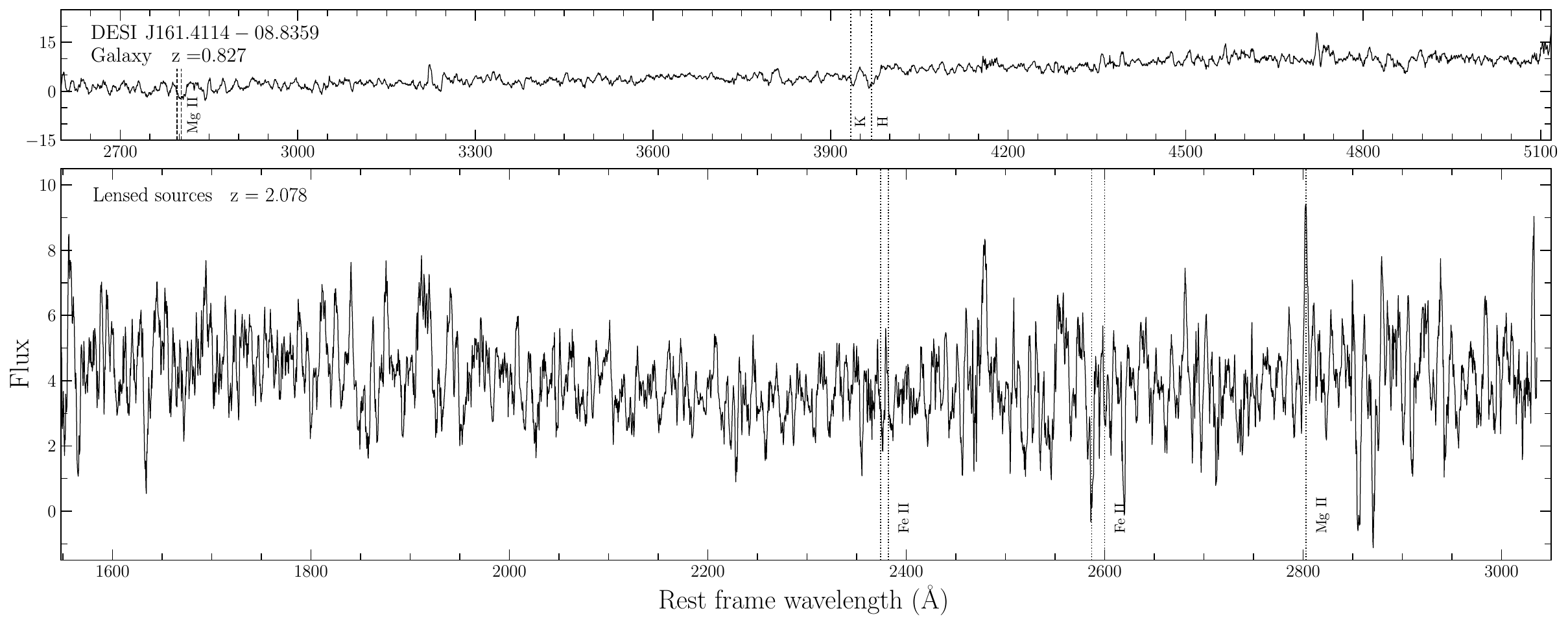}
\caption{\textit{Top:} RGB image of gravitational lens system DESI~J161.4114-8.8358 observed with MUSE. \textit{Bottom:} MUSE spectra of DESI~J161.4114-8.8358. Note that the source quality flag is $Q_z=3$. For more information on the system, see Desc. \ref{ref:lens37}.}
\label{fig:MUSEspectra37}
\end{minipage}
\end{figure*}

\begin{figure*}[!ht]
\centering
\begin{minipage}{1.0\textwidth}
\centering
\includegraphics[width=0.4\textwidth]{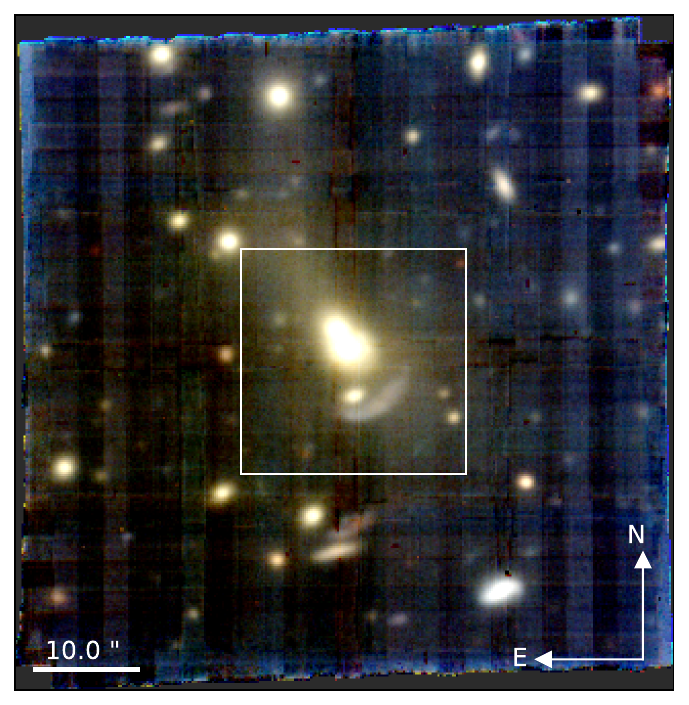}
\includegraphics[width=0.404\textwidth]{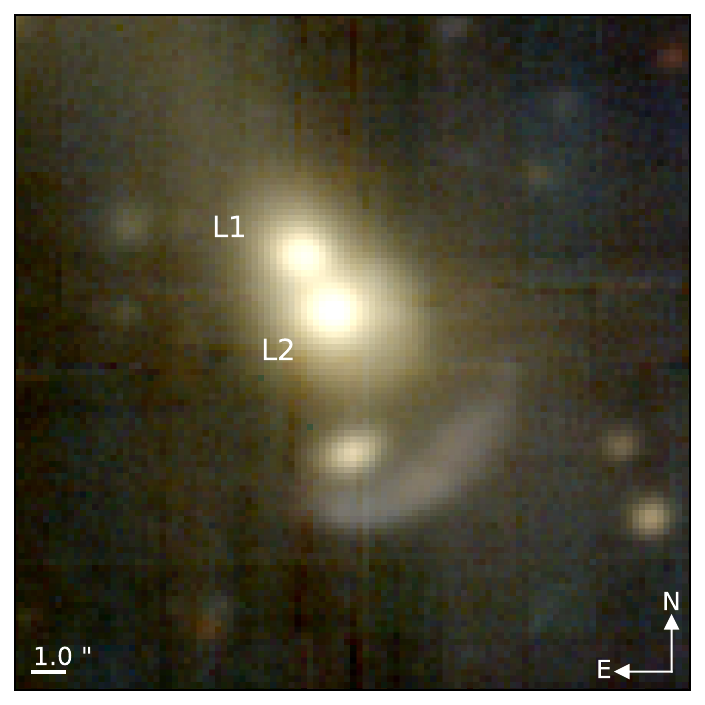}

\includegraphics[width=\textwidth]{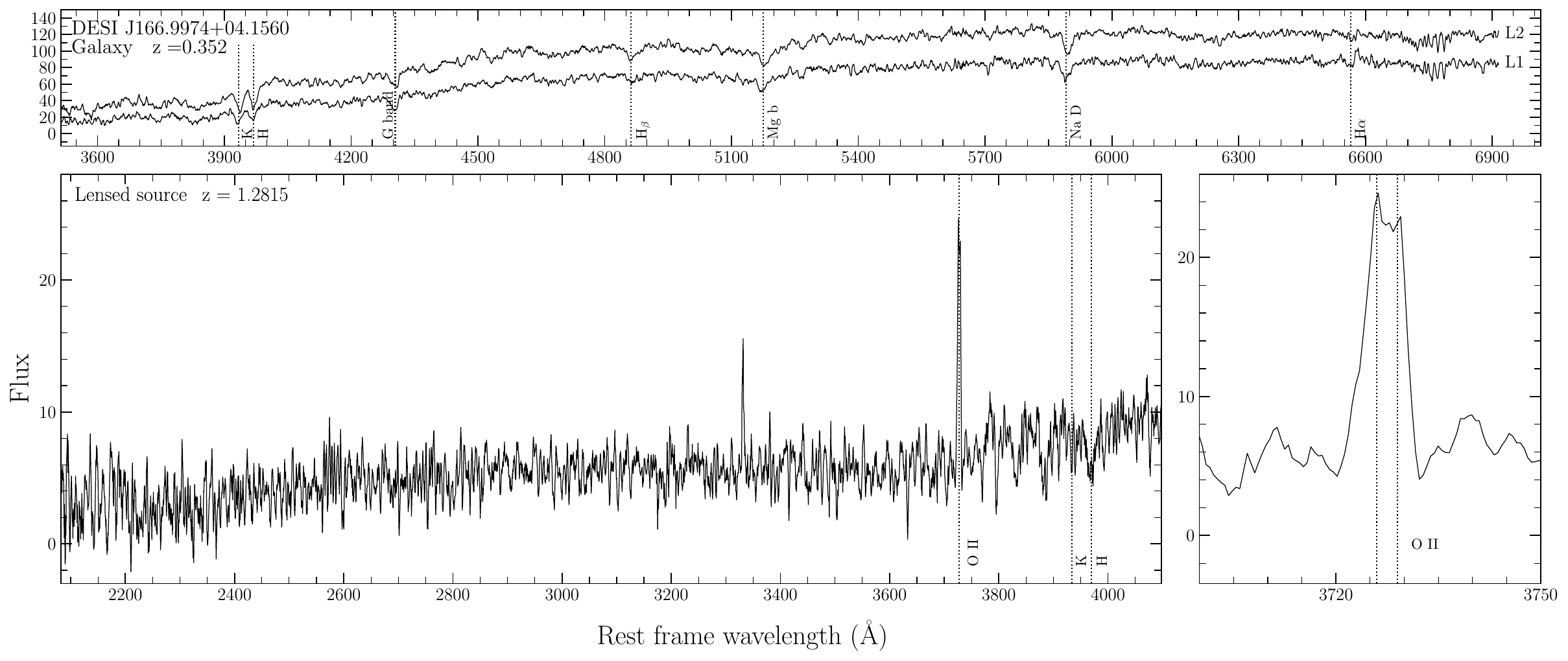}
\caption{\textit{Top:} RGB image of gravitational lens system DESI~J166.9974+04.1560 observed with MUSE. \textit{Bottom:} MUSE spectra of DESI~J166.9974+04.1560. For more information on the system, see Desc. \ref{Ref:lens38}. }
\label{fig:MUSEspectra38}
\end{minipage}
\end{figure*}

\begin{figure*}[!ht]
\centering
\begin{minipage}{1.0\textwidth}
\centering
\includegraphics[width=0.4\textwidth]{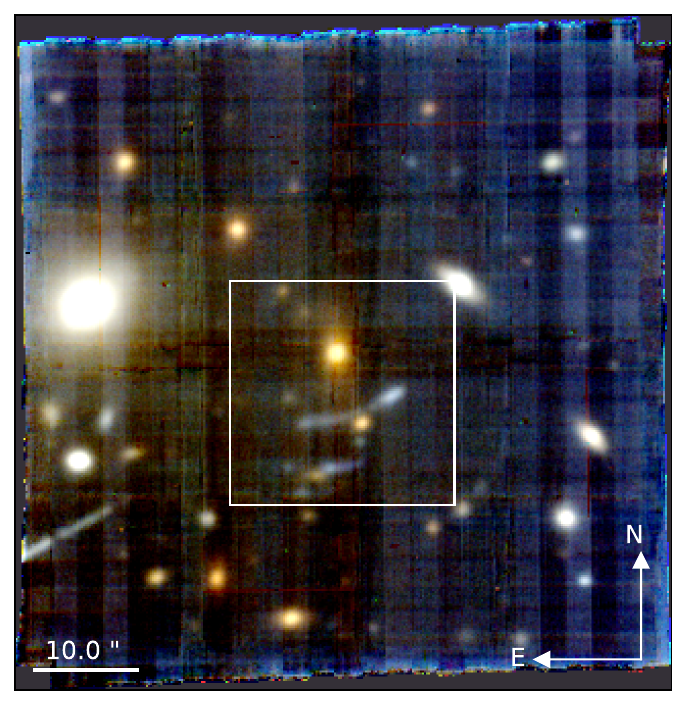}
\includegraphics[width=0.404\textwidth]{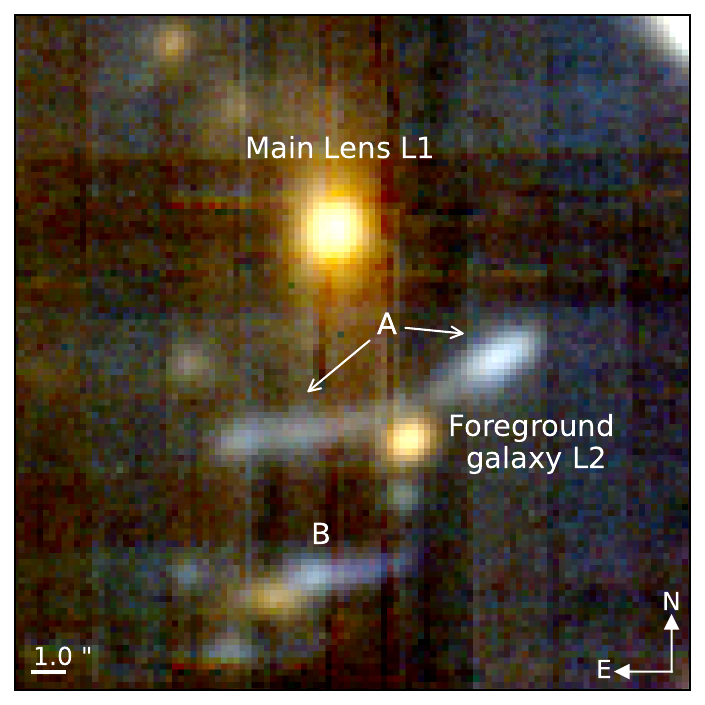}

\includegraphics[width=\textwidth]{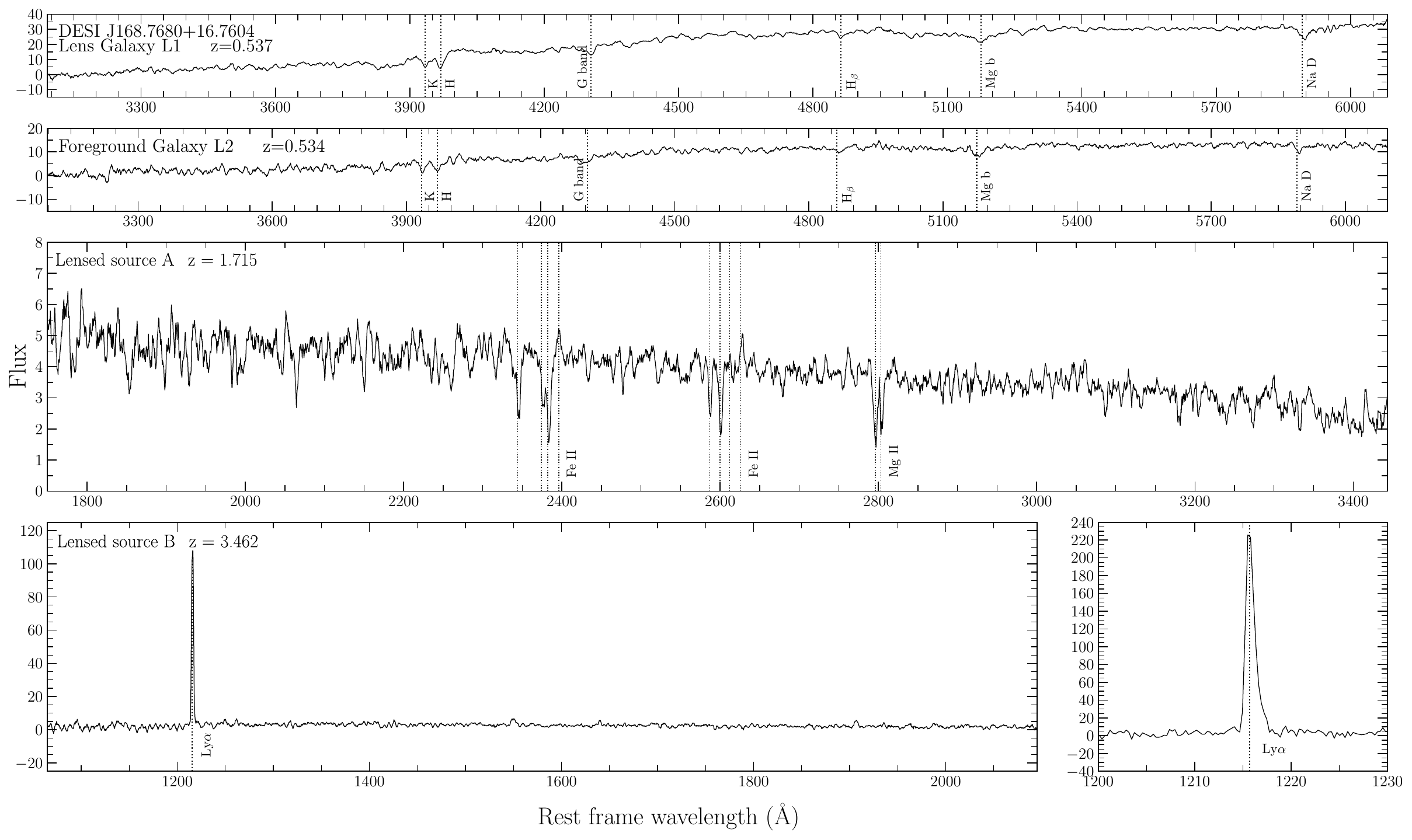}
\caption{\textit{Top:} RGB image of gravitational lens system DESI~J168.7680+16.7604 observed with MUSE. \textit{Bottom:} MUSE spectra of DESI~J168.7680+16.7604. Note that the quality flag for Source A is $Q_z=2$. For more information on the system, see Desc. \ref{Ref:lens39}.}
\label{fig:MUSEspectra39}
\end{minipage}
\end{figure*}

\begin{figure*}[!ht]
\centering

\begin{minipage}{1.0\textwidth}
\centering
\includegraphics[width=0.4\textwidth]{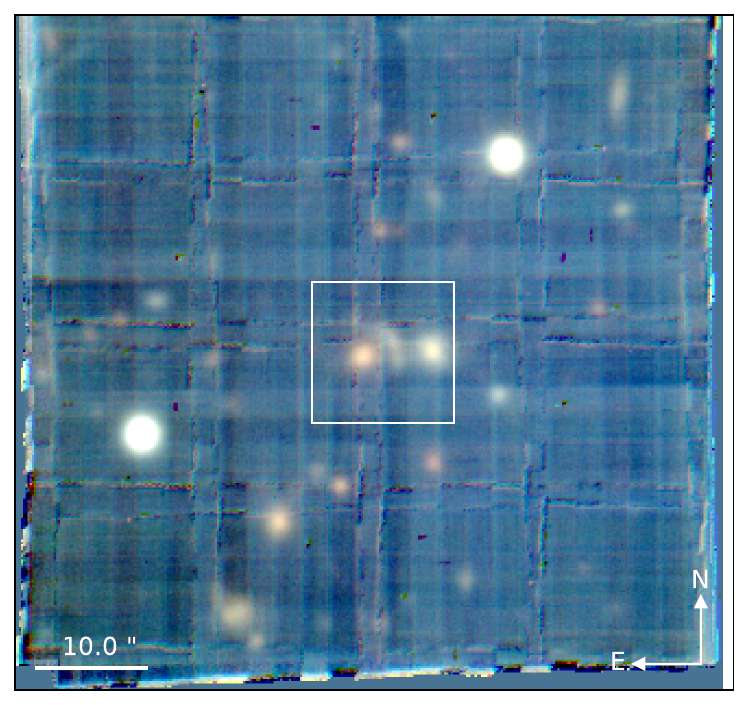}
\includegraphics[width=0.404\textwidth]{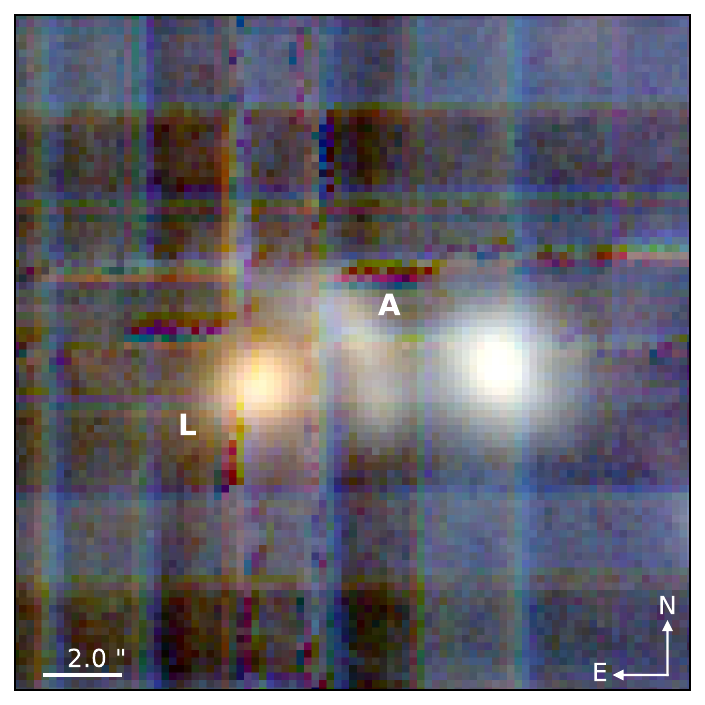}

\includegraphics[width=\textwidth]{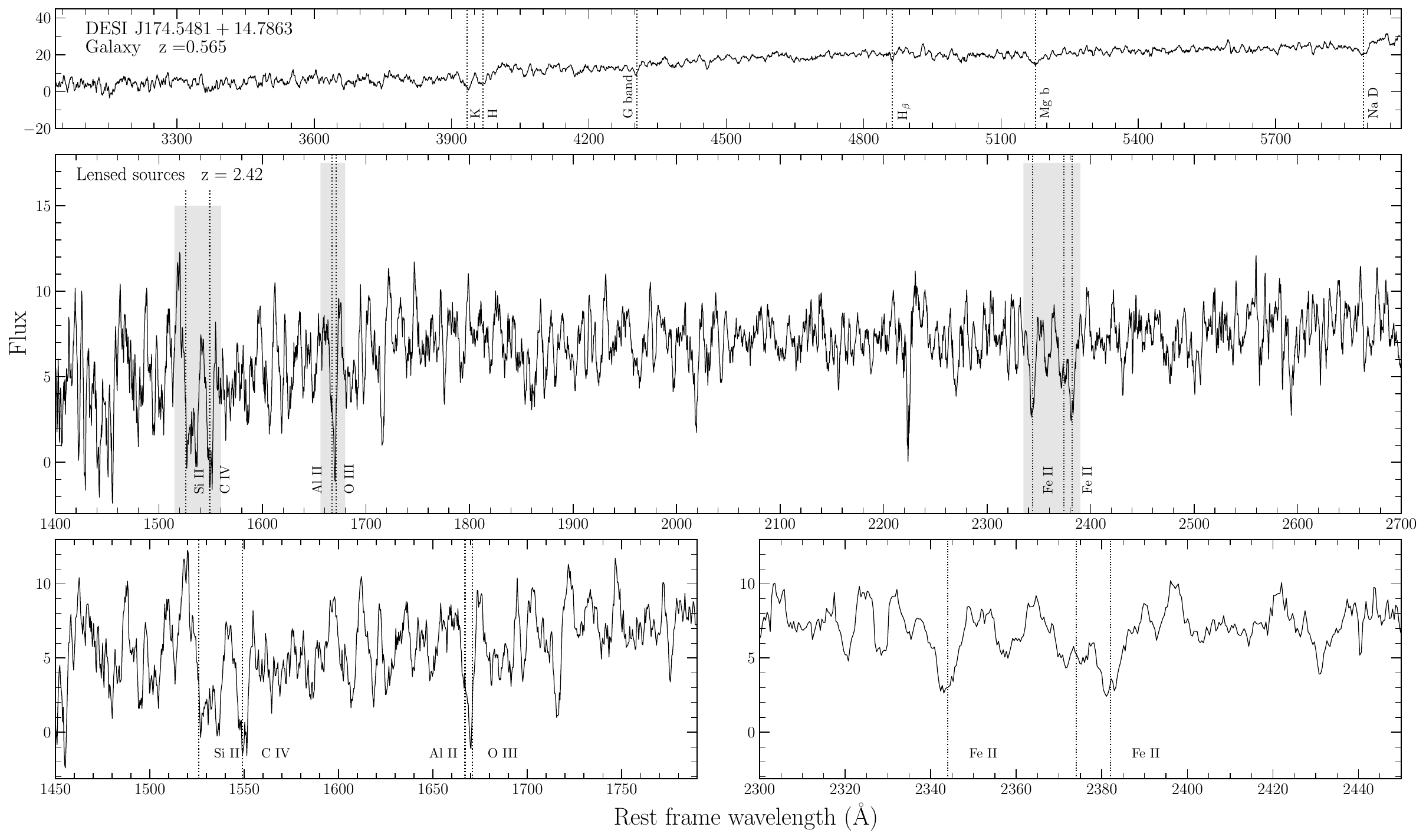}
\caption{\textit{Top:} RGB image of gravitational lens system DESI~J174.5481+14.7863 observed with MUSE. \textit{Bottom:} MUSE spectra of DESI~J174.5481+14.7863. For more information on the system, see Desc. \ref{ref:lens41}.}
\label{fig:MUSEspectra41}
\end{minipage}
\end{figure*}

\begin{figure*}[!ht]
\centering
\begin{minipage}{1.0\textwidth}
\centering
\includegraphics[width=0.4\textwidth]{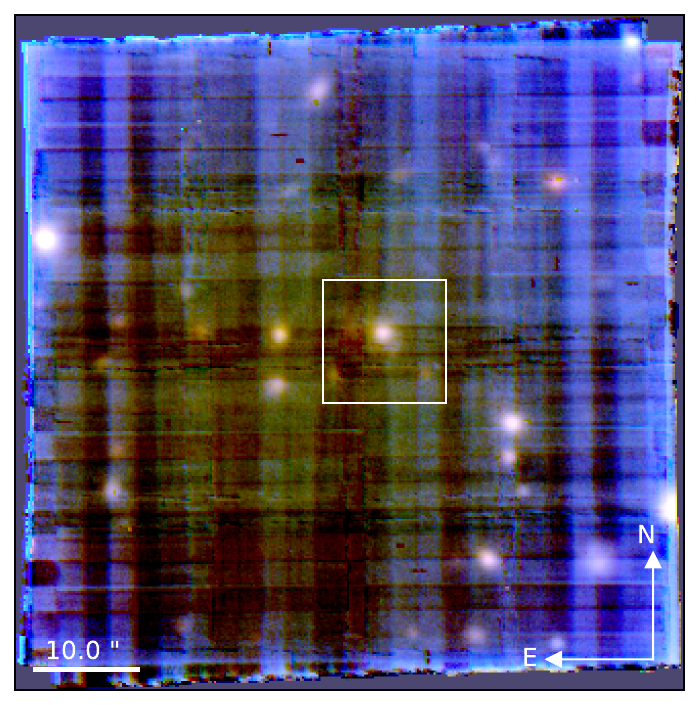}
\includegraphics[width=0.404\textwidth]{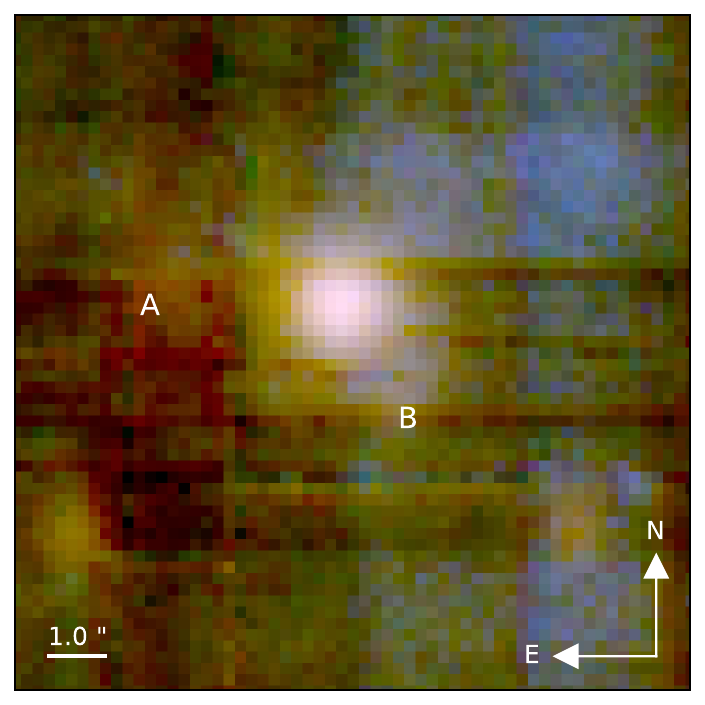}

\includegraphics[width=\textwidth]{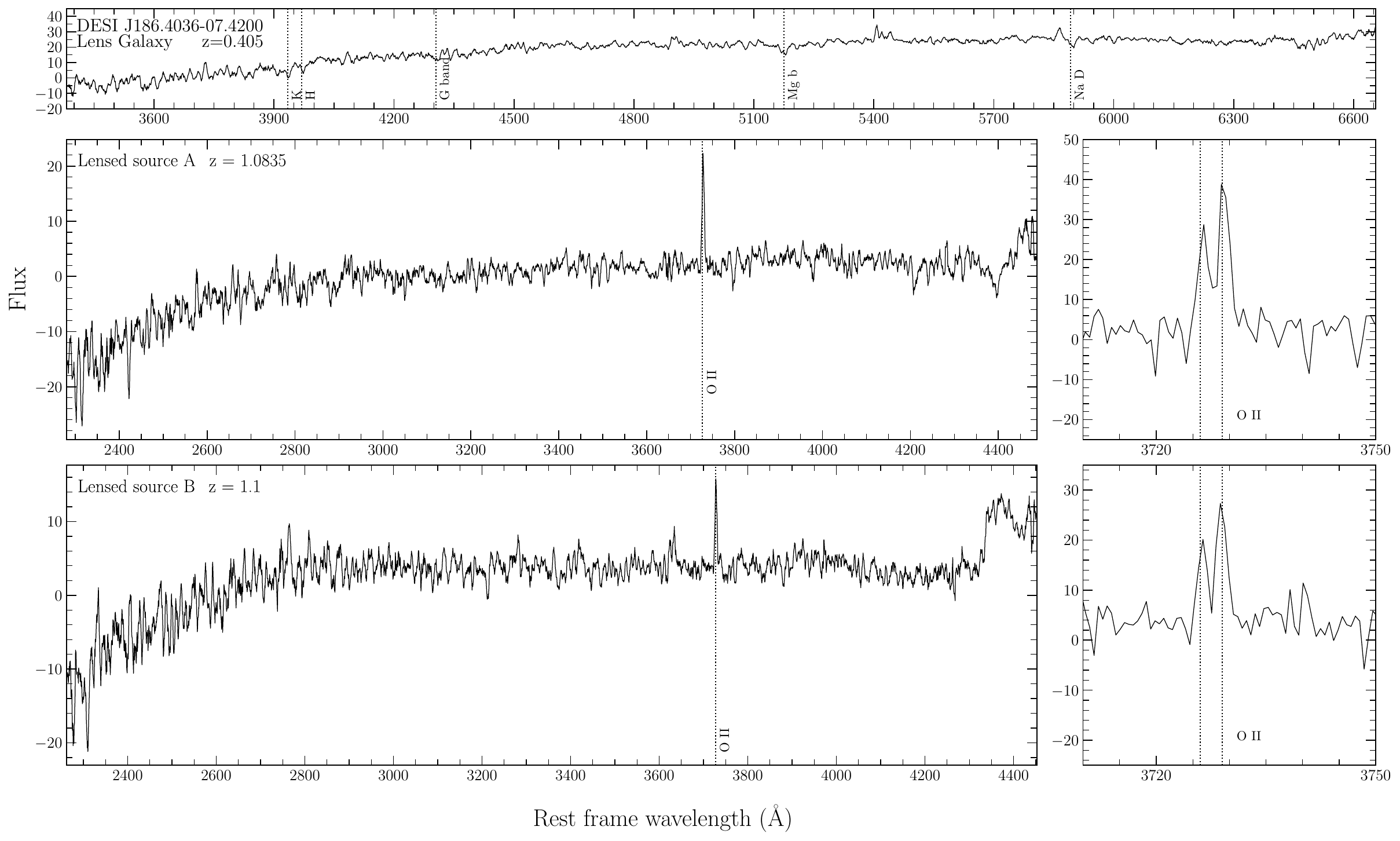}
\caption{\textit{Top:} RGB image of gravitational lens system DESI~J186.4036-07.4200 observed with MUSE. \textit{Bottom:} MUSE spectra of DESI~J186.4036-07.4200. For more information on the system, see Desc. \ref{Ref:lens106}. }
\label{fig:MUSEspectra106}
\end{minipage}
\end{figure*}

\begin{figure*}[!ht]
\centering
\begin{minipage}{1.0\textwidth}
\centering
\includegraphics[width=0.4\textwidth]{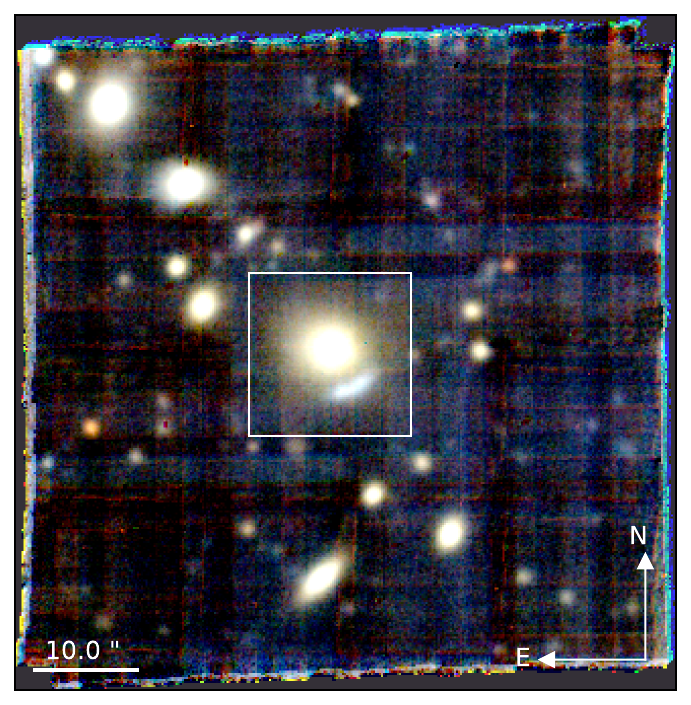}
\includegraphics[width=0.404\textwidth]{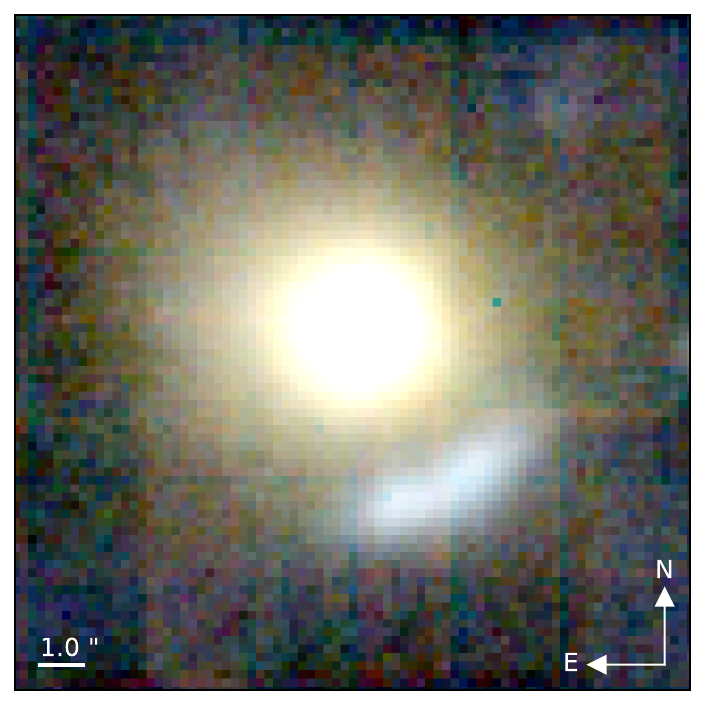}

\includegraphics[width=\textwidth]{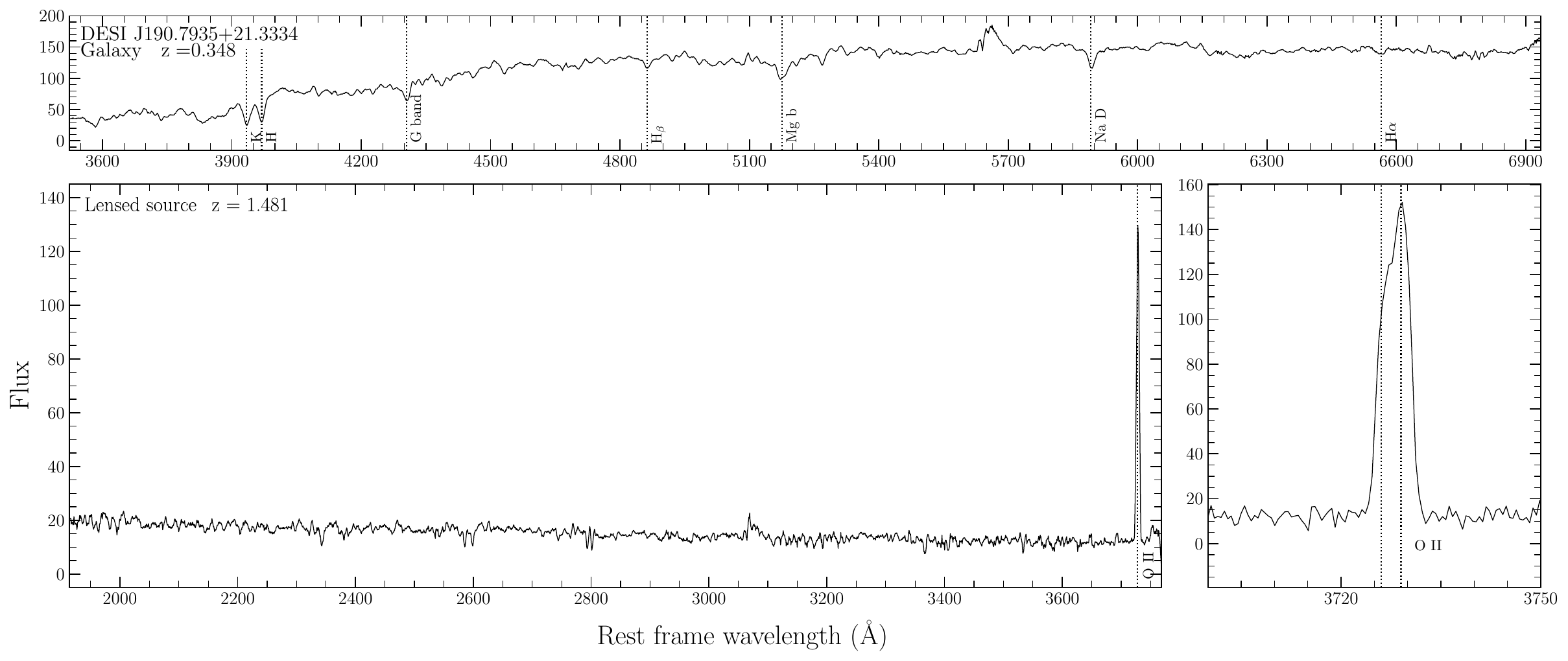}
\caption{\textit{Top:} RGB image of gravitational lens system DESI~J190.7935+21.3334 observed with MUSE. \textit{Bottom:} MUSE spectra of DESI~J190.7935+21.3334. For more information on the system, see Desc. \ref{Ref:lens47}.}
\label{fig:MUSEspectra47}
\end{minipage}
\end{figure*}

\begin{figure*}[!ht]
\centering
\begin{minipage}{1.0\textwidth}
\centering
\includegraphics[width=0.4\textwidth]{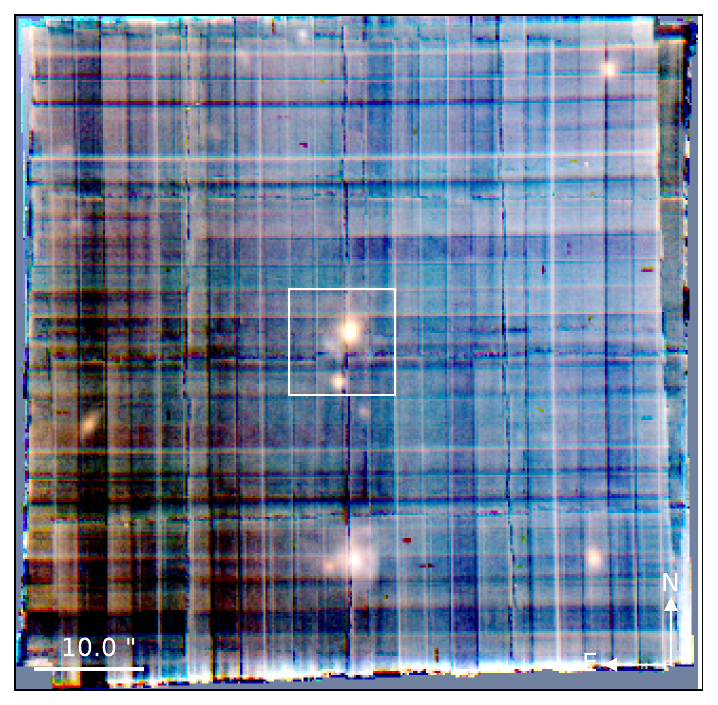}
\includegraphics[width=0.404\textwidth]{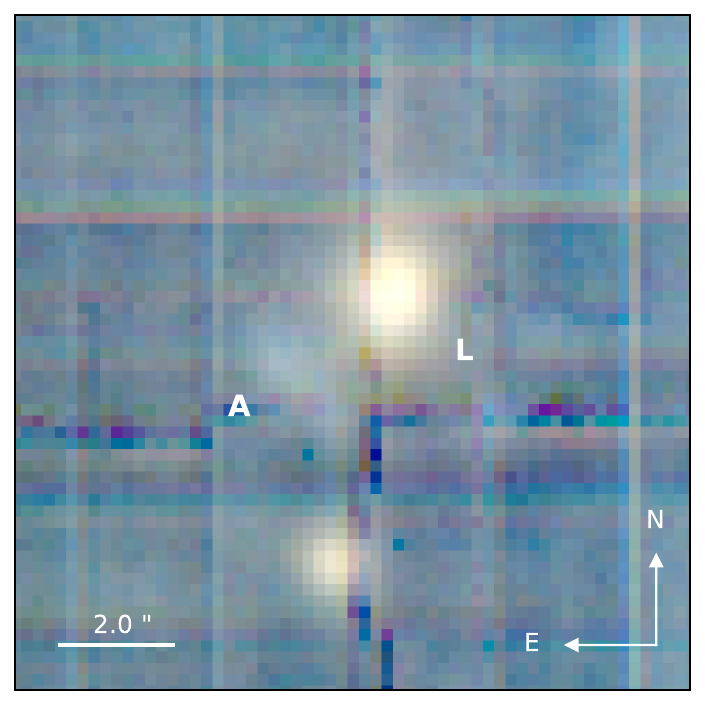}

\includegraphics[width=\textwidth]{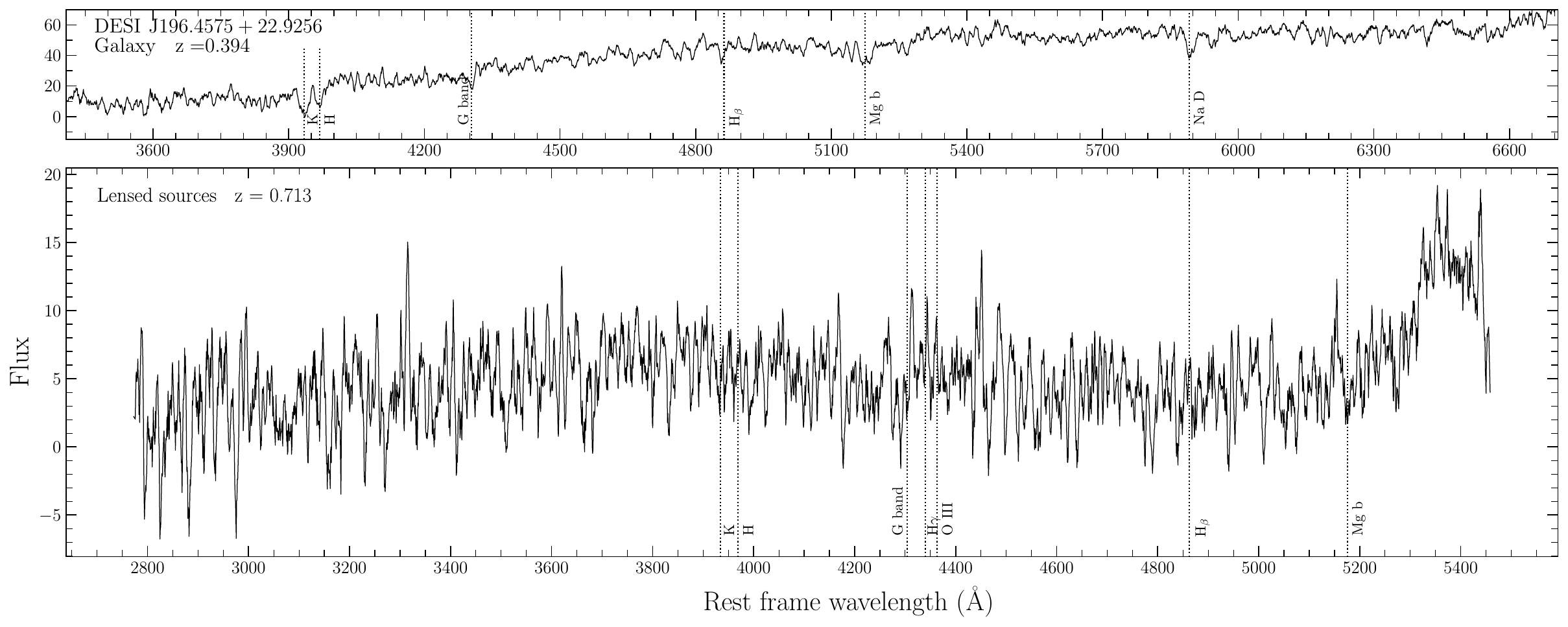}
\caption{\textit{Top:} RGB image of gravitational lens system DESI~J196.4575+22.9256 observed with MUSE. \textit{Bottom:} MUSE spectra of DESI~J196.4575+22.9256. Note that the source quality flag is $Q_z=3$. For more information on the system, see Desc. \ref{ref:lens48}. }
\label{fig:MUSEspectra48}
\end{minipage}
\end{figure*}

\begin{figure*}[!ht]
\centering
\begin{minipage}{1.0\textwidth}
\centering
\includegraphics[width=0.4\textwidth]{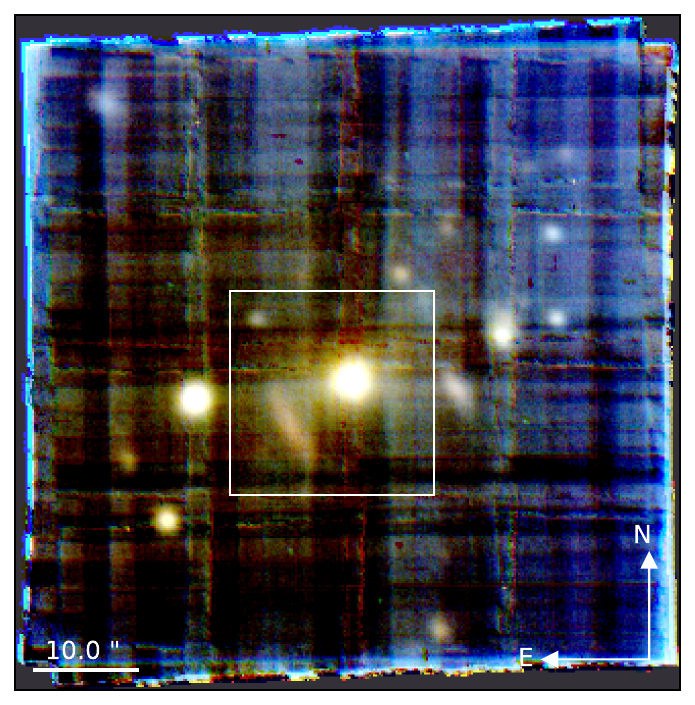}
\includegraphics[width=0.404\textwidth]{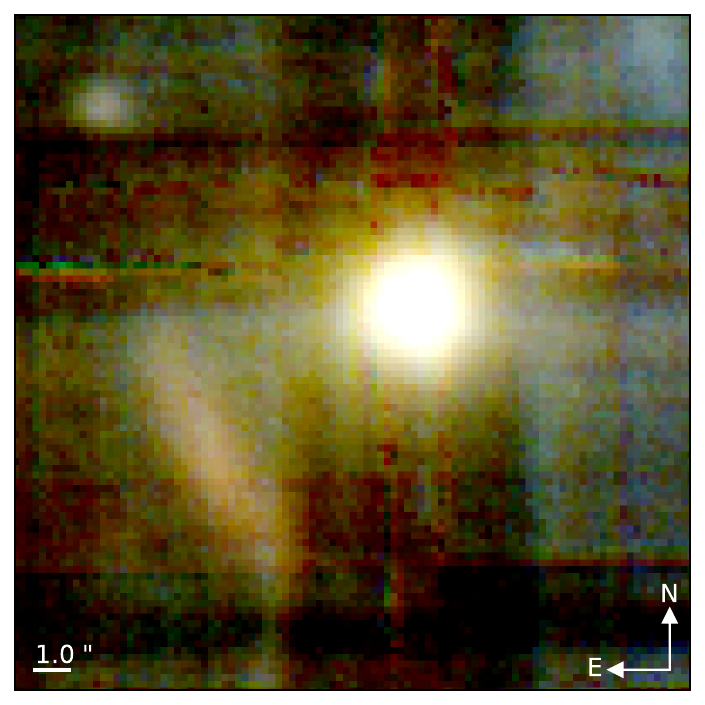}

\includegraphics[width=\textwidth]{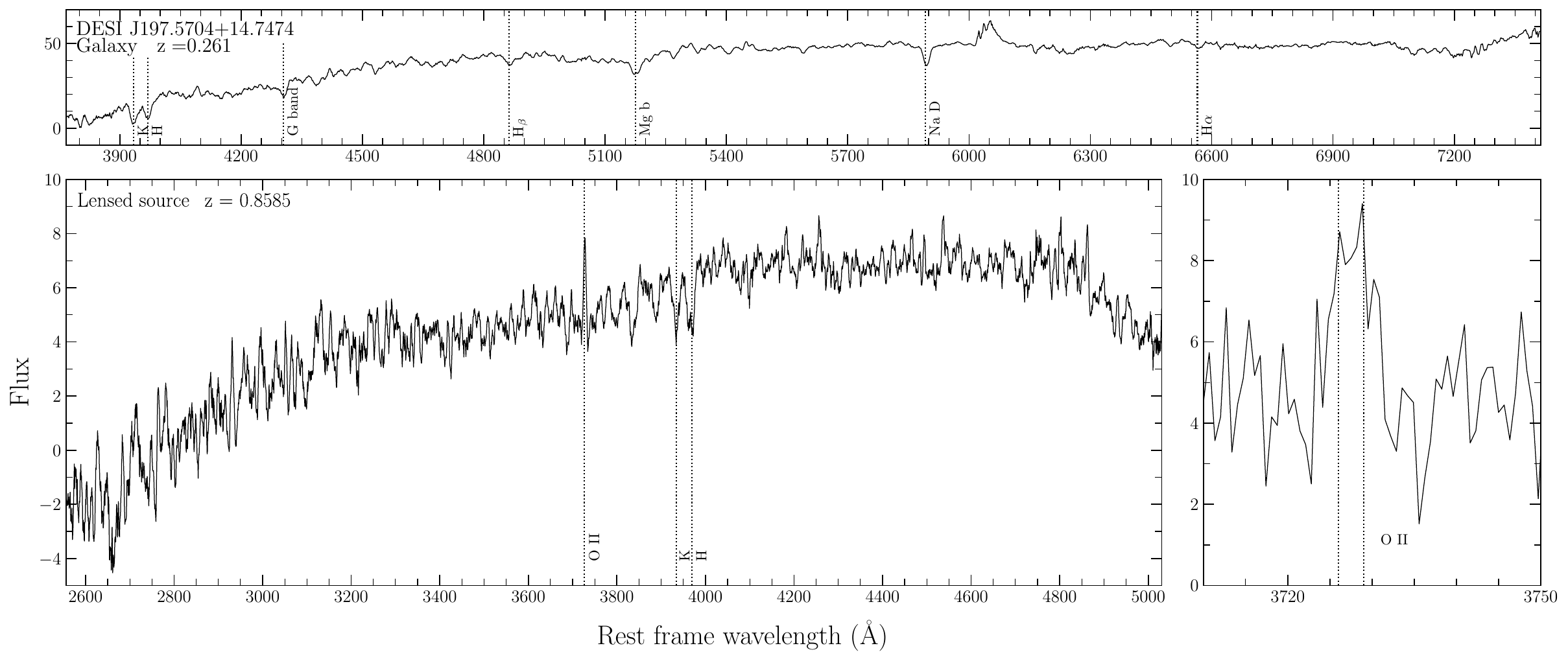}
\caption{\textit{Top:} RGB image of gravitational lens system DESI~J197.5704+14.7474 observed with MUSE. \textit{Bottom:} MUSE spectra of DESI~J197.5704+14.7474. For more information on the system, see Desc. \ref{Ref:lens49}.}
\label{fig:MUSEspectra49}
\end{minipage}
\end{figure*}

\begin{figure*}[!ht]
\centering
\begin{minipage}{1.0\textwidth}
\centering
\includegraphics[width=0.4\textwidth]{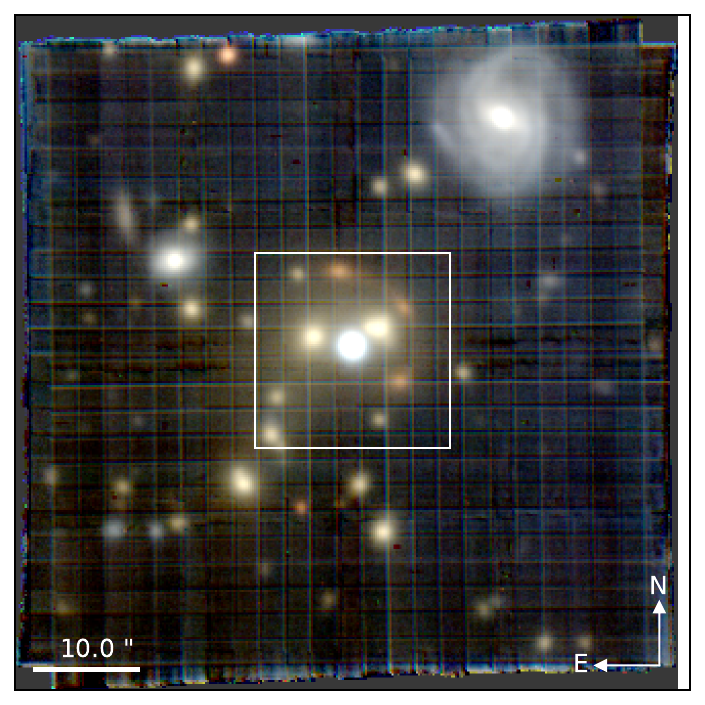}
\includegraphics[width=0.404\textwidth]{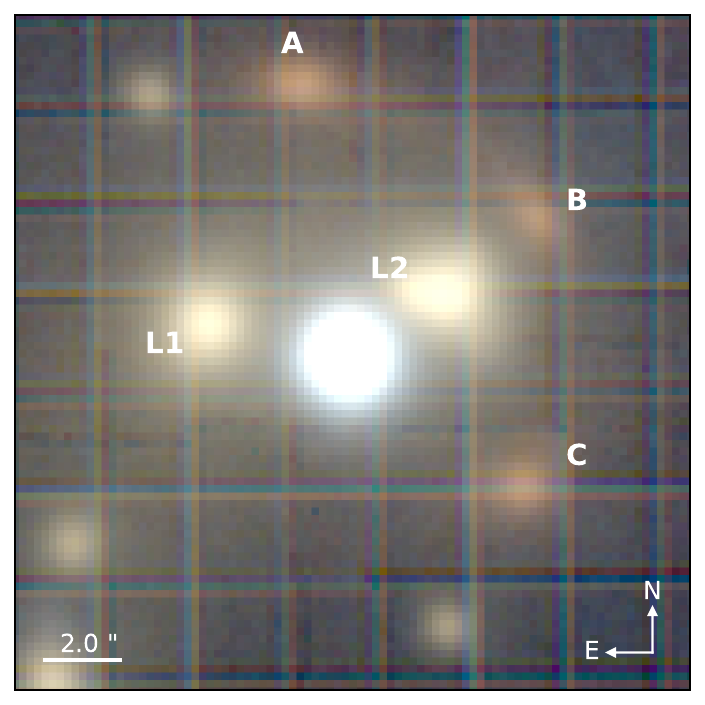}

\includegraphics[width=\textwidth]{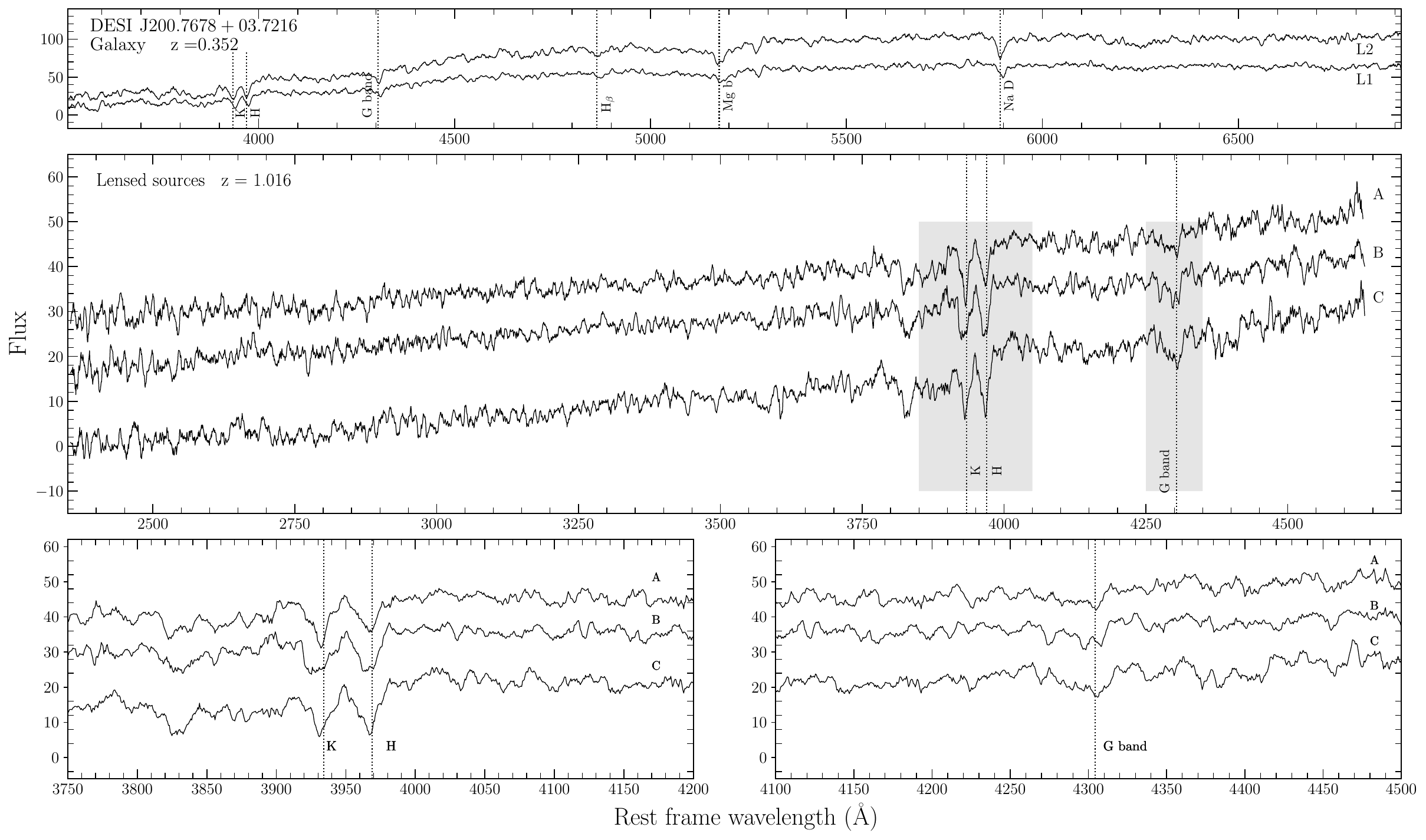}
\caption{\textit{Top:} RGB image of gravitational lens system DESI~J200.7678+03.7216 observed with MUSE. \textit{Bottom:} MUSE spectra of DESI~J200.7678+03.7216. For more information on the system, see Desc. \ref{ref:lens50}. }
\label{fig:MUSEspectra50}
\end{minipage}
\end{figure*}

\begin{figure*}[!ht]
\centering
\begin{minipage}{1.0\textwidth}
\centering
\includegraphics[width=0.4\textwidth]{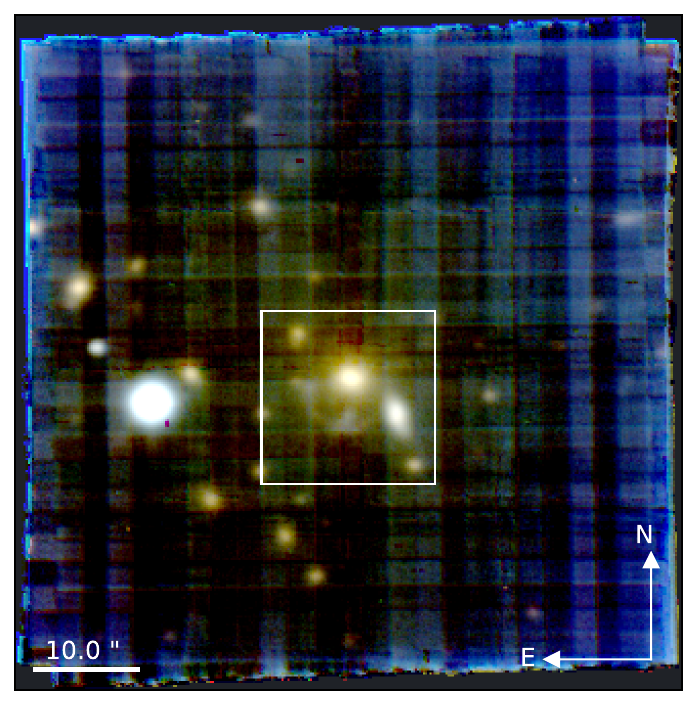}
\includegraphics[width=0.404\textwidth]{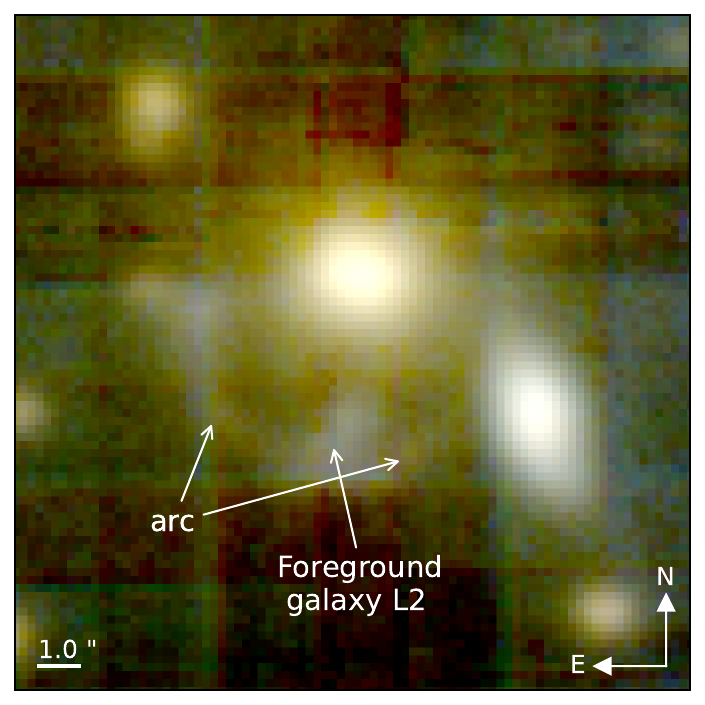}

\includegraphics[width=\textwidth]{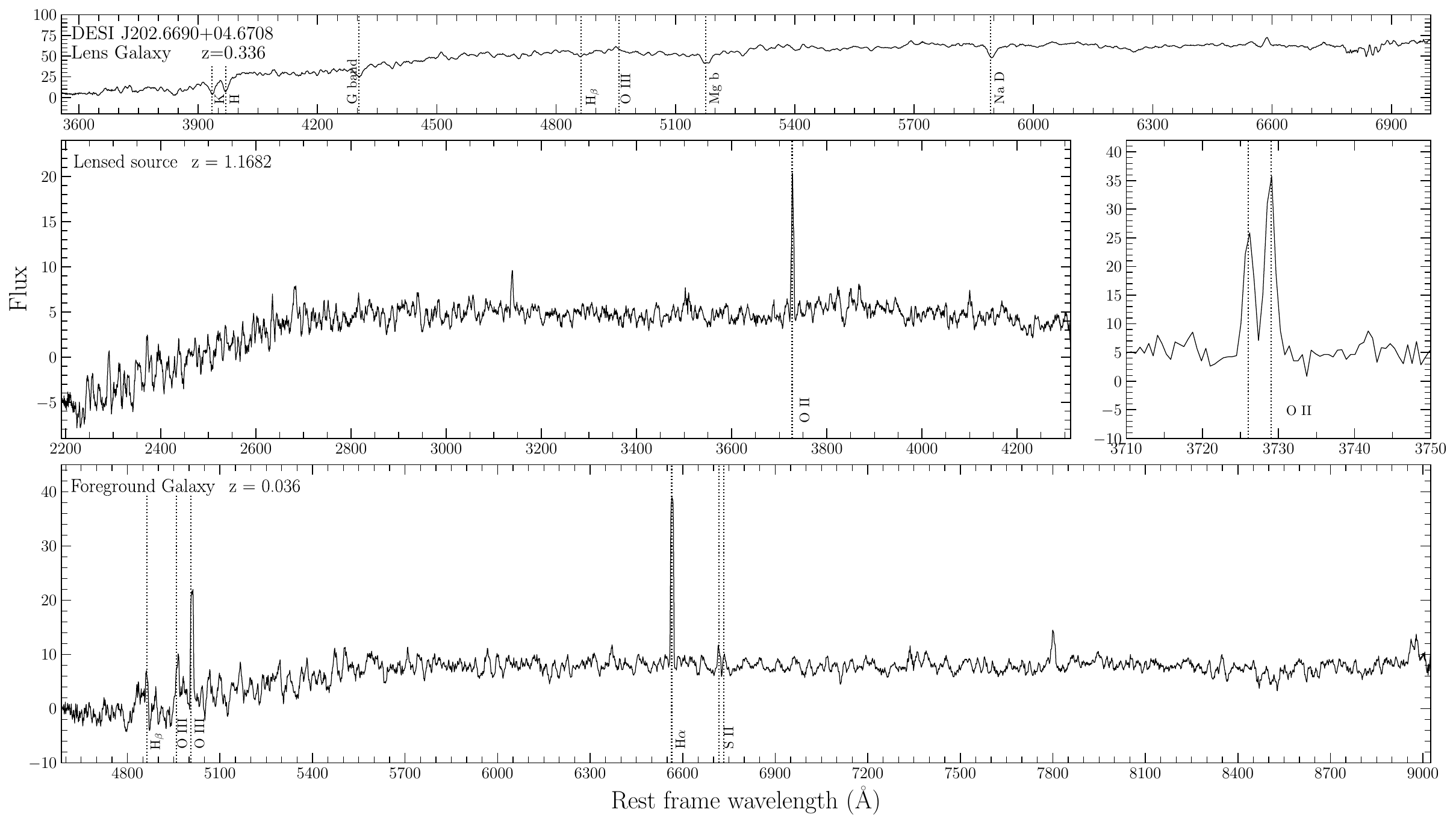}
\caption{\textit{Top:} RGB image of gravitational lens system DESI~J202.6690+04.6708 observed with MUSE. \textit{Bottom:} MUSE spectra of DESI~J202.6690+04.6708. For more information on the system, see Desc. \ref{Ref:lens119}. }
\label{fig:MUSEspectra119}
\end{minipage}
\end{figure*}

\begin{figure*}[!ht]
\centering
\begin{minipage}{1.0\textwidth}
\centering
\includegraphics[width=0.404\textwidth]{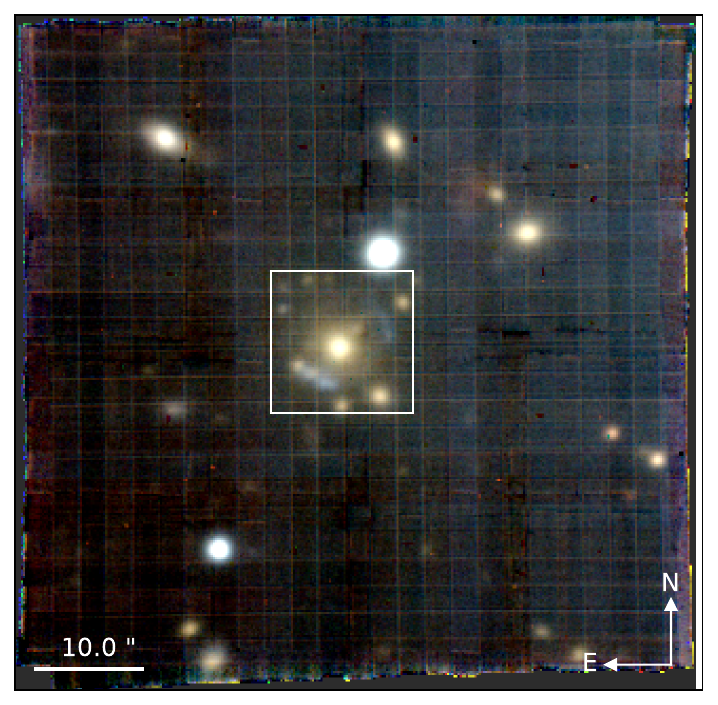}
\includegraphics[width=0.4\textwidth]{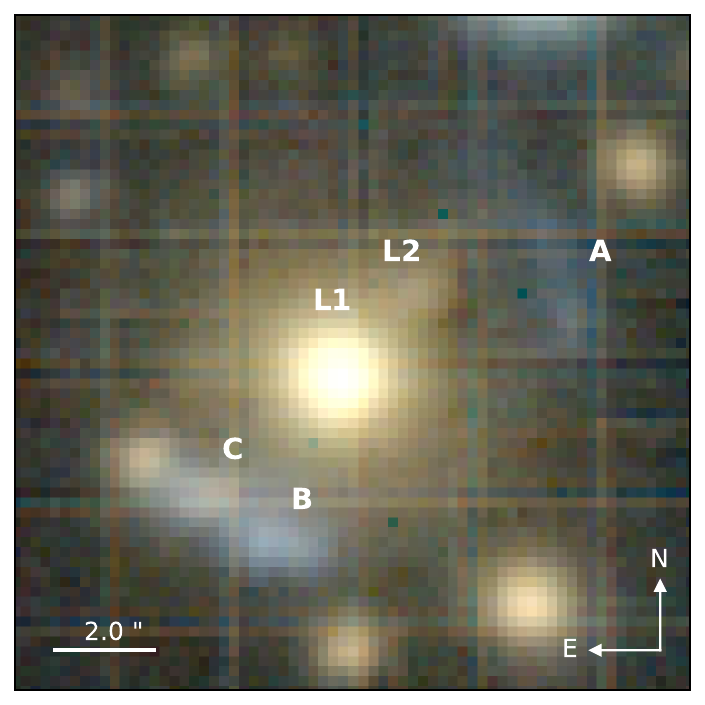}

\includegraphics[width=\textwidth]{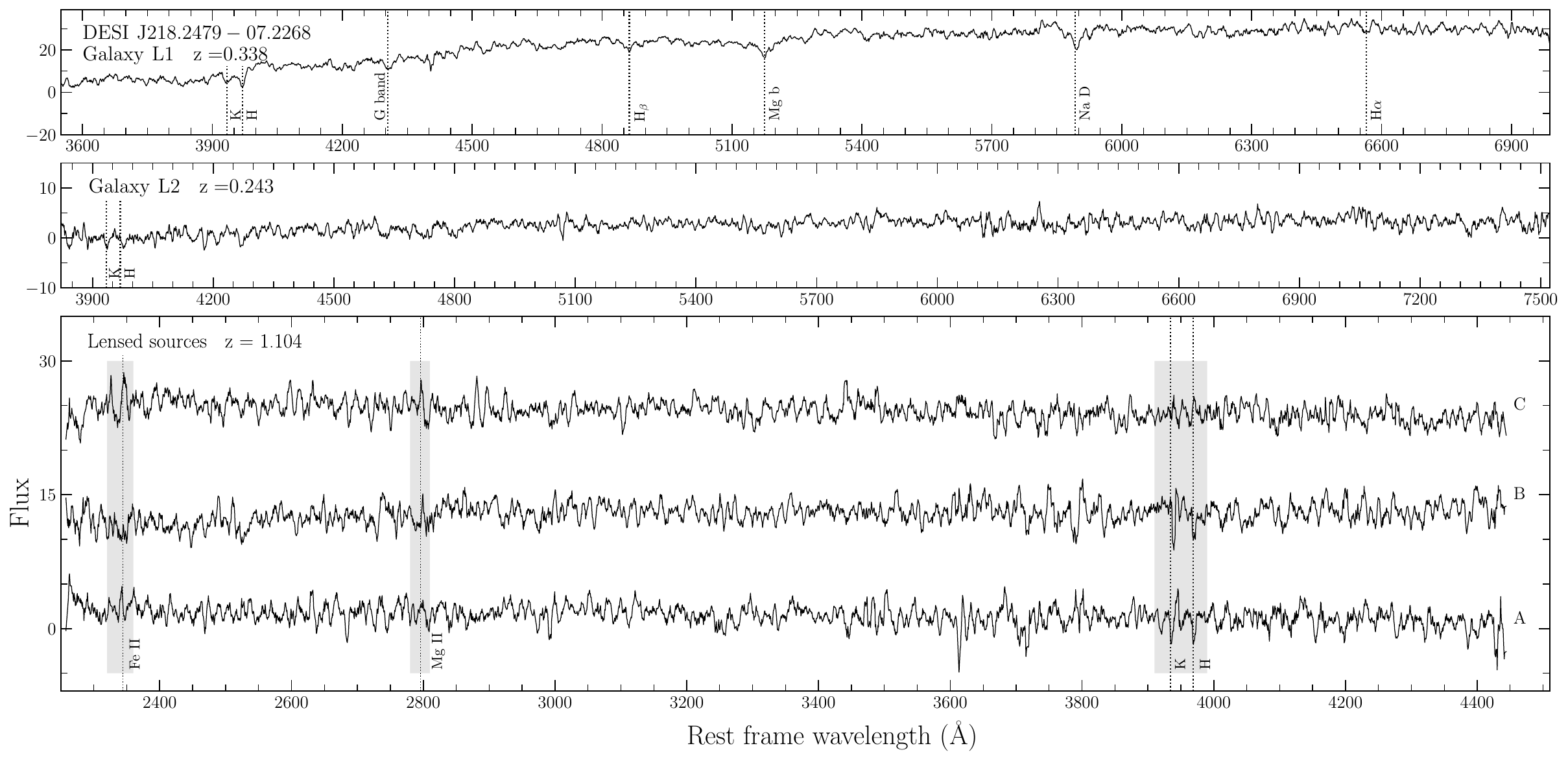}
\caption{\textit{Top:} RGB image of gravitational lens system DESI~J218.2479-7.2268 observed with MUSE. \textit{Bottom:} MUSE spectra of DESI~J218.2479-7.2268. Note that the quality flag for all sources is $Q_z=2$. For more information on the system, see Desc. \ref{ref:lens1}. }
\label{fig:MUSEspectra1}
\end{minipage}
\end{figure*}

\begin{figure*}[!ht]
\centering
\begin{minipage}{1.0\textwidth}
\centering
\includegraphics[width=0.4\textwidth]{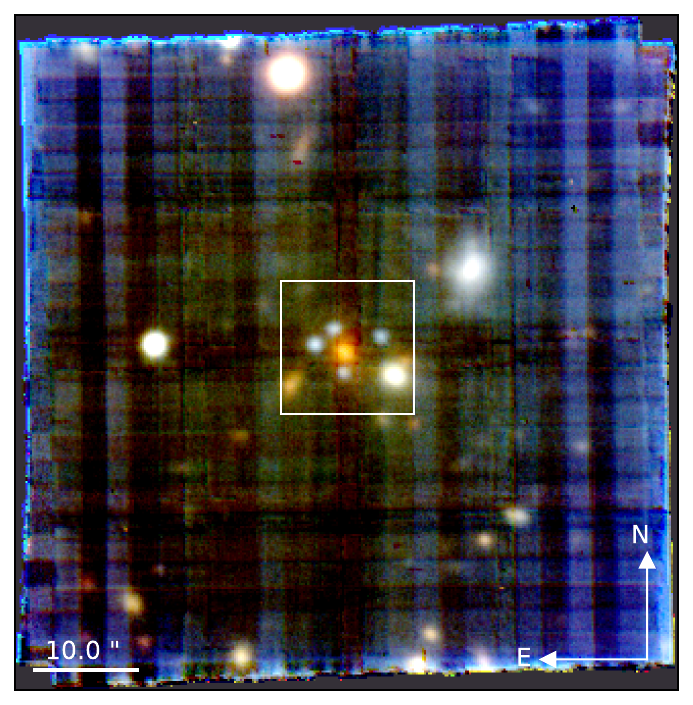}
\includegraphics[width=0.404\textwidth]{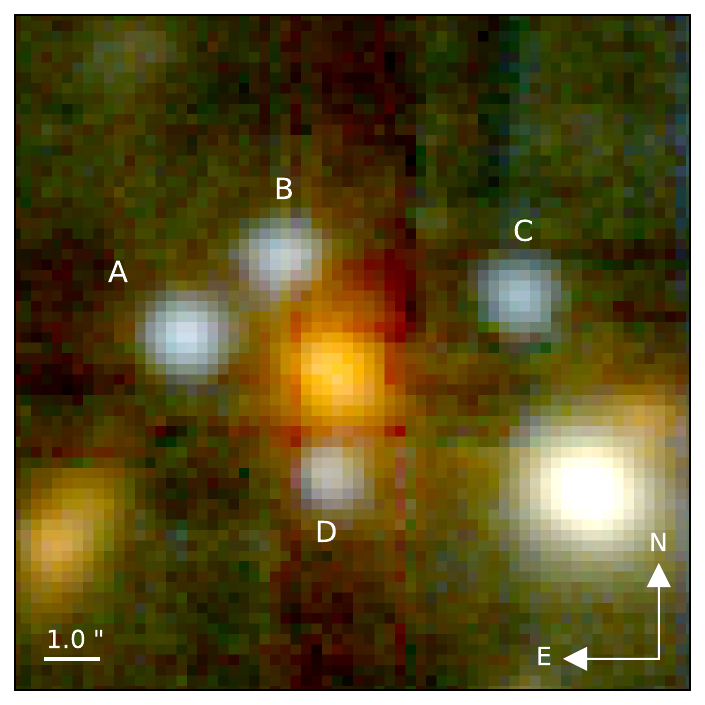}

\includegraphics[width=\textwidth]{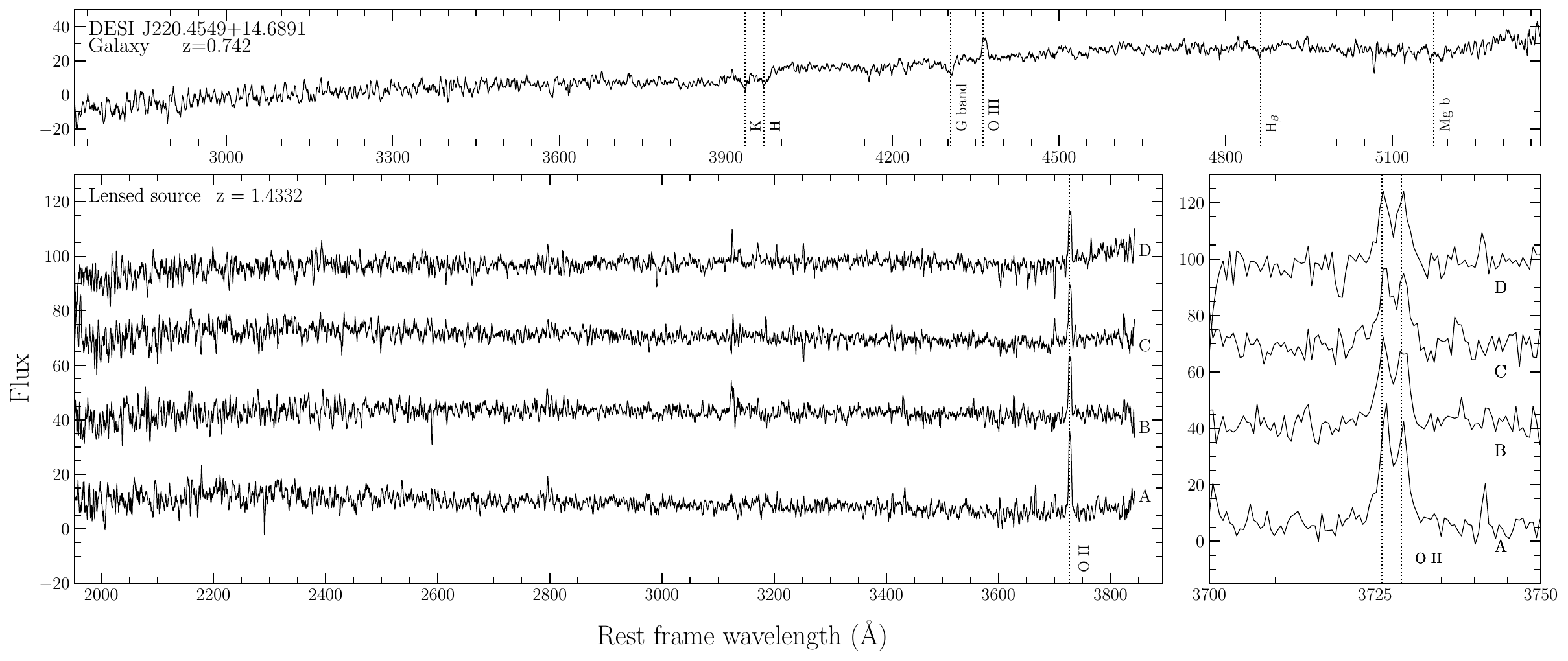}
\caption{\textit{Top:} RGB image of gravitational lens system DESI~J220.4549+14.6891 observed with MUSE. \textit{Bottom:} MUSE spectra of DESI~J220.4549+14.6891. For more information on the system, see Desc. \ref{Ref:lens52p112}.}
\label{fig:MUSEspectra52}
\end{minipage}
\end{figure*}

\begin{figure*}[!ht]
\centering

\begin{minipage}{1.0\textwidth}
\centering
\includegraphics[width=0.4\textwidth]{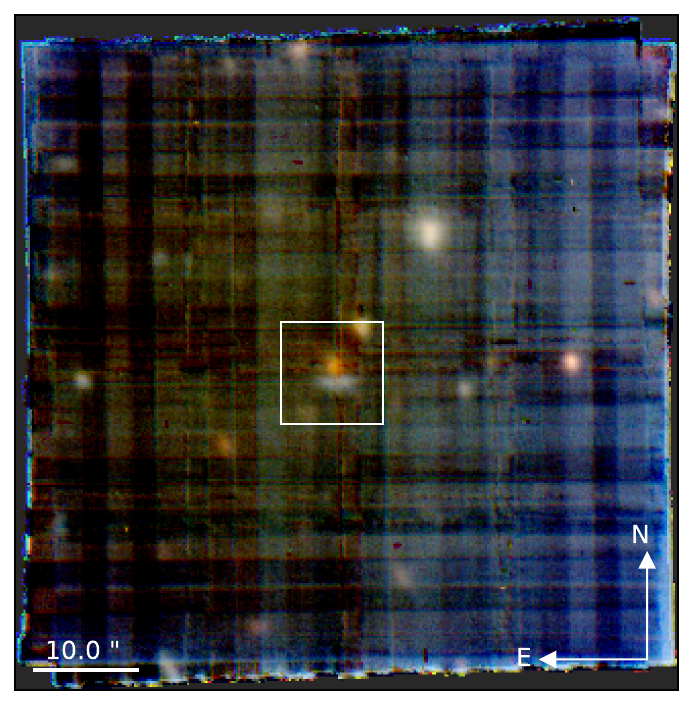}
\includegraphics[width=0.404\textwidth]{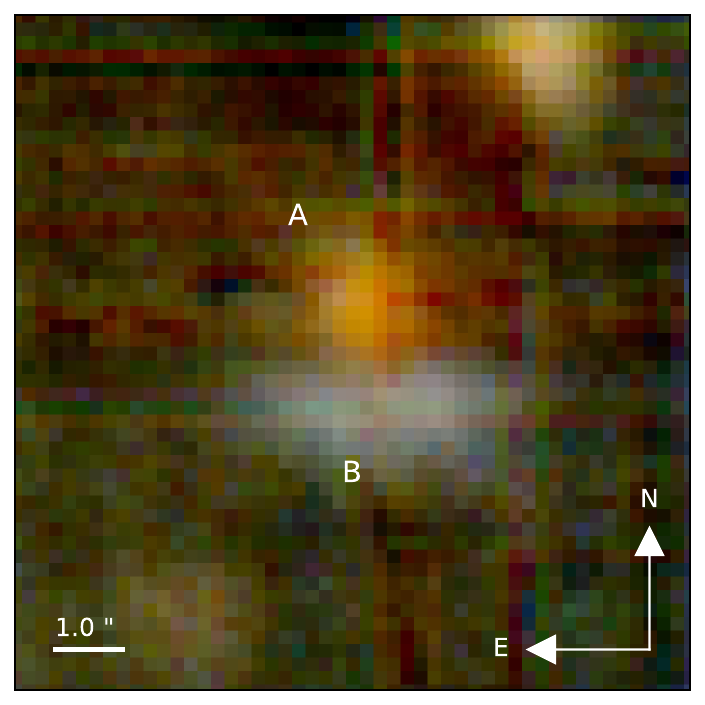}

\includegraphics[width=\textwidth]{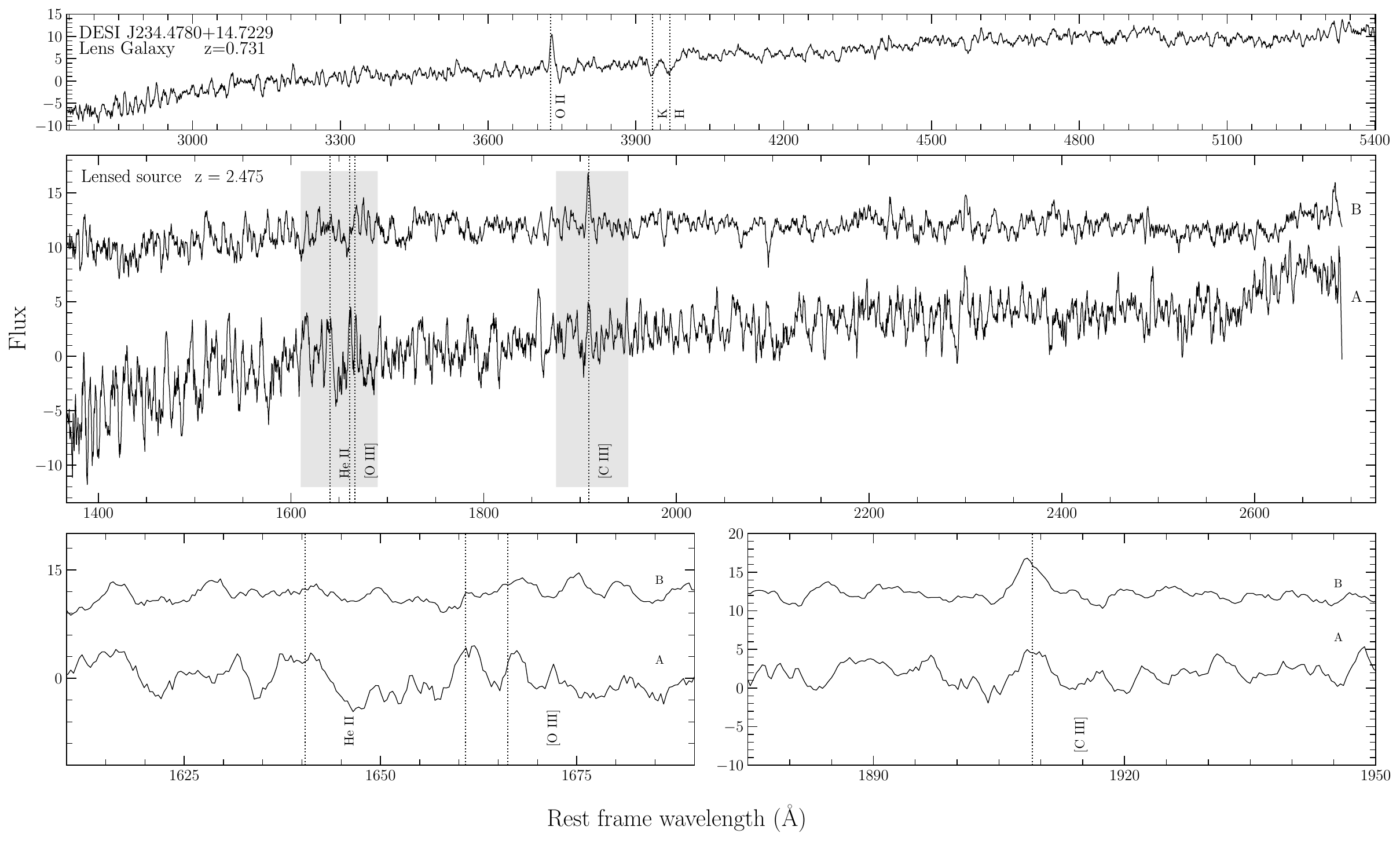}
\caption{\textit{Top:} RGB image of gravitational lens system DESI~J234.4780+14.7229 observed with MUSE. \textit{Bottom:} MUSE spectra of DESI~J234.4780+14.7229. Note that the quality flag for image A is $Q_z=3$. For more information on the system, see Desc. \ref{Ref:lens120}.}
\label{fig:MUSEspectra120}
\end{minipage}
\end{figure*}

\begin{figure*}[!ht]
\centering

\begin{minipage}{1.0\textwidth}
\centering
\includegraphics[width=0.4\textwidth]{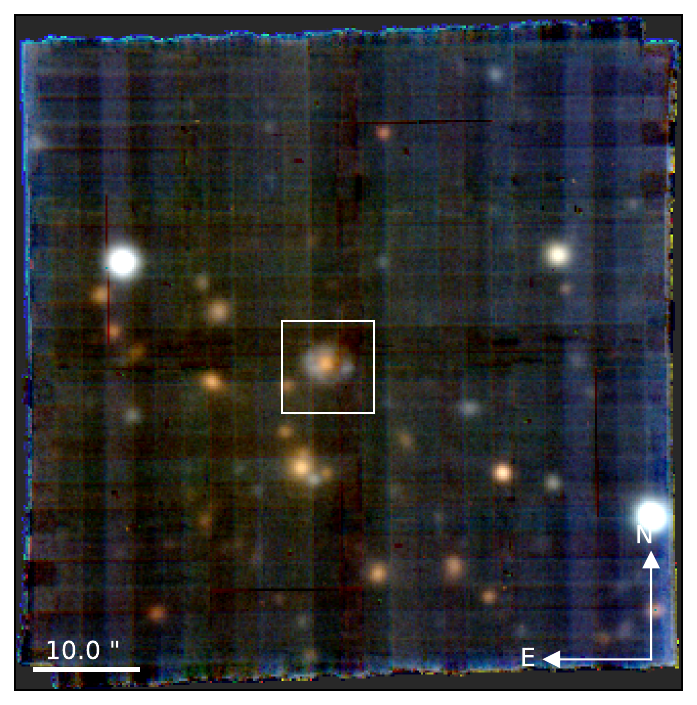}
\includegraphics[width=0.404\textwidth]{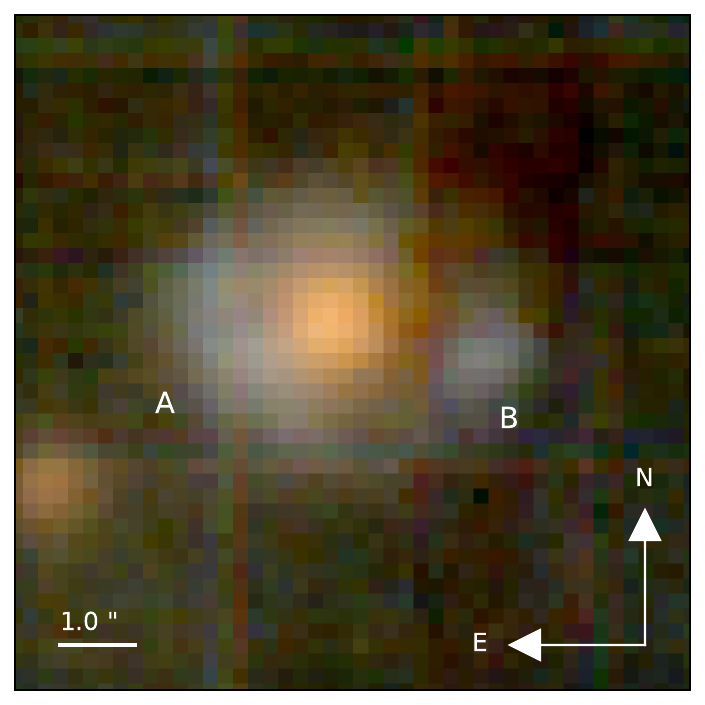}

\includegraphics[width=\textwidth]{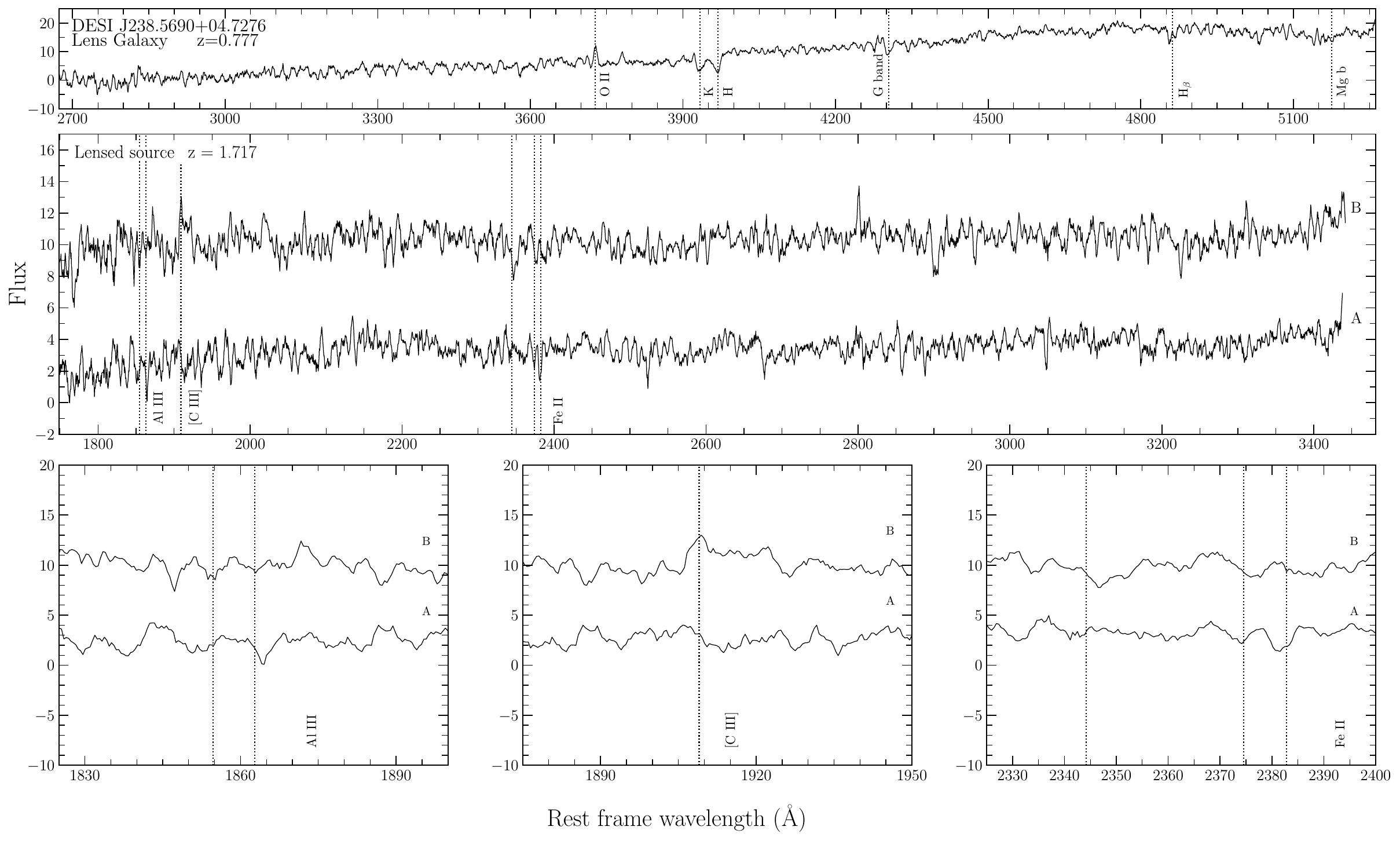}
\caption{\textit{Top:} RGB image of gravitational lens system DESI~J238.5690+04.7276 observed with MUSE. \textit{Bottom:} MUSE spectra of DESI~J238.5690+04.7276. For more information on the system, see Desc. \ref{Ref:lens122}. }
\label{fig:MUSEspectra122}
\end{minipage}
\end{figure*}

\begin{figure*}[!ht]
\centering
\begin{minipage}{1.0\textwidth}
\centering
\includegraphics[width=0.4\textwidth]{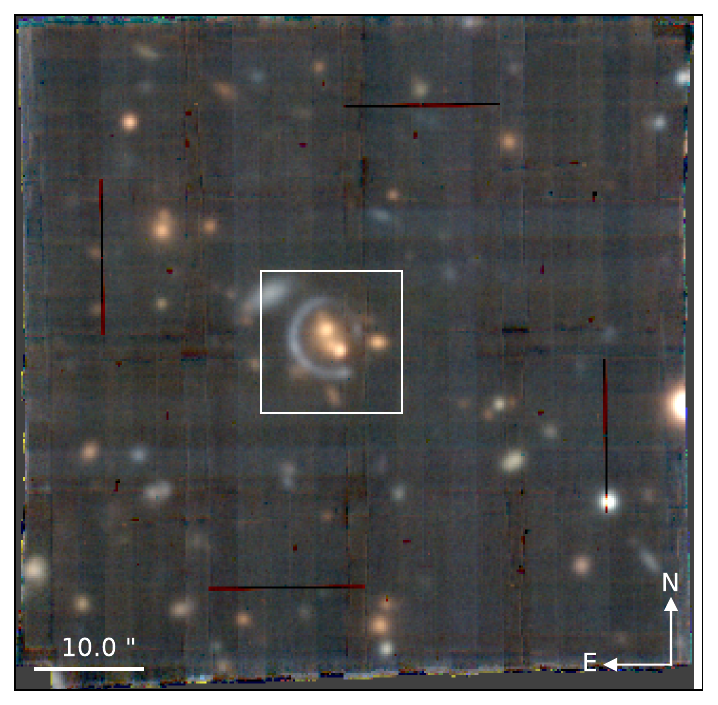}
\includegraphics[width=0.4\textwidth]{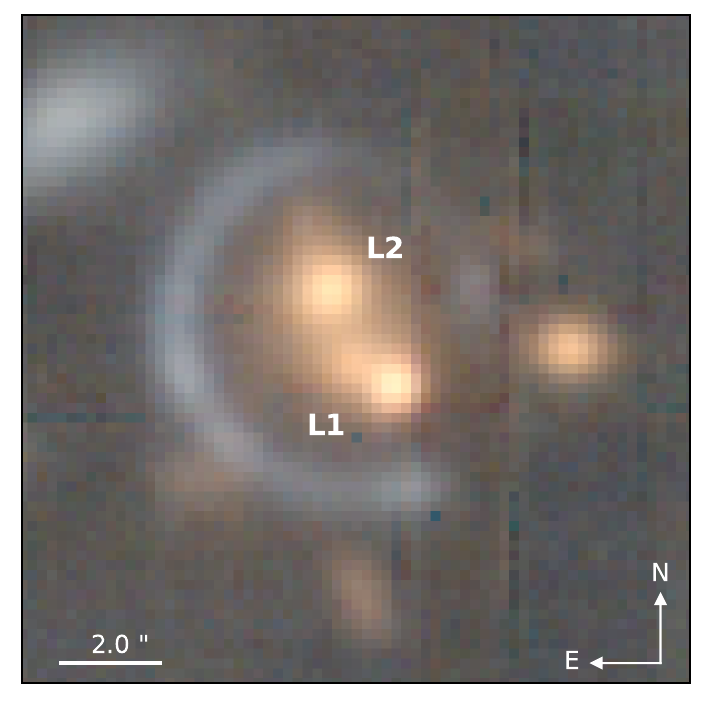}

\includegraphics[width=\textwidth]{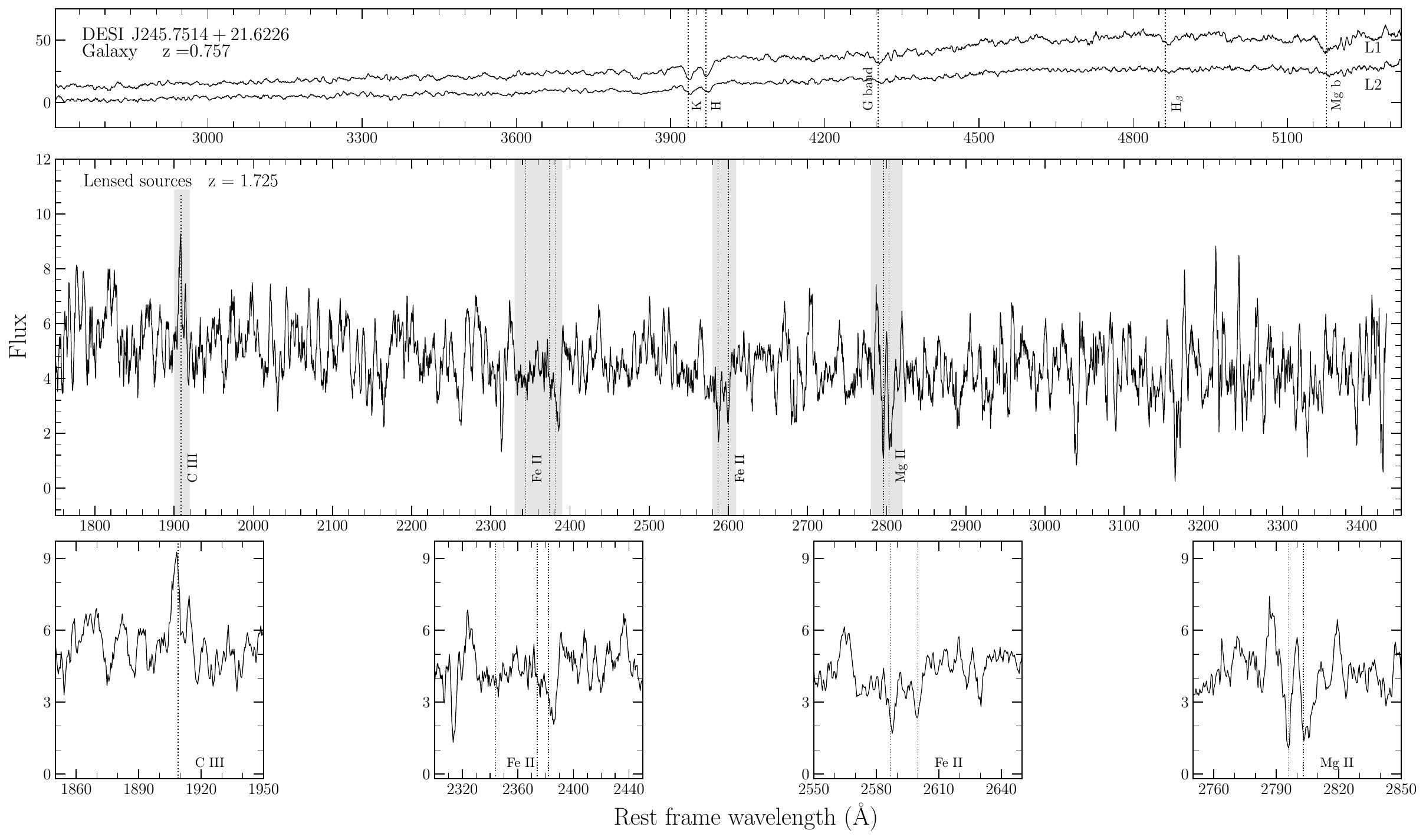}
\caption{\textit{Top:} RGB image of gravitational lens system DESI~J245.7514+21.6226 observed with MUSE. \textit{Bottom:} MUSE spectra of DESI~J245.7514+21.6226. For more information on the system, see Desc. \ref{ref:lens54}. }
\label{fig:MUSEspectra54}
\end{minipage}
\end{figure*}

\begin{figure*}[!ht]
\centering
\begin{minipage}{1.0\textwidth}
\centering
\includegraphics[width=0.27\textwidth]{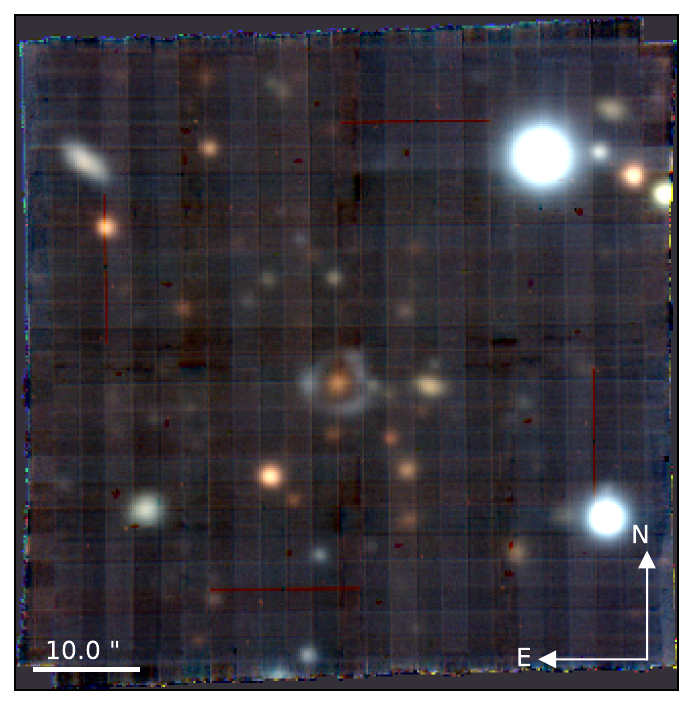}
\includegraphics[width=0.27\textwidth]{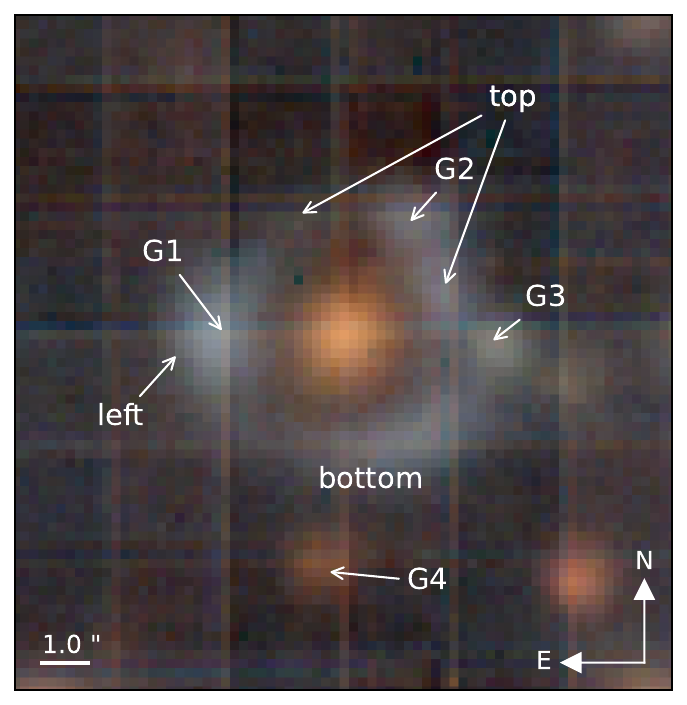}

\includegraphics[width=0.78\textwidth]{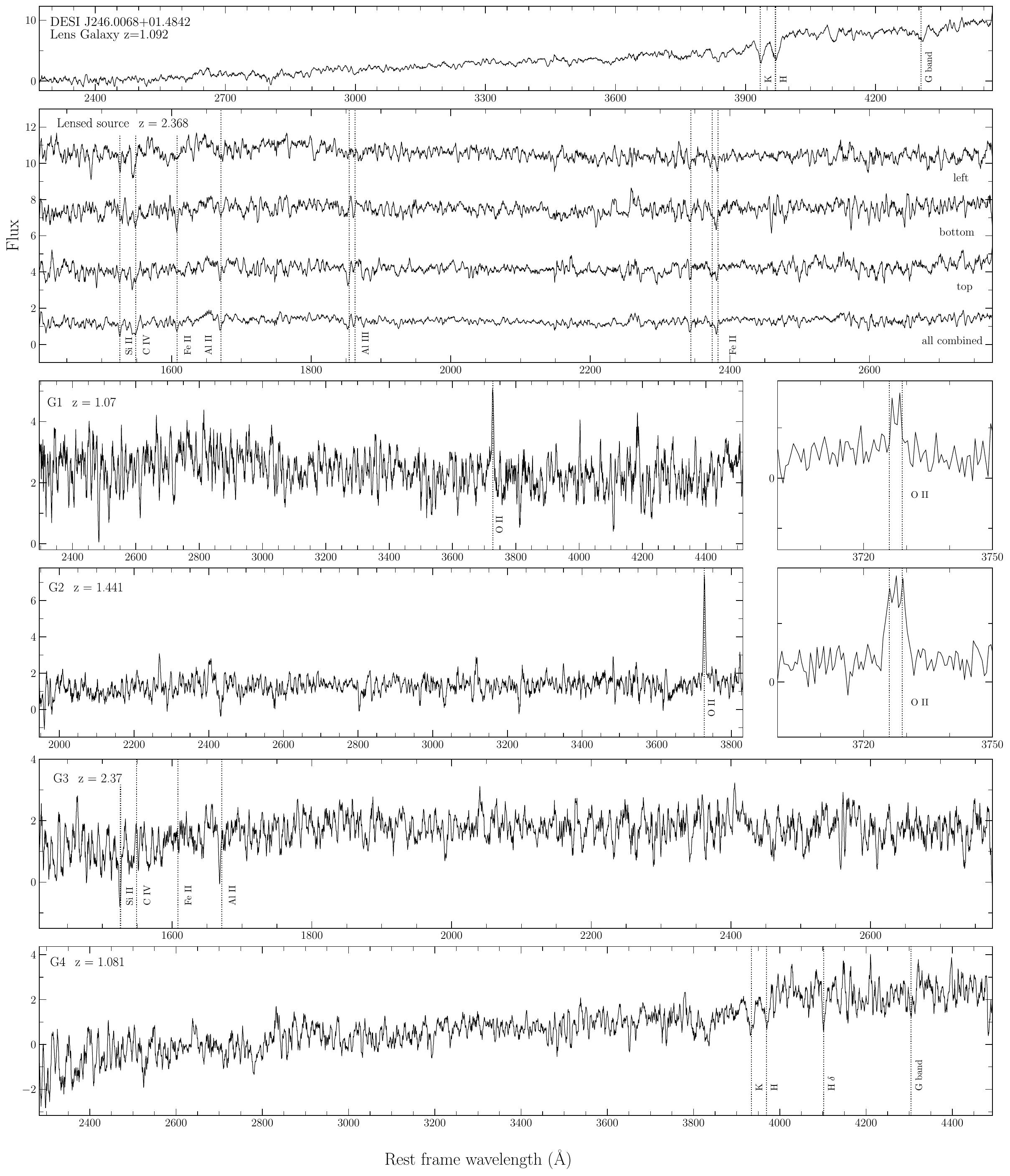}
\caption{\textit{Top:} RGB image of gravitational lens system DESI~J246.0068+01.4842 observed with MUSE. \textit{Bottom:} MUSE spectra of DESI~J246.0068+01.4842. Note that the quality flag for galaxies 1 and 3 is $Q_z=3$. For more information on the system, see Desc. \ref{ref:lens123}.}
\label{fig:MUSEspectra123}
\end{minipage}
\end{figure*}

\begin{figure*}[!ht]
\centering
\begin{minipage}{1.0\textwidth}
\centering
\includegraphics[width=0.4\textwidth]{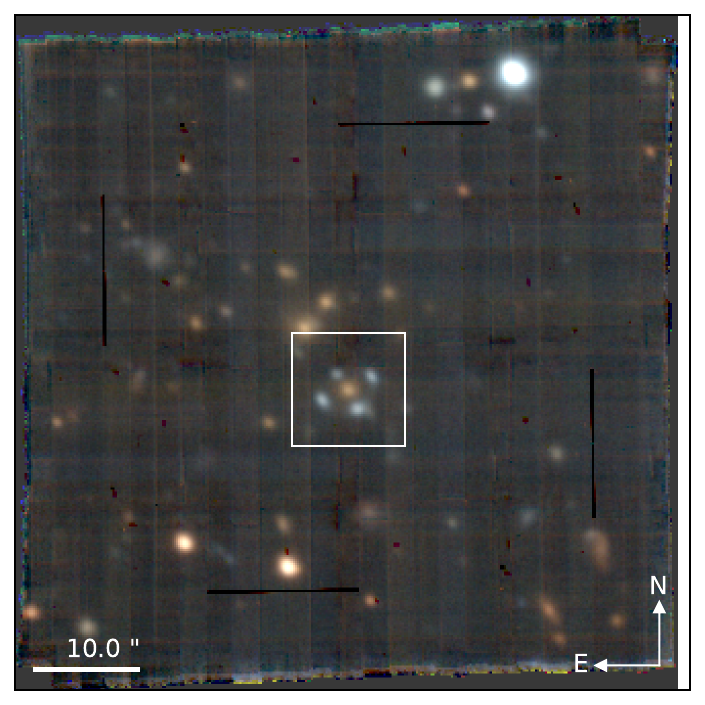}
\includegraphics[width=0.404\textwidth]{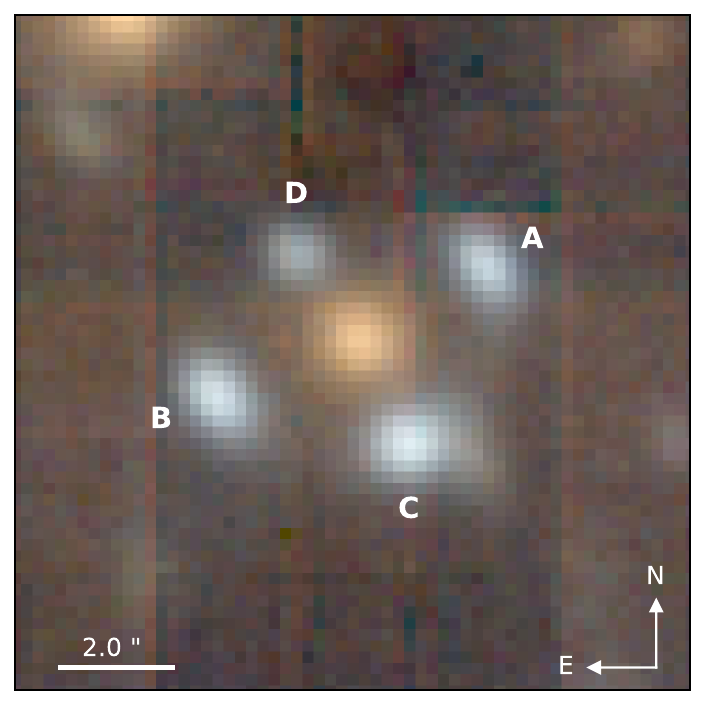}
\label{fig:MUSElens57}
\includegraphics[width=\textwidth]{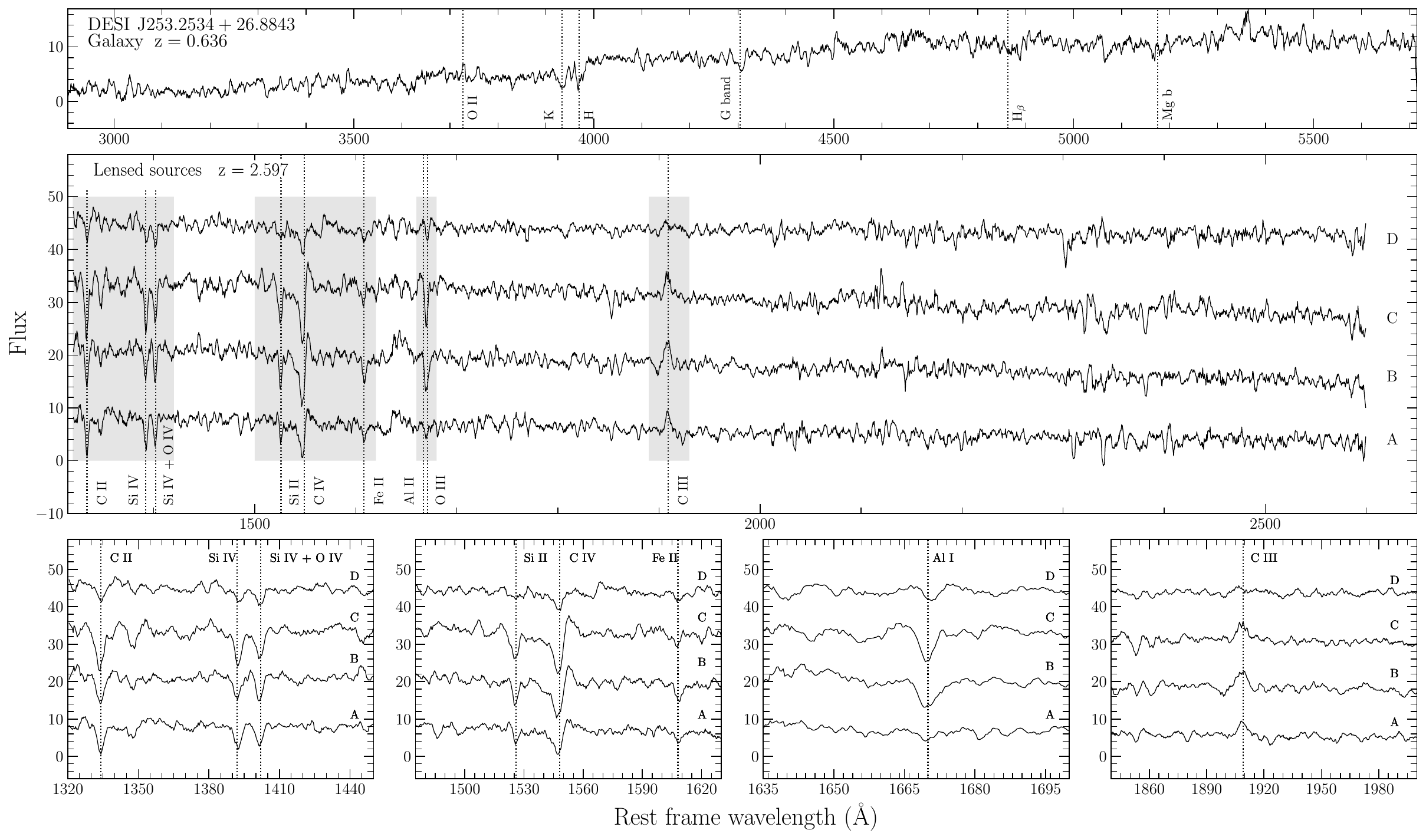}
\caption{\textit{Top:} RGB image of gravitational lens system DESI~J253.2534+26.8843 observed with MUSE. \textit{Bottom:} MUSE spectra of DESI~J253.2534+26.8843. For more information on the system, see Desc. \ref{ref:lens57}. }
\label{fig:MUSEspectra57}
\end{minipage}
\end{figure*}

\begin{figure*}[!ht]
\centering
\begin{minipage}{1.0\textwidth}
\centering
\includegraphics[width=0.4\textwidth]{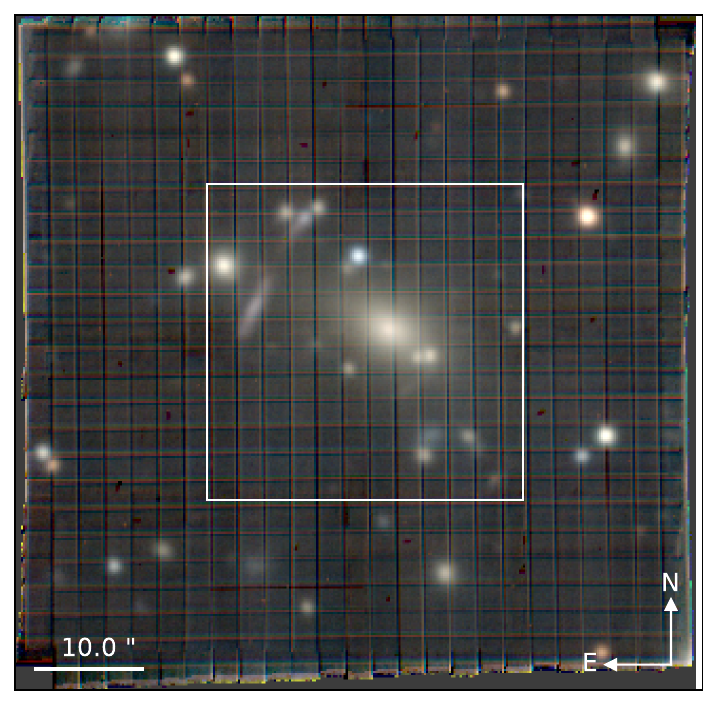}
\includegraphics[width=0.404\textwidth]{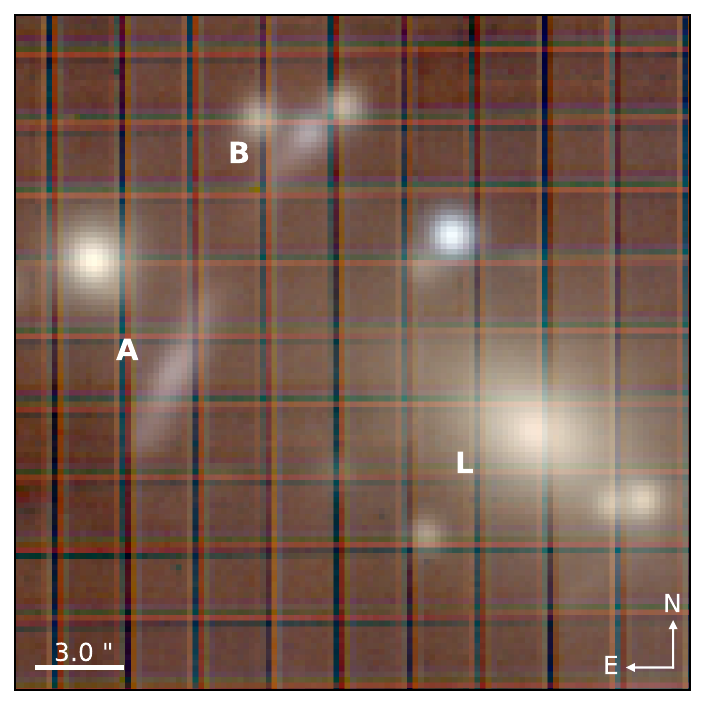}
\label{fig:MUSElens58}
\includegraphics[width=\textwidth]{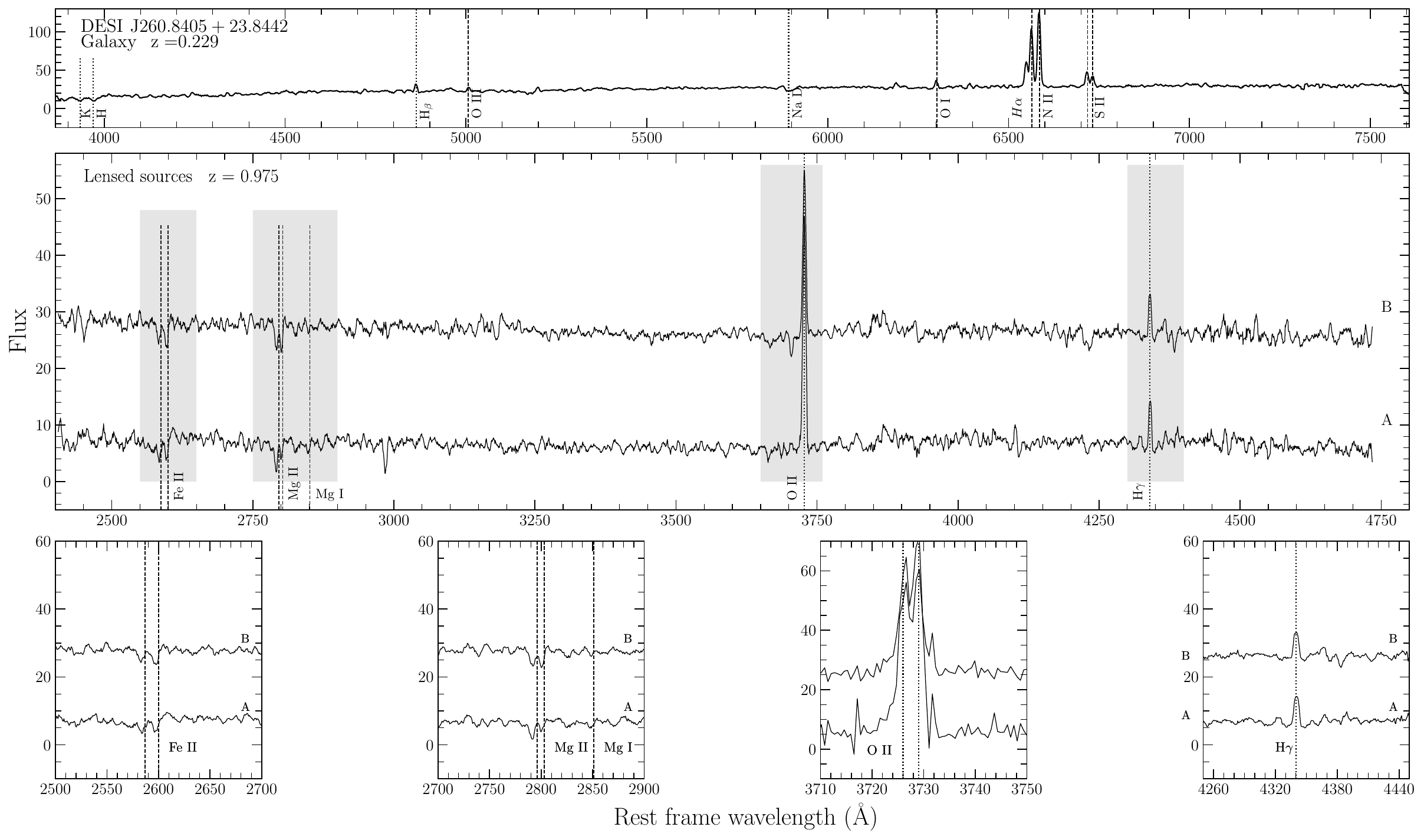}
\caption{\textit{Top:} RGB image of gravitational lens system DESI~J260.8405+23.84423 observed with MUSE. \textit{Bottom:} MUSE spectra of DESI~J260.8405+23.84423. For more information on the system, see Desc. \ref{ref:lens58}.}
\label{fig:MUSEspectra58}
\end{minipage}
\end{figure*}

\begin{figure*}[!ht]
\centering
\begin{minipage}{1.0\textwidth}
\centering
\includegraphics[width=0.4\textwidth]{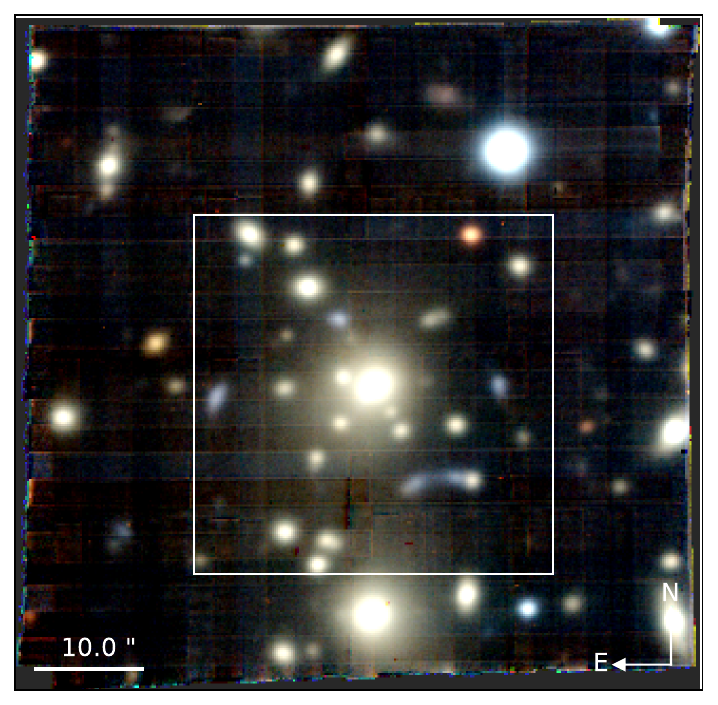}
\includegraphics[width=0.4\textwidth]{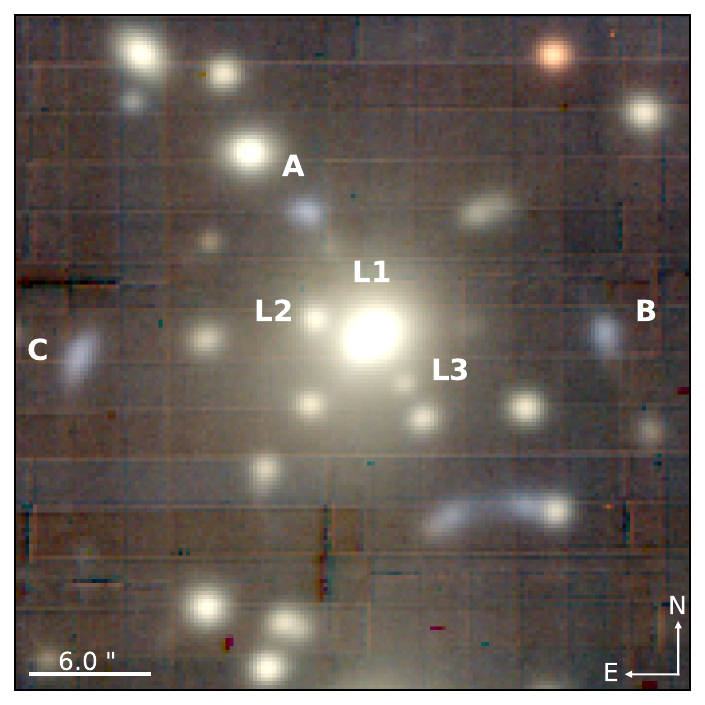}
\label{fig:MUSElens304}
\includegraphics[width=\textwidth]{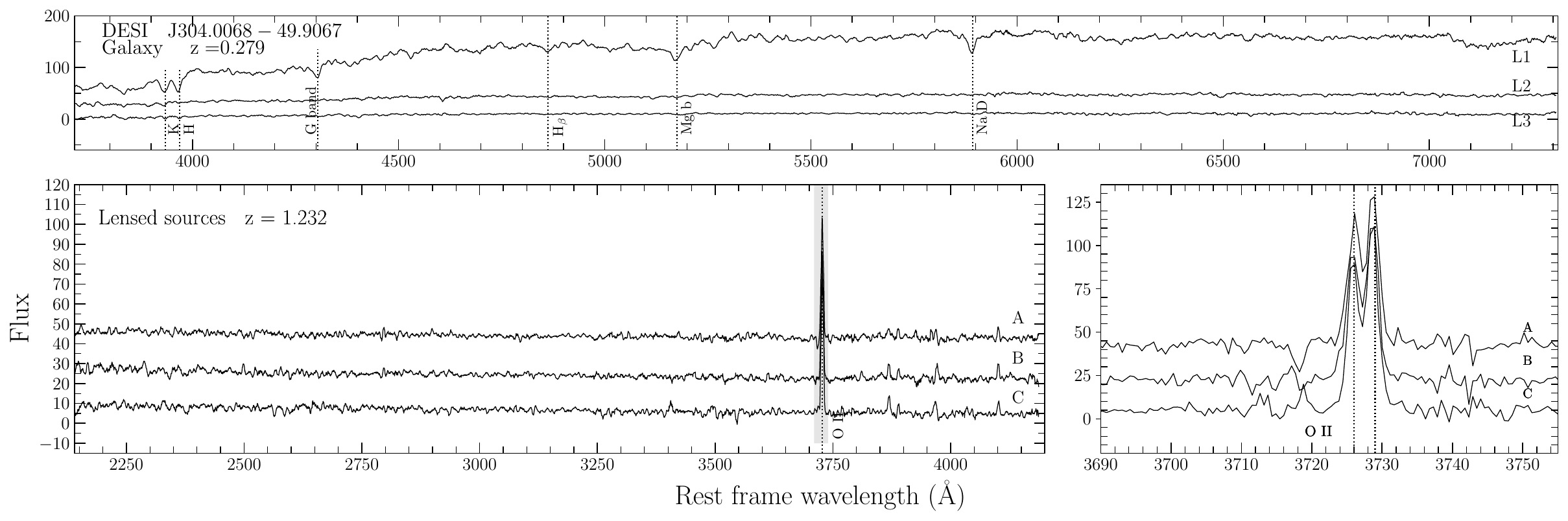}
\caption{\textit{Top:} RGB image of gravitational lens system DESI~J304.0068-49.9067 observed with MUSE. \textit{Bottom:} MUSE spectra of DESI~J304.0068-49.9067. For more information on the system, see Desc. \ref{ref:lens304}. }
\label{fig:MUSEspectra304}
\end{minipage}
\end{figure*}

\begin{figure*}[!ht]
\centering
\begin{minipage}{1.0\textwidth}
\centering
\includegraphics[width=0.4\textwidth]{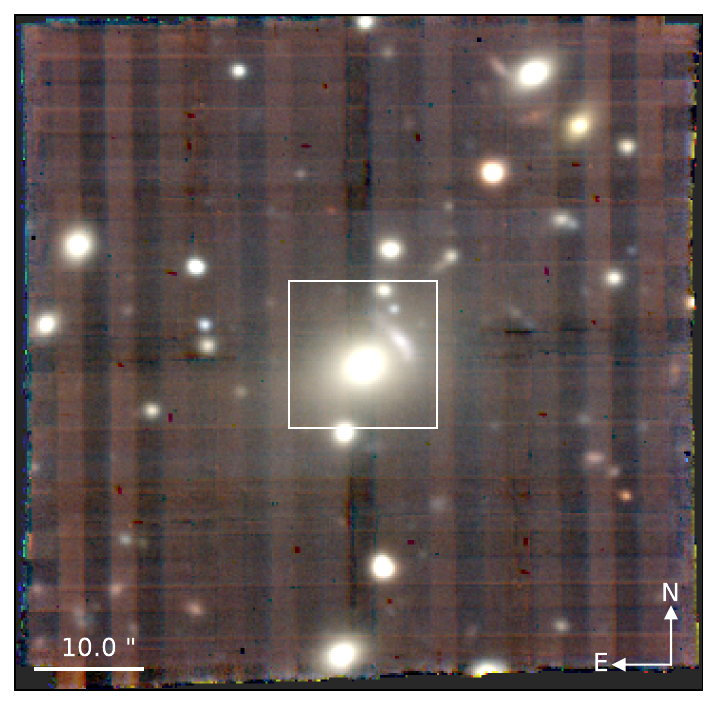}
\includegraphics[width=0.4\textwidth]{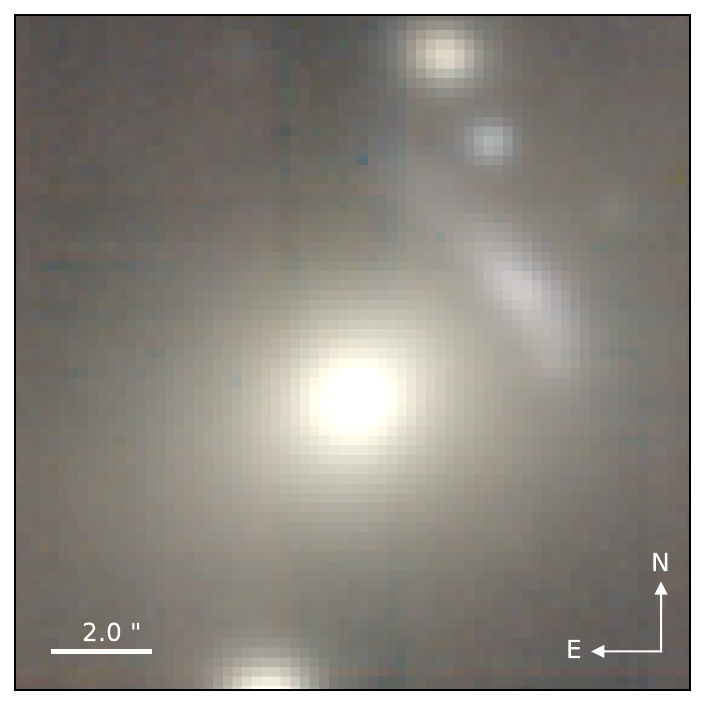}
\label{fig:MUSElens4}
\includegraphics[width=\textwidth]{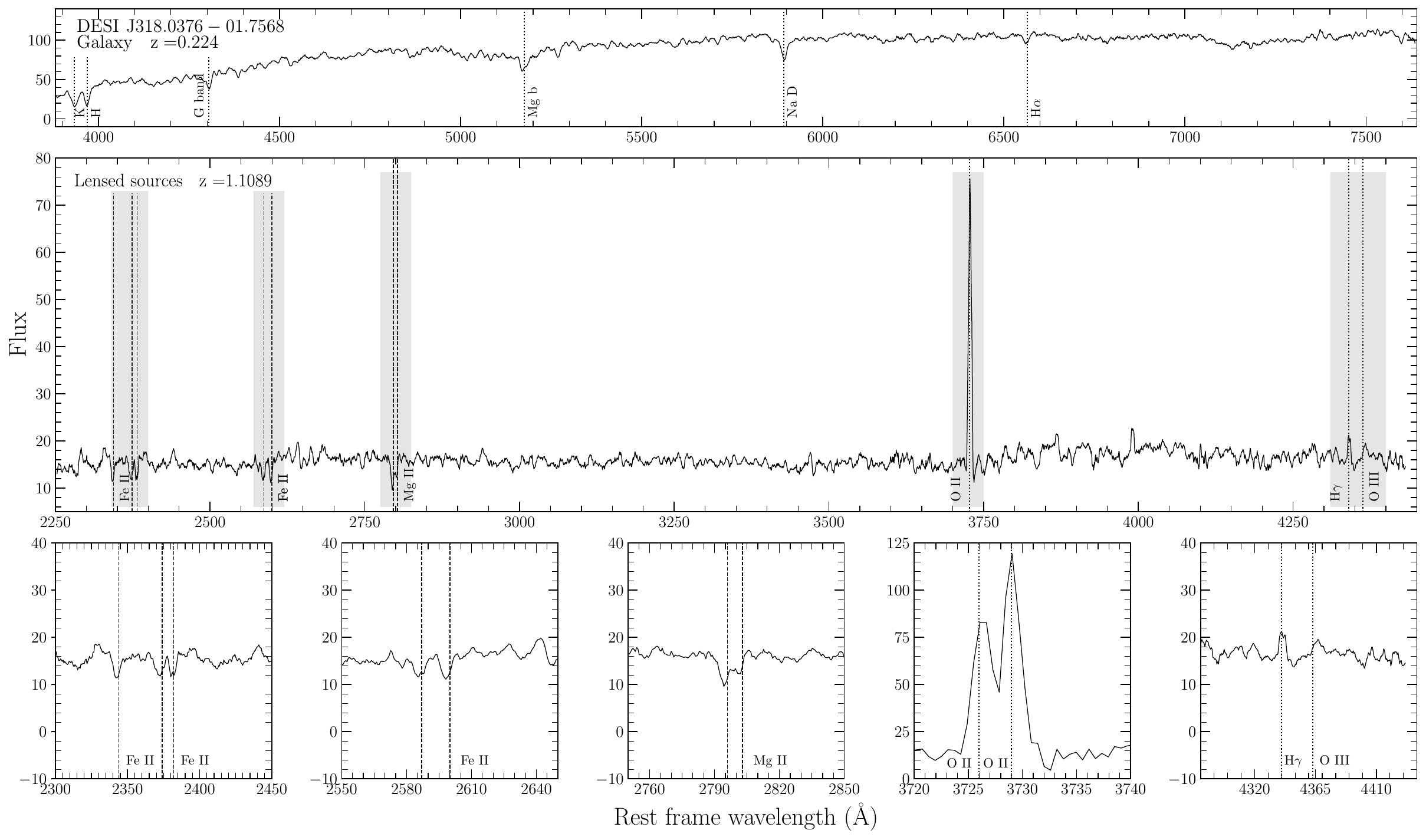}
\caption{\textit{Top:} RGB image of gravitational lens system DESI~J318.0376-01.7568 observed with MUSE. \textit{Bottom:} MUSE spectra of DESI~J318.0376-01.7568. For more information on the system, see Desc. \ref{ref:lens4}. }
\label{fig:MUSEspectra4}
\end{minipage}
\end{figure*}

\begin{figure*}[!ht]
\centering
\begin{minipage}{1.0\textwidth}
\centering
\includegraphics[width=0.4\textwidth]{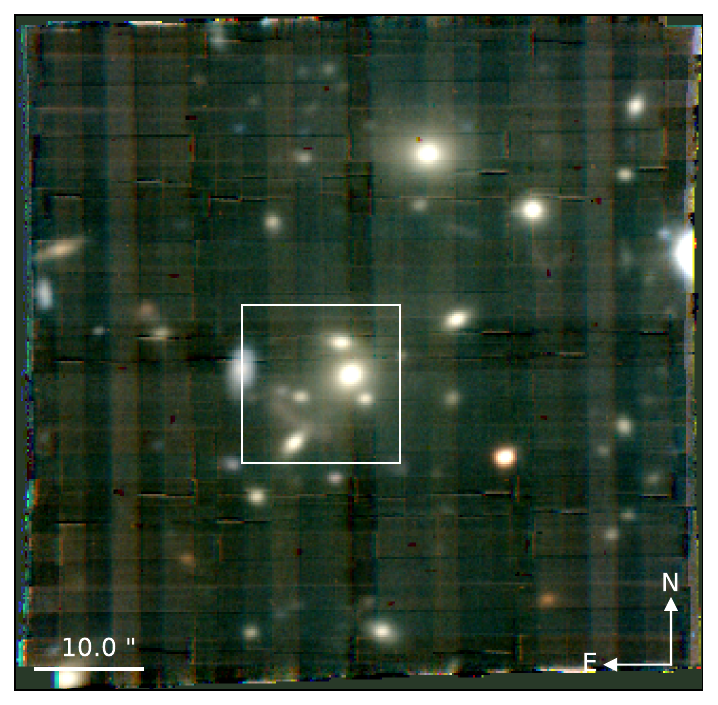}
\includegraphics[width=0.4\textwidth]{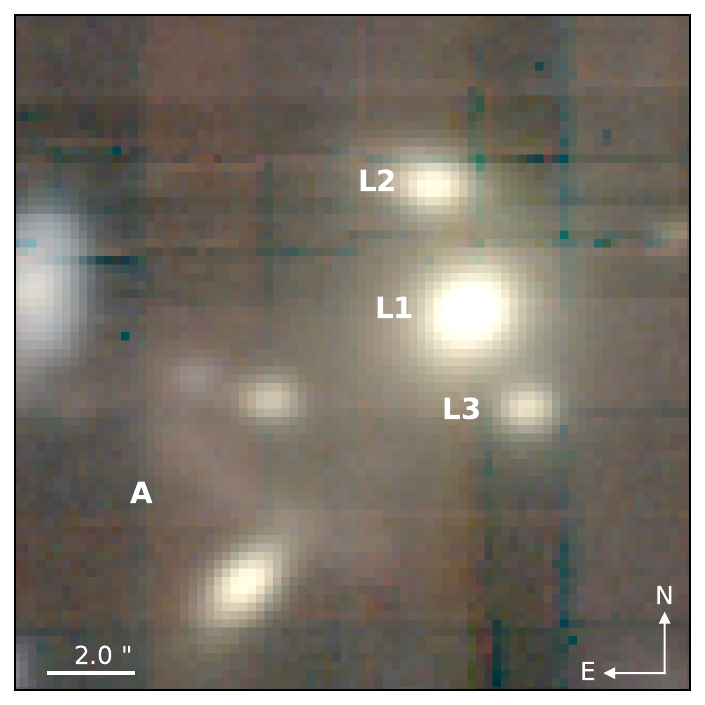}
\label{fig:MUSEspectra10_10}
\includegraphics[width=\textwidth]{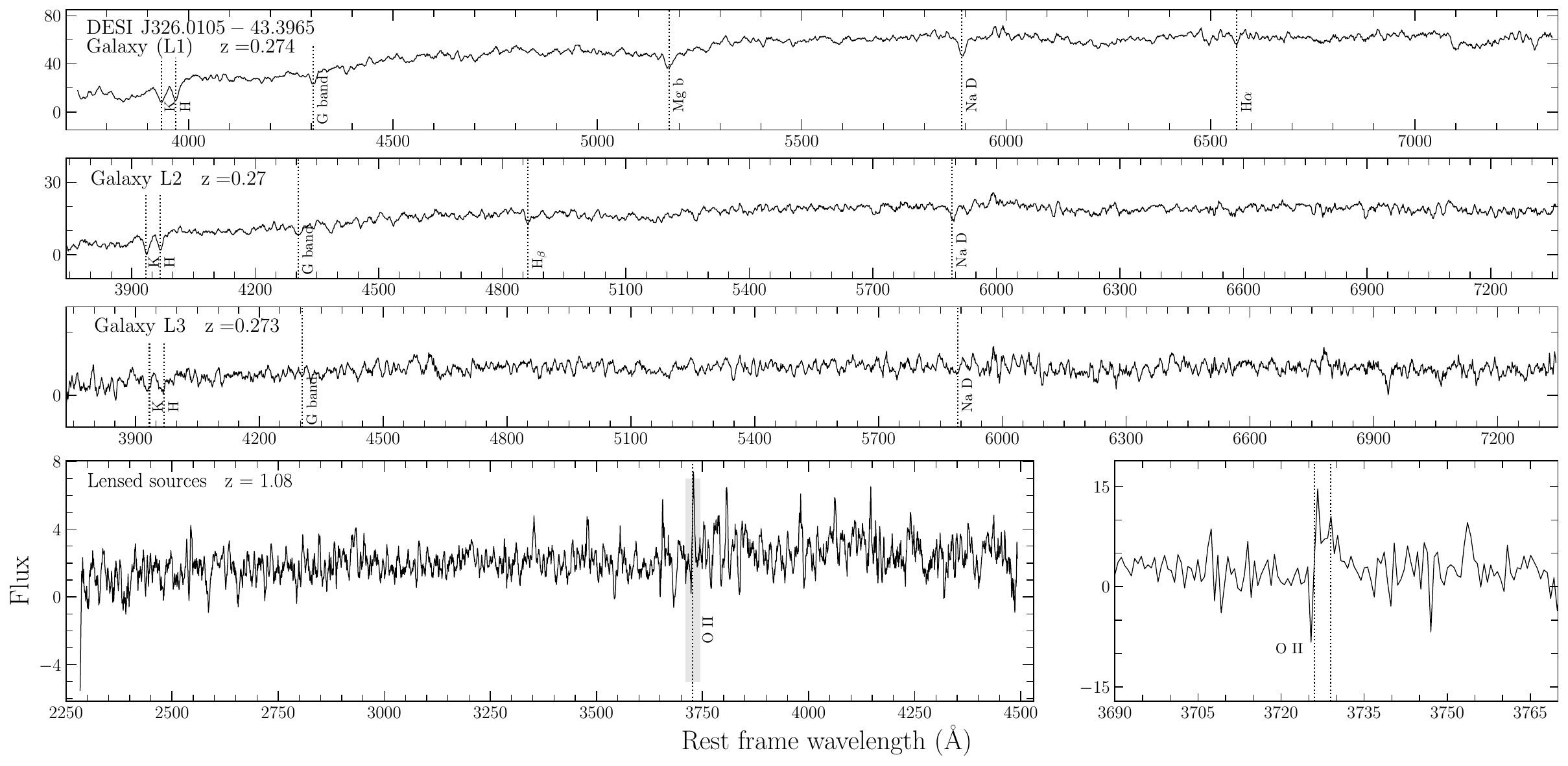 }
\caption{\textit{Top:} RGB image of gravitational lens system DESI~J326.0105-43.3965 observed with MUSE. \textit{Bottom:} MUSE spectra of DESI~J326.0105-43.3965. Note that the source quality flag is $Q_z=2$. For more information on the system, see Desc. \ref{ref:lens10}. }
\label{fig:MUSEspectra10}
\end{minipage}
\end{figure*}

\begin{figure*}[!ht]
\centering
\begin{minipage}{1.0\textwidth}
\centering
\includegraphics[width=0.4\textwidth]{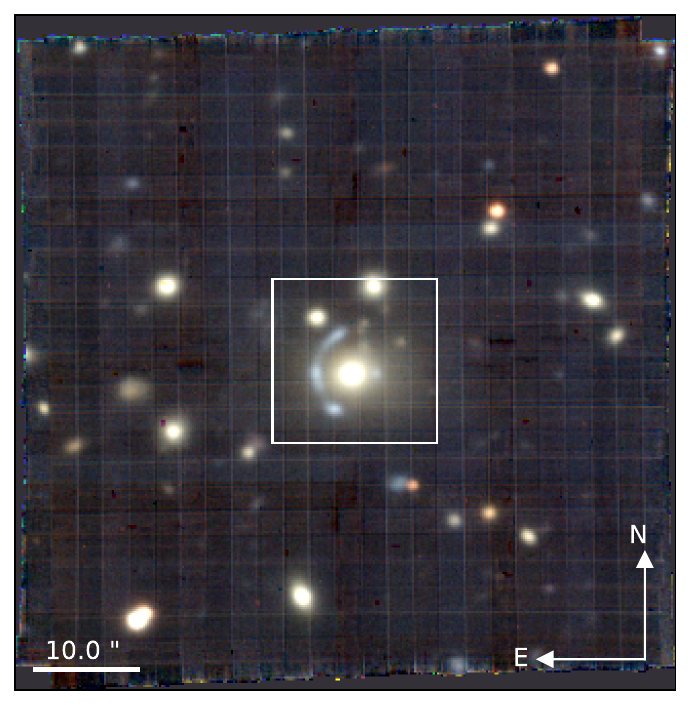}
\includegraphics[width=0.404\textwidth]{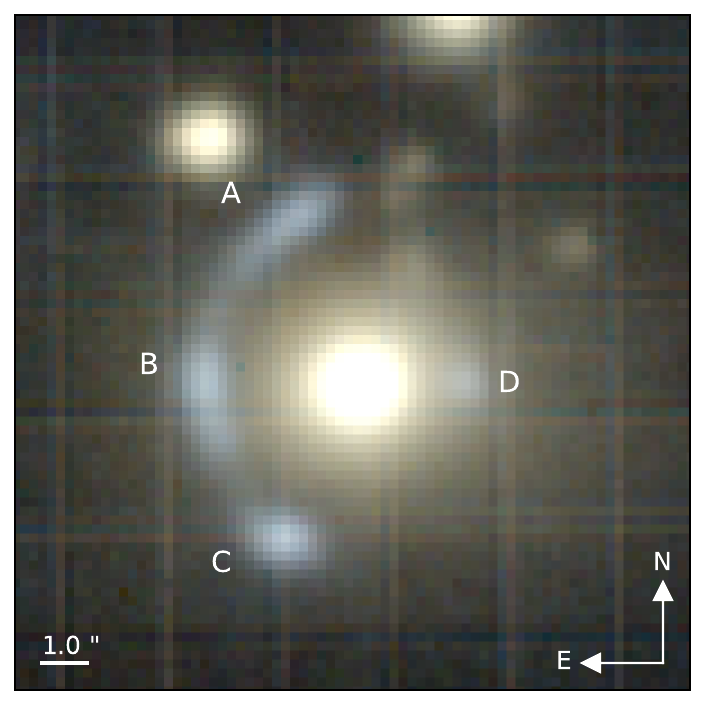}

\includegraphics[width=\textwidth]{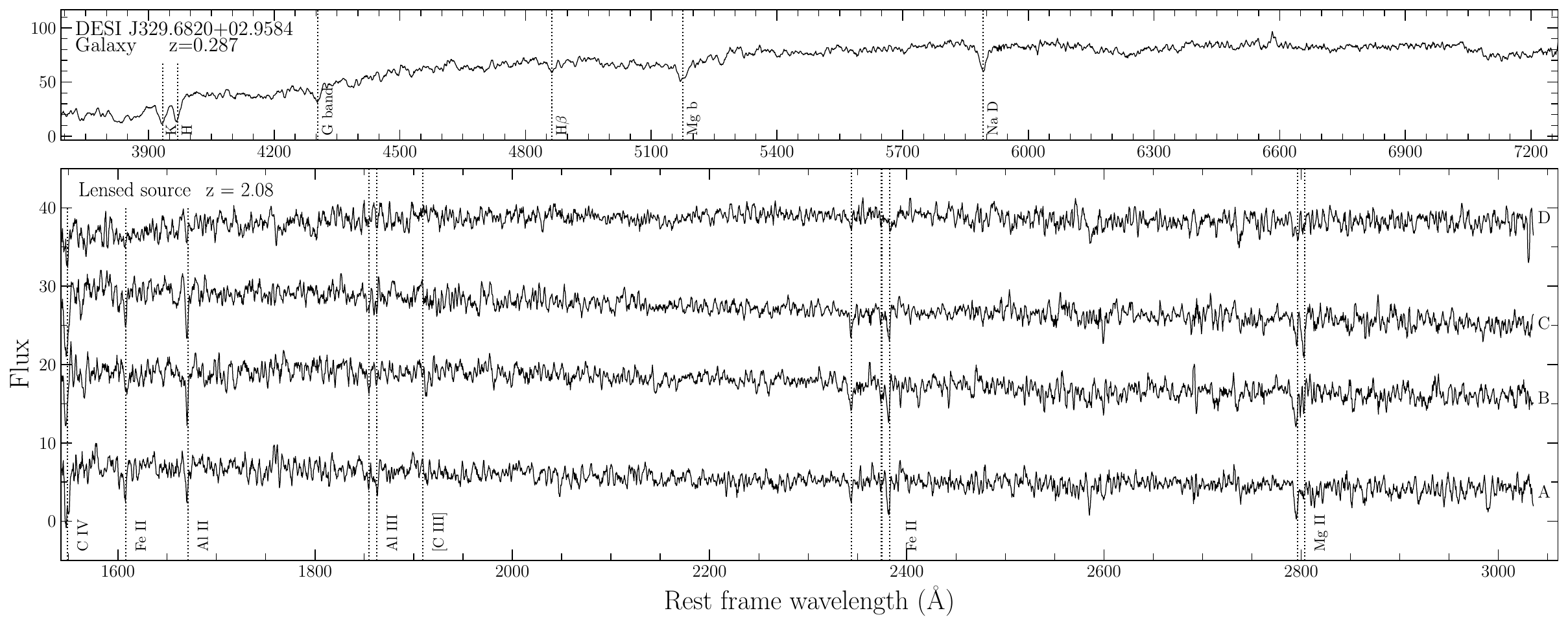}
\caption{\textit{Top:} RGB image of gravitational lens system DESI~J329.6820+02.9584 observed with MUSE. \textit{Bottom:} MUSE spectra of DESI~J329.6820+02.9584. Note that the quality flag for Source D is $Q_z=2$. For more information on the system, see Desc. \ref{Ref:lens59}.}
\label{fig:MUSEspectra59}
\end{minipage}
\end{figure*}

\begin{figure*}[!ht]
\centering
\begin{minipage}{1.0\textwidth}
\centering
\includegraphics[width=0.4\textwidth]{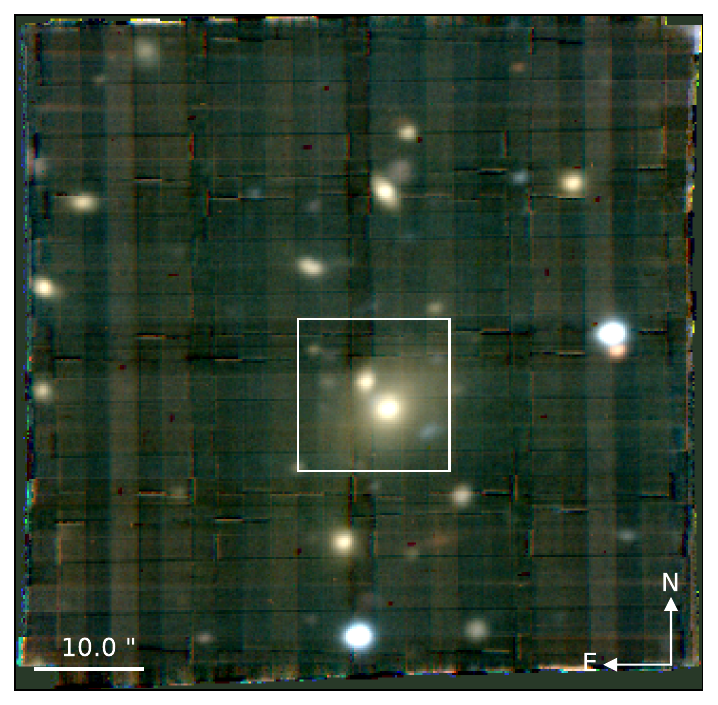}
\includegraphics[width=0.4\textwidth]{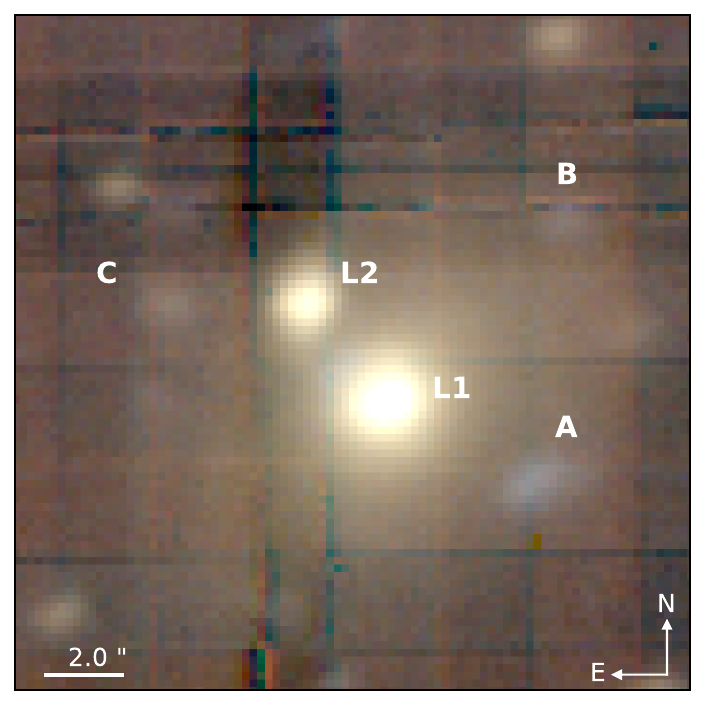}
\label{fig:MUSElens12}
\includegraphics[width=\textwidth]{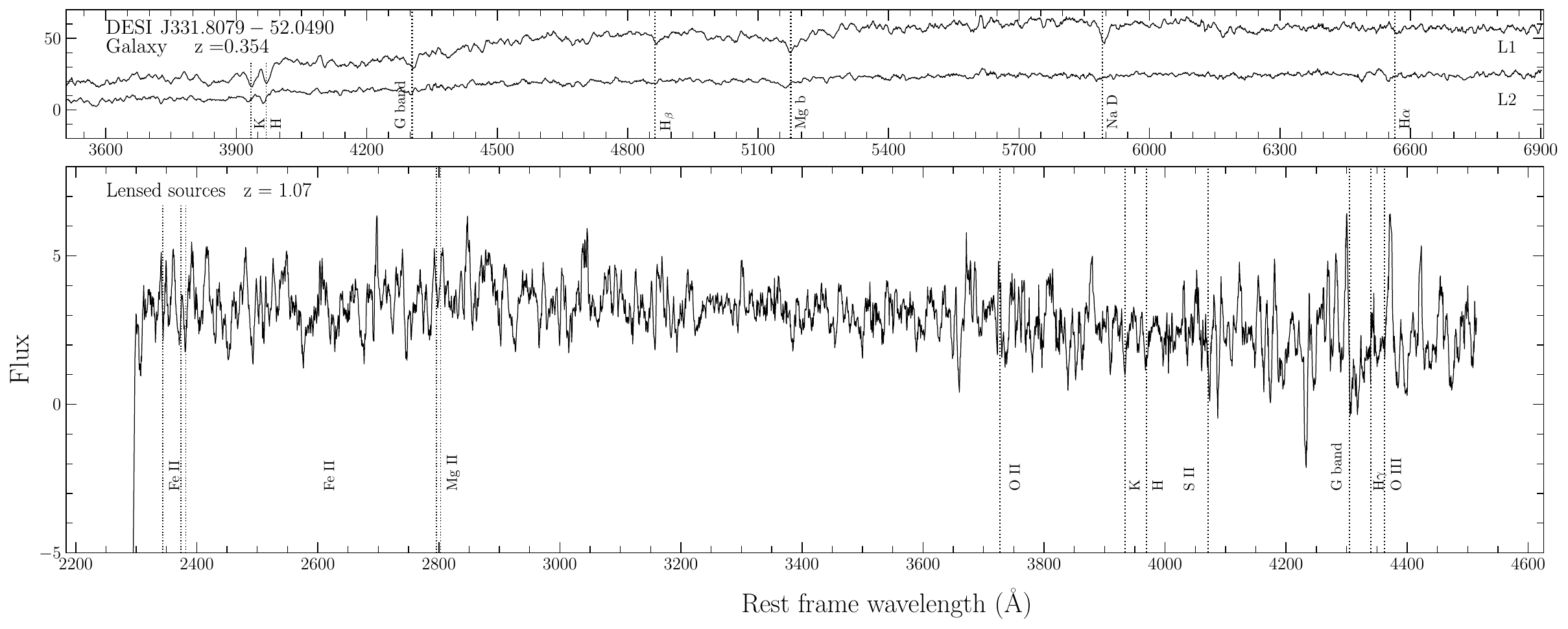}
\caption{\textit{Top:} RGB image of gravitational lens system DESI~J331.8083-52.0487 observed with MUSE. \textit{Bottom:} MUSE spectra of DESI~J331.8083-52.0487. Note that the quality flag for all sources is $Q_z=2$. For more information on the system, see Desc. \ref{ref:lens12}. }
\label{fig:MUSEspectra12}
\end{minipage}
\end{figure*}

\begin{figure*}[!ht]
\centering

\begin{minipage}{1.0\textwidth}
\centering
\includegraphics[width=0.4\textwidth]{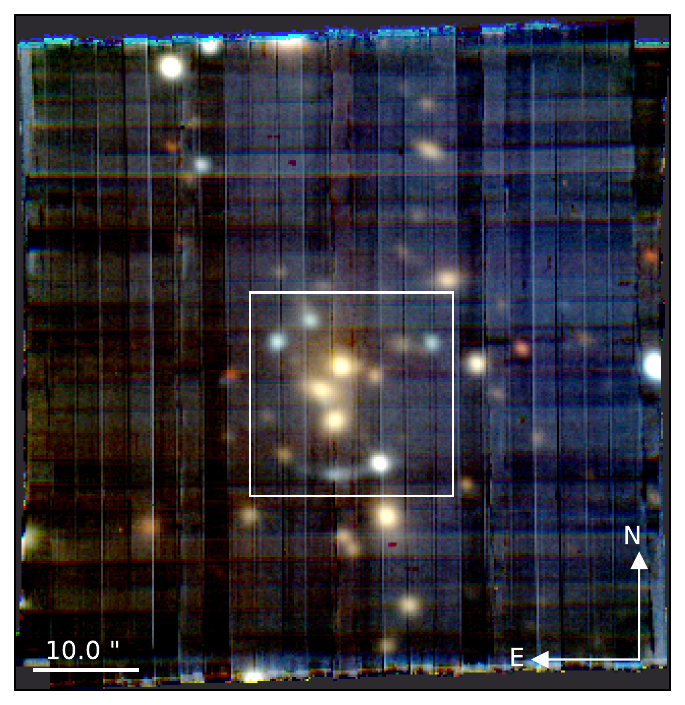}
\includegraphics[width=0.404\textwidth]{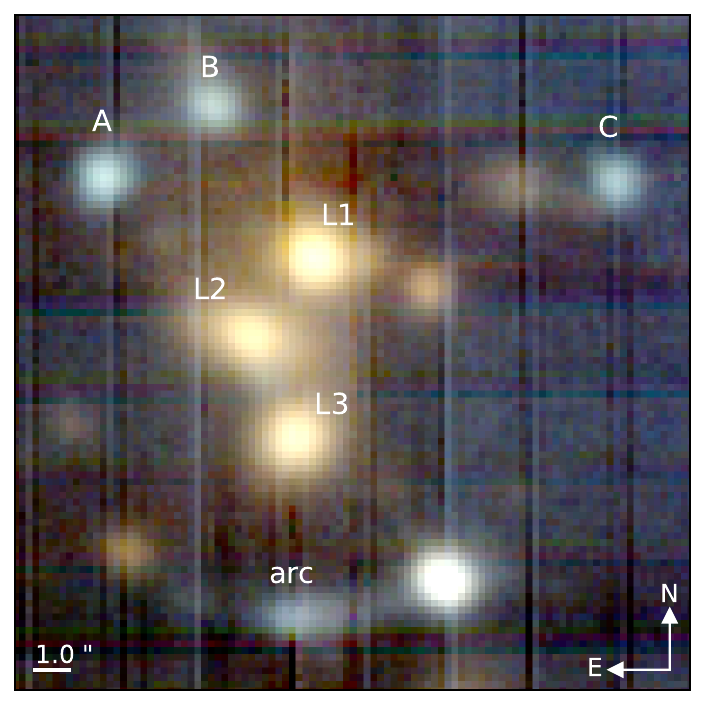}

\includegraphics[width=\textwidth]{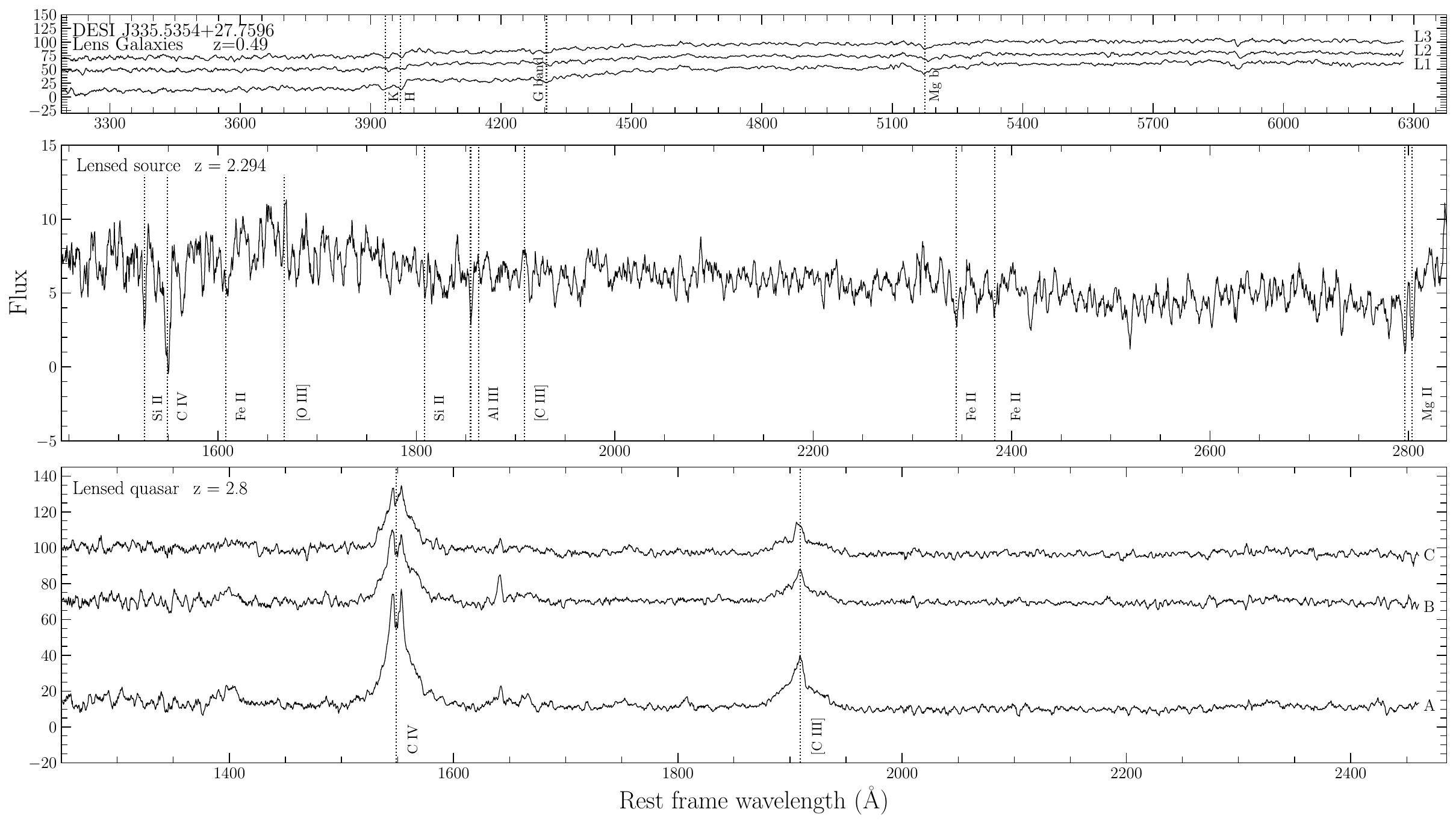}
\caption{\textit{Top:} RGB image of gravitational lens system DESI~J335.5354+27.7596 observed with MUSE. \textit{Bottom:} MUSE spectra of DESI~J335.5354+27.7596. For more information on the system, see Desc. \ref{Ref:lens64}.}
\label{fig:MUSEspectra64}
\end{minipage}
\end{figure*}

\begin{figure*}[!ht]
\centering
\begin{minipage}{1.0\textwidth}
\centering
\includegraphics[width=0.4\textwidth]{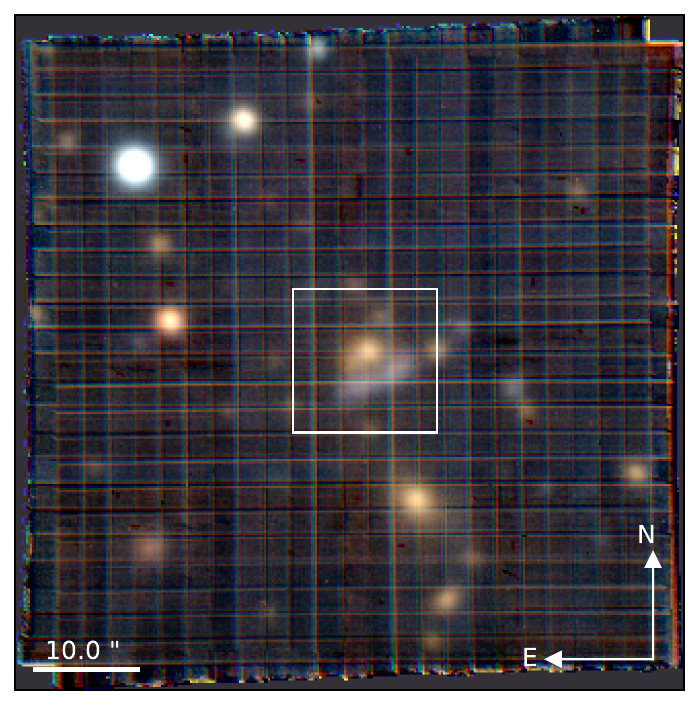}
\includegraphics[width=0.404\textwidth]{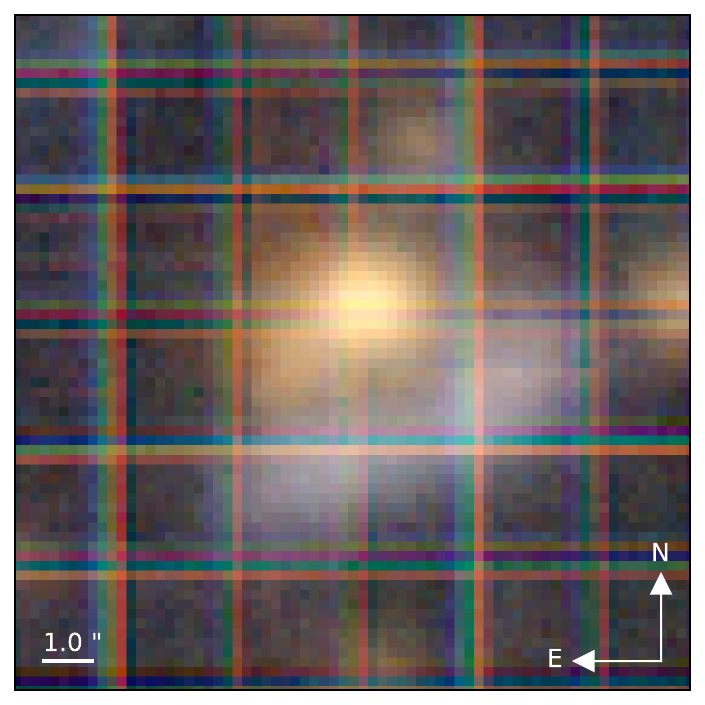}

\includegraphics[width=\textwidth]{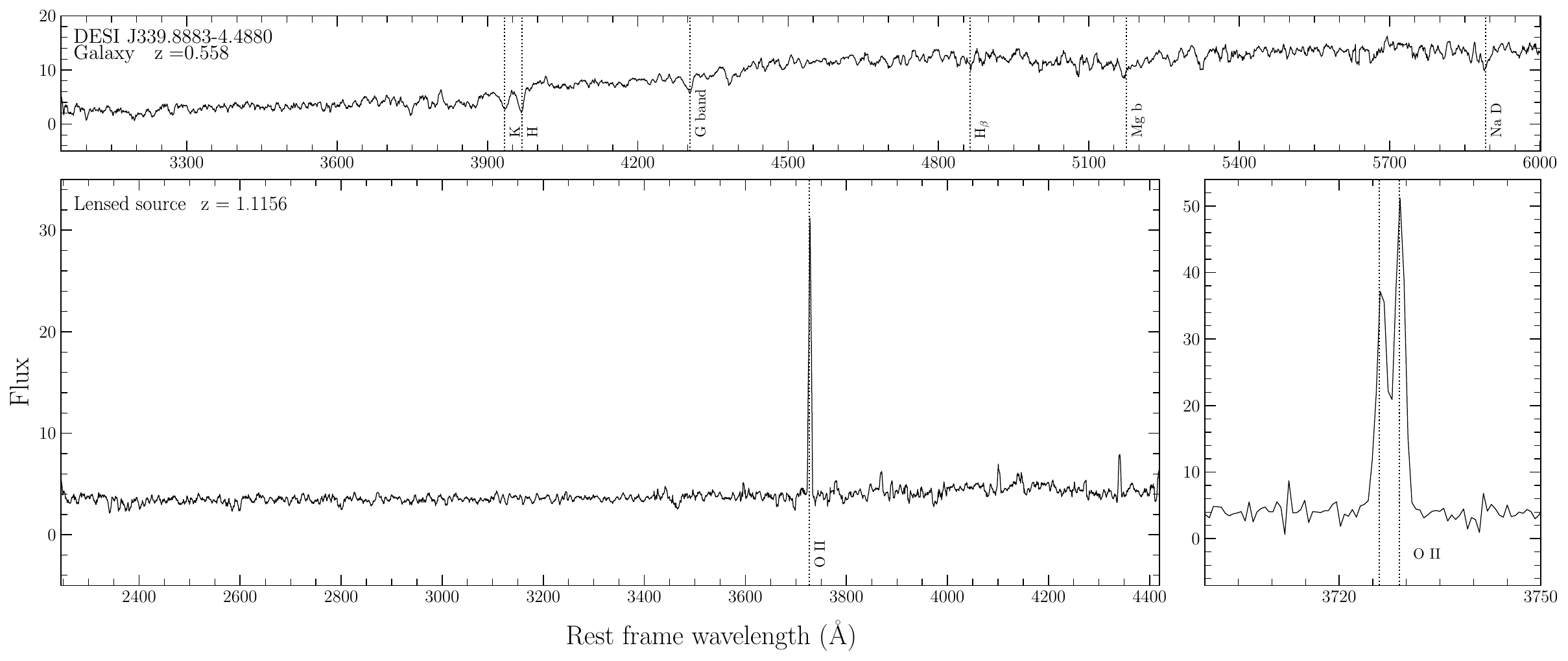}
\caption{\textit{Top:} RGB image of gravitational lens system DESI~J339.8883-4.4880 observed with MUSE. \textit{Bottom:} MUSE spectra of DESI~J339.8883-4.4880. For more information on the system, see Desc. \ref{Ref:lens61}.}
\label{fig:MUSEspectra61}
\end{minipage}
\end{figure*}

\begin{figure*}[!ht]
\centering
\begin{minipage}{1.0\textwidth}
\centering
\includegraphics[width=0.4\textwidth]{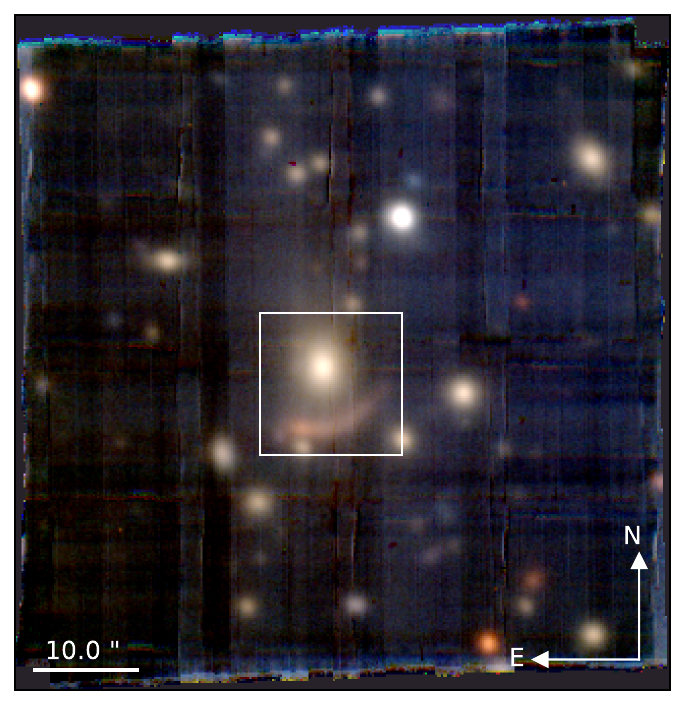}
\includegraphics[width=0.404\textwidth]{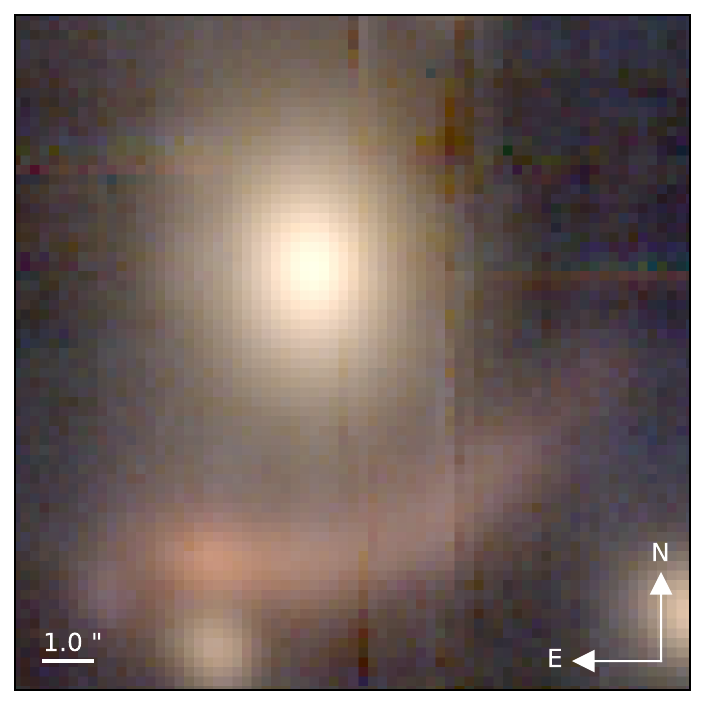}

\includegraphics[width=\textwidth]{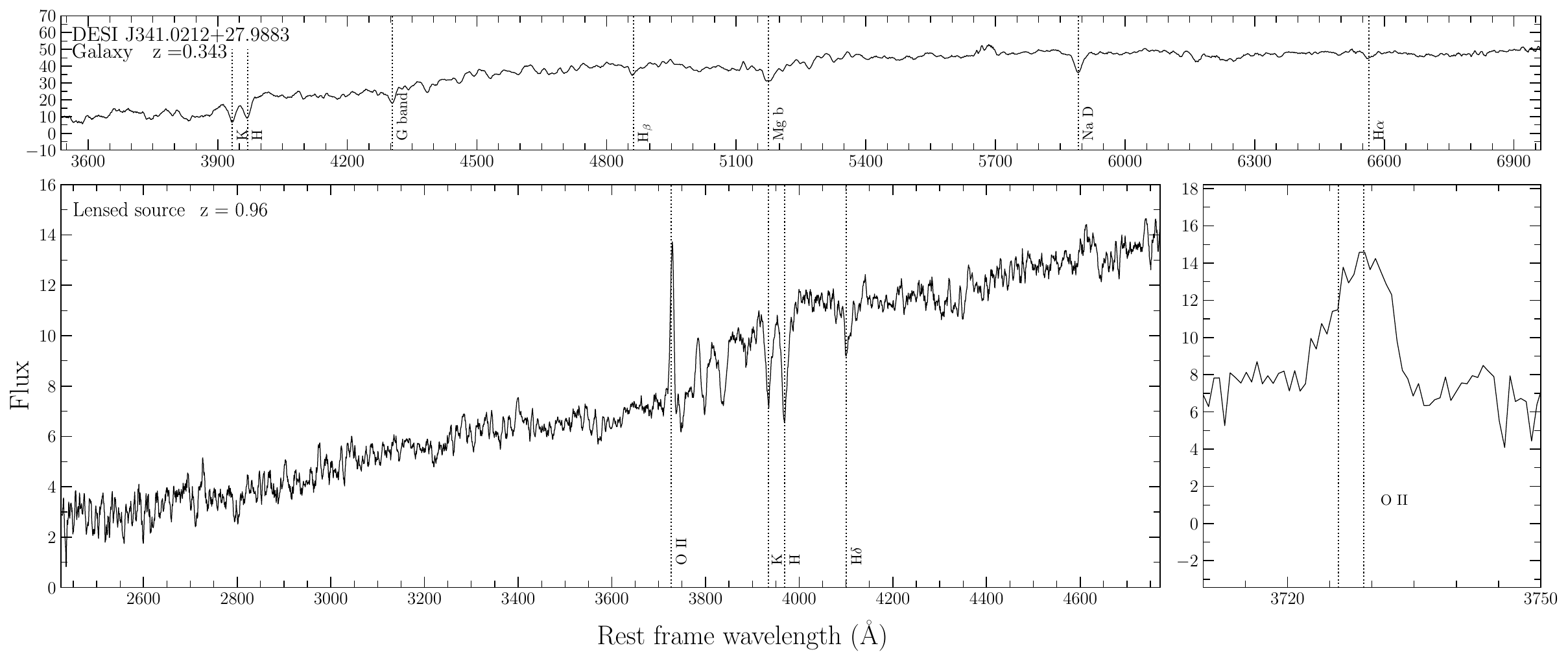}
\caption{\textit{Top:} RGB image of gravitational lens system DESI~J341.0212+27.9883 observed with MUSE. \textit{Bottom:} MUSE spectra of DESI~J341.0212+27.9883. For more information on the system, see Desc. \ref{Ref:lens65}.}
\label{fig:MUSEspectra65}
\end{minipage}
\end{figure*}

\begin{figure*}[!ht]
\centering
\begin{minipage}{1.0\textwidth}
\centering
\includegraphics[width=0.4\textwidth]{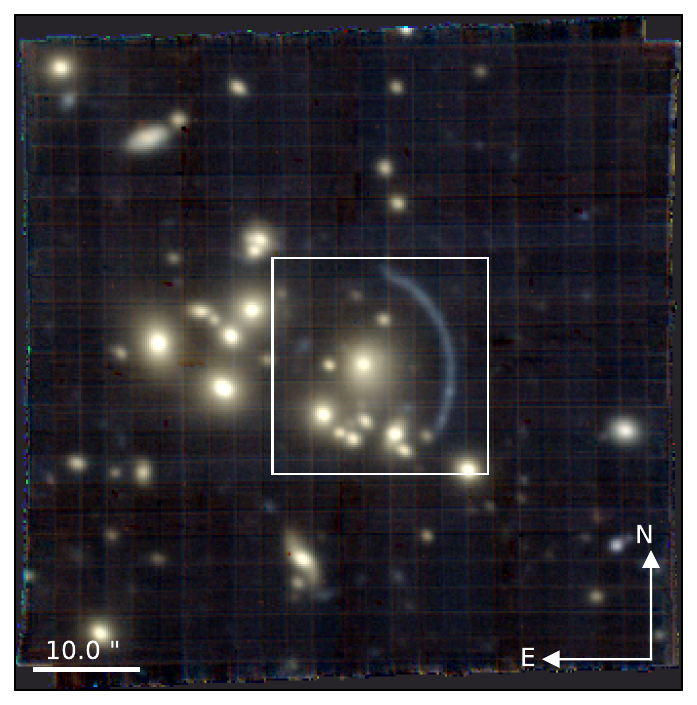}
\includegraphics[width=0.404\textwidth]{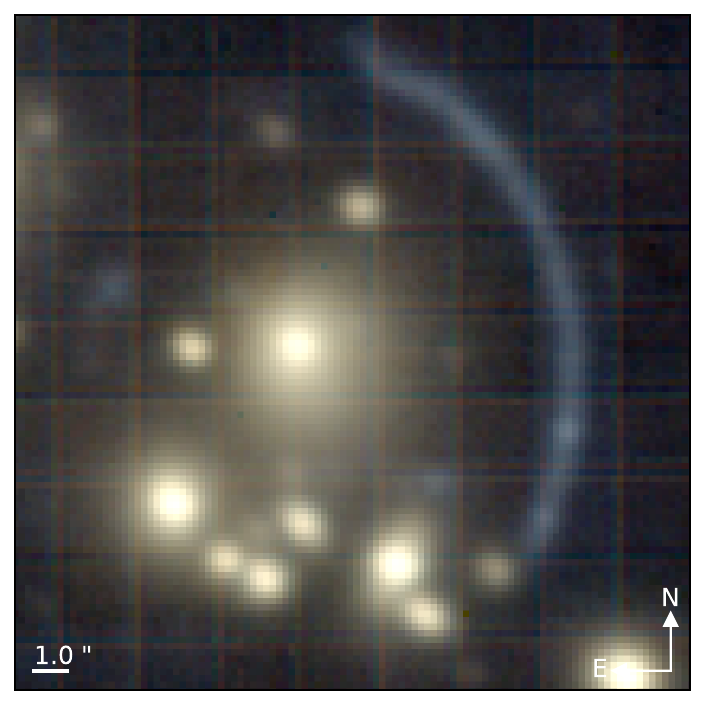}

\includegraphics[width=\textwidth]{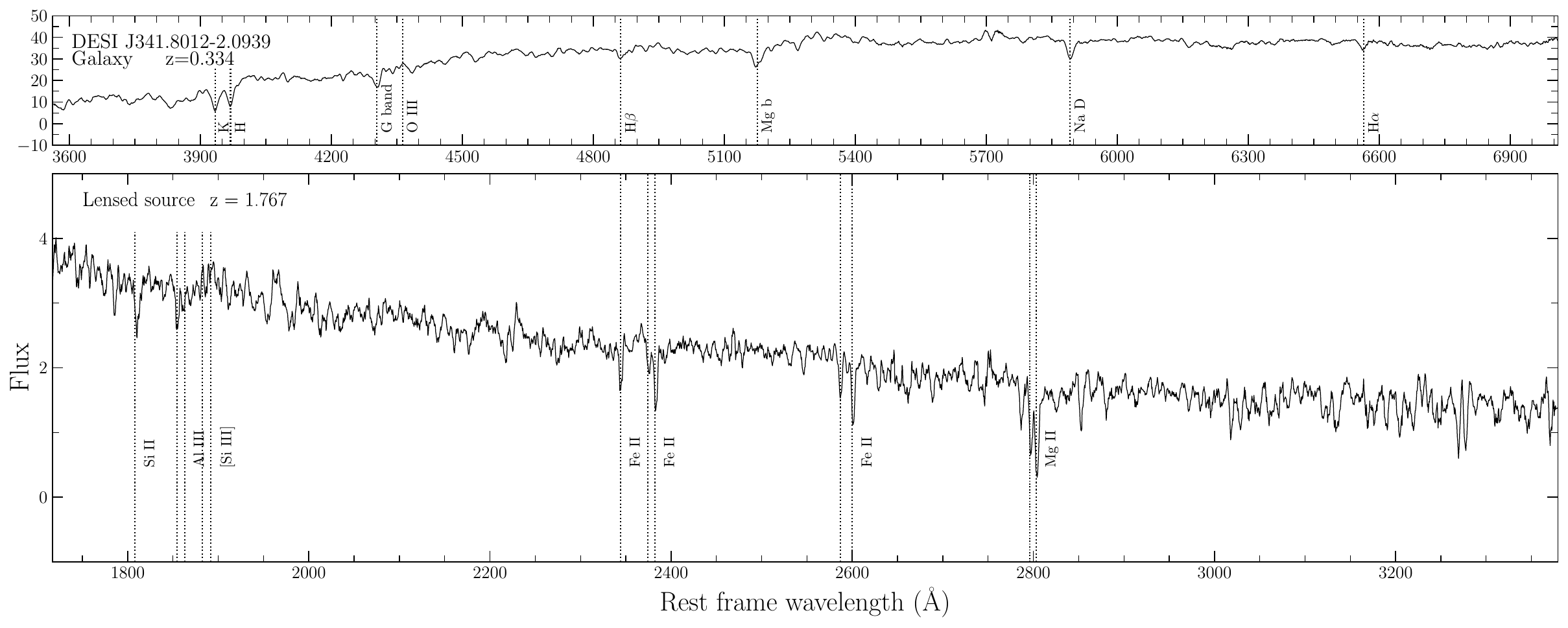}
\caption{\textit{Top:} RGB image of gravitational lens system DESI~J341.8012-2.0939 observed with MUSE. \textit{Bottom:} MUSE spectra of DESI~J341.8012-2.0939. For more information on the system, see Desc. \ref{Ref:lens63}.}
\label{fig:MUSEspectra63}
\end{minipage}
\end{figure*}

\begin{figure*}[!ht]
\centering
\begin{minipage}{1.0\textwidth}
\centering
\includegraphics[width=0.4\textwidth]{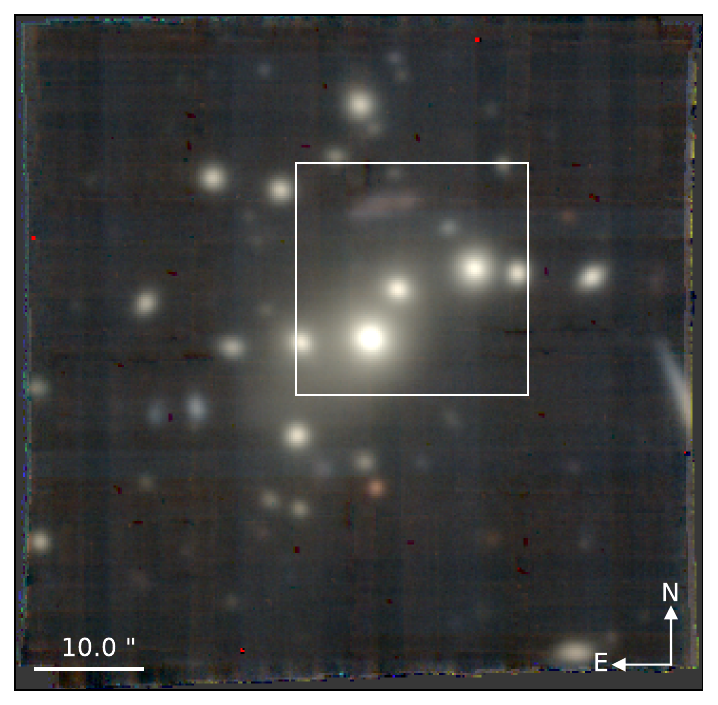}
\includegraphics[width=0.404\textwidth]{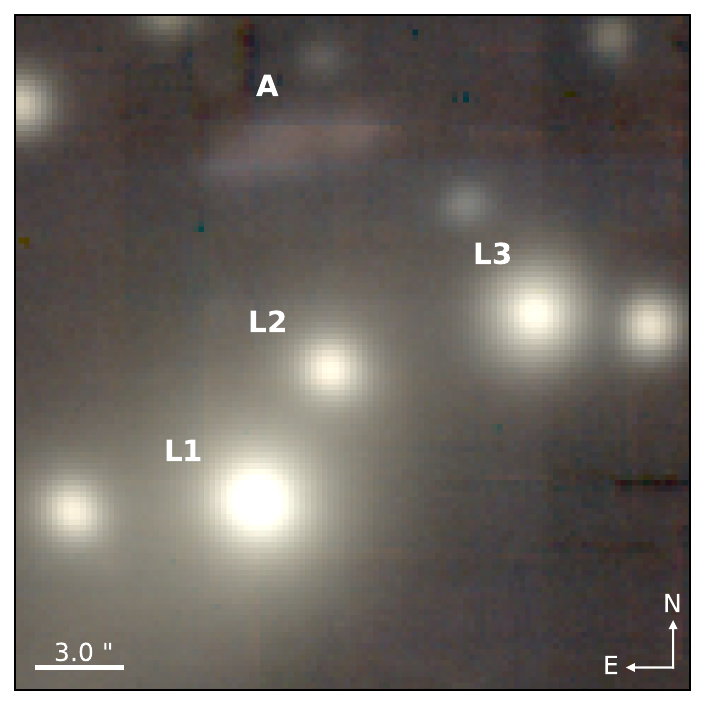}

\includegraphics[width=\textwidth]{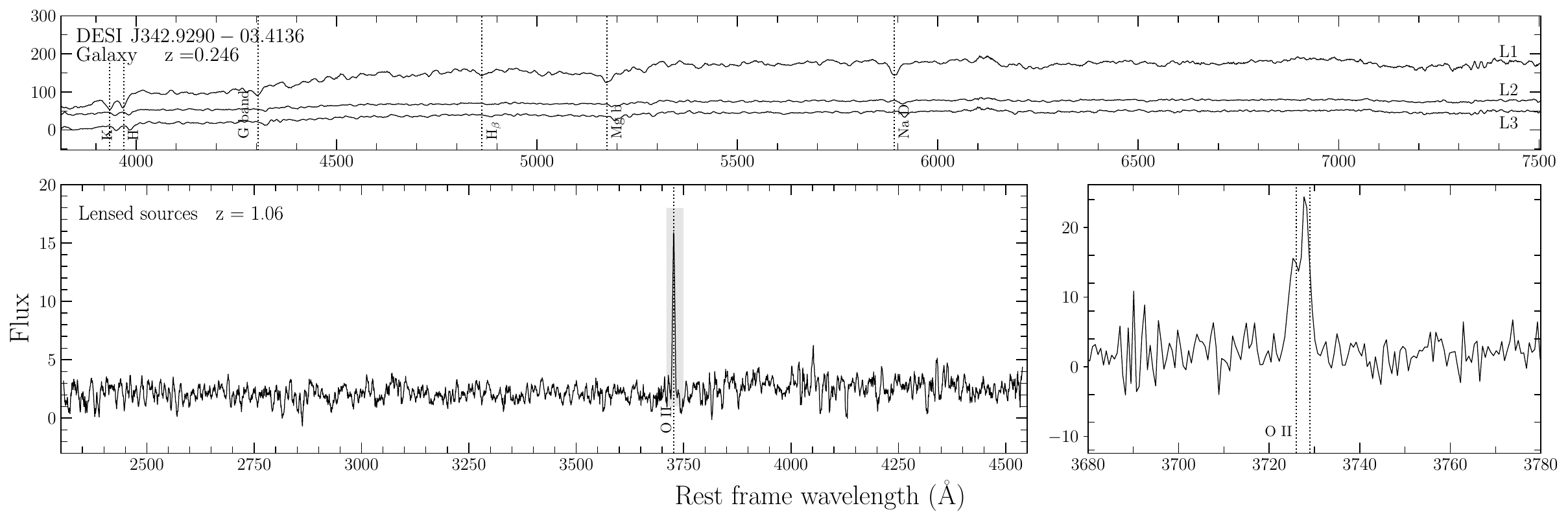}
\caption{\textit{Top:} RGB image of gravitational lens system DESI~J342.9290-03.4136 observed with MUSE. \textit{Bottom:} MUSE spectra of DESI~J342.9290-03.4136. For more information on the system, see Desc. \ref{ref:lens7}.}
\label{fig:MUSEspectra7}
\end{minipage}
\end{figure*}

\begin{figure*}[!ht]
\centering
\begin{minipage}{1.0\textwidth}
\centering
\includegraphics[width=0.4\textwidth]{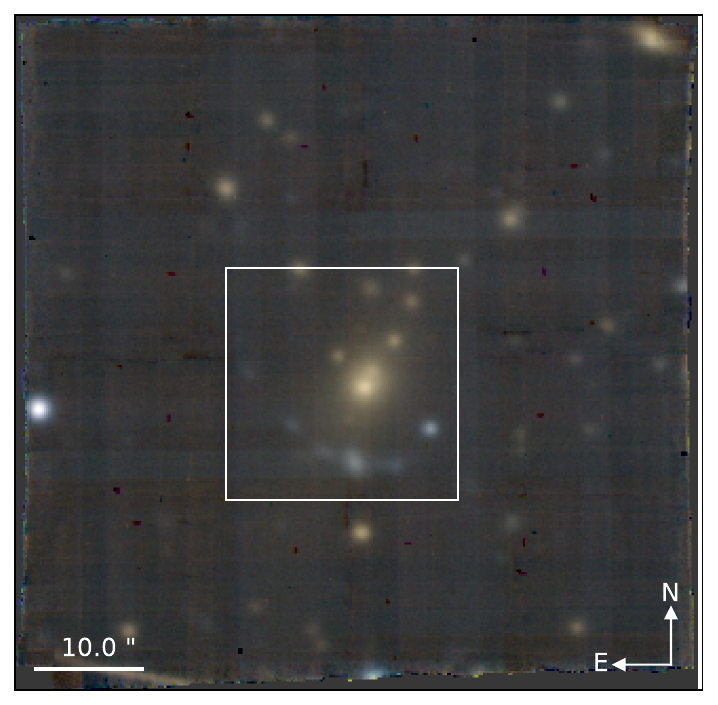}
\includegraphics[width=0.404\textwidth]{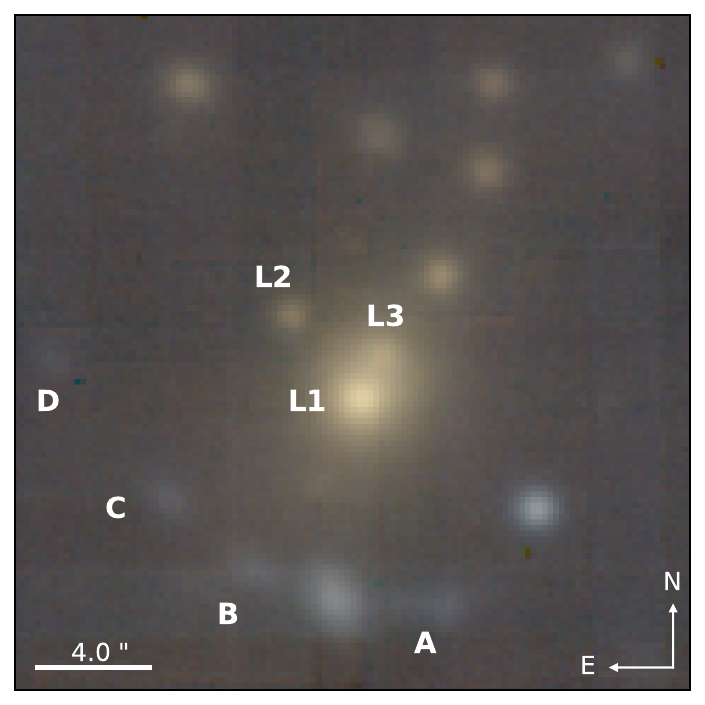}

\includegraphics[width=\textwidth]{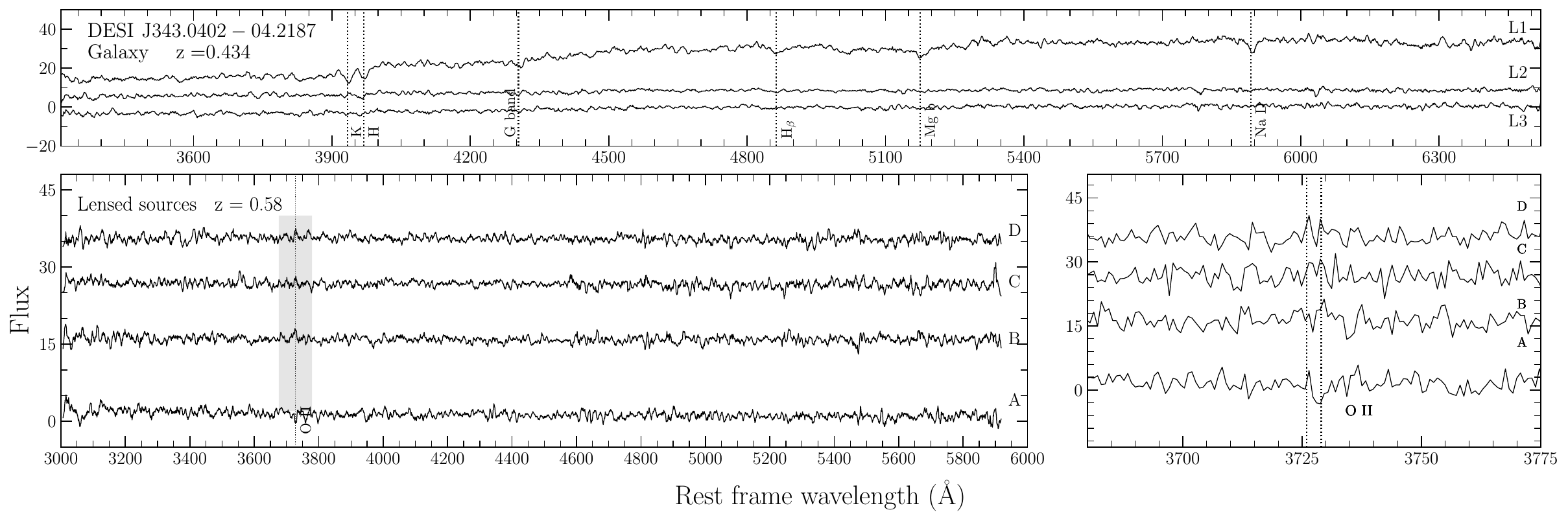}
\caption{\textit{Top:} RGB image of gravitational lens system DESI~J343.0402-04.2187 observed with MUSE. \textit{Bottom:} MUSE spectra of DESI~J343.0402-04.2187. Note that the quality flag for all sources is $Q_z=3$. For more information on the system, see Desc. \ref{ref:lens8}. }
\label{fig:MUSEspectra8}
\end{minipage}
\end{figure*}

\begin{figure*}[!ht]
\centering
\begin{minipage}{1.0\textwidth}
\centering
\includegraphics[width=0.4\textwidth]{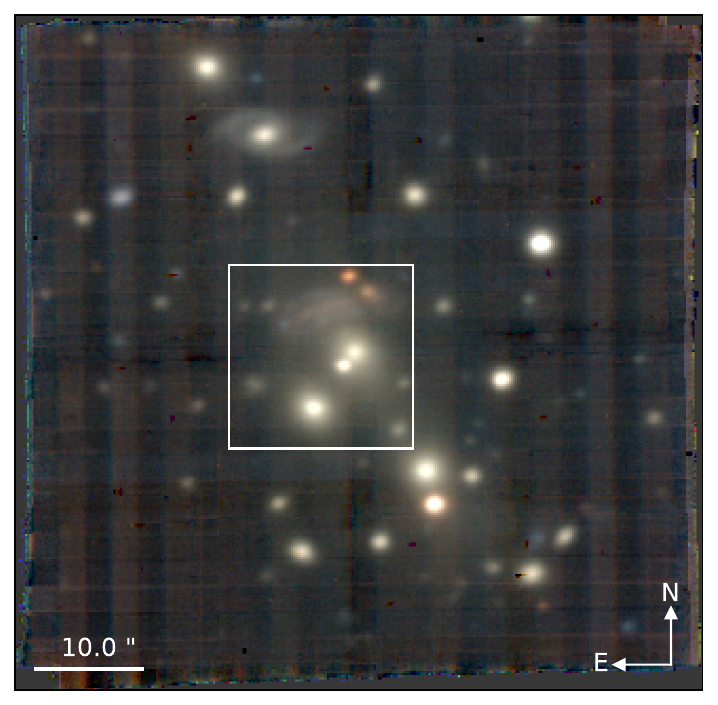}
\includegraphics[width=0.404\textwidth]{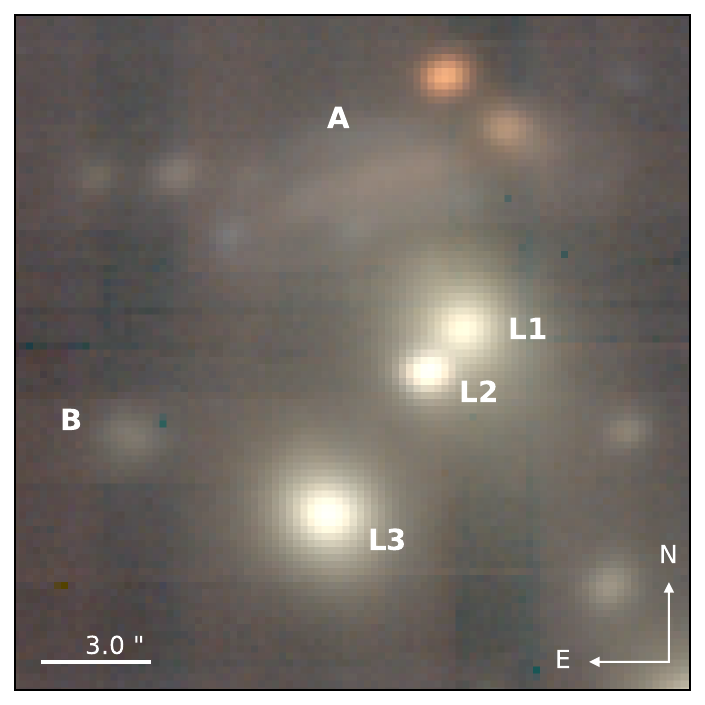}

\includegraphics[width=\textwidth]{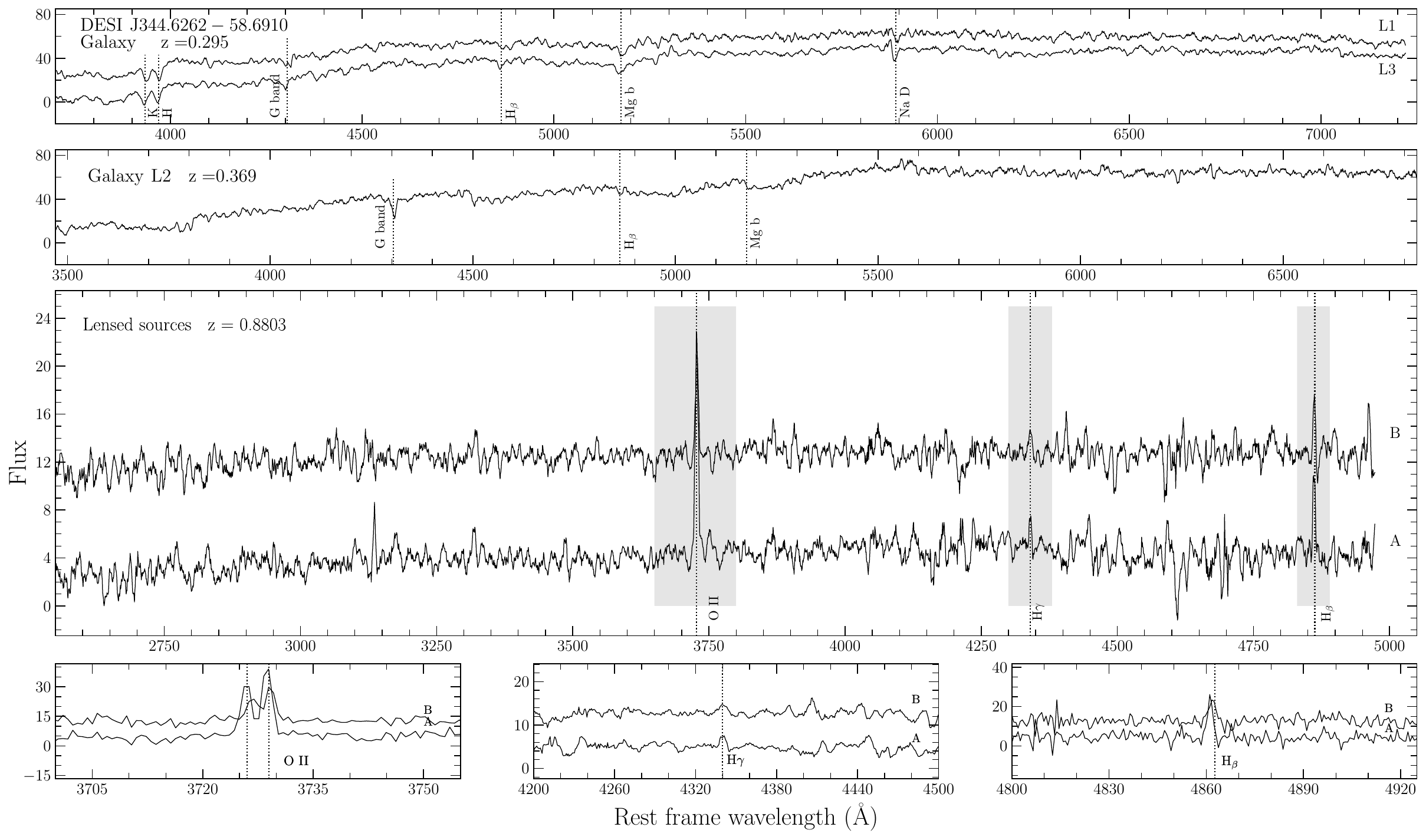}
\caption{\textit{Top:} RGB image of gravitational lens system DESI~J344.6262-58.6910 observed with MUSE. \textit{Bottom:} MUSE spectra of DESI~J344.6262-58.6910. For more information on the system, see Desc. \ref{ref:lens9}.}
\label{fig:MUSEspectra9}
\end{minipage}
\end{figure*}

\begin{figure*}[!ht]
\centering
\begin{minipage}{1.0\textwidth}
\centering
\includegraphics[width=0.4\textwidth]{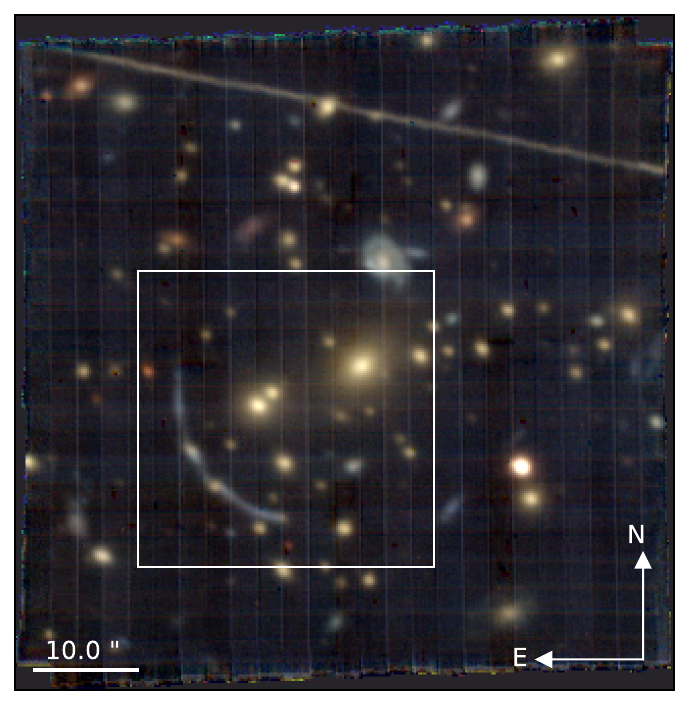}
\includegraphics[width=0.404\textwidth]{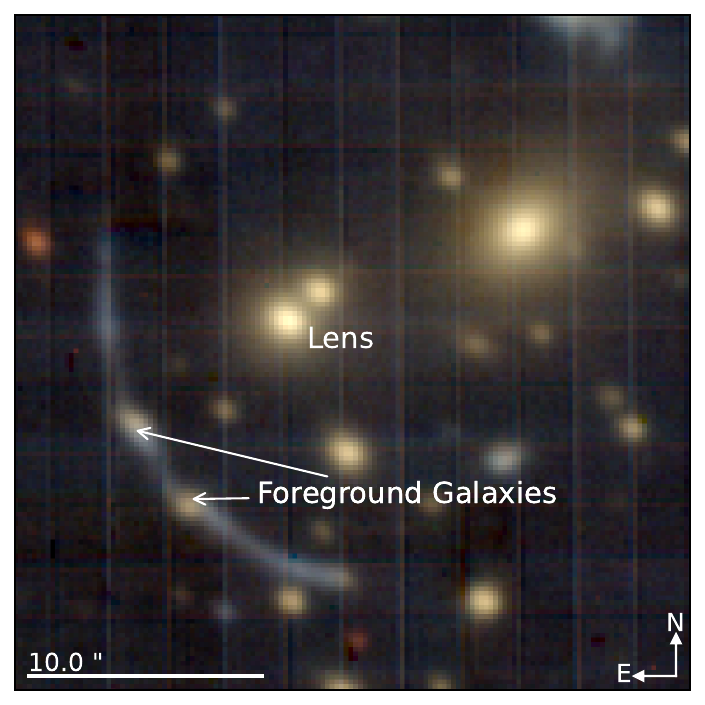}

\includegraphics[width=\textwidth]{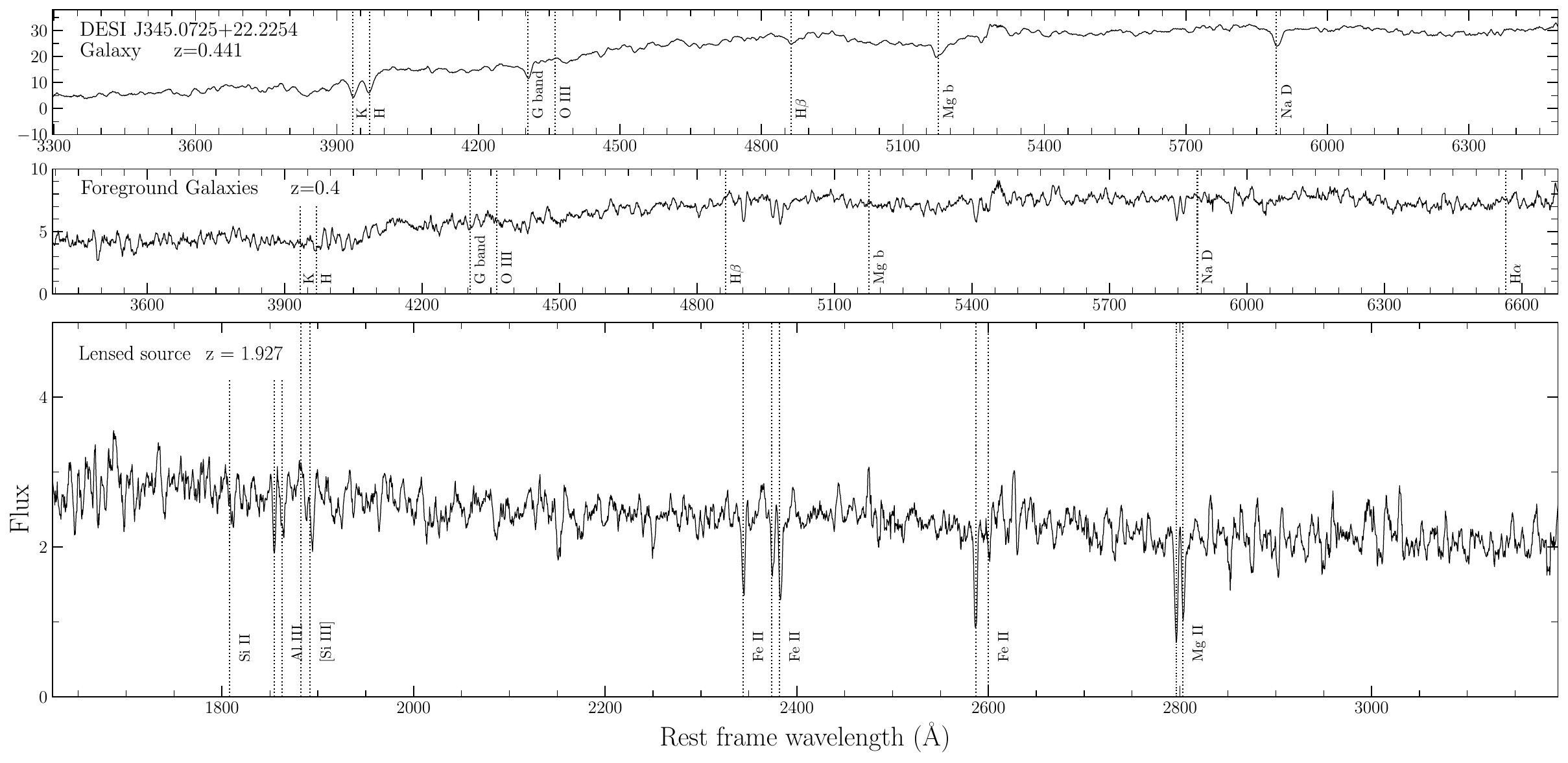}
\caption{\textit{Top:} RGB image of gravitational lens system DESI~J345.0725+22.2254 observed with MUSE. \textit{Bottom:} MUSE spectra of DESI~J345.0725+22.2254. Note that the quality flag for the foreground galaxies is $Q_z=3$. For more information on the system, see Desc. \ref{Ref:lens66}.}
\label{fig:MUSEspectra66}
\end{minipage}
\end{figure*}

\begin{figure*}[!ht]
\centering
\begin{minipage}{1.0\textwidth}
\centering
\includegraphics[width=0.4\textwidth]{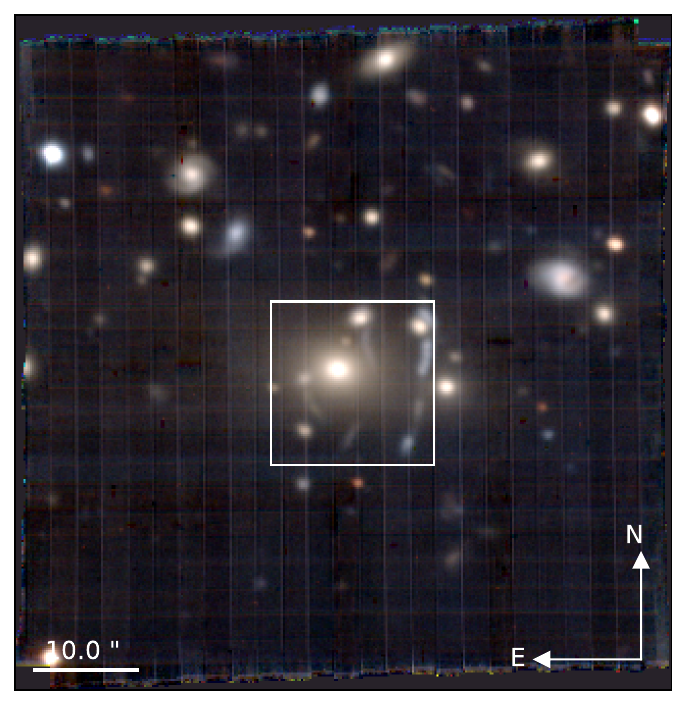}
\includegraphics[width=0.404\textwidth]{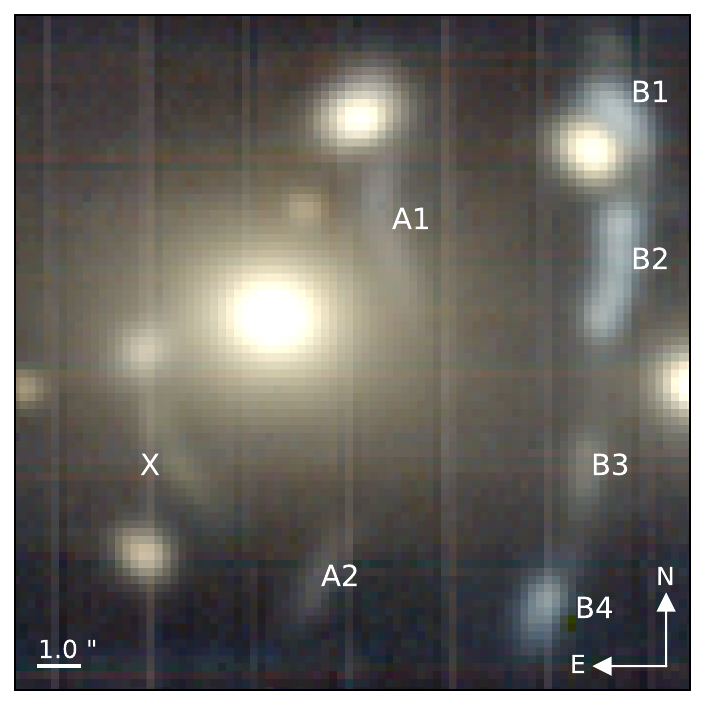}

\includegraphics[width=\textwidth]{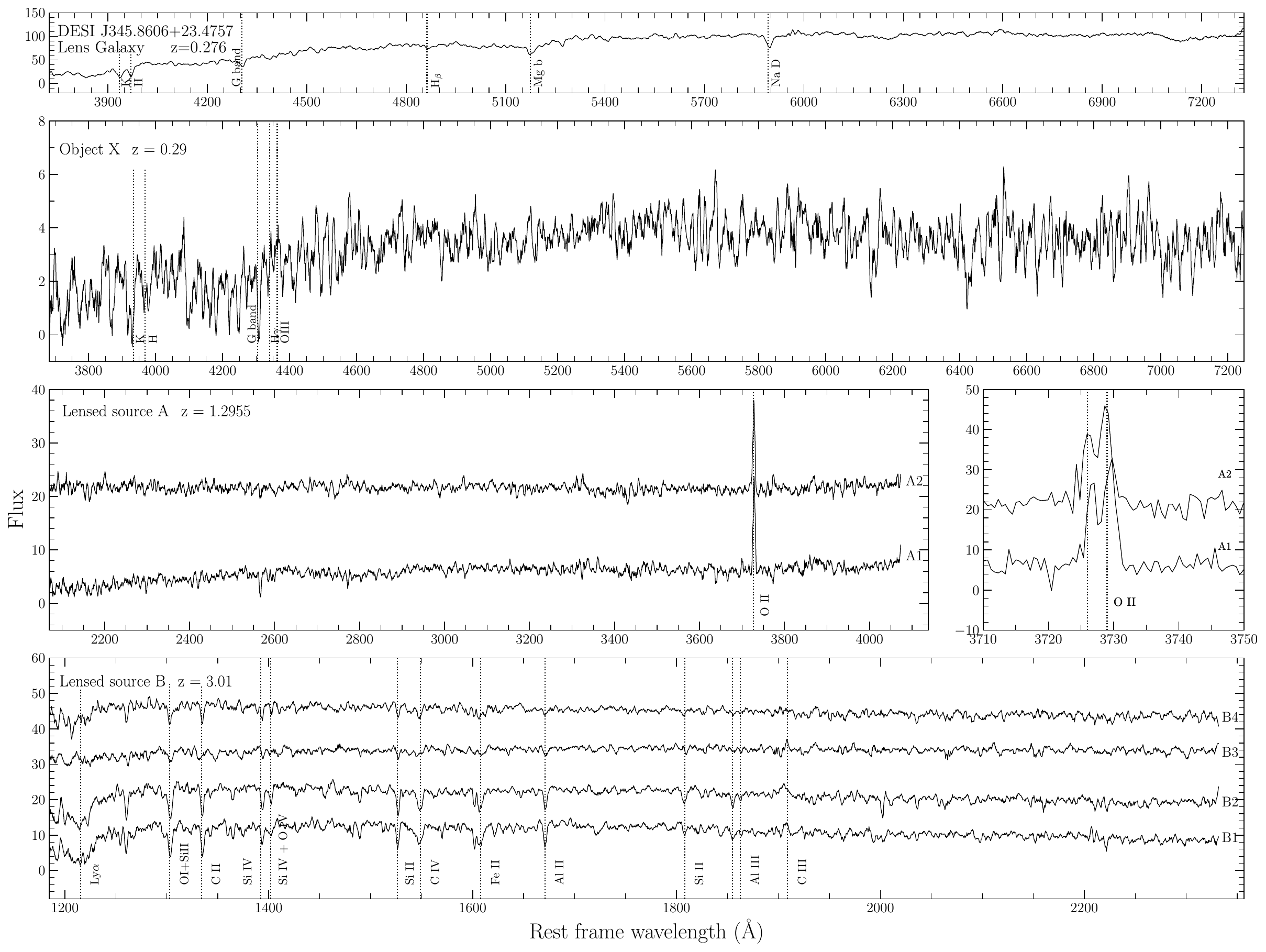}
\caption{\textit{Top:} RGB image of gravitational lens system DESI~J345.8606+23.4757 observed with MUSE. \textit{Bottom:} MUSE spectra of DESI~J345.8606+23.4757. Note that the quality flag for Object X is $Q_z=3$. For more information on the system, see Desc. \ref{Ref:lens67}.}
\label{fig:MUSEspectra67}
\end{minipage}
\end{figure*}


\begin{figure*}[!ht]
\centering
\begin{minipage}{1.0\textwidth}
\centering
\includegraphics[width=1\textwidth]{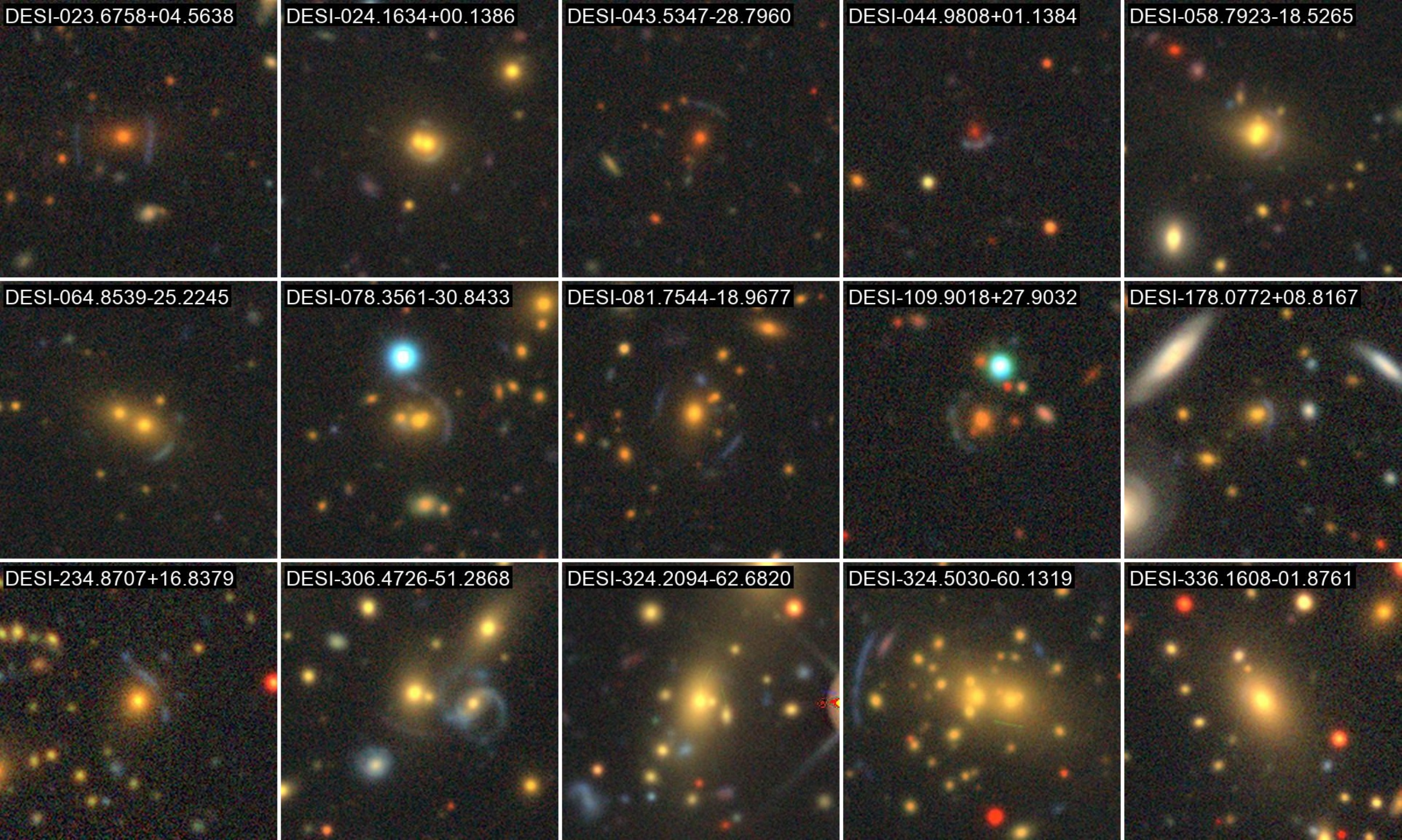}
\caption{DESI Legacy Survey cut-outs for the systems listed in Table~\ref{tab:lens_only_systems}, in which only the foreground lens galaxy redshift could be measured. Each panel shows a $50\times50$ arcsec field centered on the lens galaxy, arranged in a $5\times3$ mosaic.}
\label{fig:lens_only_systems_mosaic}
\end{minipage}
\end{figure*}

\begin{figure*}[!ht]
\centering
\begin{minipage}{1.0\textwidth}
\centering
\includegraphics[width=1\textwidth]{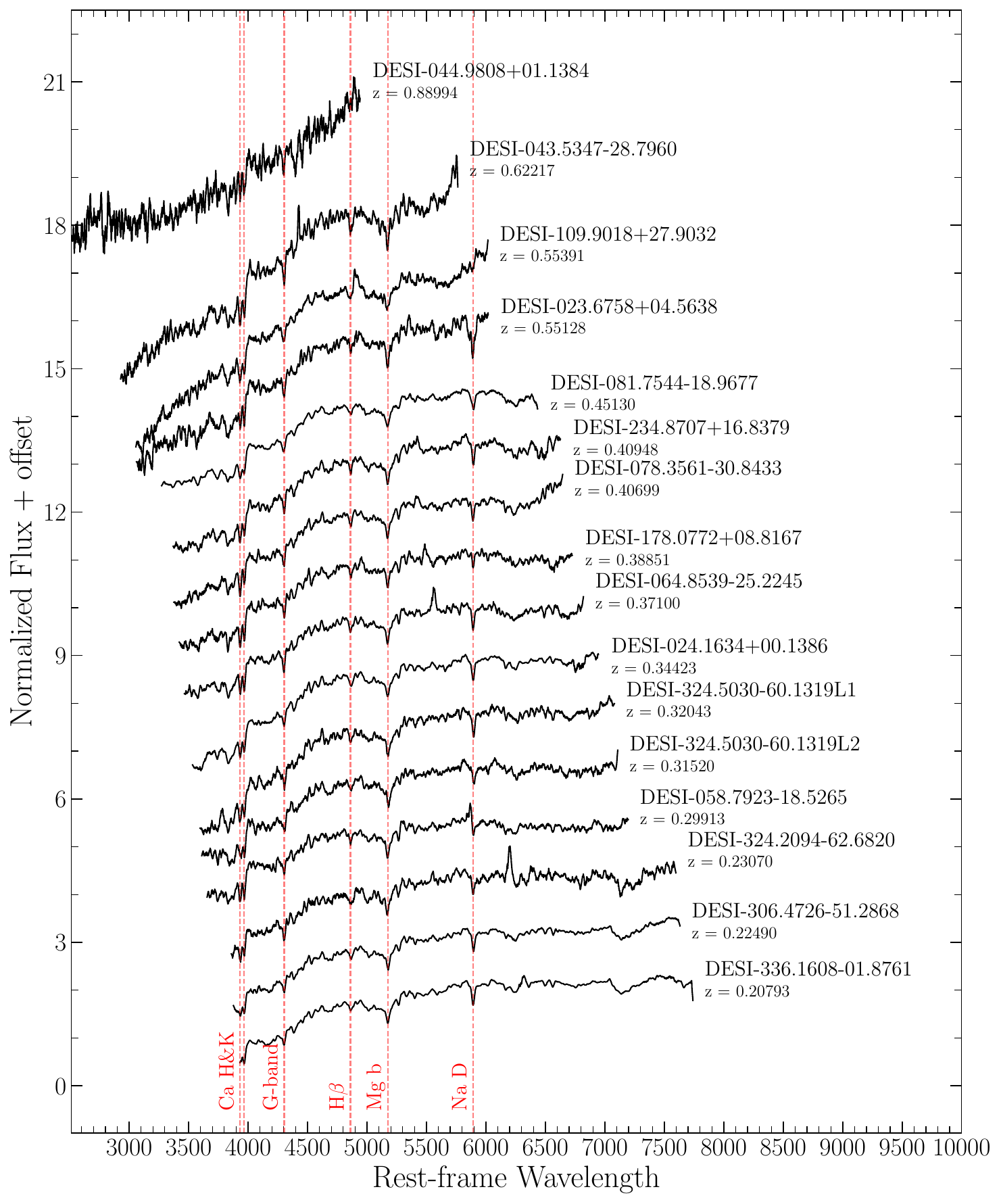}
\caption{Rest-frame spectra of the systems listed in Table~\ref{tab:lens_only_systems}, for which only the redshift of the foreground lens galaxy could be determined. The spectra are shifted to the lens redshift. Prominent absorption features, particularly the Ca H \& K lines, are visible and were used for the redshift measurements.}
\label{fig:lens_only_systems_spectra}
\end{minipage}
\end{figure*}


\begin{figure*}[!ht]
\centering
\begin{minipage}{1.0\textwidth}
\centering
\includegraphics[width=0.4\textwidth]{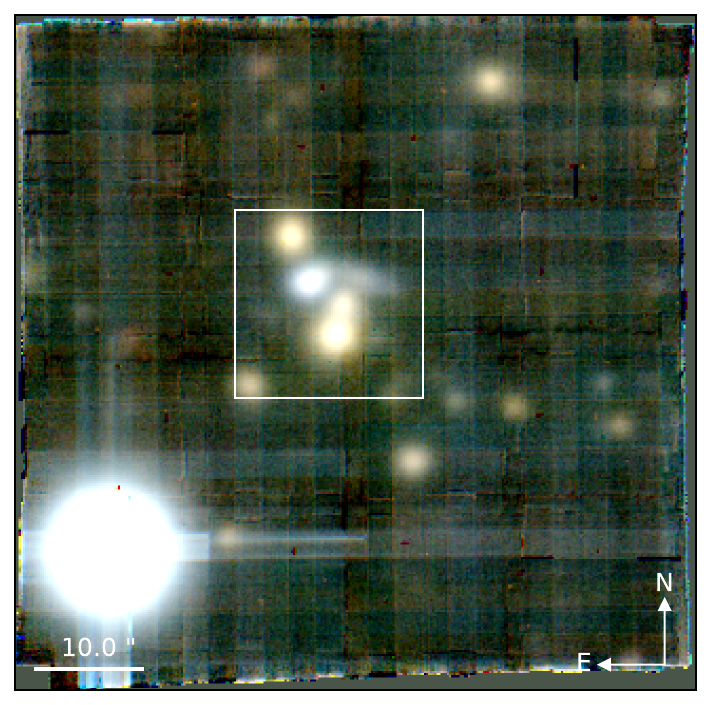}
\includegraphics[width=0.404\textwidth]{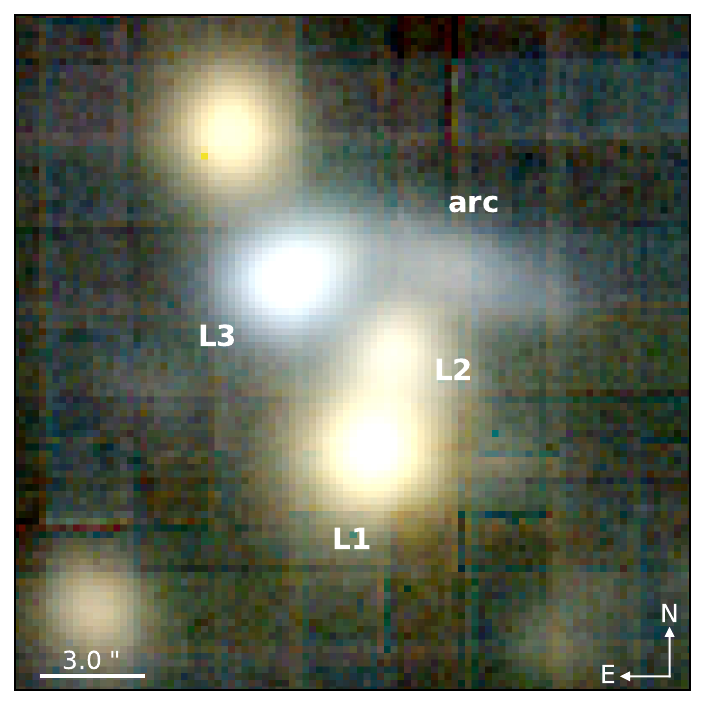}
\includegraphics[width=\textwidth]{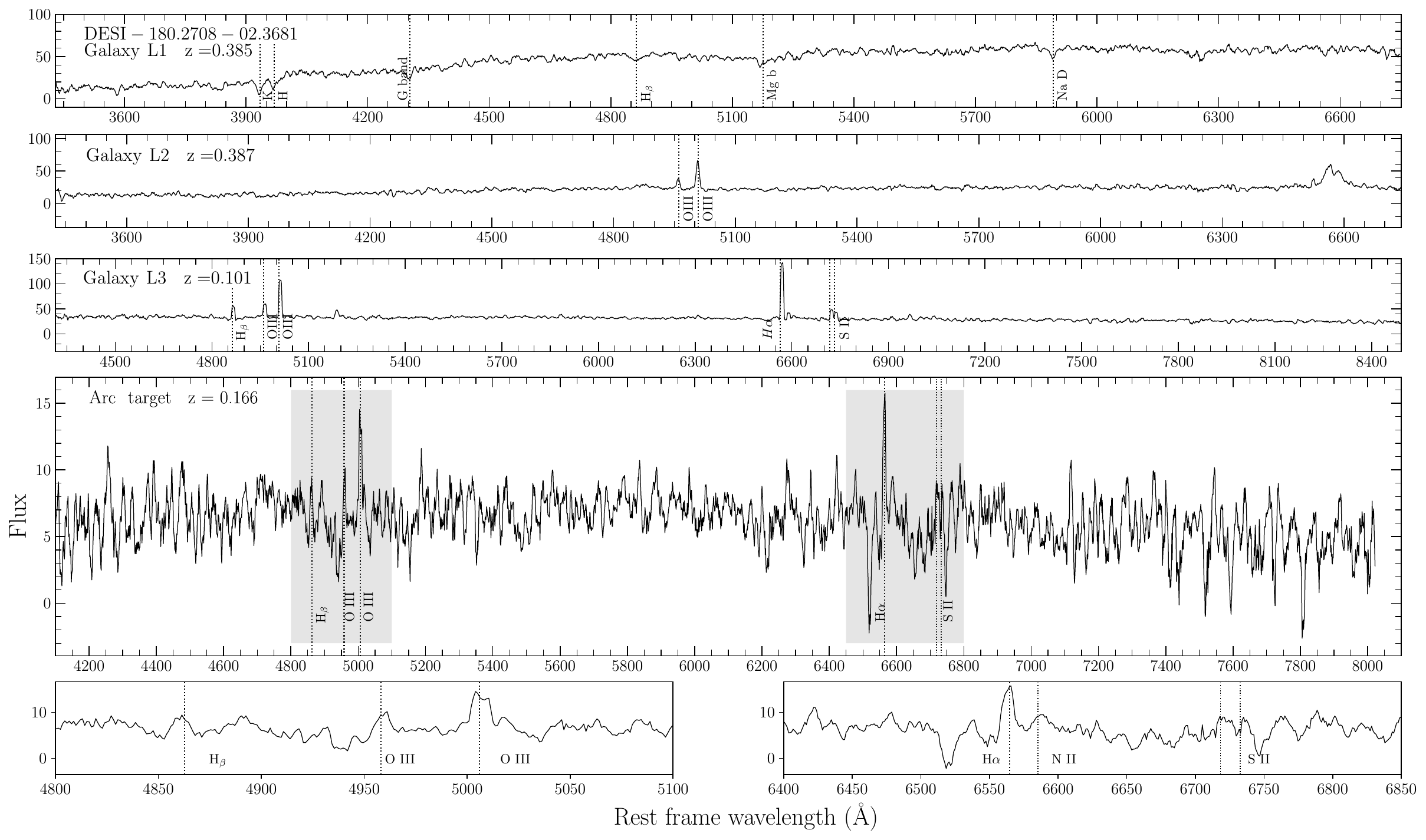}
\caption{\textit{Top:} RGB image of the confirmed non-gravitational lens DESI~J180.2707-2.3681 observed with MUSE. \textit{Bottom:} MUSE spectra of DESI~J180.2707-2.3681. For more information on the system, see Desc. \ref{ref:lens45}.}
\label{fig:MUSEspectra45}
\end{minipage}
\end{figure*}

\begin{figure*}[!ht]
\centering
\begin{minipage}{1.0\textwidth}
\centering
\includegraphics[width=0.4\textwidth]{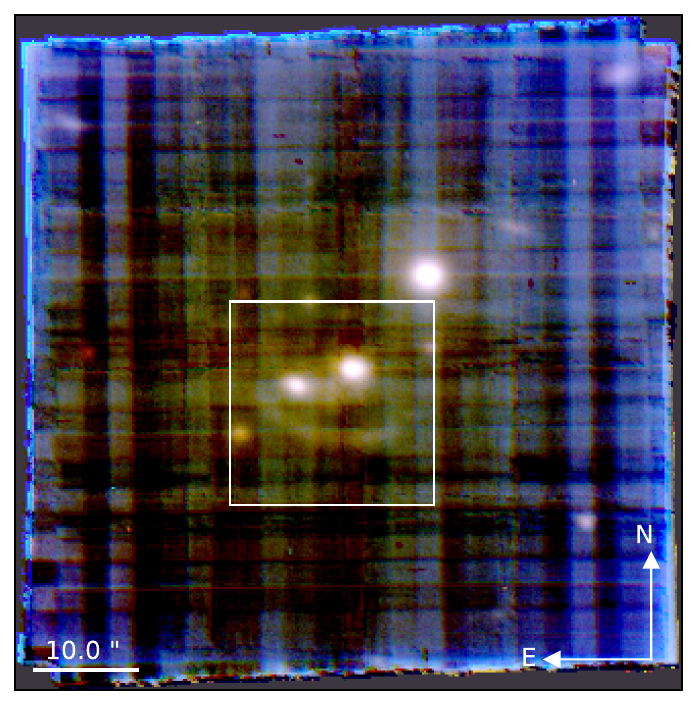}
\includegraphics[width=0.404\textwidth]{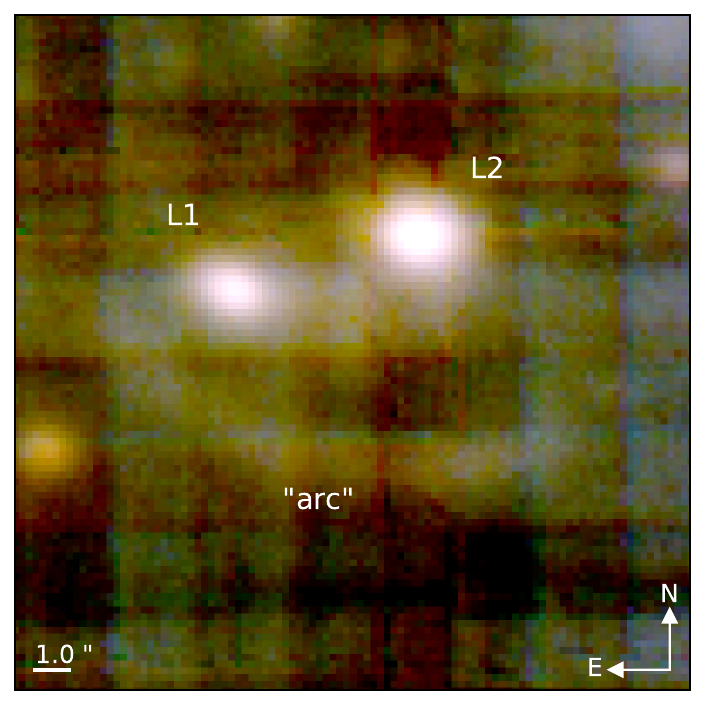}
\includegraphics[width=\textwidth]{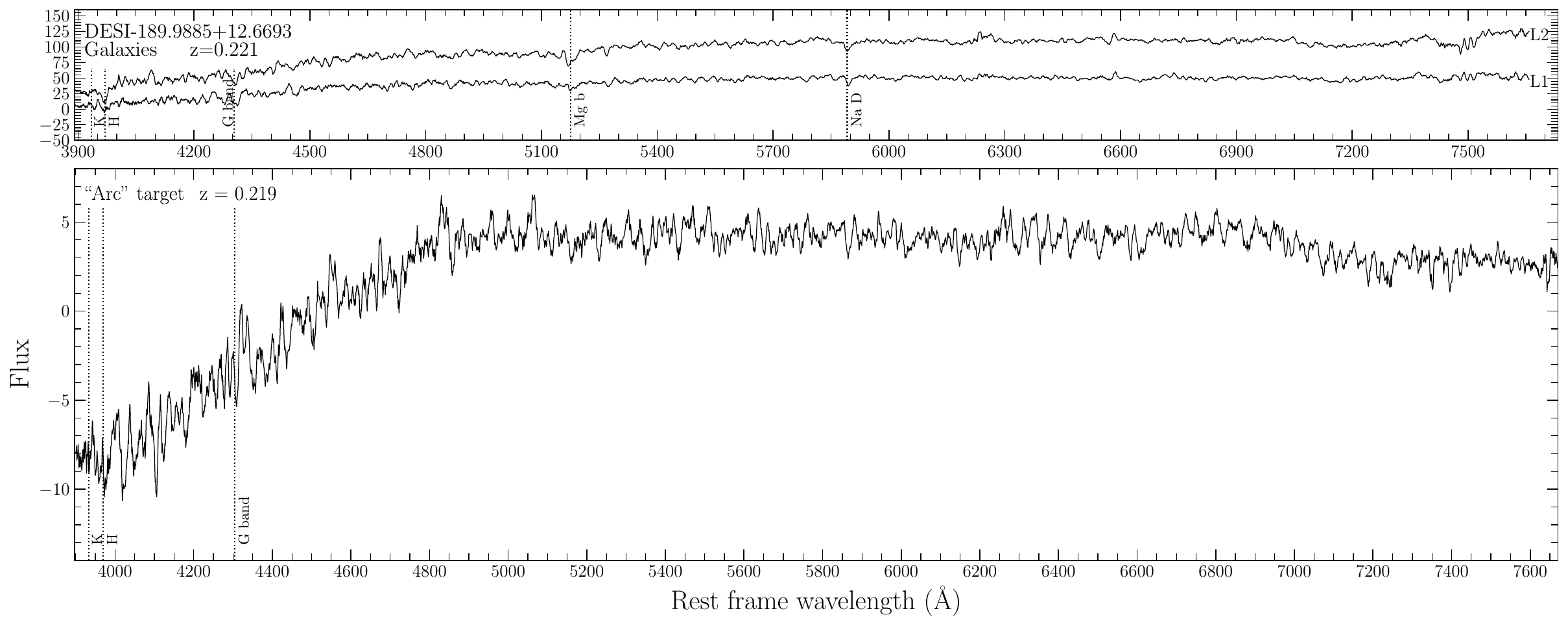}
\caption{\textit{Top:} RGB image of the confirmed non-gravitational lens DESI~J189.9885+12.6693 observed with MUSE. \textit{Bottom:} MUSE spectra of DESI~J189.9885+12.6693. For more information on the system, see Desc. \ref{ref:lens118}.}
\label{fig:MUSEspectra118}
\end{minipage}
\end{figure*}

\begin{figure*}[!ht]
\centering
\begin{minipage}{1.0\textwidth}
\centering
\includegraphics[width=0.4\textwidth]{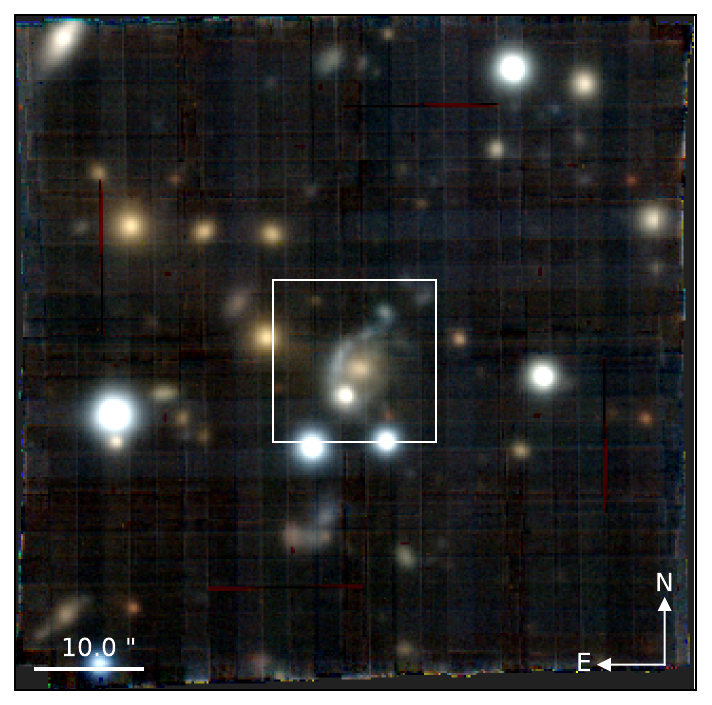}
\includegraphics[width=0.404\textwidth]{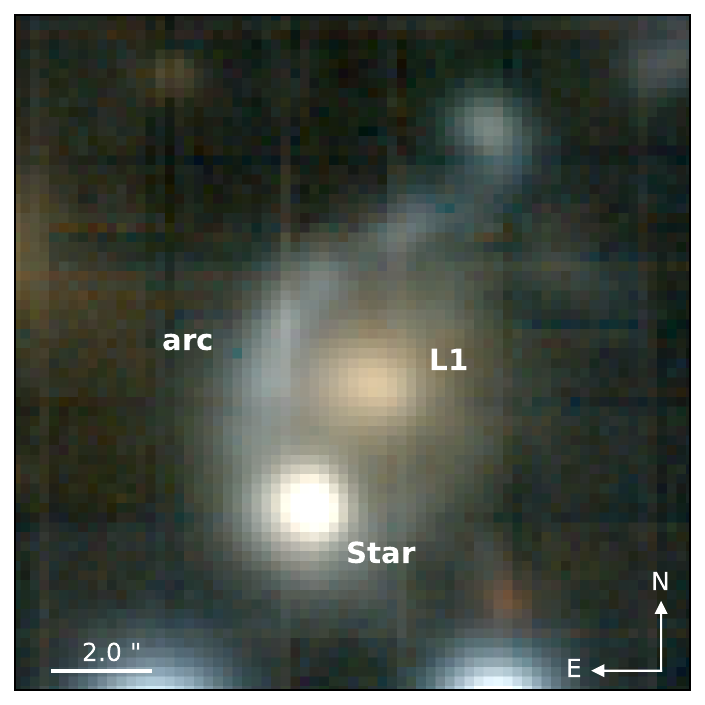}
\includegraphics[width=\textwidth]{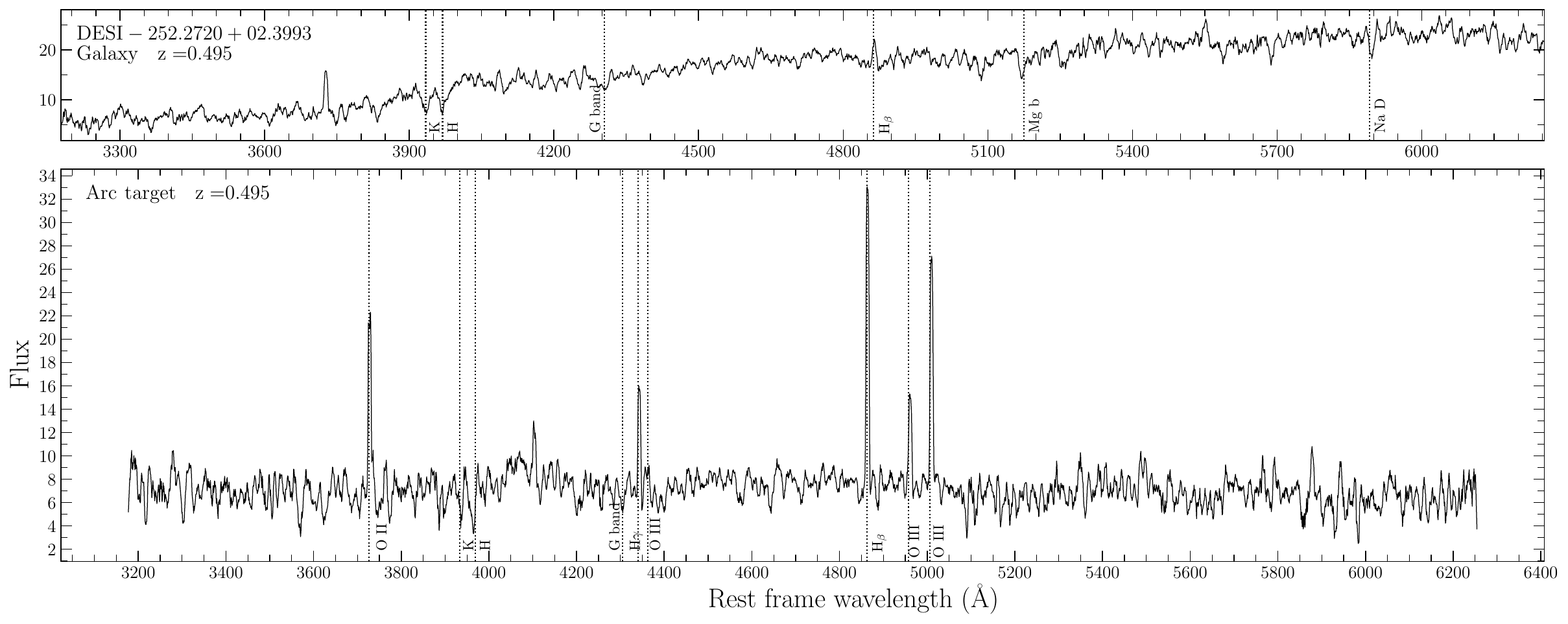}
\caption{\textit{Top:} RGB image of the confirmed non-gravitational lens DESI~J252.2720+02.3993 observed with MUSE. \textit{Bottom:} MUSE spectra of DESI~J252.2720+02.3993. For more information on the system, see Desc. \ref{ref:lens56}. }
\label{fig:MUSEspectra56}
\end{minipage}
\end{figure*}

\begin{figure*}[!ht]
\centering
\begin{minipage}{1.0\textwidth}
\centering
\includegraphics[width=0.4\textwidth]{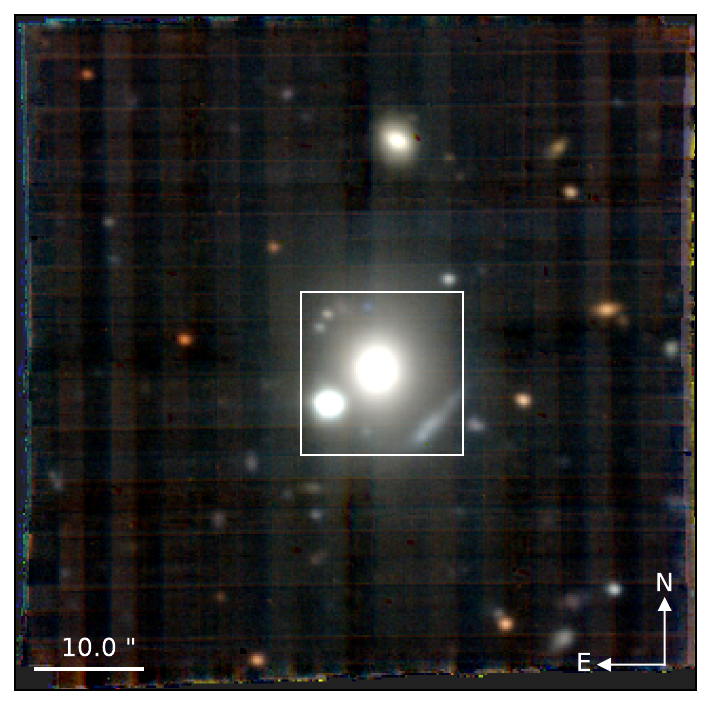}
\includegraphics[width=0.404\textwidth]{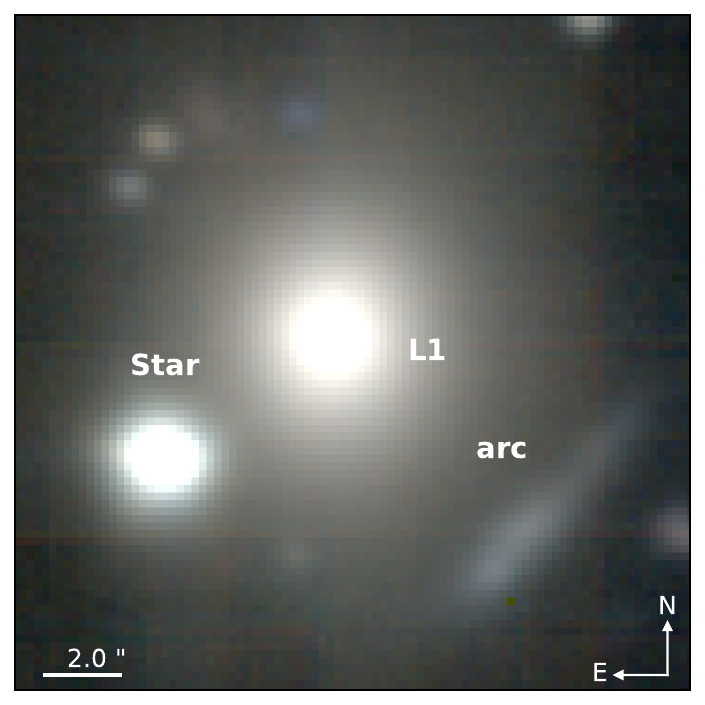}
\includegraphics[width=\textwidth]{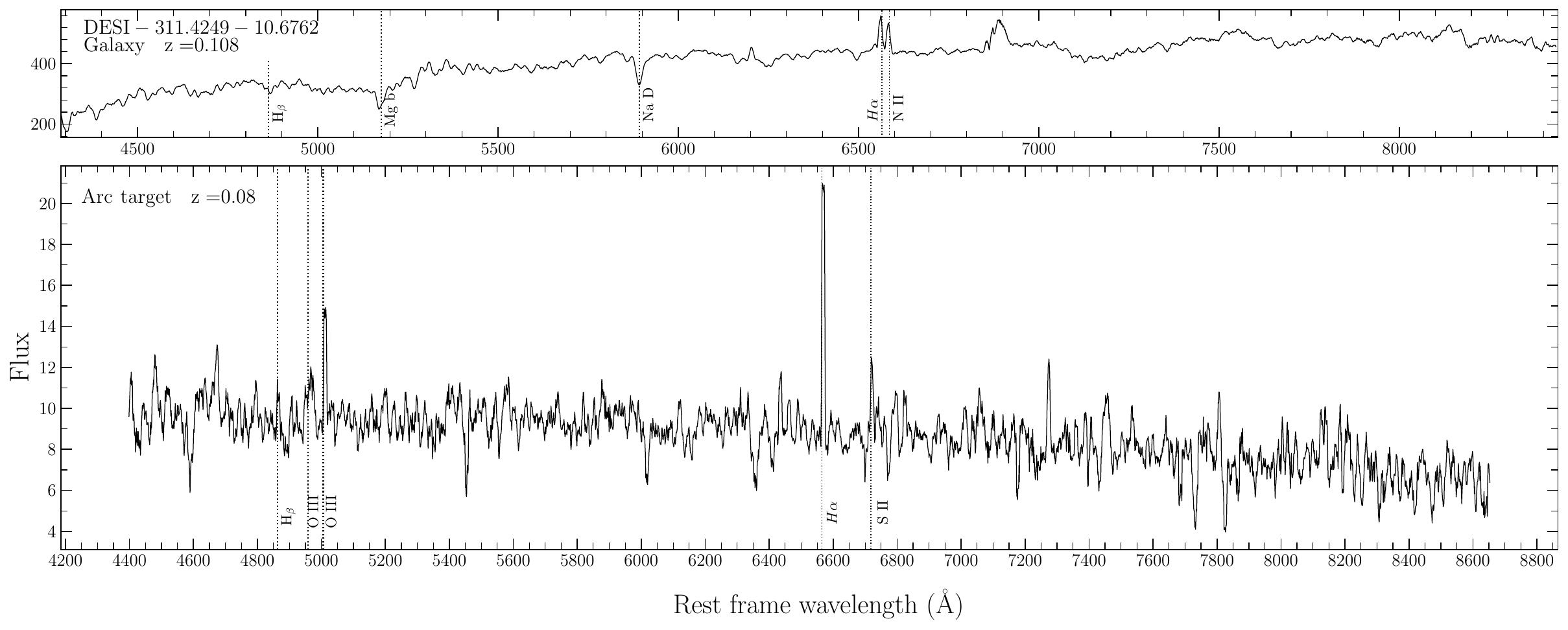}
\caption{\textit{Top:} RGB image of the confirmed non-gravitational lens DESI~J311.4249-10.6762 observed with MUSE. \textit{Bottom:} MUSE spectra of DESI~J311.4249-10.6762. For more information on the system, see Desc. \ref{ref:lens11}.}
\label{fig:MUSEspectra11}
\end{minipage}
\end{figure*}

\begin{figure*}[!ht]
\centering
\begin{minipage}{1.0\textwidth}
\centering
\includegraphics[width=0.4\textwidth]{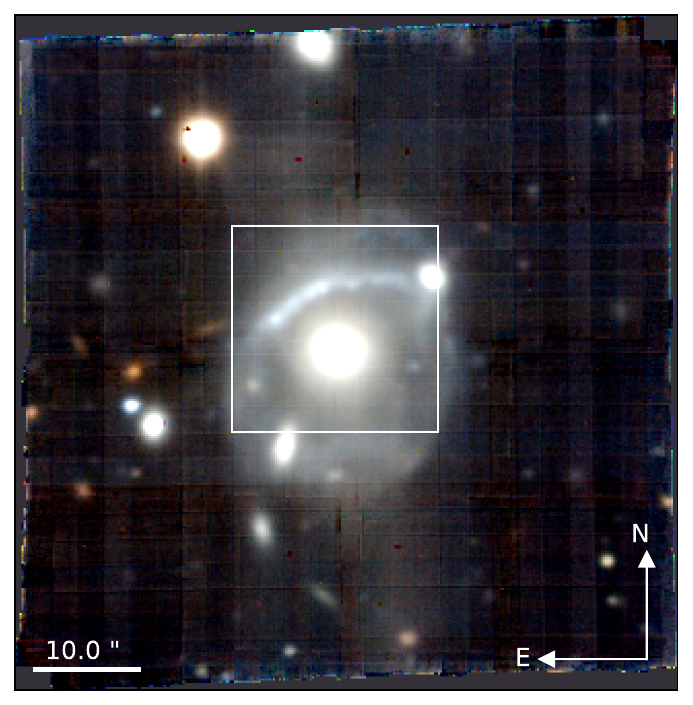}
\includegraphics[width=0.404\textwidth]{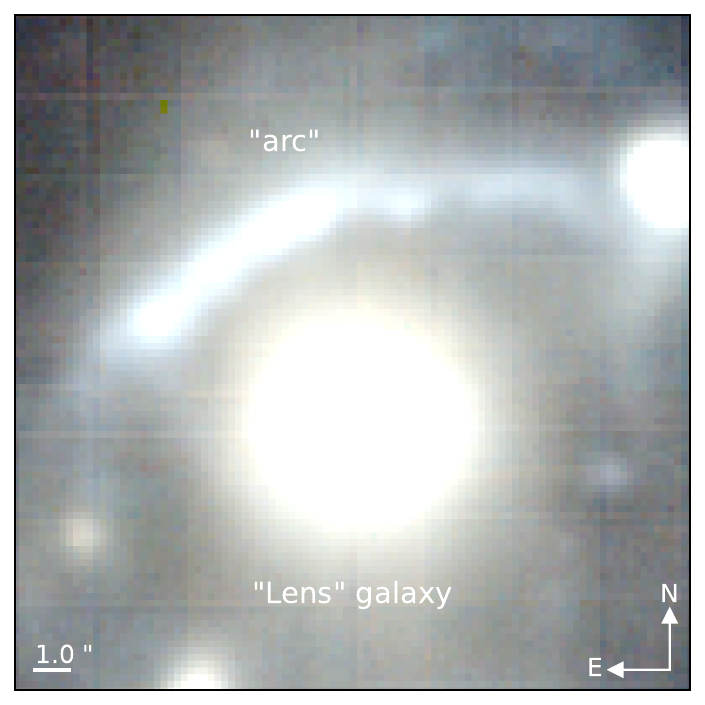}
\includegraphics[width=\textwidth]{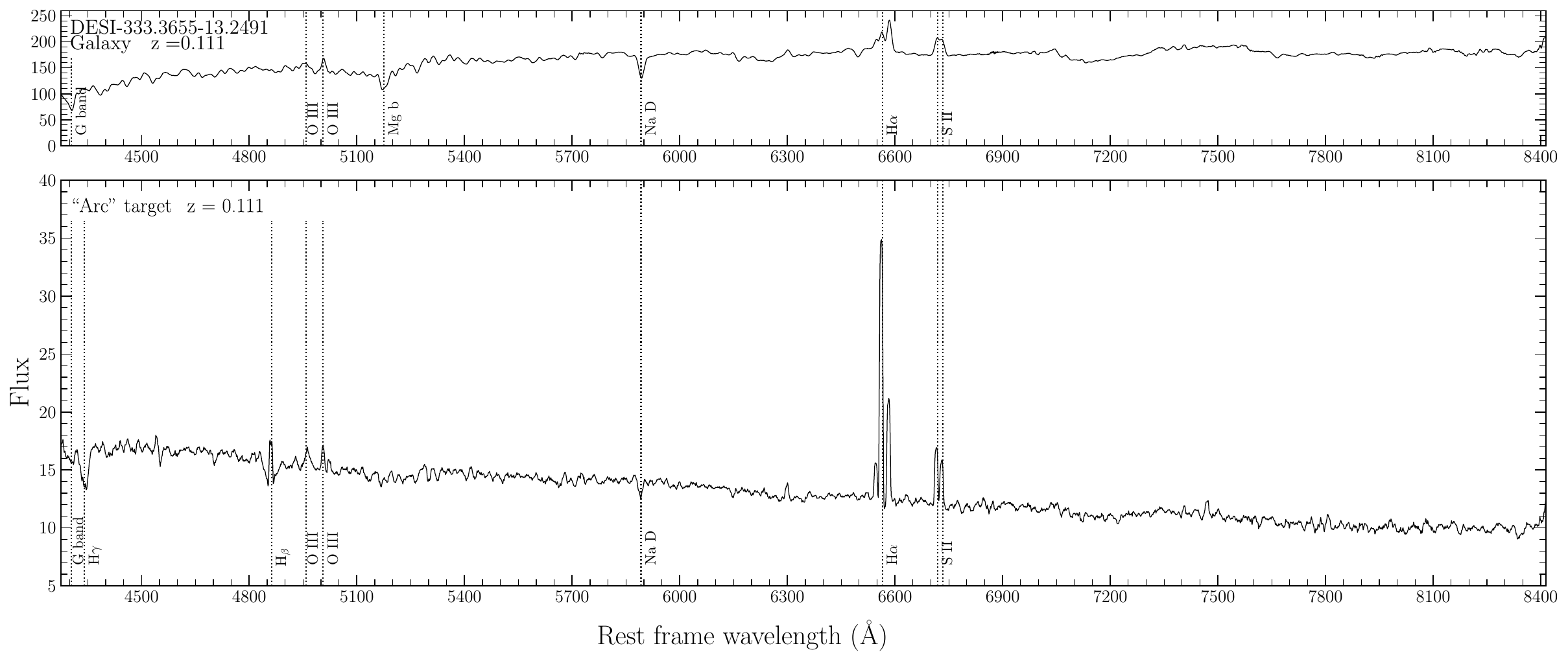}
\caption{\textit{Top:} RGB image of the confirmed non-gravitational lens DESI~J333.3655-13.2491 observed with MUSE. \textit{Bottom:} MUSE spectra of DESI~J333.3655-13.2491. For more information on the figure, see Desc. \ref{ref:lens60}.}
\label{fig:MUSEspectra60}
\end{minipage}
\end{figure*}

\begin{figure*}[!ht]
\centering
\begin{minipage}{1.0\textwidth}
\centering
\includegraphics[width=0.4\textwidth]{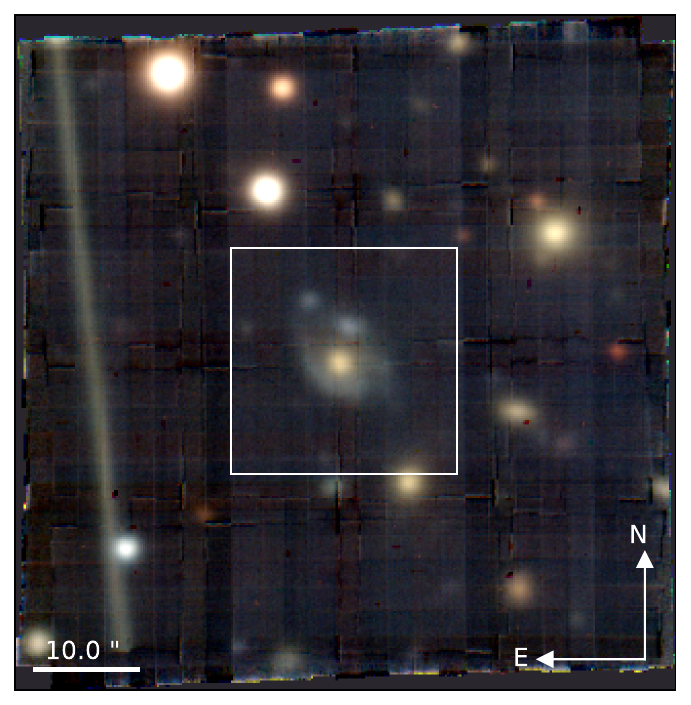}
\includegraphics[width=0.404\textwidth]{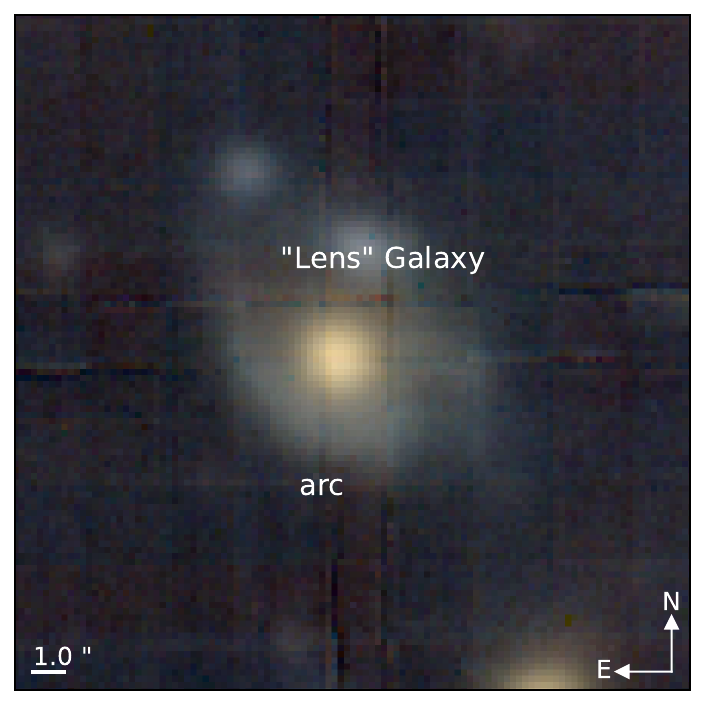}
\includegraphics[width=\textwidth]{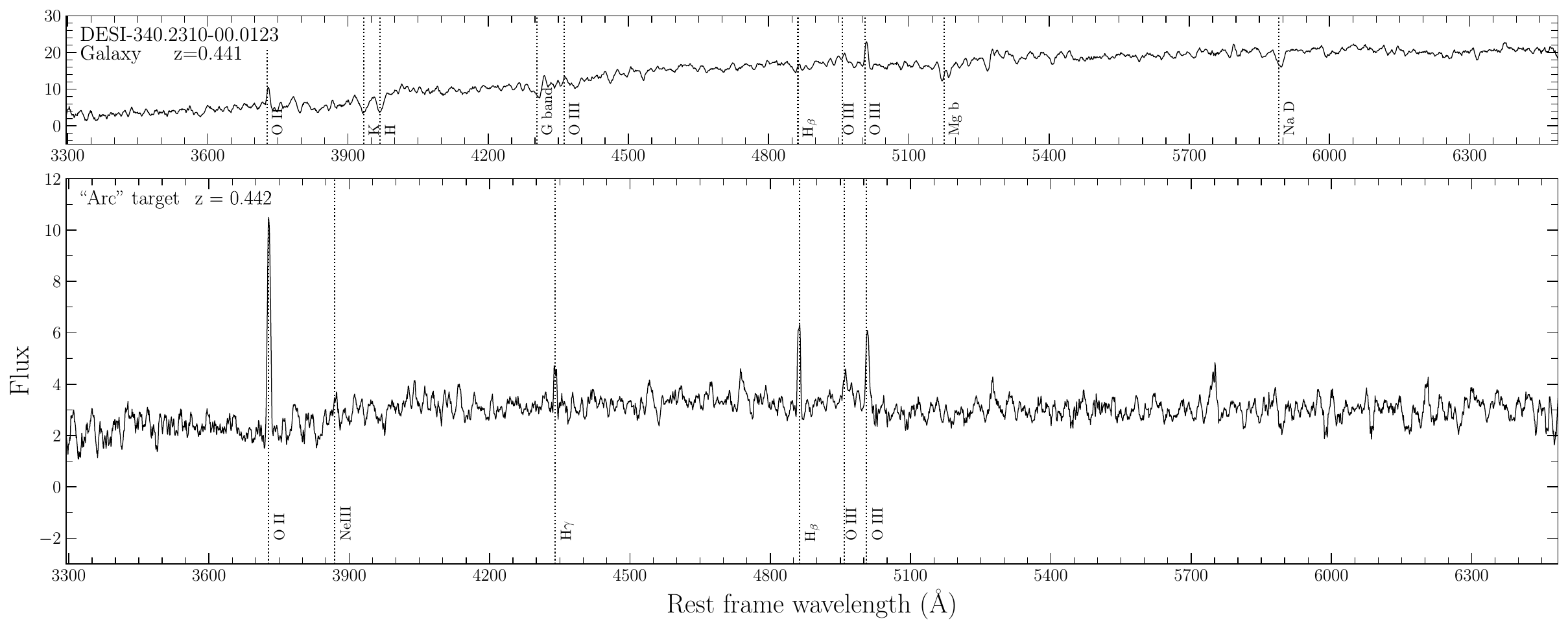}
\caption{\textit{Top:} RGB image of the confirmed non-gravitational lens DESI~J340.2310-00.0123 observed with MUSE. \textit{Bottom:} MUSE spectra of DESI~J340.2310-00.0123. For more information on the figure, see Desc. \ref{ref:lens62}.}
\label{fig:MUSEspectra62}
\end{minipage}
\end{figure*}



\begin{figure}
    \centering
    \includegraphics[width=\textwidth]{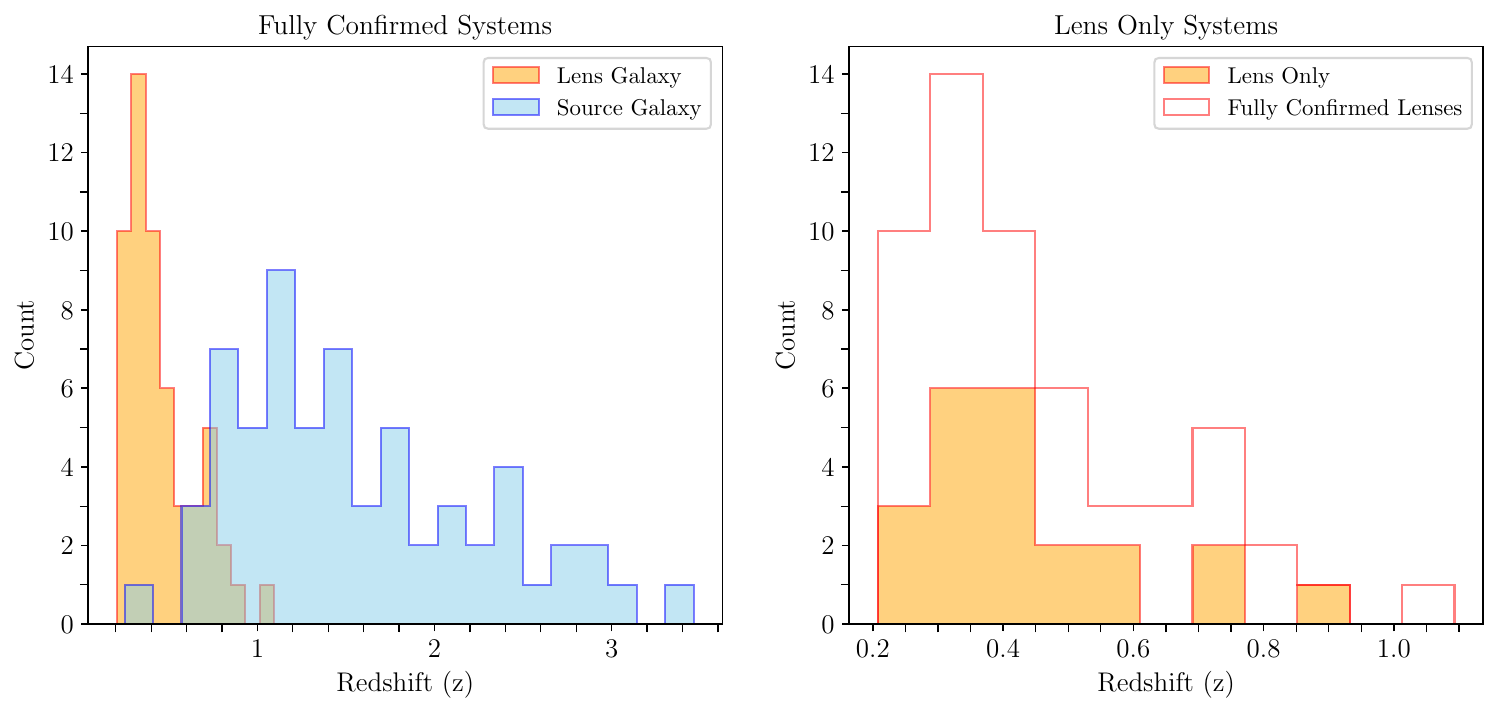}
    \caption{\textit{Left:} Histogram of all lens and source redshifts for fully confirmed systems. 
    \textit{Right:} Redshifts for lens only systems.}
     \label{fig:lens-source-redshift-histogram}
\end{figure}

\begin{figure}   
  \centering
    \includegraphics[width=0.75\textwidth]{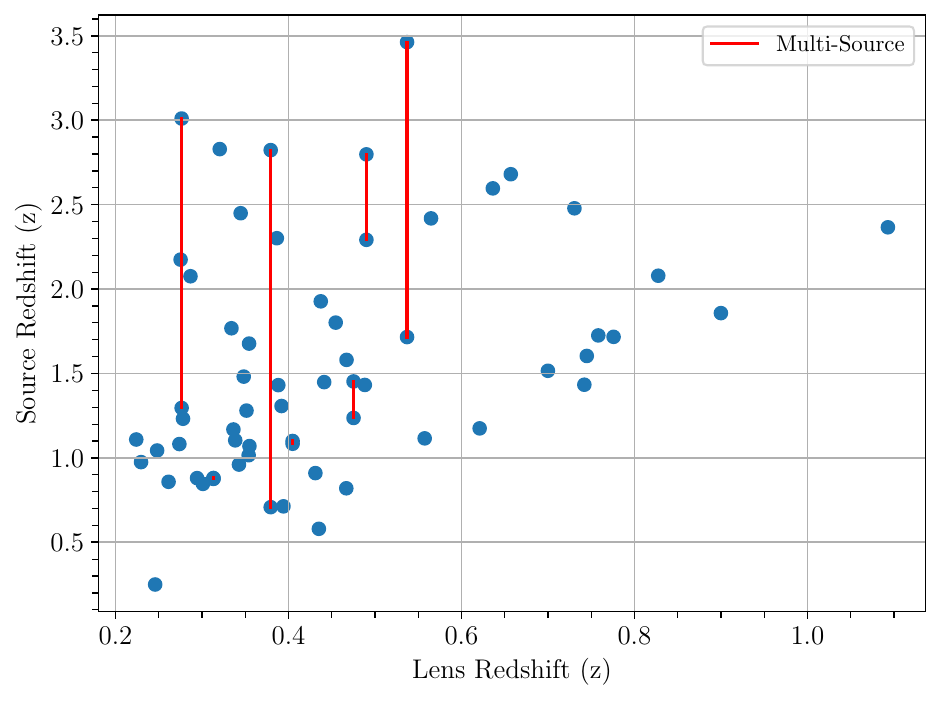}
    \caption{Scatter plot of lens vs source redshifts for all fully confirmed systems.}
    \label{fig:lens-source-redshift-scatter}
\end{figure}




\end{document}